\documentclass{pas}

\usepackage{multirow}
\RequirePackage{aas-macros}
\usepackage{amsmath}
\usepackage[nopatch]{microtype}
\usepackage{booktabs}
\usepackage{subcaption}
\usepackage{natbib}
\usepackage{hyperref}
\usepackage{morefloats}

\begin{document}

\lefttitle{Publications of the Astronomical Society of Australia}
\righttitle{T. O'Beirne}

\jnlPage{1}{4}
\jnlDoiYr{2021}
\doival{10.1017/pasa.xxxx.xx}

\articletitt{Research Paper}

\title{WALLABY pilot survey: properties of H\,{\sc i}-selected dark sources and low surface brightness galaxies}

\author{\sn{T.} \gn{O'Beirne}$^{1,2,3,4}$, \sn{L.} \gn{Staveley-Smith}$^{4,3}$, \sn{V. A.} \gn{Kilborn}$^{1,3}$, \sn{O. I.} \gn{Wong}$^{2,4,3}$, \sn{T.} \gn{Westmeier}$^{4,3}$, \sn{M. E.} \gn{Cluver}$^{1}$, \sn{K.} \gn{Bekki}$^{4}$, \sn{N.} \gn{Deg}$^{5}$, \sn{H.} \gn{Dénes}$^{6}$, \sn{B.-Q.} \gn{For}$^{4,3}$, \sn{K.} \gn{Lee-Waddell}$^{4,2,7}$, \sn{C.} \gn{Murugeshan}$^{8}$, \sn{K.} \gn{Oman}$^{9}$, \sn{J.} \gn{Rhee}$^{4}$, \sn{A. X.} \gn{Shen}$^{2}$ and \sn{E. N.} \gn{Taylor}$^{1}$}

\affil{$^1$Centre for Astrophysics and Supercomputing, Swinburne University of Technology, Hawthorn, Victoria 3122, Australia, $^2$CSIRO Space \& Astronomy, PO Box 1130, Bentley WA 6102, Australia, $^3$ARC Centre of Excellence for All-Sky Astrophysics in 3 Dimensions (ASTRO-3D), $^4$International Centre for Radio Astronomy Research (ICRAR), The University of Western Australia, 35 Stirling Highway, Crawley, WA 6009, Australia, $^5$Department of Physics, Engineering Physics, and Astronomy, Queen's University, Kingston ON K7L 3N6, Canada, $^6$School of Physical Sciences and Nanotechnology, Yachay Tech University, Hacienda San José S/N, 100119, Urcuquí, Ecuador, $^7$International Centre for Radio Astronomy Research (ICRAR), Curtin University, Bentley, WA 6102, Australia, $^8$CSIRO Space \& Astronomy, PO Box 76, Epping, NSW 1710, Australia, and $^9$Centre for Extragalactic Astronomy, Physics Department, Durham University, South Road, Durham DH1 3LE, United Kingdom.}

\corresp{T. O'Beirne, Email: tamsynobeirne.astro@gmail.com}


\history{(Received xx xx xxxx; revised xx xx xxxx; accepted xx xx xxxx)}

\begin{abstract}
We examine the optical counterparts of the 1829 neutral hydrogen (H\,{\sc i}) detections in three pilot fields in the Widefield ASKAP L-band Legacy All-sky  {Blind} surveY (WALLABY) using data from the Dark Energy Spectroscopic Instrument (DESI) Legacy Imaging Surveys DR10. We find that $17$~per~cent (315) of the detections are optically low surface brightness galaxies (LSBGs; mean $g$-band surface brightness within 1 $ R_e$ of $> 23$ mag arcsec$^{-2}$) and $3$~per~cent (55) are optically `dark'. We find that the gas-rich WALLABY LSBGs have low star formation efficiencies, and have stellar masses spanning five orders of magnitude, which highlights the diversity of properties across our sample.  $75$~per~cent of the LSBGs and all of the dark H\,{\sc i} sources had not been catalogued prior to WALLABY. We examine the optically dark sample of the WALLABY pilot survey to verify the fidelity of the catalogue and investigate the implications for the full survey for identifying dark H\,{\sc i} sources. We assess the H\,{\sc i} detections without optical counterparts and identify 38 which pass further reliability tests. Of these, we find that 13 show signatures of tidal interactions. The remaining 25 detections have no obvious tidal origin, 
 {so are candidates for isolated galaxies with high H\,{\sc i} masses, but low stellar masses and star-formation rates. Deeper H\,{\sc i} and optical follow-up observations are required to verify the true nature of these dark sources.}
\end{abstract}


\maketitle

\section{Introduction}

The deep, high-resolution radio and optical surveys that have been developed in recent years have begun to allow us to probe the previously unreachable low surface brightness Universe. Historically, we have been oblivious to much of our Universe due to the brightness of the sky background and the limitations of instruments. \cite{Disney:87} argued that the galaxies that we observe should be thought of as `icebergs' or `crouching giants'. That is, what we observe above the sky background is not a reliable indicator of what lies beneath. Insignificant dwarf elliptical galaxies could just be the tip of giant low surface brightness spirals, such as Malin 1 \citep{Bothun:87}. Although the existence of low surface brightness galaxies (LSBGs) is nothing new \citep[e.g.][]{Sandage:84,Impey:88}, it is only recently that they have been shown to make up a significant fraction of the galaxy census, with multiple populations of both low surface brightness and optically dark sources being found, including ultra diffuse galaxies \citep[UDGs, e.g.][]{VanDokkum:15, Koda:15, Leisman:17,ManceraPina:20,For:23,Gannon:24} and neutral atomic hydrogen (H\,{\sc i}) clouds without or with extremely faint optical counterparts \citep[e.g.][]{Kilborn:06,Matsuoka:12,Cannon:15,jozsa:21,Wong:21,obeirne:24}. As deeper optical observations are obtained we are able to improve our understanding of these sources. Many dark H\,{\sc i} clouds identified in the Arecibo Legacy Fast ALFA (ALFALFA) survey have since been revealed to host stellar counterparts \citep[e.g.][]{Du:24,Jones:24}. With current and future H\,{\sc i} and optical surveys such as the Widefield ASKAP L-band Legacy All-sky  {Blind} surveY \citep[WALLABY,][]{Koribalski:20} and the Dark Energy Spectroscopic Instrument (DESI) Legacy Imaging Surveys data release 10 \citep[hereafter referred to as the Legacy Survey;][]{Dey:19}, we are entering an era where these LSBGs can be studied in large numbers for the first time.

At the extreme end of the spectrum of LSBGs lie dark galaxies: dark matter haloes that lack stars. Dark galaxy candidates are optically dark sources with surface brightnesses below the sensitivity limits of current optical telescopes. Additionally, they are predicted to have significant H\,{\sc i} gas \citep{Jimenez:97,Jimenez:20}. They could be key to reconciling Lambda Cold Dark Matter ($\Lambda$CDM) simulations with observations, in particular resolving the `Missing Satellites Problem' \citep{Kauffmann:93,Moore:99,Klypin:99} and the `Too Big To Fail' problem \citep{BoylanKolchin:11,Papastergis:16}.   Hydrodynamical simulations have shown that $\Lambda$CDM can actually reproduce the stellar mass function of observed satellites in the Local Group \citep{Sawala:16}. However, reionisation and feedback from supernovae and stellar winds play a key role in suppressing star formation. In effect, this makes a proportion of the dark matter subhaloes invisible to our optical telescopes.  \cite{Lee:24} study the formation and evolution of dark galaxies using the IllustrisTNG cosmological hydrodynamical simulation. They predict that, at the present epoch (z = 0), dark galaxies are predominantly located in void regions and higher spin parameters than luminous galaxies. This work selects for relatively low mass dark matter haloes with $M\sim10^{9}$ M$_{\odot}$. The theoretical work done by \cite{Jimenez:20} shows that we do not expect to find large numbers of dark galaxies above a halo mass of $\sim10^{10}$ M$_{\odot}$. They predict that the number density of dark galaxies with halo masses $>3\times10^{10}$ M$_{\odot}$ is only $10^{-6}$ Mpc$^{-3}$, so large volumes will be required in order to identify them. In their models, \cite{Benitez-Llambay:20} find that all haloes with $M_{200}>5\times10^9$ M$_{\odot}$ should host a luminous galaxy, and a population of starless gaseous haloes should exist with masses between $10^6$ M$_{\odot}$ and $5\times10^9$ M$_{\odot}$. In these haloes, gas is expected to be in thermal equilibrium with the ultraviolet background radiation and in hydrostatic equilibrium in the gravitational potential of the halo. Additionally, it has been postulated that dark minihaloes could host some type of compact gas cloud, such as ultra compact high velocity clouds \citep[UCHVC;][]{Adams:13}  and REionization-Limited H\,{\sc i} Clouds \citep[RELHICs;][]{Benitez-llambay:17}.

Looking at the low surface brightness Universe can increase our understanding of galaxy formation and evolution. The gas-star formation cycle plays a key role in the life-cycle of galaxies, centred on the balance between gas inflows from the intergalactic medium, its consumption through star formation, and outflows \citep{Kennicutt:12,Lilly:13}. The H\,{\sc i} in a galaxy acts as a reservoir for star formation. If this gas is stripped then the star formation in a galaxy can become quenched. This can occur as a result of star formation feedback, as well as through ram pressure stripping and tidal interactions \citep[ {e.g.}][]{Cortese:21}. Alternatively, isolation can also be responsible for low star formation. Giant low surface brightness spirals are usually found in low density environments and rarely found to be interacting with other systems \citep{Das:13}. A lack of mergers and tidal interactions combined with massive dark matter halos may allow for increased stability with a slow, steady accretion of gas, avoiding large star-forming events. We can look for internal and external factors that cause star formation to be suppressed in these low surface brightness galaxies.  H\,{\sc i}-rich dark galaxies could give us insights into the early stages of galaxy formation, giving us a chance to study the pristine conditions of the very first galaxies.

Yet even now, true dark galaxies have been shown to be extremely rare in observations. \cite{Xu:23} claim that their recent detection of FAST J0139+4328 with the Five-hundred-meter Aperture Spherical radio Telescope \citep[FAST;][]{Jiang:19} is the first isolated dark galaxy detected in the local Universe. Their optical imaging, however, is limited to the shallow Panoramic Survey Telescope and Rapid Response System data \citep[Pan-STARRS;][]{Chambers:16},  {which has a surface brightness limit of 24 mag arcsec$^{-2}$ \citep{Sola:22}}.  This source's status as a dark galaxy is not confirmed \citep{Benitez-llambay:24,Karunakaran:24}. Prior to this candidate, there have in fact been several other promising potential dark galaxies \citep[e.g.][]{Kilborn:00,Kent:10,Bilek:20,Wong:21}. 

While both are devoid of stars, it is useful to distinguish between dark galaxies, which have a primordial origin, and dark clouds, which are debris from interactions. This includes tidal debris, such as the dark clouds in \cite{Taylor:22, Jozsa:22}, and debris from ram pressure stripping, such as the dark cloud in the Virgo cluster studied by \cite{Oosterloo:05}.  {Dark galaxies reside in dark matter haloes and are stable to the effects of harassment \citep{Taylor:16}, while dark clouds are dark matter poor and transient in comparison.} It is often difficult to confirm the formation mechanism of a dark H\,{\sc i} source, as was the case with VIRGOHI21. This source was initially identified as a dark galaxy candidate \citep{Davies:04,Minchin:05} before later being revealed to favour an interaction based origin \citep{Bekki:05,Haynes:07,Duc:08}.

Although dark galaxies and dark clouds are invisible to optical instruments, they are detectable by radio telescopes if they are sufficiently gas-rich, and consequently  H\,{\sc i} is an excellent probe of the optically low surface brightness Universe.  H\,{\sc i}  often extends well beyond the optical disc, making it an exceedingly useful tracer of environment and galaxy evolution. The extended H\,{\sc i} is susceptible to  {ram pressure stripping and tidal forces}, causing it to be significantly impacted by the environment in which it resides \citep[e.g.][]{Oosterloo:05,LeeWaddell:14}.  Large area, untargeted H\,{\sc i} surveys are the key to identifying low surface brightness galaxies and dark galaxy candidates in large numbers. The H\,{\sc i} Parkes All Sky Survey \citep[HIPASS,][]{Barnes:01,Doyle:05}, however, was unable to confirm any dark galaxies due to poor angular resolution and source confusion. In the Arecibo Legacy Fast ALFA \citep[ALFALFA,][]{Giovanelli:05} survey less than $2$~per~cent of sources were missing optical counterparts \citep{Haynes:18}. Many of these had a tidal origin \citep[e.g.][]{Leisman:16}, and a few dark galaxy candidates were followed up \citep[e.g.][]{Kent:10,Janowiecki:15}. By performing an optical search for LSBGs in the region covered by the Arecibo H\,{\sc i} Strip Survey, \cite{Trachternach:06} demonstrate how optical and H\,{\sc i} surveys sample different parts of LSBG population to complement each other, finding that LSBGs are expected to make up $>30$~per~cent of the local galaxy number density.  With its  {improved} angular resolution and sensitivity, WALLABY has the potential to detect dark galaxy candidates and extremely low surface brightness galaxies in large numbers by being able to better locate the origin of the emission and better separate emission from other nearby objects in denser group and cluster environments. The fast survey speed will also allow more of these rare objects to be detected in the large volume covered. Two dark H\,{\sc i} clouds have already been identified in the pre-pilot survey observations \citep{Wong:21}.

The paper is structured as follows. Section~\ref{sec:surveys} outlines the surveys used in our analysis, and Section~\ref{sec:methods} presents the methods used to identify the dark and low surface brightness sources and calculate the H\,{\sc i} and multiwavelength properties. Section~\ref{sec:results} presents the dark H\,{\sc i} sources that we identify and the global properties that they and the LSBGs possess though several scaling relations. Finally, Section~\ref{sec:discussion} discusses our results and Section~\ref{sec:conclusion} summarises our findings. Throughout this work we use velocity in the optical convention ($v=cz$) and we adopt the AB magnitude convention. We assume a cosmology with $H_{0}=70$ km~s$^{-1}$ Mpc$^{-1}$ and $\Omega_{m,0} = 0.3$.

\section{Surveys}
\label{sec:surveys}

\subsection{WALLABY}
\label{sec:surveys.WALLABY} 

WALLABY \citep{Koribalski:20} is an H\,{\sc i} survey being conducted with the Australian Square Kilometre Array Pathfinder \citep[ASKAP;][]{Hotan:21}. The phase 1 pilot survey targeted three 60 deg$^{2}$ fields around the Hydra Cluster, Norma Cluster and the NGC 4636 galaxy group in a redshift range of $z<0.08$ \citep{Westmeier:22,Deg:22}. The phase 2 pilot survey targeted the three additional fields around the NGC 5044 group, NGC 4808 group and the Vela Cluster \citep{Murugeshan:24}. In this study we have made use of the Hydra, NGC 4808 and NGC 5044 fields as they are free from bright continuum sources and consequently have the most reliable detections.  Additionally, these fields are not near the Galactic plane and overlap with the Legacy Survey. The full survey is currently underway and will observe approximately $1.4\pi$ sr of the sky in 8832 hours over the next few years. It is expected to detect $\sim210,000$ galaxies out to a redshift of $z\approx0.1$ across the majority of the southern hemisphere. WALLABY has better angular resolution and sensitivity than previous wide-field H\,{\sc i} surveys, such as ALFALFA and HIPASS, with a 30~arcsec beam. The survey has a frequency range of 1295.5-1439.5~MHz and a channel resolution of 4 km~s$^{-1}$. The phase 1 data has a target noise level of 1.6~mJy~beam$^{-1}$ per channel, which corresponds to a $5\sigma$ column density of $8.6 \times 10^{19} (1+z)^4$ cm$^{-2}$ across 20 km~s$^{-1}$. The Hydra field has a slightly higher noise level of 1.85 mJy beam$^{-1}$ per channel.  The phase 2 data has an observed median rms noise of 1.7 mJy beam$^{-1}$ per channel, which corresponds to a 5$\sigma$ H\,{\sc i} column density sensitivity of $ \sim 9.1 \times 10^{19}(1 + z)^{4}$ cm$^{-2}$ across 20 km~s$^{-1}$.  Version 2 of the Source Finding Application \citep[SoFiA;][]{Serra:15,Westmeier2:21} is used by the WALLABY team for source identification. This allows for many H\,{\sc i} properties to be included in the WALLABY catalogues, including the integrated fluxes and central frequencies of the sources. 

\subsection{DESI Legacy Imaging Surveys}
\label{sec:legacy}

The Legacy Survey is a combination of three public projects: the Dark Energy Camera Legacy Survey (DECaLS), the Beijing-Arizona Sky Survey \citep[BASS;][]{Zou:17}, and the Mayall z-band Legacy Survey (MzLS). The Legacy Survey is primarily being conducted with the 4 m Blanco Telescope at the Cerro Tololo Inter-American Observatory. The Legacy Survey provides imaging in the $g$, $r$, $i$ and $z$-bands over 20,000 square degrees with seeing on the order of 1 arcsec.  Data release 10 has $3\sigma$ limiting surface brightnesses of 29.8, 29.4, 27.7 and 28.0 mag arcsec$^{-2}$ in the $g$, $r$, $i$ and $z$-bands respectively, measured in $10 \times 10$ arcsec boxes as measured by \cite{obeirne:24} following the depth definition by \cite{ROman:20}.  These deep optical images allow extremely faint optical counterparts to WALLABY H\,{\sc i} detections to be found and dark sources to be identified. The next deeper optical photometric survey across a comparably wide field will be the Legacy Survey of Space and Time \citep[LSST;][]{Ivezi:19ApJ}.

\subsection{GALEX}

The Galaxy Evolution Explorer \citep[GALEX;][]{Martin:05} provides imaging in the near ultraviolet (NUV; 1770-2730 \AA) and the far ultraviolet (FUV; 1350-1780 \AA). The surveys include the All-sky Imaging Survey (AIS), a Medium Imaging Survey (MIS) of 1000 deg$^{2}$ and a Deep Imaging Survey (DIS) of 100 deg$^{2}$ , with depths of m$_{\rm AB}\sim$20.5, m$_{\rm AB}\sim$23 and m$_{\rm AB}\sim$25 respectively. The resolution of the NUV and FUV images are 4.2 arcsec and 5.3 arcsec. We use the GALEX images to estimate star formation rates in Section~\ref{sec:phot-SFR}. As GALEX does not have complete sky coverage in the WALLABY fields that we have used, we find that $5$~per~cent of our dark and low surface brightness sample was not observed by GALEX.

\subsection{WISE}

In addition to calculating the star formation rates of the LSBGs from the GALEX UV emission, we also investigated the infrared emission from the Wide-field Infrared Survey Explorer \citep[WISE;][]{Wright:10}. WISE provides data over the entire sky in four mid-infrared bands: W1 (3.4 $\mu$m),  W2 (4.6 $\mu$m), W3 (12 $\mu$m) and W4 (22 $\mu$m). Unfortunately, we were only able to measure the infrared photometry for 19 of the LSBGs due to the sensitivity limits of the WISE data.


\section{Methods}
\label{sec:methods}

\subsection{H\,{\sc i} properties}
\label{sec:HI_prop}

In this work we make use of the H\,{\sc i} properties provided in the WALLABY catalogues, including central frequency and integrated flux. We apply the statistical flux corrections following the methods outlined in \cite{Westmeier:22} and \cite{Murugeshan:24} for the phase 1 and 2 pilot data respectively.  {This is done to account for the flux deficit of the faint WALLABY sources compared to the flux that would be recovered by single dish observations.}  {We calculate the H{\sc i} mass from the integrated flux measurement, propagating the flux uncertainty to calculate the uncertainty in the H{\sc i} mass. The true H{\sc i} mass error will be dominated by systematic errors, such as the partial detection of a galaxy, source confusion and distance uncertainty. Throughout this work we use the luminosity distances approximated by the Hubble law using the heliocentric velocities of the H{\sc i} detections. The uncertainty in the H{\sc i} velocities (approximately 4 km s$^{-1}$) is small compared with the uncertainties in typical optical velocity estimates. Approximately half of the previously catalogued sources in the Hydra and NGC 5044 fields are in the 6dF Galaxy Survey \citep[6dFGS;][]{Jones:09}, and approximately half of the previously catalogued sources in the NGC 4808 field are in the Sloan Digital Sky Survey \citep[SDSS;][]{Abazajian:09}. These have average quoted velocity errors of 46 km s$^{-1}$ and 12 km s$^{-1}$ respectively. The median offset between our H{\sc i} velocities and the corresponding optical velocities is 20 km s$^{-1}$.}

In addition to galaxy properties provided in the WALLABY catalogues, we measure the sizes of the H\,{\sc i} discs ($D_{\rm HI}$) from the integrated intensity (moment 0) maps. $D_{\rm HI}$ defined as the major axis of the 1 M$_{\odot}$ pc$^{-2}$ isodensity contour. To measure this, we model the H\,{\sc i} moment 0 map as a 2-dimensional Gaussian.  { Commensurate with the signal-to-noise ratio of the resolved detections, the uncertainty is estimated as 10 arcsec converted to a physical size at the angular diameter distance of the source.} A beam smearing correction is also applied to the H\,{\sc i} diameter using Equations (\ref{eq:DHIcorr}) to (\ref{eq:Dbeam}):
\begin{equation}
    \sigma_{\rm galaxy} = \sqrt{\sigma_{\rm model}^2 - \sigma_{\rm beam}^2},
    \label{eq:DHIcorr}
\end{equation}
\begin{equation}
    \sigma_{\rm beam} = \frac{30"}{2\sqrt{2\ln(2)}},
\end{equation}
\begin{equation}
    D_{\rm HI} = 2\sqrt{\ln(A) \times 2 \sigma_{\rm galaxy}^2},
    \label{eq:Dbeam}
\end{equation}
where $ \sigma_{\rm galaxy}$ is the major axis standard deviation of the deconvolved galaxy, $\sigma_{\rm model}$ is the major axis standard deviation of the 2-dimensional Gaussian model of the moment 0 map (which is the real galaxy convolved with the WALLABY beam) and $\sigma_{\rm beam}$ is the standard deviation of the beam. $D_{\rm HI}$ is the H\,{\sc i} diameter in arcsec and $A$ is the amplitude of the Gaussian model in M$_{\odot}$ pc$^{-2}$. Equation (\ref{eq:Dbeam}) is the equation for a 1-dimensional Gaussian along the major axis. This deconvolution method is valid because the WALLABY beam is circular (30 arcsec). Due to the resolution limitations  {we do not correct for the effect of inclination on optical depth.}  This could lead to an over estimation of $D_{HI}$ for any edge on galaxies. To measure $D_{\rm HI}$ we require $\sigma_{\rm galaxy} > 2\sigma_{\rm beam}$ to ensure the galaxy is sufficiently well resolved.  Unfortunately $89.7$~per~cent of the dark and low surface brightness sample were not sufficiently resolved to meaningfully measure $D_{\rm HI}$.  Additionally, we find that $0.3$~per~cent of the sample were not well-modelled by a Gaussian and $0.8$~per~cent  were too diffuse to reach a density of 1 M$_{\odot}$ pc$^{-2}$, and consequently do not have a $D_{\rm HI}$ measurement.

\subsection{Photometry}

\subsubsection{Sérsic Modelling}
\label{sec:sersic}

To measure photometric properties in the Legacy Survey and GALEX images, we create models of the galaxies using the python package \textsc{AstroPhot} \citep{Stone:23}. We model the galaxies using a Sérsic profile \citep{Graham:05}, given by: 
\begin{equation}
    I(R) = I_{e} \exp \left[ -b_{n} \left( \left( \frac{R}{R_{e}} \right)^{1/n} - 1 \right) \right]
    \label{eq:sersic}
\end{equation}
where $I(R)$ is the brightness profile as a function of semi-major axis, $R$ is the semi-major axis length, $I_{e}$ is the brightness at the half-light radius $R_{e}$, $n$ is the Sérsic index that controls the shape of the profile, and $b_{n}$ is a function of $n$ that is not involved in the fit. \textsc{AstroPhot} fits seven parameters: the centre coordinates, position angle, axis ratio,  $I_{e}$, $n$ and $R_{e}$. These fits allow us to obtain photometric properties, including total fluxes and central surface brightnesses, and to create meaningful apertures using the effective radii, position angles and axis ratios.

 {Consistent multiband photometry for extended diffuse galaxies is a challenge. Using a model optimises the signal to noise, and a consistent model across the bands allows for consistent photometry, and consequently good colours.} \textsc{AstroPhot} is a powerful tool, handling point spread functions (PSFs), multiple sources (e.g. foreground stars) and joint fitting in multiple bands.  For the Legacy Survey images, we model the PSF ourselves using \textsc{AstroPhot}. To do this, we fit a Moffat PSF profile to five stars in each of the image cutouts. All galaxy models take the respective PSFs into account, and additionally galaxies with bright foreground stars are modelled together with the point source models of the foreground stars. We also make use of the joint modelling function of \textsc{AstroPhot} to model multiple bands together, allowing all parameters except $I_{e}$ to be fit together in both bands. The $g$- and $i$-band images are modelled together, as are the NUV and FUV images. An example of the Sérsic modelling is shown in Appendix~\ref{appen:sersic}.

There are limitations to our identification of LSBGs. We model all of our galaxies using a Sérsic profile, which may not reflect the range of properties present in all galaxies and has implicit limitations \citep[e.g.][]{Trujillo:01}. Across the 1829 galaxies in the three WALLABY fields looked at in this study,  6.3~per~cent of the sources are not well modelled by a Sérsic profile.  This includes galaxies with foreground stars that were not able to be removed accurately as they saturated the detector.  A further $0.8$~per~cent had missing coverage in the Legacy Survey image, and $4.4$~per~cent LSBGs did not have reliable models of the UV emission. Moreover, the proportion of LSBG sources in our sample could be underestimated. This is because a proportion of the H\,{\sc i} detections contained more than one optical source. These detections could contain galaxies currently undergoing interactions within a shared HI envelope, two galaxies at different redshifts, or could alternatively be the result of the limited angular resolution of WALLABY compared to the Legacy Survey. As the individual H\,{\sc i} properties could not be determined, these sources were not included in the sample. Similarly, the H\,{\sc i} properties could not be accurately determined for the galaxies that had their H\,{\sc i} emission split across multiple detections by SoFiA. Overall, $12.1$~per~cent of the 1829 WALLABY galaxies were not modeled because of these reasons. Future work will include studying the properties and distributions of galaxy pairs and groups with interacting H\,{\sc i} gas.

\subsubsection{Stellar Mass}
\label{sec:mstar}

To calculate the stellar mass of the LSBGs, we use the relation between stellar mass to light ratio ($\frac{\Upsilon^*}{\rm M_{\odot}/L_{\odot}}$) and $g-i$ colour from \cite{Du:20}:
\begin{equation}
    \log_{10}(\Upsilon^{*}) = a + b({g-i}) ,
    \label{Mstar_lsb}
\end{equation}
where $a=-1.152$ and $b=1.328$ are the coefficients. {We account for Galactic extinction using the correction method outlined in \cite{Yuan:13}, adopting the $R(a)$ values from \cite{Schlegel:98} and the $E(B-V)$ values from \cite{Schlafly:11}. We use the mean \cite{Schlafly:11} $E(B-V)$ values\footnote{available at \url{https://irsa.ipac.caltech.edu/applications/DUST/}}. Additionally, we apply $k$-corrections following \cite{Chilingarian:10}.} While the $r$\nobreakdash-band Legacy Survey images are deeper than the $i$-band images, almost a third of the low surface brightness sources had not been observed by the Legacy Survey in the $r$-band at the time of writing. Consequently, for consistency the $g-i$ colour is used to calculate the stellar mass to light ratio for all sources. \cite{Du:20} derive this relationship from their sample of LSBGs selected from the $\alpha$.40 H\,{\sc i} survey \citep{Haynes:11} and SDSS. LSBGs have been shown to have low star formation rates and low stellar mass densities \citep{Burkholder:01,Lei:18}. These distinct properties suggest that LSBGs could have different formation and evolutionary histories compared to high surface brightness galaxies, and different stellar populations have very different spectral energy distributions. This leads to different stellar mass-to-light ratio scaling relations, for example, \cite{Du:20} show that the relation from \cite{Bell:03} overestimates the stellar mass for their population of LSBGs. Hence, it is important that we use a stellar mass to light ratio scaling relation derived specifically for LSBGs. The mass-to-light ratio can be converted to a stellar mass using the Equations (\ref{eq:MLR}), (\ref{eq:Lum}) and (\ref{eq:absMag}):
\begin{equation}
    \Upsilon^{*} = \frac{M_{*}}{L},
    \label{eq:MLR}
\end{equation}
where $M_{*}$ is the stellar mass, and $L$ is the luminosity. The luminosity can be calculated from the $g$-band magnitude.
\begin{equation}
    \frac{L}{L_{\odot}} = 10^{-0.4(M_g - M_{g, \mathrm{solar}})},
    \label{eq:Lum}
\end{equation}
\begin{equation}
    M_g = m_g - 5\log\left(\frac{D_L}{\rm 10 pc}\right),
    \label{eq:absMag}
\end{equation}
where $M_g$ is the $g$-band absolute magnitude and $M_{g, \mathrm{solar}}=5.05$~M$_{\odot}$ is the absolute solar magnitude in the $g$-band \citep{Willmer:18}. The uncertainty associated with the stellar mass is calculated by propagating the uncertainty in the $g$-band flux measurement from the Sérsic modelling and the 0.24 dex scatter in the $\log_{10}(\Upsilon^{*})$ from \cite{Du:20}.

\begin{figure}
    \centering
    \includegraphics[width=0.45\textwidth]{ 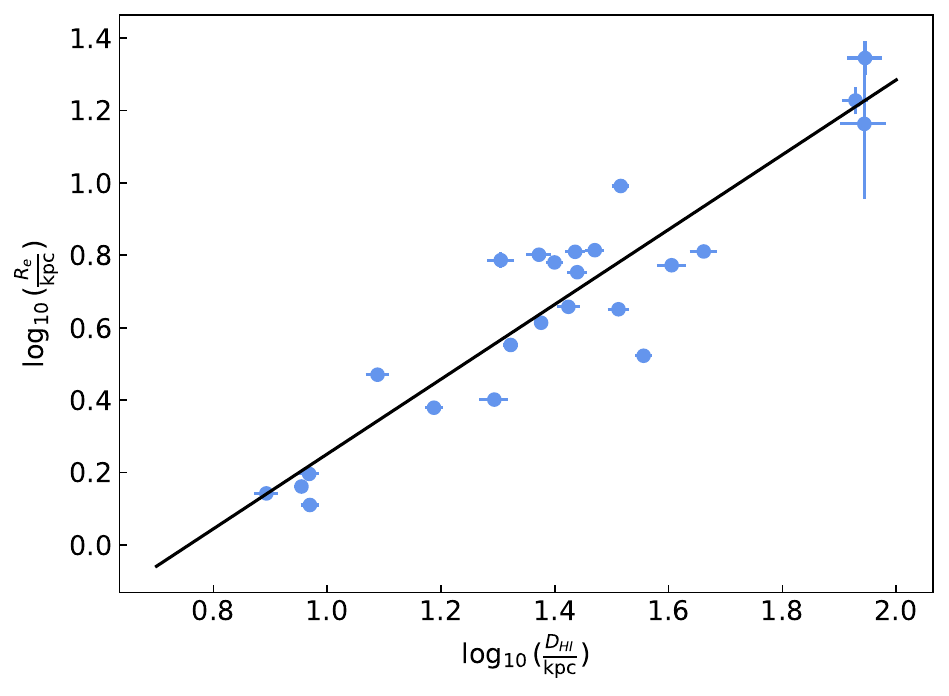}
    \caption{The effective radius ($R_{e}$) as a function of H\,{\sc i} size ($D_{\rm HI}$) for the well-resolved LSBGs. The black line shows the best-fitting line:  $\log_{10}\left(\frac{R_{e}}{\rm kpc}\right) = (1.0\pm0.1) \times \log_{10}\left(\frac{D_{\rm HI}}{\rm kpc}\right) - (0.8\pm0.1)$.}
    \label{fig:RE_DHI}
\end{figure}

We measure the $3\sigma$ upper limits on the stellar masses of the dark sources using an aperture on the Legacy Survey images within the location of the H\,{\sc i} detection. To create a meaningful aperture, we study the correlation between H\,{\sc i} size and effective radius for the resolved LSBGs, as shown in Figure~\ref{fig:RE_DHI}. We find that this relation has a best-fitting line defined by $\log_{10}\left(\frac{R_{e}}{\rm kpc}\right) = (1.0\pm0.1) \times \log_{10}\left(\frac{D_{\rm HI}}{\rm kpc}\right) - (0.8\pm0.1)$. For the dark sources we use this relation to create a circular aperture corresponding to their measured H\,{\sc i} size. For the purpose of choosing an appropriate aperture to measure the upper limits, we measure $D_{\rm HI}$ for poorly resolved dark sources with $\sigma_{\rm galaxy} < 2\times \sigma_{\rm beam}$ without correcting for the beam, noting that 4 of the 55 dark sources still do not have H\,{\sc i} size measurements, as 3 were too diffuse and for 1 source the Gaussian model did not converge.

\subsubsection{Star Formation Rates}
\label{sec:phot-SFR}

Newly formed high mass stars release hot UV radiation. This radiation from young stars is absorbed by dust and re-emitted in the mid- and far-infrared; this reprocessed emission can therefore be used as a SFR indicator. The SFR can be derived from the WISE W3 luminosity after correcting for the contribution of old stellar populations to the W3 band; this Rayleigh-Jeans emission is estimated using the W1 luminosity \citep{Cluver:17}. However, as the LSBGs do not have strong W3 infrared emission, we calculate their star formation rates (SFRs) solely from the GALEX ultra-violet (UV) images. The UV traces young massive stars and hence is often used as an indicator of SFR, however it is susceptible to dust extinction which must be accounted for. The SFRs are calculated following the method of \cite{Hao:11} using the total fluxes measured in the NUV and FUV bands. First we apply a Galactic reddening correction using Equations (\ref{eq:mnuv_corr}) to (\ref{eq:beta}):
\begin{equation}
    m_{\rm NUV,corr} = m_{\rm NUV} - \alpha
    \label{eq:mnuv_corr}
\end{equation}
\begin{equation}
    m_{\rm FUV,corr} = m_{\rm FUV} - \beta
    \label{eq:mfuv_corr}
\end{equation}
\begin{equation}
    \alpha = 8.36 E(B-V) + 14.3 (E(B-V))^{2} - 82.8 (E(B-V))^{3}
\end{equation}
\begin{equation}
    \beta = 10.47 E(B-V) + 8.59 (E(B-V))^{2} - 82.8 (E(B-V))^{3}
    \label{eq:beta}
\end{equation}

where $m_{\rm NUV,corr}$ and $m_{\rm FUV,corr}$ are the NUV and FUV magnitudes after applying the Galactic reddening correction to the measured NUV and FUV magnitudes ($m_{\rm NUV}$ and $m_{\rm FUV}$), and $\alpha$ and $\beta$ are functions of $E(B-V)$. {Next we apply the $k$-correction following \cite{Chilingarian:10}.} Then we apply an attenuation correction using Equations (\ref{eq:mfuv_atten}) and (\ref{eq:gamma}):
\begin{equation}
    m_{\rm FUV,atten} = m_{\rm FUV,corr} - \gamma
    \label{eq:mfuv_atten}
\end{equation}
\begin{equation}
    \gamma = 3.83 (m_{\rm FUV,corr} - m_{\rm NUV,corr} - 0.022)
    \label{eq:gamma}
\end{equation}
where $m_{\rm FUV,atten}$ is the attenuation-corrected FUV magnitude. Finally, the SFR is calculated using Equation (\ref{eq:SFR_GALEX}):
\begin{equation}
    \log_{10}\left(\frac{\rm SFR}{{\rm M}_{\odot} {\rm yr}^{-1}}\right) = \log_{10}\left(\frac{\rm L_{FUV}}{\rm erg~s^{-1}}\right) - 43.35
    \label{eq:SFR_GALEX}
\end{equation}
where $L_{FUV}$ is the reddening- and attenuation-corrected FUV luminosity.  {However, if $FUV-NUV\le0$, the internal dust attenuation correction can not be applied, as the $\gamma$ correction parameter becomes unphysical.} The uncertainty in $\log_{10}\left(\frac{\rm SFR}{\rm M_{\odot} yr^{-1}}\right)$ is taken to be $\pm0.115$ dex from the scatter of the relation in Equation (\ref{eq:SFR_GALEX}) \citep{Hao:11}. $3\sigma$ upper limits on the SFRs for the LSBGs that were not detected in the FUV are calculated from elliptical apertures with semi-major axes of two times the effective radius from the optical Sérsic models (or as estimated in Section~\ref{sec:mstar} for the dark sources).

\subsection{Source Classification and Reliability}
\label{sec:reliability}

Galaxies are classified as LSBGs if they have a mean $g$-band surface brightness fainter than 23 mag arcsec$^{-2}$ within 1 effective radius (as calculated in the Sérsic model) and are classified as dark H\,{\sc i} sources if they have no visible counterpart  {in the $g$-band image} and no Sérsic model could be made at the H\,{\sc i} coordinates of the Legacy Survey image.  {The optical properties of the dark sources are further analysed by coadding all four bands of Legacy Survey images and convolving with a boxcar kernel with a size of 2.6 arcsec by 2.6 arcsec to degrade the resolution and enhance the surface brightness sensitivity, greater enabling the detection of diffuse emission. These images are shown in Appendix \ref{append:coadd}, and one `dark' source may have evidence of a potential optical counterpart (see Section \ref{sec:disc-dgc}).} Additionally, we inspect the dark H\,{\sc i} sources for signatures that suggest a tidal origin. Sources that have asymmetric H{\sc i} features or lie within the virial radius of a neighbouring galaxy are  {potentially} tidal debris. We fit a Sérsic model to the neighbouring galaxies in NASA/IPAC Extragalactic Database (NED) that are within $\pm 300$ km~s$^{-1}$ and measure their effective radii.  We use the relation $R_{e} = 0.015R_{200}$ from \cite{Kravtsov:13} to estimate the virial radius from the effective radius.

\begin{figure*}
     \centering
       \begin{subfigure}[b]{0.48\textwidth}
         \centering
         \includegraphics[width=\textwidth]{ 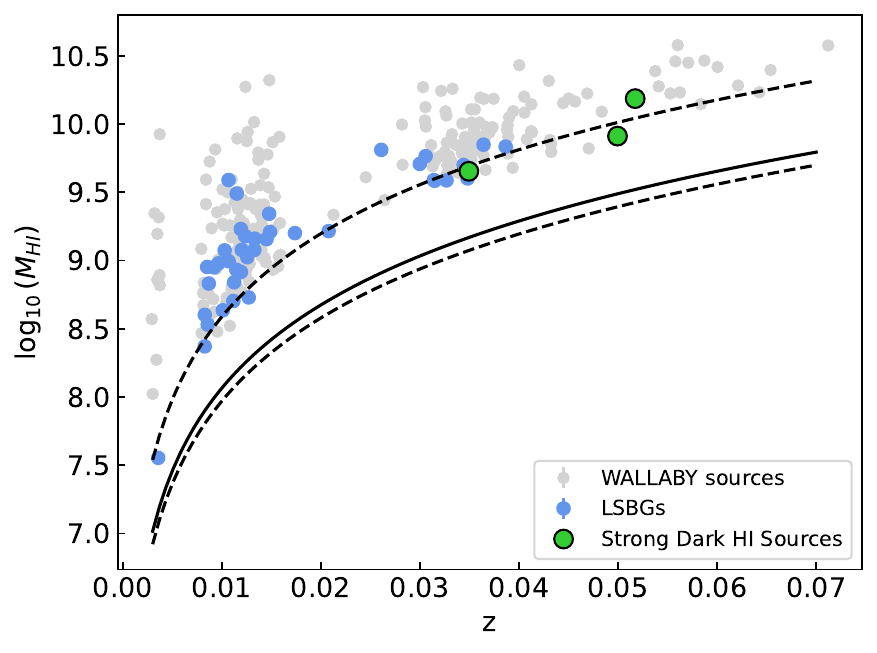}
         \caption{Phase 1 Data}
         \label{fig:mhi_z1}
     \end{subfigure}
     \begin{subfigure}[b]{0.48\textwidth}
         \centering
         \includegraphics[width=\textwidth]{ 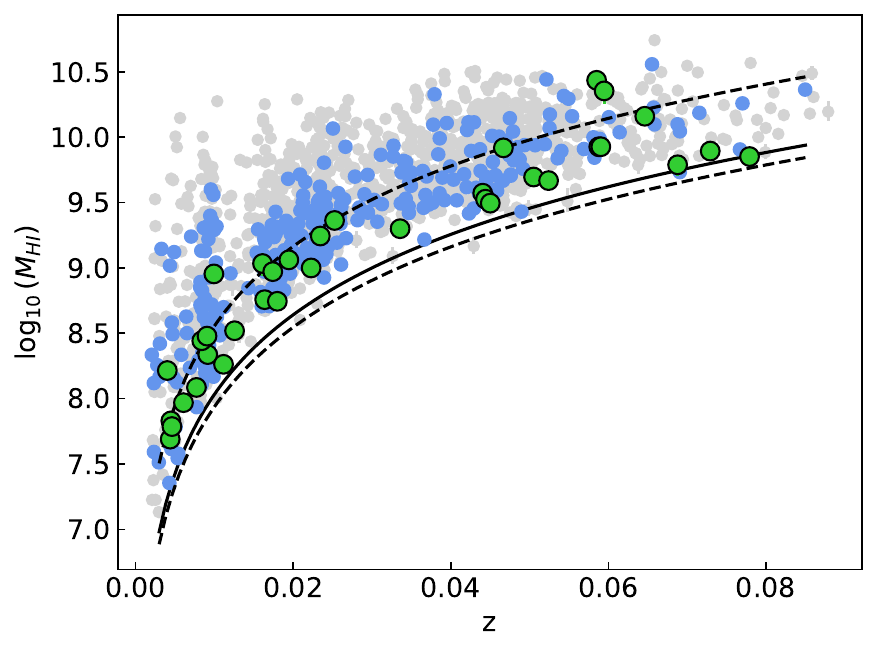}
         \caption{Phase 2 Data}
         \label{fig:mhiz2}
     \end{subfigure}
    \caption{The H\,{\sc i} mass as a function of redshift for the LSBGs (blue) and strong candidate dark source detections (green) compared to the rest of the detections in the WALLABY fields (grey).  The phase 1 data and phase 2 data are plotted separately in panels (a) and (b) respectively as the detection limits of the pilot surveys differ. The $5\sigma$ detection limit shown by the solid black line is calculated using a line width of 1 MHz.  {The $5\sigma$ detection limits shown by the dashed black lines are calculated using the minimum and maximum line widths.}}
        \label{fig:mhi_z}
\end{figure*}

It is essential to consider the reliability of the detection of the dark H\,{\sc i} sources, as the false positive rate of the SoFiA implementation in the WALLABY pipeline still requires further investigation. False positives could arise from a number of factors, including the presence of interference and residual continuum emission in the data cube. Nevertheless, all of the dark H\,{\sc i} sources presented in this work have passed the quality checks required to be published in the WALLABY catalogues \citep{Westmeier:22,Murugeshan:24}.  {There are, however, notes in the catalogue to caution users against trusting the reliability of these sources without further analysis.} To distinguish strong dark source candidates from  {uncertain} detections (requiring follow-up observations),  {we consider the strength of the detection with respect to the survey detection limit} and we inspect the unmasked H\,{\sc i} cubes and spectra to assess the strength of the detection compared to the noise level in the cube.  Figure~\ref{fig:mhi_z} shows the H\,{\sc i} mass plotted against the redshift for the dark H\,{\sc i} sources and LSBGs with respect to all of the WALLABY detections in the fields used. The phase 1 and 2 pilot data are plotted separately as the phase 2 survey has a lower noise level  {than the Hydra field}. The limiting detectable H\,{\sc i} mass is dependent on the velocity range, so the detection limit for each dark source is calculated by integrating the median local RMS noise level (of 1.85 mJy in the Hydra field and 1.7 mJy in the phase 2 cubes) over the $w_{20}$. The $5\sigma$ detection limit line shown in the figure is calculated assuming a line width of 1 MHz (211 km~s$^{-1}$) for consistency with the WALLABY pilot data releases, however, it is important to note that this width is not representative of all galaxies and the $w_{20}$ emission line widths in our sample range from 0.09 MHz to 1.55 MHz (19 km~s$^{-1}$ to 327 km~s$^{-1}$). Seven of the dark sources have H\,{\sc i} masses less than a factor of two above the detection limit (for their given $w_{20}$ width). Dark sources lying close to the detection limit could help to explain why large numbers of dark candidates have not been detected in previous H\,{\sc i} surveys, however it also suggests that these seven dark sources are less reliable detections  {and could simply be false positives caused by noise or artefacts in the data.}

We identify a total of 38 strong candidate sources (leaving 17  {uncertain} H\,{\sc i} detections). All of the strong dark source candidates have a  {peak} signal-to-noise (SNR) ratio $>4.9$, where the SNR is defined as the peak flux density divided by the rms provided in the WALLABY catalogues. Furthermore, we check for dark H\,{\sc i} sources that could be impacted by satellite radio frequency interference (RFI), whether from sidelobes and haromonics of radionavigation satellites below 1293 MHz, or satellites that use frequencies within the clean mid band \citep[1293 to 1437 MHz;][]{Lourencco:24}. We identify three candidates (WALLABY J$130119+053553$, WALLABY J$132709-163509$ and WALLABY J$132238-204726$) that could be affected by RFI based on their central frequencies. All of these had already been flagged as  {uncertain} detections, which is a testament to the accuracy of our classification.  Three of the strong H\,{\sc i} source candidates have a Parkes H\,{\sc i} detection within 15.5 arcmin (size of HIPASS beam) of the WALLABY coordinates: WALLABY J131244-155218/HIPASS J1312-15, WALLABY J131928-123828/HIPASS J1319-12 \citep{Barnes:01}, and WALLABY J132825-253528/HIDEEP J1329-2533 \citep{Minchin:03}. Deep follow-up H\,{\sc i} observations, such as with MeerKAT  { \citep[e.g.][]{Namumba:21,Maccagni:24,Zabel:24} }, are still required to definitively confirm the detections of the dark H\,{\sc i} sources. 
%


\section{Results}
\label{sec:results}

\subsection{Source Identification}
\label{sec:source_ident}

Table~\ref{tab:classes} shows the number of WALLABY sources in each category for each pilot survey field used. In total we find 315 LSBGs, 38 strong dark H\,{\sc i} detections, and 17  {uncertain} dark H\,{\sc i} detections. Of the 38 strong dark H\,{\sc i} detections, 13 have signatures that suggest they could be tidal debris (1 from the Hydra field, 1 from the NGC 4808 field and 11 from the NGC 5044 field). Table~\ref{tab:tidal} presents the sources that have clear tidal features. 

\begin{figure*}
     \centering
       \begin{subfigure}[b]{0.32\textwidth}
         \centering
         \includegraphics[width=\textwidth]{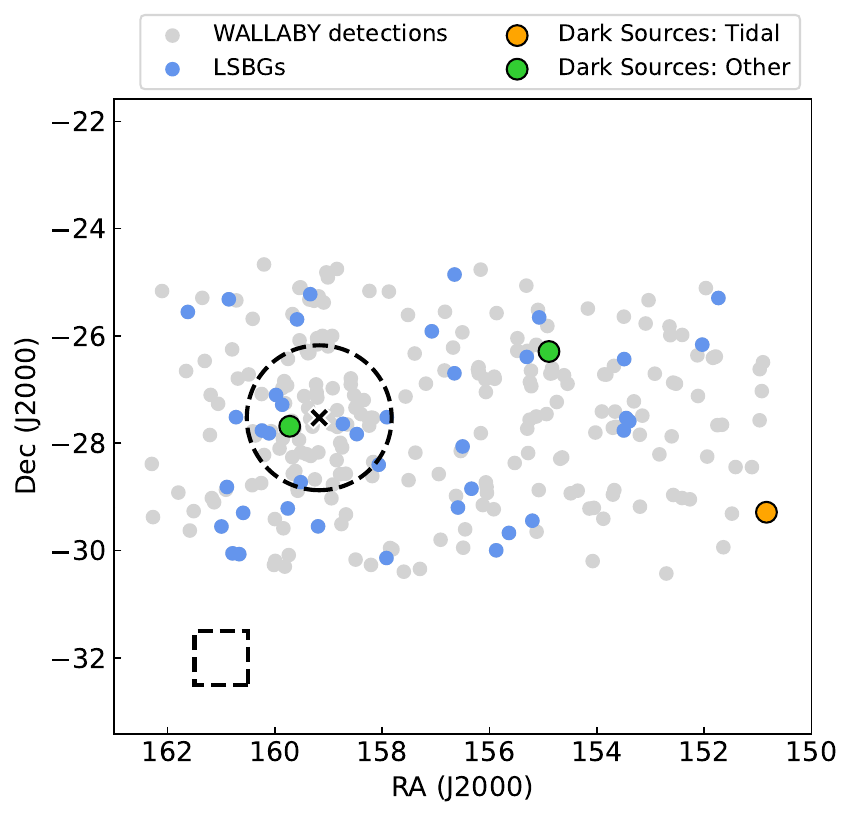}
         \caption{Hydra field}
         \label{fig:hydra}
     \end{subfigure}
     \begin{subfigure}[b]{0.3\textwidth}
         \centering
         \includegraphics[width=\textwidth]{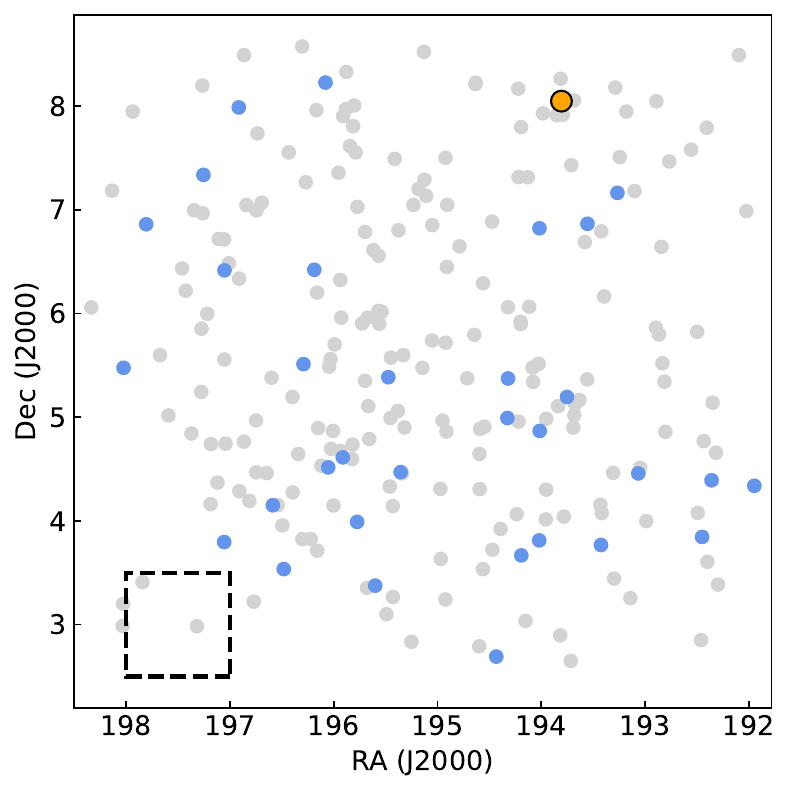}
         \caption{NGC 4808 field}
         \label{fig:ngc4808}
     \end{subfigure}
     \begin{subfigure}[b]{0.32\textwidth}
         \centering
         \includegraphics[width=\textwidth]{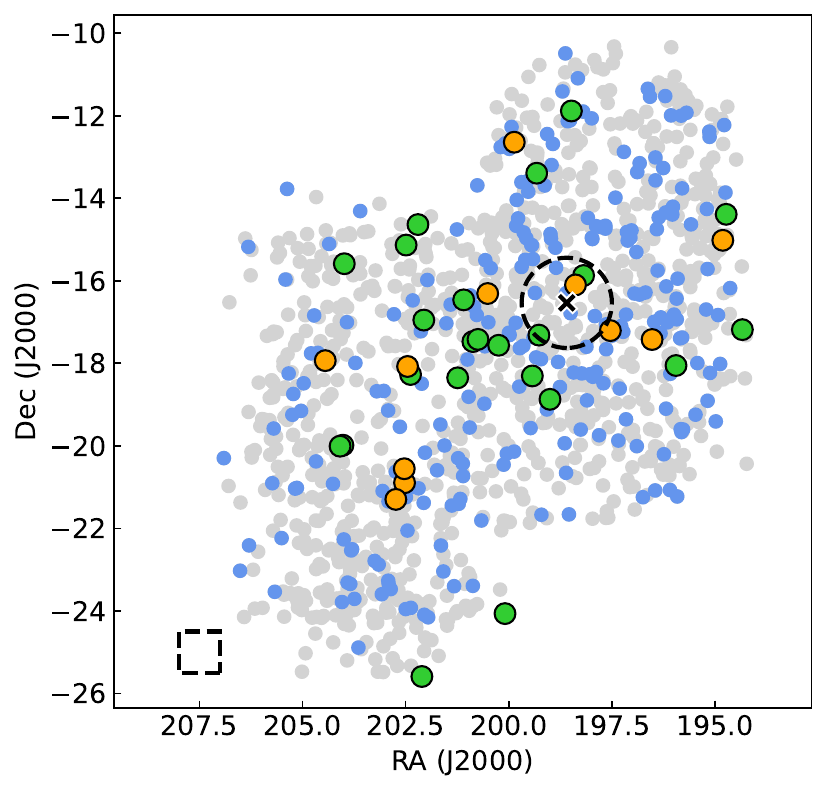}
         \caption{NGC 5044 field }
         \label{fig:ngc5044}
     \end{subfigure}
    \caption{The location of the LSBGs (blue), dark H\,{\sc i} sources with tidal features (orange) and the rest of the dark H\,{\sc i} sources (green) with respect to the rest of the WALLABY sources (grey).  {A 1 deg$^{2}$ box is shown in the lower left corner for scale.} The Hydra cluster and virial radius is shown in panel (a) and the NGC 5044 group and $r_{500}$ overdensity radius is shown in panel (c). }
    \label{fig:loc}
\end{figure*}

Table~\ref{tab:DGC} contains the properties of the strong dark H\,{\sc i} detections. This includes the WALLABY field, WALLABY name, right ascension, declination, central velocity, luminosity distance, $w_{50}$, $w_{20}$, H\,{\sc i} size, H\,{\sc i} mass, stellar mass $3\sigma$ upper limit, SFR $3\sigma$ upper limit,  {peak} signal-to-noise ratio and whether the source has signatures to suggest that it is a tidal remnant.  Throughout this section, the  {uncertain} dark H\,{\sc i} detections are not included in Figures \ref{fig:loc} to \ref{fig:mhi mstar sfr}, however their properties are presented in Table \ref{tab:unreliable} in Appendix \ref{append:weak}. The H\,{\sc i} contours overlaid on optical images, moment 1 maps and unmasked spectra of the strong and  {uncertain} dark H\,{\sc i} detections are shown in Appendices \ref{append:strong} and \ref{append:weak} respectively.

\begin{table}
    \small
    \setlength{\tabcolsep}{4pt} 
    \centering
    \caption{The number of WALLABY dark H\,{\sc i} sources and LSBGs }
    \label{tab:classes}
    \begin{tabular}{lcccc}
        \hline
        Field & Hydra & NGC 4808 & NGC 5044 & Total \\
        \hline
        Strong dark H\,{\sc i} detections & 3 & 1 & 34 & 38 \\
         {Uncertain} dark H\,{\sc i} detections & 2 & 2 & 13 & 17 \\
        LSBGs  & 39 & 32 & 244 & 315 \\
        \hline
        Total WALLABY sources & 272 & 231 & 1326 & 1829\\
        \hline
    \end{tabular}
\end{table}

\begin{table*}
    \centering
    \caption{The strong dark H\,{\sc i} detections that show evidence that suggests they could be tidal remnants. }
    \label{tab:tidal}
    \begin{tabular}{cccccl}
    \hline
        WALLABY Name & Field & Separation  & Separation  & $\Delta v$ & Comment \\
         &  &  (arcsec) &  (kpc) & (km s$^{-1}$) \\
        \hline
         J100321-291708 & Hydra & 317 & 320 & 81 & Within the virial radius of WISEA J100257.09-291041.3 \\
         J125513+080246 & NGC 4808 & 154 & 29 & 36 & Within the virial radius of NGC 4795 \\
         J125915-150108 & NGC 5044 & 122 & 10 & 144 & Within the virial radius of NGC 4856, consists of 2 clouds with  \\
         &  &   &  & &  additional WALLABY detection to the east of NGC 4856 \\
         J131928-123828 & NGC 5044 & 77 & 13 & 286 & Within the virial radius of NGC 5077, asymmetric H\,{\sc i} distribution   \\
         J131331-160600 & NGC 5044 & 280 & 52 & 19 & Within the virial radius of MCG -03-34-020 \\
         J133006-205341 & NGC 5044 & 321 & 113 & 25 & Asymmetric HI, within the virial radius of WISEA J132943.49-205512.9\\
         J132202-161829 & NGC 5044 & 130 & 62 & 15 & Within the virial radius of WISEA J132153.70-162049.0 \\
         J130606-172523 & NGC 5044 & 31 & 28 & 76 & Within the virial radius of WISEA J130604.17-172549.4, 2 interacting\\
         &  &   &  & &  galaxies to the south \\
         J132948-180438 & NGC 5044 & 399 & 37 & 157 & Within the virial radius of NGC 5170 \\
         J133008-203319 & NGC 5044 & 81 & 32 & 298 & Within the virial radius of ESO 576-G067 \\
         J133747-175606 & NGC 5044 & 286 & 27 & 30 & Within the virial radius of NGC 5247, stream of 3 clouds  \\
         J133057-211755 & NGC 5044 & 255 & 93 & 240 & Within the virial radius of WISEA J133039.64-211911.9, asymmetric H\,{\sc i}, \\
         &  &   &  & & 2 additional WALLABY detections within 50 kpc and 500 km s$^{-1}$   \\
          {J131009-171227} & NGC 5044 & 51 & 26 & 43 & Within the virial radius of WISEA J131007.86-171139.6 \\
         \hline
    \end{tabular}
    
\end{table*}

\begin{table*}
    \centering
    \caption{Properties of the strong dark source detections. From left to right the columns are: WALLABY field, WALLABY name, right ascension, declination, central velocity, luminosity distance, emission line width at half maximum, $w_{20}$ emission line width,  H\,{\sc i} size (major axis at 1 M$_{\odot}$ pc$^{-2}$ contour level), H\,{\sc i} mass,  stellar mass $3\sigma$ upper limit, SFR $3\sigma$ upper limit,  {peak} signal-to-noise ratio, and whether the source has evidence to suggest it could be a tidal remnant. }
    \label{tab:DGC}
    \setlength{\tabcolsep}{3pt}
    \begin{tabular}{llcccccccccccc}
    \hline
       Field &  Name  & RA (J2000) & DEC (J2000) & $\frac{cz}{\rm km s^{-1}}$ & $\frac{d_L}{\rm Mpc}$ & $\frac{w_{50}}{\rm km~s^{-1}}$ & $\frac{w_{20}}{\rm km~s^{-1}}$ & $\frac{D_{\rm HI}}{\rm kpc}$ & $\log(\frac{M_{HI}}{\rm M_{\odot}})$ & $\log(\frac{M_{*}}{\rm M_{\odot}})$ & $\log(\frac{SFR}{\rm M_{\odot} yr^{-1}})$ & SNR & Tidal\\
       \hline
& J101934-261721 & 154.892 & -26.289 & 14981 & 222.0 & 44.8 & 103.9 &    & 9.91$\pm$0.04 & $<5.47$ & $<-1.73$ & 10.9 & No \\
Hydra & J103853-274100 & 159.724 & -27.684 & 10478 & 153.6 & 54.6 & 81.4 &    & 9.65$\pm$0.04 & $<6.81$ & $<-2.87$ & 11.1 & No \\
& J100321-291708 & 150.839 & -29.286 & 15516 & 230.2 & 96.6 & 105.2 &    & 10.19$\pm$0.05 & $<7.49$ & $<-2.35$ & 9.2 & Yes \\
        \hline
        
NGC 4808 & J125513+080246 & 193.805 & 8.046 & 2749 & 39.5 & 39.2 & 97.0 &    & 8.34$\pm$0.07 & $<4.65$ & $<-4.32$ & 6.5 & Yes \\
        \hline
& J125721-171102 & 194.339 & -17.184 & 4834 & 69.9 & 45.7 & 72.2 &    & 9.03$\pm$0.09 & $<5.84$ & $<-2.63$ & 5.1 & No \\
& J125855-142319 & 194.731 & -14.389 & 19384 & 290.3 & 95.0 & 129.5 &    & 10.16$\pm$0.08 & $<5.77$ & $<-2.20$ & 6.3 & No \\
& J130347-180311 & 195.946 & -18.053 & 17712 & 264.2 & 69.8 & 115.2 &    & 9.93$\pm$0.07 & $<5.98$ & $<-1.51$ & 7.1 & No \\
& J131244-155218 & 198.187 & -15.872 & 2983 & 42.9 & 21.3 & 55.0 & 22.2$\pm2.0$ & 8.95$\pm$0.03 & $<4.06$ &   & 14.2 & No \\
& J131355-115301 & 198.483 & -11.884 & 15152 & 224.6 & 71.6 & 114.2 &    & 9.69$\pm$0.06 & $<5.68$ & $<-2.56$ & 7.3 & No \\
& J131600-185222 & 199.002 & -18.873 & 17633 & 263.0 & 69.1 & 137.7 &    & 9.93$\pm$0.05 & $<6.01$ & $<-2.38$ & 9.0 & No \\
& J131704-171858 & 199.269 & -17.316 & 1825 & 26.2 & 63.2 & 100.6 & 4.6$\pm1.3$ & 7.97$\pm$0.07 & $<4.22$ & $<-4.55$ & 7.2 & No \\
& J131717-132332 & 199.323 & -13.392 & 13216 & 195.0 & 87.0 & 130.9 &    & 9.57$\pm$0.07 & $<5.91$ & $<-2.58$ & 6.9 & No \\
& J131743-181822 & 199.429 & -18.306 & 4916 & 71.1 & 71.0 & 100.0 &    & 8.76$\pm$0.08 & $<4.95$ & $<-3.54$ & 6.0 & No \\
& J132022-240400 & 200.092 & -24.067 & 17551 & 261.7 & 108.0 & 151.8 &    & 10.43$\pm$0.07 & $<6.42$ & $<-2.29$ & 6.5 & No \\
& J132059-173347 & 200.246 & -17.563 & 2320 & 33.3 & 93.4 & 107.4 &    & 8.09$\pm$0.06 & $<3.88$ & $<-3.38$ & 7.2 & No \\
& J132259-172513 & 200.748 & -17.420 & 13322 & 196.6 & 94.0 & 128.3 &    & 9.53$\pm$0.08 & $<5.14$ & $<-2.67$ & 5.8 & No \\
& J132328-172821 & 200.871 & -17.473 & 3770 & 54.3 & 69.4 & 121.8 &    & 8.52$\pm$0.07 & $<3.78$ & $<-2.99$ & 6.6 & No \\
& J132422-162744 & 201.094 & -16.462 & 21876 & 329.5 & 26.7 & 89.2 &    & 9.89$\pm$0.07 & $<4.90$ & $<-2.01$ & 6.9 & No \\
& J132457-182105 & 201.239 & -18.352 & 3349 & 48.2 & 65.3 & 85.0 &    & 8.26$\pm$0.07 & $<5.29$ & $<-3.86$ & 6.7 & No \\
& J132814-165706 & 202.060 & -16.952 & 10070 & 147.4 & 103.6 & 109.4 &    & 9.30$\pm$0.08 & $<4.90$ & $<-2.29$ & 5.8 & No \\
NGC 5044 & J132825-253528 & 202.107 & -25.591 & 17837 & 266.1 & 67.9 & 113.0 &    & 10.35$\pm$0.10 & $<7.28$ & $<-1.86$ & 4.9 & No \\
& J132848-143813 & 202.202 & -14.637 & 20634 & 309.9 & 85.9 & 90.8 &    & 9.79$\pm$0.09 & $<5.45$ & $<-1.37$ & 5.6 & No \\
& J132931-181615 & 202.381 & -18.271 & 1320 & 18.9 & 51.3 & 123.6 &    & 7.69$\pm$0.06 & $<3.41$ & $<-5.21$ & 7.8 & No \\
& J132957-150800 & 202.489 & -15.134 & 13504 & 199.4 & 46.8 & 73.1 &    & 9.49$\pm$0.07 & $<5.24$ & $<-3.42$ & 6.6 & No \\
& J133556-153510 & 203.987 & -15.586 & 23372 & 353.2 & 70.7 & 81.9 &    & 9.85$\pm$0.07 & $<5.23$ & $<-1.35$ & 6.3 & No \\
& J133604-195904 & 204.020 & -19.985 & 6683 & 97.0 & 21.8 & 80.6 &    & 9.00$\pm$0.07 & $<6.73$ & $<-3.15$ & 6.9 & No \\
& J133621-200033 & 204.090 & -20.009 & 15723 & 233.4 & 64.4 & 130.5 &    & 9.67$\pm$0.07 & $<5.71$ & $<-1.91$ & 6.7 & No \\

& J125915-150108 & 194.813 & -15.019 & 1209 & 17.3 & 109.1 & 159.4 &    & 8.21$\pm$0.04 & $<3.39$ & $<-4.62$ & 11.1 & Yes \\
& J130606-172523 & 196.526 & -17.423 & 14002 & 207.0 & 128.5 & 203.5 &    & 9.92$\pm$0.05 & $<6.43$ & $<-1.66$ & 8.5 & Yes \\
& J131009-171227 & 197.538 & -17.208 & 7576 & 110.3 & 106.8 & 308.0 &    & 9.36$\pm$0.05 & $<5.17$ & $<-2.01$ & 9.7 & Yes \\
& J131331-160600 & 198.383 & -16.100 & 2722 & 39.1 & 45.0 & 88.3 & 9.8$\pm1.9$ & 8.48$\pm$0.05 & $<4.15$ & $<-4.37$ & 9.1 & Yes \\
& J131928-123828 & 199.867 & -12.641 & 2520 & 36.2 & 217.9 & 248.4 & 8.8$\pm1.7$ & 8.44$\pm$0.06 & $<3.69$ &   & 8.3 & Yes \\
& J132202-161829 & 200.512 & -16.308 & 7033 & 102.2 & 44.7 & 111.7 &    & 9.24$\pm$0.05 & $<4.40$ & $<-1.83$ & 9.0 & Yes \\
& J132948-180438 & 202.453 & -18.077 & 1344 & 19.3 & 116.7 & 144.1 &    & 7.83$\pm$0.05 & $<3.34$ & $<-5.12$ & 8.4 & Yes \\
& J133006-205341 & 202.527 & -20.895 & 5223 & 75.6 & 48.8 & 113.7 &    & 8.97$\pm$0.06 & $<4.60$ & $<-2.24$ & 7.7 & Yes \\
& J133008-203319 & 202.533 & -20.555 & 5836 & 84.6 & 17.9 & 83.9 &    & 9.06$\pm$0.08 & $<6.66$ & $<-3.29$ & 6.3 & Yes \\
& J133057-211755 & 202.738 & -21.299 & 5398 & 78.1 & 17.7 & 33.4 &    & 8.74$\pm$0.06 & $<4.82$ & $<-2.75$ & 7.5 & Yes \\
& J133747-175606 & 204.449 & -17.935 & 1386 & 19.9 & 54.2 & 69.7 &    & 7.79$\pm$0.05 &   &   & 9.7 & Yes \\
        \hline
    \end{tabular}
\end{table*}

Figure \ref{fig:loc} shows the location of the LSBGs and dark H\,{\sc i} sources with respect to the rest of the WALLABY detections in the three fields. The Hydra cluster and virial radius \citep[$r_{200}=1.44$ Mpc projected onto the sky at 61 Mpc;][]{Reiprich:02,Reynolds:22} and NGC 5044 and $r_{500}$ overdensity radius \citep[$r_{500}=620$ kpc projected onto the sky at 33 Mpc;][]{Osmond:04} are also shown. There does not appear to be a preferential distribution of the dark H\,{\sc i} sources within each field. Most have been detected in the NGC 5044 field, as this field covers a larger area than the other two  {(NGC 5044 covers $\sim 120$ deg$^2$, Hydra covers $\sim 60$ deg$^2$ and NGC 4808 covers $\sim 30$ deg$^2$)}.

\subsection{Global Galaxy Properties}
\label{sec:scal_rel}

\begin{figure}
     \centering
         \includegraphics[width=0.45\textwidth]{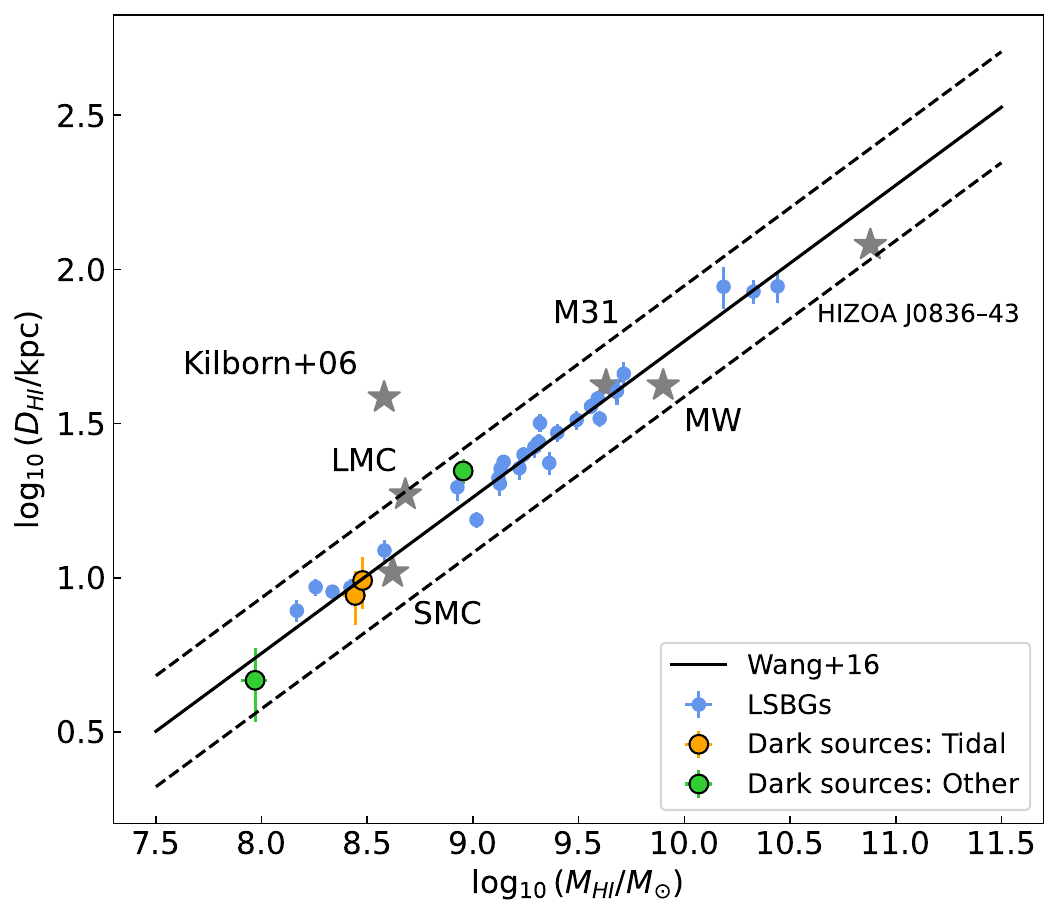}
         \label{fig:sizemass_LSB}
    \caption{The H\,{\sc i} size-mass relation for the well resolved LSBGs (blue), dark sources with tidal features (orange) and the rest of the dark sources (green). The H\,{\sc i} size ($D_{\rm HI})$ is the size of the semi-major axis at 1 M$_{\odot}$ pc$^{-2}$. The relation and 3$\sigma$ scatter from Wang et al. (2016) is shown by the solid and dotted lines respectively.}
    \label{fig:sizemass}
\end{figure}

Figure~\ref{fig:sizemass} shows the H\,{\sc i} size-mass relation with the resolved WALLABY LSBGs and dark H\,{\sc i} sources. The H\,{\sc i} size is defined as the major axis of the 1 M$_{\odot}$ pc$^{-2}$ isodensity contour. As discussed in Section~\ref{sec:HI_prop}, the H\,{\sc i} size was only able to be measured for the $\sim 9.2$~per~cent of sources that were sufficiently well resolved. For comparison, the best-fitting relation and $3\sigma$ scatter for the galaxies studied by \cite{Wang:16} are shown. The relationship extends over a large range of H\,{\sc i} masses, from $\sim10^{5.5}$ M$_{\odot}$ all the way to $\sim10^{11}$ M$_{\odot}$. Notable galaxies have been marked by star symbols. The sample studied by \cite{Wang:16} contains galaxies across a range of morphologies and environments, and yet all lie on the same $D_{\rm HI}-M_{HI}$ relation with a small $1\sigma$ scatter of $\sim0.06$ dex. Some (almost) dark H\,{\sc i} sources are known to lie above this relation, such as the dark cloud found in \cite{Kilborn:06}. The WALLABY LSBGs and dark H\,{\sc i} sources follow this relation remarkably well. The $1\sigma$ scatter of the LSBGs and dark H\,{\sc i} detections are only 0.04 dex and 0.08 dex respectively. Hence, despite having extreme optical properties, our well-resolved LSBGs and strong dark H\,{\sc i} sources appear to have typical H\,{\sc i} sizes. We emphasise here that whilst the well-resolved dark sources are consistent with this relation, we are still yet to confirm whether they are genuine H\,{\sc i} detections. The tightness of the H\,{\sc i} size-mass correlation is thought to arise from a constant average H\,{\sc i} surface density across different galaxy morphologies \citep{Broeils:97}, and environmental stripping processes, such as ram pressure stripping, that cause gas disc truncation have been shown not to impact this relation \citep{Stevens:19}. The H\,{\sc i} surface density has been shown to be regulated by the conversion of H\,{\sc i} to molecular hydrogen and star formation, however \cite{Wang:16} find that these two factors cannot be the only or major drivers of the H\,{\sc i} size-mass relation.  This is consistent with our LSBGs and dark sources following the relation, despite their lack of significant star formation, to which we turn our attention to next.

\begin{figure*}[h!]
     \centering
  
     \begin{subfigure}[b]{0.48\textwidth}
         \centering
         \includegraphics[width=\textwidth]{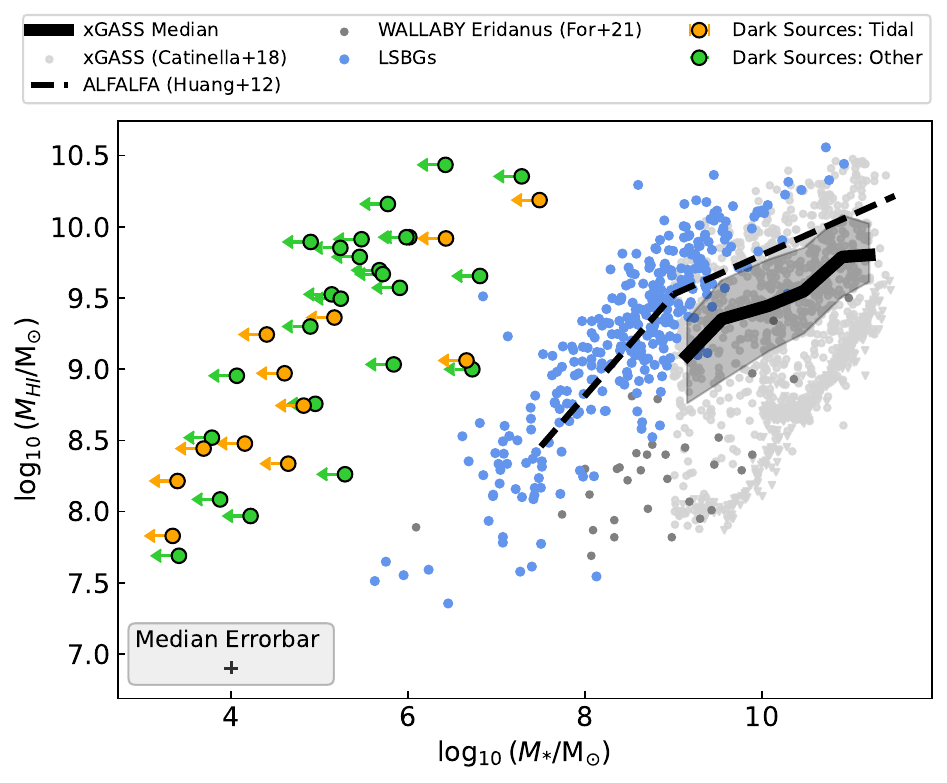}
         \caption{ }
         \label{fig:mhimstar}
     \end{subfigure}
     \hfill
     \begin{subfigure}[b]{0.465\textwidth}
         \centering
         \includegraphics[width=\textwidth]{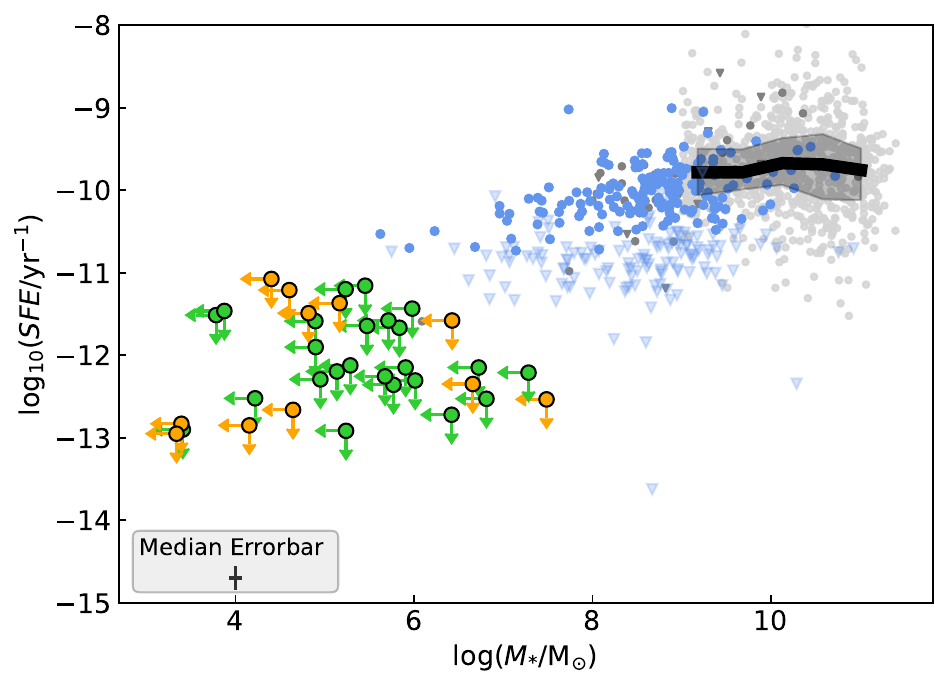}
         \caption{ }
         \label{fig:mhiSFE}
     \end{subfigure}
     \hfill

     \begin{subfigure}[b]{0.465\textwidth}
         \centering
         \includegraphics[width=\textwidth]{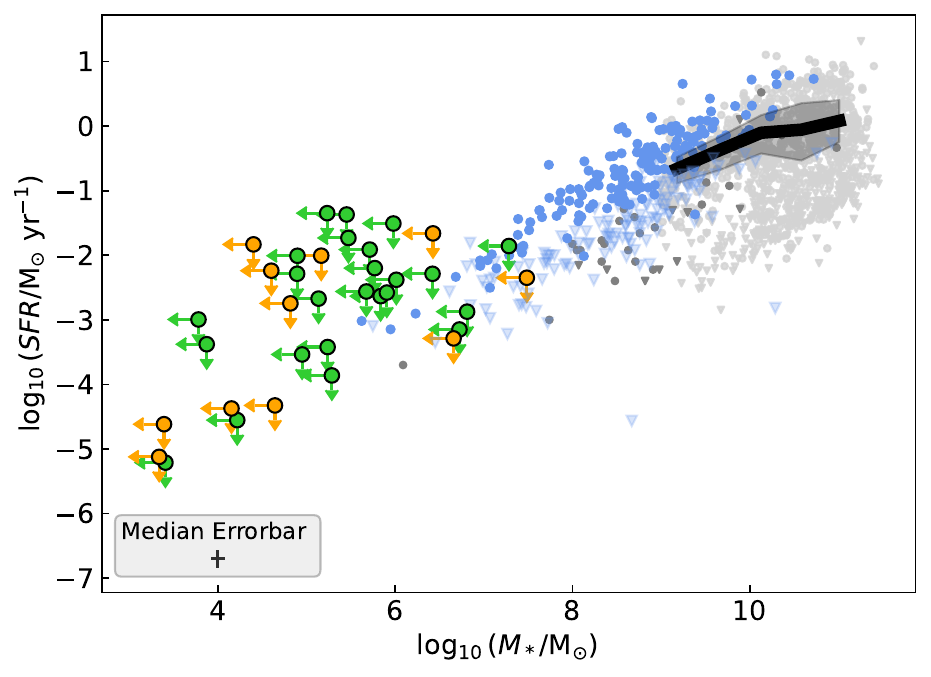}
         \caption{ }
         \label{fig:mstar_sfr}
     \end{subfigure}
     \hfill
     \begin{subfigure}[b]{0.465\textwidth}
         \centering
         \includegraphics[width=\textwidth]{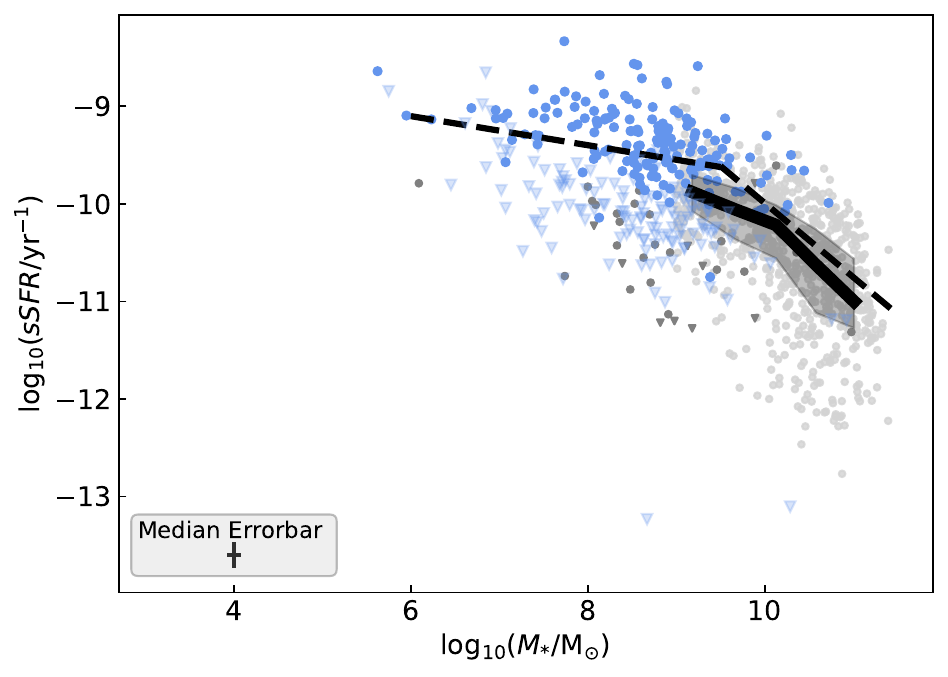}
         \caption{ }
         \label{fig:mstar_ssfr}
     \end{subfigure}

    \caption{Gas and stellar property comparisons with the LSBGs (blue), dark sources with tidal features (orange) and the rest of the dark sources (green). The xGASS galaxies are shown in light grey, with the rolling median and interquartile range given by the black line and shaded region. The scaling relations of the ALFALFA galaxies are shown by the dashed lines. The WALLABY Eridanus galaxies are shown in dark grey. Upper limits are denoted by triangular symbols or arrows. The plots show (a) the H\,{\sc i} mass ($M_{HI}$) against the stellar mass ($M_*$), (b) the star formation efficiency (SFE) against $M_*$ (c) the star formation rate (SFR) against $M_*$, and (d) the specific star formation rate (sSFR) against $M_*$.}
    \label{fig:mhi mstar sfr}
\end{figure*}

Figure~\ref{fig:mhi mstar sfr} presents H\,{\sc i} mass, stellar mass, SFR and star formation efficiency (SFE) scaling relations. We use the GALEX Arecibo SDSS Survey \citep[xGASS;][]{Catinella:18} galaxies as our control sample to compare with our LSBG and dark H\,{\sc i} sources. xGASS is a gas fraction- and volume-limited H\,{\sc i} survey of galaxies selected by stellar mass and redshift, minimising the effect of detection limit bias. The xGASS galaxies are shown by light grey markers, and the rolling median (excluding non-detections) and interquartile range are shown by the black line and shaded region. The dashed lines show the scaling relations from ALFALFA \citep{Huang:12}, another H\,{\sc i} selected sample, for comparison. Additionally, we compare the WALLABY pre-pilot observation Eridanus galaxies from \cite{For:21}, shown by dark grey markers. This sample consists of 55 H\,{\sc i} detections, 43 of which are part of the Eridanus supergroup. It has a large fraction of H\,{\sc i} deficient galaxies and their distorted H\,{\sc i} morphologies suggest the presence of ongoing tidal interactions \citep{Wang:22}. The LSBGs in our sample are shown in blue, with the median errorbars shown in the bottom left of the plot. The upper limits of the dark H\,{\sc i} sources with tidal signatures and the other dark H\,{\sc i} sources are denoted by the orange and green markers respectively.

Figure~\ref{fig:mhimstar} shows the stellar mass plotted against the H\,{\sc i} mass. Our LSBGs have large H\,{\sc i} masses compared to the xGASS sample within the overlapping stellar mass range. This is a combination of selection effects: our sample of LSBGs are, by definition, low surface brightness and so are expected to have low stellar mass, while at the same time have enough H\,{\sc i} to have been detected in WALLABY.  {The LSBGs are not as gas-deficient as the Eridanus sources, and} tend to follow the trend of the ALFALFA galaxies for $\log_{10}\left(\frac{M_*}{\rm M_{\odot}}\right) < 9$, while those with $\log_{10}\left(\frac{M_*}{\rm M_{\odot}}\right) > 9$ seem to have higher H\,{\sc i} masses. There are, however, several LSBGs with $\log_{10}\left(\frac{M_*}{\rm M_{\odot}}\right) < 9$ that deviate from the ALFALFA scaling relation towards the parameter space occupied by the  upper limits of the dark H\,{\sc i} sources.  The stellar masses of the LSBGs extend over a large range due to the varied nature of the sample, spanning from dwarfs all the way to large diffuse galaxies. There are 15 LSBGs with very high stellar masses ( $\log_{10}\left(\frac{M_*}{\rm M_{\odot}}\right) > 10$). They are all fairly distant sources at redshifts between 0.034 and 0.078, with large effective radii ($7.4 < R_e < 22.1 $ kpc ). 12 of the 15 show regular rotation in their moment 1 maps.  

Figure~\ref{fig:mhiSFE} presents the SFE (the ratio of the SFR and the H\,{\sc i} mass) as a function of stellar mass. The SFEs of the LSBGs tend to lie below the xGASS median, and the distribution is relatively flat across the stellar mass range  \citep[which is consistent with the findings of ][]{Wong:16}. Figures~\ref{fig:mstar_sfr} and \ref{fig:mstar_ssfr} show the stellar mass against the SFR and specific SFR (sSFR; the ratio of the SFR and the stellar mass). The LSBGs tend to have lower stellar masses than the xGASS galaxies, but within the range of stellar masses spanned by the xGASS sample, most of the WALLABY LSBGs have SFRs that lie above the xGASS median. The sSFR of the LSBGs tend to be larger than those of xGASS across all stellar masses, and follow the trend of the ALFALFA galaxies. Altogether, Figure~\ref{fig:mhi mstar sfr} illustrates that the WALLABY LSBGs are a distinct population from the xGASS galaxies by selection, and that most, but not all, tend to follow the trend of (H\,{\sc i} selected) ALFALFA galaxies. The stellar mass and SFR of the LSBGs is an extention of the high surface brightness population. However, once the H\,{\sc i} is taken into consideration, the completely different mode of galaxy evolution becomes clear, with the large reservoirs of gas used sparingly for their mass.  The stellar mass, SFE and SFR upper limits of the dark H\,{\sc i} sources highlight the parameter space that is limited by the depth of our current surveys.


\section{Discussion}
\label{sec:discussion}

\subsection{Low Surface Brightness Galaxies}
\label{sec:disc-lsb}

We find that LSBGs make up a significant proportion of the gas-rich galaxy population, with $17$~per~cent of the 1829 WALLABY detections used in this study having low surface brightness optical counterparts. Even this proportion may underestimate the true number of LSBGs due to sensitivity limitations. Using the cosmological hydrodynamical simulation Horizon-AGN \citep{Dubois:14}, \cite{Martin:19} predicted that LSBGs (mean $r$-band surface brightness within 1 $R_e$ that is $ > 23$ mag arcsec$^{-2}$) are expected to make up $47$~per~cent of galaxies with $M_{*}>10^8$ M$_{\odot}$, and $85$~per~cent of galaxies with $M_{*}>10^7$ M$_{\odot}$.  Of the 315 LSBGs that we identified, $75$~per~cent were not catalogued in NED (within 10 arcsec of the optical source centres). This highlights both the power of H\,{\sc i} surveys like WALLABY for identifying  {gas-rich} low surface brightness sources  {that may be missed by photometric studies}, and the lack of deep optical imaging in the southern hemisphere.  Combined with new deep optical surveys such as the Legacy Survey, we are finally entering an era where these extreme sources can be studied in large numbers for the first time, and foreshadows what could be achieved with LSST.

 {There were no matches between the LSBGs and UDGs in the Hydra cluster catalogued by \cite{LaMarca:22} and the WALLABY LSBGs. This does not necessarily imply that the cluster LSBGs and UDGs are completely devoid of H{\sc i}, but rather that their H{\sc i} masses may lie below the WALLABY detection threshold \citep[WALLABY has a $5\sigma$ H{\sc i} mass sensitivity of $\sim5.5\times10^8(D/100$ Mpc$)^2$ M$_{\odot}$ for point sources,][]{Murugeshan:24}. \cite{For:23} had similar results searching for H{\sc i}-detections of UDGs candidates from the Systematically Measuring Ultra-diffuse Galaxies survey \citep[SMUDGes;][]{Zaritsky:22} in the Eridanus supergroup using the WALLABY pre-pilot data. They found 6 UDGs that were undetected in H{\sc i} and only one H{\sc i}-bearing low surface brightness dwarf galaxy.}

In addition to calculating the SFR of the LSBGs from the GALEX UV emission, we also investigated the infrared emission from WISE.  Of the 315 LSBGs, 190 were detected in GALEX FUV, while only 19 had a W3 detection in WISE. Of all the LSBGs that were detected in W3, none had a sufficiently high W3 luminosity to calculate a SFR after subtraction of the stellar continuum using the W1 luminosity correction.  This lack of dust, even in sources with UV-detected star formation, suggests that the interstellar medium conditions of low surface brightness sources are relatively dust-poor compared to more massive and luminous galaxies (up to the sensitivity limits of the WISE data). 

Our sample of LSBGs spans a large range in stellar masses ($5\times10^5$ M$_{\odot}$ $<M_*<1\times10^{11}$ M$_{\odot}$) and SFR ($9\times10^{-5}$ M$_{\odot}$ yr$^{-1}$ $<$SFR$<6$ M$_{\odot}$ yr$^{-1}$). This shows that the LSBGs exhibit considerable diversity, from irregular dwarfs to massive spirals. This large sample of LSBGs with significant H\,{\sc i} content in a range of environments suggests that a low surface brightness cannot always be the result of a loss of gas and early quenching, as has been suggested for cluster ultra-diffuse galaxies \citep{VanDokkum:15}. Internal mechanisms that suppress significant star formation in these galaxies must also exist, such as supernova feedback \citep[e.g.][]{DiCintio:17}, and high spin \citep[e.g.][]{Leisman:17} which has been directly linked to galaxies with high gas fractions \citep{ManceraPina:21b}.

\subsection{Dark H\,{\sc i} Sources}
\label{sec:disc-dgc}

In this work, we aim to investigate the nature of the dark H\,{\sc i} sources in the WALLABY pilot catalogues. While they both lack optical counterparts, it is useful to differentiate between  {isolated dark sources} and dark tidal clouds, as they have different formation mechanisms and consequently different properties.  {There is a possibility that a small number of the isolated dark sources could be higher H\,{\sc i} mass counterparts of primordial dark galaxies.} Using the IllustrisTNG cosmological hydrodynamical simulations, \cite{Lee:24} find that in the early Universe, dark galaxies initially tend to form in less dense regions. Their star formation is suppressed by heating from cosmic reionisation and a lack of mergers and interactions. They are predicted to form in dark matter haloes with high spin parameters \citep{Jimenez:20} and be stable to the effects of harassment \citep{Taylor:16}.  { REionization-Limited H\,{\sc i} Clouds \citep[RELHIC;][]{Benitez-llambay:17}  are a type of dark galaxy predicted by simulations. They are starless, low mass dark matter haloes that host gas which is almost completely ionised but with small (H\,{\sc i} size $<1$~kpc ), round neutral cores.  The gas is in hydrostatic equilibrium with the gravitational potential of the dark matter halo and in thermal equilibrium with the ionizing UV background. Some observed ultra compact high velocity clouds \citep[UCHVC;][]{Adams:13} are consistent with RELHICs.} On the other hand, dark tidal clouds lack dark matter as they are formed from stripped gas in galaxy interactions \citep[e.g.][]{Duc:08}  {and exist over shorter timescales}. 13 of the strong H\,{\sc i} detections have signatures that suggest they are tidal debris or dark tidal dwarf galaxies (as discussed in Section \ref{sec:source_ident}).  {The other 25 dark sources are either artefacts of an unknown nature or extreme LSBGs (with optical counterparts beyond the limits of the Legacy Survey). These dark galaxy candidates} that we have identified are not consistent with UCHVCs or RELHICs as their H\,{\sc i} sizes and velocity widths are too large.

\begin{figure*}
     \centering
       \begin{subfigure}[b]{0.49\textwidth}
         \centering
         \includegraphics[width=\textwidth]{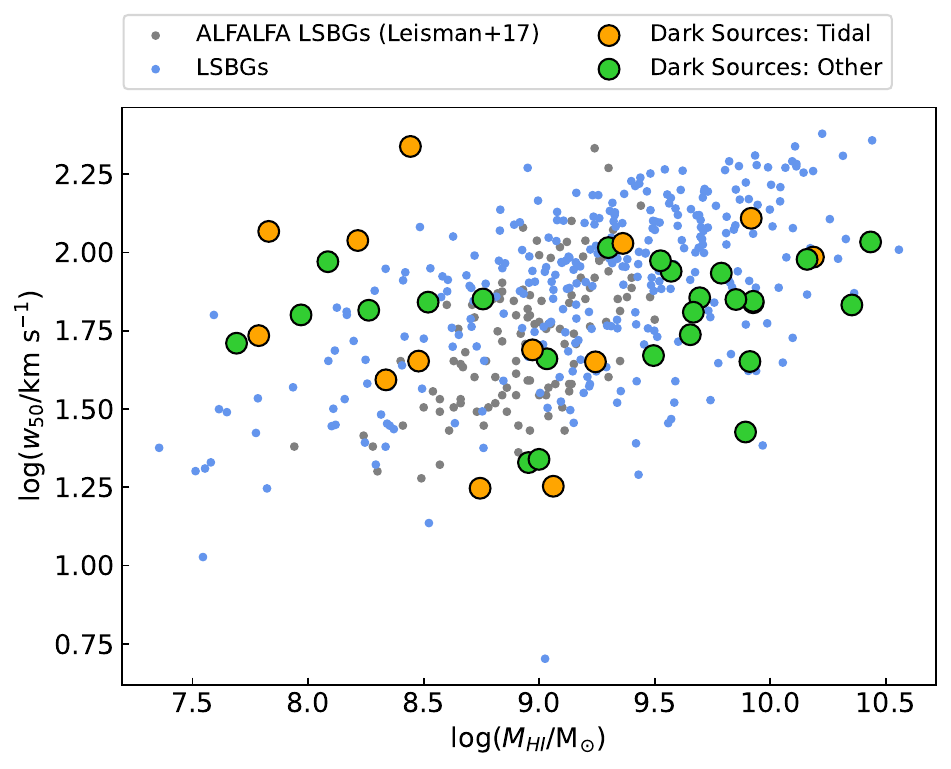}
         \caption{ }
         \label{fig:w50mhi}
     \end{subfigure}
     \begin{subfigure}[b]{0.49\textwidth}
         \centering
         \includegraphics[width=\textwidth]{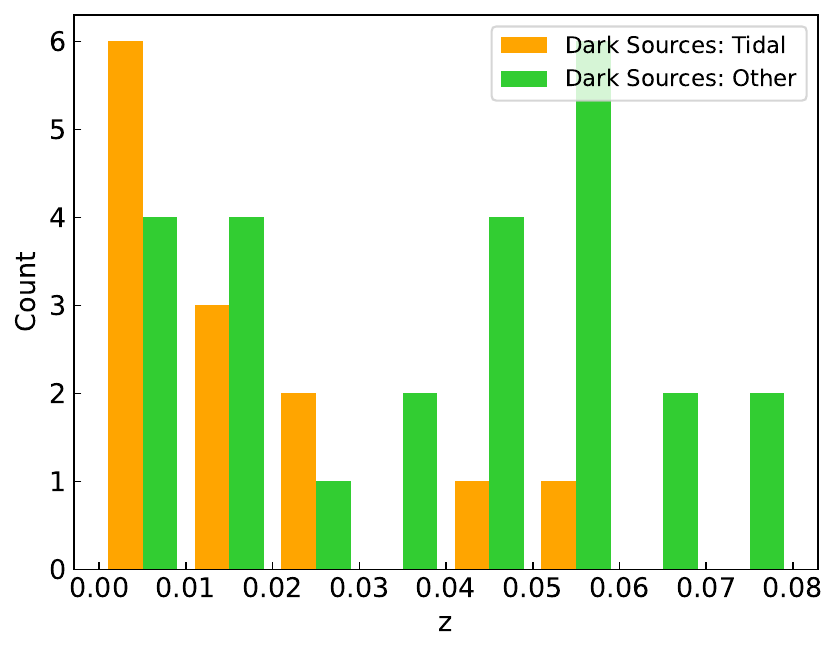}
         \caption{ }
         \label{fig:z}
     \end{subfigure}
    \caption{ Panel (a) shows the H\,{\sc i} mass against $w_{50}$ emission line width for the WALLABY LSBGs (blue), ALFALFA almost dark sources (grey),  dark tidal sources (orange) and other  {candidate} dark sources (green). Panel (b) presents the histogram of the redshifts of the dark tidal sources and the other dark sources. }
    \label{fig:disc}
\end{figure*}

Figure \ref{fig:w50mhi} presents the $w_{50}$ emission line width against the H\,{\sc i} mass for the LSBGs, dark tidal sources and the other  {candidate} dark sources. Additionally, the almost dark galaxies from the ALFALFA survey \citep{Leisman:17} are shown for comparison. The dark sources span the same parameter space as the LSBGs, with the dark tidal sources preferentially distributed in the lower H\,{\sc i} mass region. The ALFALFA almost dark sources span the same range of $w_{50}$ values and have a narrower H\,{\sc i} mass range. Figure \ref{fig:z} highlights that we are only able to detect dark tidal cloud candidates at lower redshifts, suggesting that at higher redshifts they may be subject to source confusion, or be too low mass to be detectable.   Figure~\ref{fig:mhimstar} highlights that all the  {candidate} dark H\,{\sc i} sources,  {assuming they are genuine detections},  {would have to} have considerable H\,{\sc i} mass  {($>4.9\times10^7$ M$_{\odot}$)}, despite their low stellar mass upper limits. All of the  {candidates would} have $\frac{M_{HI}}{M_{*}}>187$, and $89$~per~cent of the dark sources have $\frac{M_{HI}}{M_{*}}>500$. The sensitivities of these new optical and H\,{\sc i} surveys are pushing us to consider the limits of what we define as a galaxy.

Unfortunately, none of the  {candidate} dark sources were sufficiently well resolved both spatially and spectrally to have kinematic models generated from the WALLABY Kinematic Analysis Proto-Pipeline \citep[WKAPP;][]{Deg:22,Murugeshan:24}, and consequently meaningful dynamical mass estimates could not be made. Higher resolution observations of the dark  {candidates, such as with MeerKAT, would allow us to determine which are artefacts of an unknown nature, which are tidal debris of unknown origin, and which may be `failed' galaxies.} Many models and simulations predict dark galaxies with halo masses of order $<10^9$ M$_{\odot}$ \citep[e.g.][]{Benitez-llambay:17,Jimenez:20,Benitez-Llambay:20,Lee:24}. In contrast, 17 of the 25 strong dark H\,{\sc i} detections without tidal features have H\,{\sc i} masses $>10^{9}$ M$_{\odot}$. If these dark galaxy candidates are revealed to be genuine detections by follow-up H\,{\sc i} observations, this indicates a significant gap in our current understanding of galaxy formation. On the other hand, deeper optical observations could reveal faint stellar counterparts to many of the high-mass candidates.  {In fact, although no optical counterpart can be easily seen in the $g$-band image of WALLABY J131244-155218 (Figure \ref{fig:faint}), an extremely faint optical counterpart is just visible in the co-added image of the $g$, $r$, $i$ and $z$-bands (convolved with a boxcar kernel with a size of 2.6 arcsec by 2.6 arcsec), as shown in Figure \ref{fig:coadd}.}  {Co-added images of all the dark sources are presented in Appendix \ref{append:coadd}.}

\begin{figure}
    \centering
    \includegraphics[width=\linewidth]{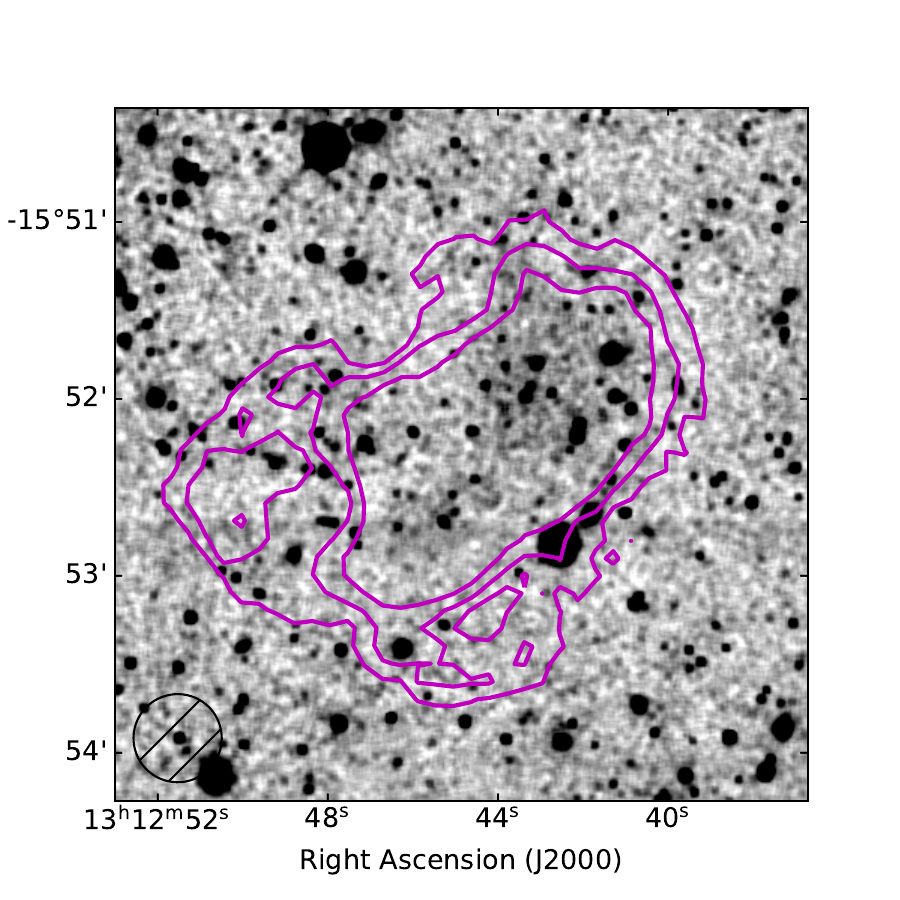}
    \caption{Co-added $g$, $r$, $i$ and $z$-band image convolved with a boxcar kernel of WALLABY J131244-155218. H\,{\sc i} contours ([0.1,0.5,1]$\times 10^{20}$ cm$^{-2}$) are overlaid and the WALLABY beam is shown in the lower left corner. Although this appears to be a dark source from the $g$-band image, a faint optical counterpart is visible in the co-added image.}
    \label{fig:coadd}
\end{figure}

Very few of the ALFALFA H\,{\sc i} sources were found to be dark galaxy candidates \citep[][]{Haynes:11,Janowiecki:15,Cannon:15}. While we are yet to confirm whether all of the dark WALLABY H\,{\sc i} sources are genuine detections, the discrepancy between the number of dark sources detected by the two surveys could arise from the different survey properties. While ALFALFA and WALLABY have similar sensitivities, the significantly better angular resolution of WALLABY means that we are better able to localise the H\,{\sc i} emission and thus reduce source confusion.  {The 55 dark WALLABY detections are likely to be a mix of artefacts (especially the uncertain detections shown in Appendix \ref{append:weak}), tidal debris, extreme LSBGs and dark galaxy candidates. }


\section{Summary and Conclusions}
\label{sec:conclusion}

We have presented 315 LSBGs identified in the WALLABY pilot data. To measure the photometry of the optical and ultraviolet observations, we use the Python package \textsc{AstroPhot} to fit Sérsic models to the galaxies.  {These fits are done consistently across multiple bands and are designed to give consistent, SNR-optimised photometry.} The LSBGs are defined by a mean $g$-band surface brightnesses within 1 $R_e$ fainter than 23 mag arcsec$^{-2}$, and the faintest LSBG in our sample has a mean $g$-band surface brightness within 1 $R_e$ of 25.7 mag arcsec$^{-2}$. All of our dark H\,{\sc i} sources and $75$~per~cent of our LSBGs had not been catalogued prior to WALLABY (within 10 arcsec of the optical source centres), highlighting both the extreme nature of these sources and the lack of multiwavelength coverage in the southern hemisphere.

In addition to the LSBGs, we find 55 H\,{\sc i}  {detections} without optical counterparts  {in the deepest observations available}. We investigate the nature of  {these candidate} dark H\,{\sc i} sources in the WALLABY pilot catalogues.  We assess their reliability, and identify 38 to be strong  {candidates}. Of these, we find that 13 show signatures of tidal remnants, while the other 25  {are isolated, so could be fainter LSBGs, genuinely dark galaxies, or artefacts of an unknown nature.}  A large proportion of the non-tidal dark sources have large H\,{\sc i} masses  {(if they are genuine detections)}, with 68 per~cent having H\,{\sc i} masses $> 10^9$ M$_{\odot}$. This is in conflict with simulations, which predict an abundance of lower mass dark galaxies.  {If the dark sources are revealed to be genuine dark galaxy candidates} by follow-up H\,{\sc i} observations, this indicates a significant gap in our current understanding of galaxy formation. On the other hand, deeper optical observations could reveal faint stellar counterparts to many of the high mass candidates. While we are yet to confirm whether all of the dark WALLABY sources are genuine detections, the discrepancy between the number of dark sources detected by ALFALFA and WALLABY may arise from the different survey properties. Although ALFALFA and WALLABY have similar sensitivities, the significantly better angular resolution of WALLABY means that we are better able to localise the H\,{\sc i} emission and thus reduce source confusion.

We use scaling relations to study the global galaxy properties of our dark and low surface brightness sample. Both the WALLABY LSBGs and dark sources (that are sufficiently well resolved) follow the H\,{\sc i} size-mass relation remarkably well. Hence, despite having extreme optical properties, our LSBGs and dark H\,{\sc i} sources do in fact have typical H\,{\sc i}  {galaxy} properties. The WALLABY LSBGs have high H\,{\sc i} masses for their stellar masses when compared with the xGASS galaxies due to selection effects. On the other hand, they do have similar H\,{\sc i} masses to the (H\,{\sc i}-selected) ALFALFA sample for $\log_{10}\left(\frac{M_*}{\rm M_{\odot}}\right) < 9$, while those with $\log_{10}\left(\frac{M_*}{\rm M_{\odot}}\right) > 9$ seem to have higher H\,{\sc i} masses. The sSFRs of the LSBGs follow the trend of the ALFALFA galaxies across all stellar masses, and tend to be larger than those of the xGASS galaxies. We find that the WALLABY LSBGs have low SFEs, and have stellar masses spanning five orders of magnitude, which highlights the varied morphologies across our sample, ranging from tiny dwarf galaxies to large ultra-diffuse galaxies.  The stellar mass and star formation rate upper limits of the dark sources illustrate the unexplored parameter space that is limited by the sensitivity of current surveys.

Assuming that the non-tidal dark sources are not preferentially distributed with respect to the environment, across the 1.4$\pi$ steradians to be covered by the full WALLABY survey we can expect to detect $\sim570$ isolated dark sources. We have highlighted the challenges that the full survey will face with respect to distinguishing true dark sources from false positive detections. To confirm the reliability of our WALLABY dark sources, follow-up H\,{\sc i} observations with a high-resolution and high-sensitivity instrument, such as MeerKAT  {or the upcoming Square Kilometre Array}, is required. Additionally, deep optical imaging could help push down to lower surface brightness levels to reveal whether the dark sources are failed galaxies with very little stellar content, or perhaps none at all. Our understanding of galaxy formation and evolution is coupled to the galaxies that are visible  {(in the optical wavelengths)} in past and current surveys. This work provides a glimpse into the future of studying the H\,{\sc i}-rich optically faint Universe and the potential for interplay between radio and multiwavelength observations in the upcoming Square Kilometre Array era.


\section*{Acknowledgements}





 {We are grateful to the anonymous referee for their useful feedback and suggestions.} We thank Barbara Catinella for her contributions that have improved this manuscript.

This scientific work uses data obtained from Inyarrimanha Ilgari Bundara / the Murchison Radio-astronomy Observatory. We acknowledge the Wajarri Yamaji People as the Traditional Owners and native title holders of the Observatory site. CSIRO’s ASKAP radio telescope is part of the Australia Telescope National Facility (https://ror.org/05qajvd42). Operation of ASKAP is funded by the Australian Government with support from the National Collaborative Research Infrastructure Strategy. ASKAP uses the resources of the Pawsey Supercomputing Research Centre. Establishment of ASKAP, Inyarrimanha Ilgari Bundara, the CSIRO Murchison Radio-astronomy Observatory and the Pawsey Supercomputing Research Centre are initiatives of the Australian Government, with support from the Government of Western Australia and the Science and Industry Endowment Fund.

WALLABY acknowledges technical support from the Australian SKA Regional Centre (AusSRC).

Parts of this research were supported by the Australian Research Council Centre of Excellence for All Sky Astrophysics in 3 Dimensions (ASTRO 3D), through project number CE170100013. 

This investigation has made use of the NASA/IPAC Extragalactic Database (NED) which is operated by the Jet Propulsion Laboratory, California Institute of Technology, under contract with the National Aeronautics and Space Administration, and NASA's Astrophysics Data System.

The Legacy Surveys consist of three individual and complementary projects: the Dark Energy Camera Legacy Survey (DECaLS; Proposal ID \#2014B-0404; PIs: David Schlegel and Arjun Dey), the Beijing-Arizona Sky Survey (BASS; NOAO Prop. ID \#2015A-0801; PIs: Zhou Xu and Xiaohui Fan), and the Mayall z-band Legacy Survey (MzLS; Prop. ID \#2016A-0453; PI: Arjun Dey). DECaLS, BASS and MzLS together include data obtained, respectively, at the Blanco telescope, Cerro Tololo Inter-American Observatory, NSF’s NOIRLab; the Bok telescope, Steward Observatory, University of Arizona; and the Mayall telescope, Kitt Peak National Observatory, NOIRLab. Pipeline processing and analyses of the data were supported by NOIRLab and the Lawrence Berkeley National Laboratory (LBNL). The Legacy Surveys project is honored to be permitted to conduct astronomical research on Iolkam Du’ag (Kitt Peak), a mountain with particular significance to the Tohono O’odham Nation.

NOIRLab is operated by the Association of Universities for Research in Astronomy (AURA) under a cooperative agreement with the National Science Foundation. LBNL is managed by the Regents of the University of California under contract to the U.S. Department of Energy.

This project used data obtained with the Dark Energy Camera (DECam), which was constructed by the Dark Energy Survey (DES) collaboration. Funding for the DES Projects has been provided by the U.S. Department of Energy, the U.S. National Science Foundation, the Ministry of Science and Education of Spain, the Science and Technology Facilities Council of the United Kingdom, the Higher Education Funding Council for England, the National Center for Supercomputing Applications at the University of Illinois at Urbana-Champaign, the Kavli Institute of Cosmological Physics at the University of Chicago, Center for Cosmology and Astro-Particle Physics at the Ohio State University, the Mitchell Institute for Fundamental Physics and Astronomy at Texas A\&M University, Financiadora de Estudos e Projetos, Fundacao Carlos Chagas Filho de Amparo, Financiadora de Estudos e Projetos, Fundacao Carlos Chagas Filho de Amparo a Pesquisa do Estado do Rio de Janeiro, Conselho Nacional de Desenvolvimento Cientifico e Tecnologico and the Ministerio da Ciencia, Tecnologia e Inovacao, the Deutsche Forschungsgemeinschaft and the Collaborating Institutions in the Dark Energy Survey. The Collaborating Institutions are Argonne National Laboratory, the University of California at Santa Cruz, the University of Cambridge, Centro de Investigaciones Energeticas, Medioambientales y Tecnologicas-Madrid, the University of Chicago, University College London, the DES-Brazil Consortium, the University of Edinburgh, the Eidgenossische Technische Hochschule (ETH) Zurich, Fermi National Accelerator Laboratory, the University of Illinois at Urbana-Champaign, the Institut de Ciencies de l’Espai (IEEC/CSIC), the Institut de Fisica d'Altes Energies, Lawrence Berkeley National Laboratory, the Ludwig Maximilians Universitat Munchen and the associated Excellence Cluster Universe, the University of Michigan, NSF’s NOIRLab, the University of Nottingham, the Ohio State University, the University of Pennsylvania, the University of Portsmouth, SLAC National Accelerator Laboratory, Stanford University, the University of Sussex, and Texas A\&M University.

BASS is a key project of the Telescope Access Program (TAP), which has been funded by the National Astronomical Observatories of China, the Chinese Academy of Sciences (the Strategic Priority Research Program “The Emergence of Cosmological Structures” Grant \# XDB09000000), and the Special Fund for Astronomy from the Ministry of Finance. The BASS is also supported by the External Cooperation Program of Chinese Academy of Sciences (Grant \# 114A11KYSB20160057), and Chinese National Natural Science Foundation (Grant \# 12120101003, \# 11433005).

The Legacy Survey team makes use of data products from the Near-Earth Object Wide-field Infrared Survey Explorer (NEOWISE), which is a project of the Jet Propulsion Laboratory/California Institute of Technology. NEOWISE is funded by the National Aeronautics and Space Administration.

The Legacy Surveys imaging of the DESI footprint is supported by the Director, Office of Science, Office of High Energy Physics of the U.S. Department of Energy under Contract No. DE-AC02-05CH1123, by the National Energy Research Scientific Computing Center, a DOE Office of Science User Facility under the same contract; and by the U.S. National Science Foundation, Division of Astronomical Sciences under Contract No. AST-0950945 to NOAO.

KAO acknowledges support by the Royal Society through a Dorothy Hodgkin Fellowship (DHF/R1/231105).

\section*{Data Availability}
The WALLABY source catalogue and associated data products (e.g. cubelets, moment maps, integrated spectra, radial surface density profiles) are available online through the CSIRO ASKAP Science Data Archive (CASDA) and the Canadian Astronomy Data Centre (CADC). All source and kinematic model data products are mirrored at both locations. Links to the data access services and the software tools used to produce the data products as well as documented instructions and example scripts for accessing the data are available from the WALLABY Data Portal (\url{https://wallaby-survey.org/data/}).  {The photometric properties and additional H{\sc i} properties measured in this study are available on request.}

\bibliographystyle{apj}
\bibliography{example2}

\appendix

\section{Sersic Models}
\label{appen:sersic}

\begin{figure*}
	\centering
		\begin{subfigure}[b]{0.99\textwidth}
			\centering
			\includegraphics[width=\textwidth]{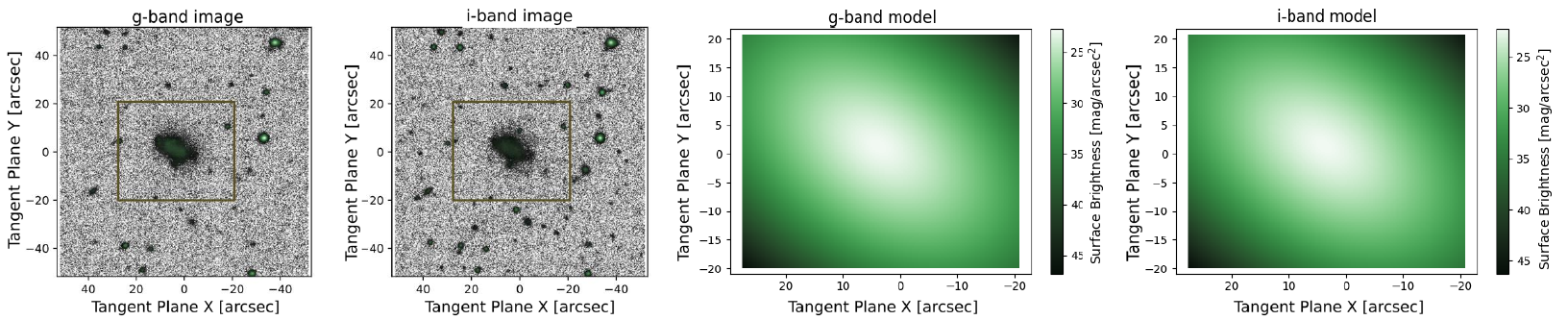}
			\caption{ }
		\end{subfigure}
		\begin{subfigure}[b]{0.99\textwidth}
			\centering
			\includegraphics[width=\textwidth]{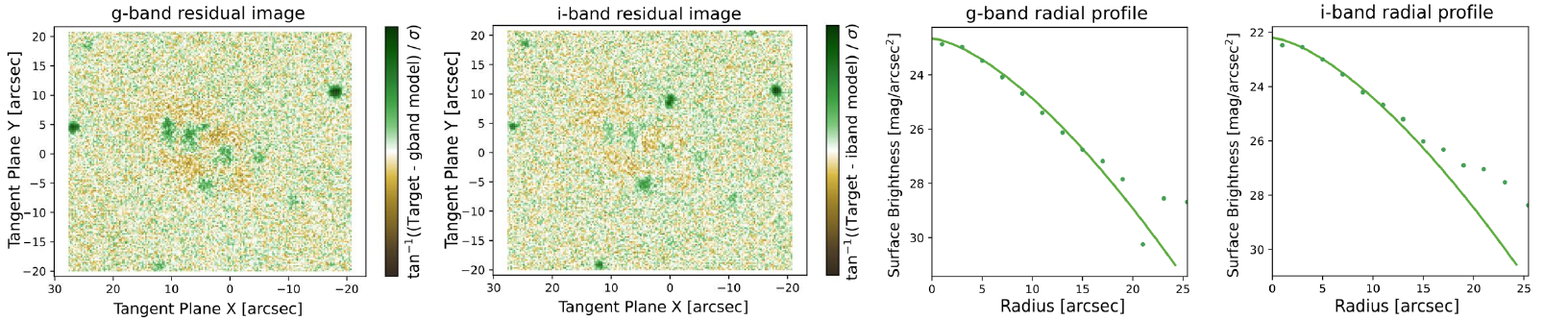}
			\caption{ }
		\end{subfigure}
	\caption{Sérsic fits to the g and i band Legacy Survey images of LSBG WALLABY J102113-262325. The left two panels of (a) are the optical images, with the box identifying the area to be modelled. The right two panels of (a) are the Sérsic models in each of the two bands. The left two panels of (b) are the residuals from the target image subtract the model. The right two panels are the radial surface brightness profiles. The points show the median of pixel values at a given radius of the image and the lines show the fitted models.}
    \label{fig:sersic1}
\end{figure*}

\begin{figure*}
	\centering
		\begin{subfigure}[b]{0.99\textwidth}
			\centering
			\includegraphics[width=\textwidth]{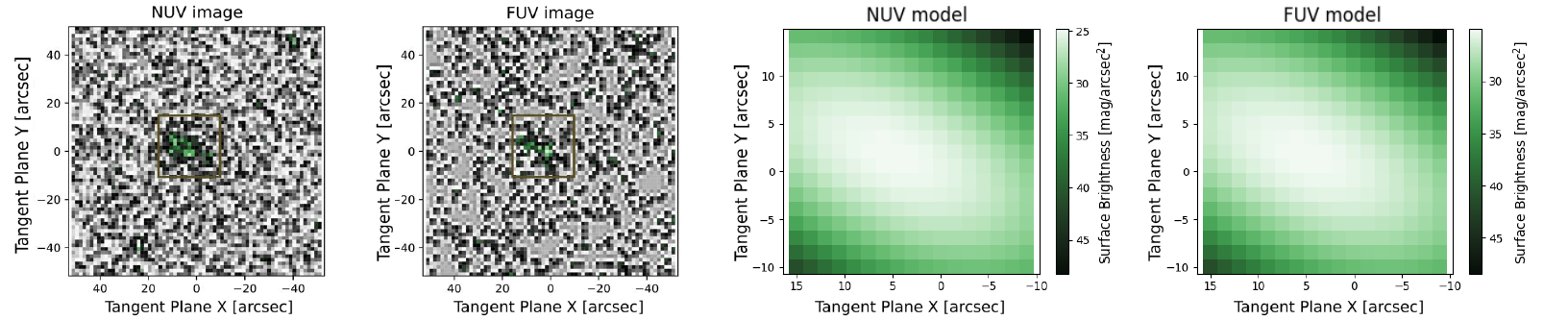}
			\caption{ }
		\end{subfigure}
		\begin{subfigure}[b]{0.99\textwidth}
			\centering
			\includegraphics[width=\textwidth]{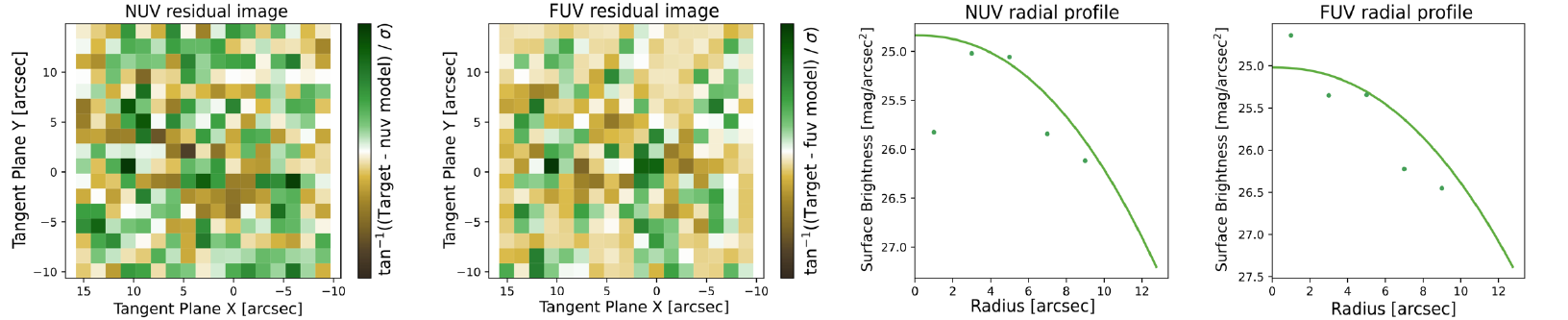}
			\caption{ }
		\end{subfigure}
	\caption{Sérsic fits to the NUV and FUV band GALEX images of LSBG WALLABY J102113-262325. The left two pannels of (a) are the UV images, with the box identifying the area to be modelled. The right two panels of (a) are the Sérsic models in each of the two bands. The left two panels of (b) are the residuals from the target image subtract the model. The right two panels are the radial surface brightness profiles. The points show the median of pixel values at a given radius of the image and the lines show the fitted models. }
    \label{fig:sersic2}
\end{figure*}

In this section, we use WALLABY J102113-262325 to illustrate an example of Sérsic fits to a LSBG. As discussed in Section~\ref{sec:sersic}, we use Sérsic models to measure the photometric properties of the galaxies. The g-band and i-band Legacy Survey images are modelled together and are shown in Figure~\ref{fig:sersic1}, and the NUV and FUV GALEX images are modelled together and are shown in Figure~\ref{fig:sersic2}. The parameters measured from the models are shown in Table~\ref{tab:sersic3}. From these properties, we estimated the mean g-band surface brightness within 1 effective radius to be 23.3 mag arcsec$^{-2}$, the stellar mass of this LSBG to be $3\times10^7$ M$_{\odot}$ and the SFR to be 0.013 M$_{\odot}$ yr$^{-1}$.

\begin{table}
    \centering
        \caption{Properties of the LSBG WALLABY J102113-262325 measured from the Sérsic models. The optical parameters are measured from the g and i band Legacy Survey images, and the UV parameters are measured from the NUV and FUV GALEX images. The properties presented in this table are: $m_{\lambda}$ the total apparent magnitude in the respective bands, $R_e$ the effective radius, $PA$ the position angle, $q$ the axis ratio, and $n$ the Sérsic index.}
        \label{tab:sersic3}
    \begin{tabular}{ccc}
    \hline
       Wavelength & Property  & Value \\
       \hline
       & $m_g$ & 17.8 mag \\
       & $m_i$ & 17.3 mag \\
      Optical & $R_e$ & 1.1 kpc \\
       & $PA$ & 124$^{\circ}$ \\
       & $q$ & 0.63 \\
       & $n$ & 0.67 \\
        \hline
        & $m_{NUV}$ & 19.6 mag \\
       & $m_{FUV}$ & 19.8 mag \\
     UV  & $R_e$ & 1.2 kpc \\
       & $PA$ & 115$^{\circ}$ \\
       & $q$ & 0.50 \\
       & $n$ & 0.45 \\
        \hline
    \end{tabular}
\end{table}


\section{Strong Dark Source Detections}
\label{append:strong}

\begin{figure*}
	\centering
		\begin{subfigure}[b]{0.24\textwidth}
			\centering
			\includegraphics[width=\textwidth]{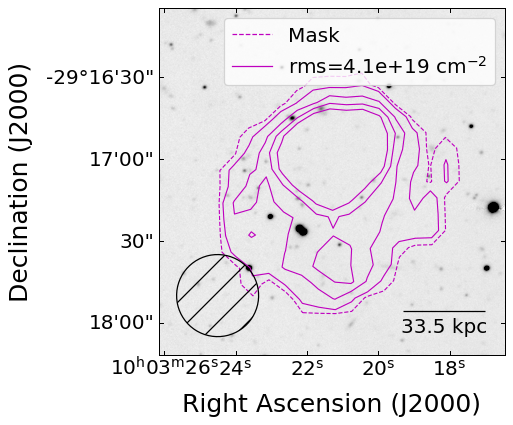}
			\caption{ }
		\end{subfigure}
		\begin{subfigure}[b]{0.24\textwidth}
			\centering
			\includegraphics[width=\textwidth]{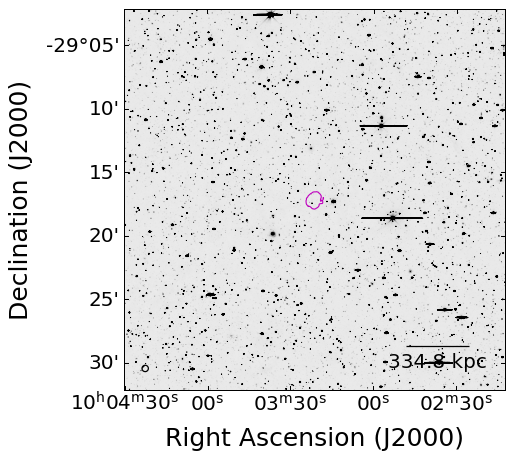}
			\caption{ }
		\end{subfigure}
		\begin{subfigure}[b]{0.26\textwidth}
			\centering
			\includegraphics[width=\textwidth]{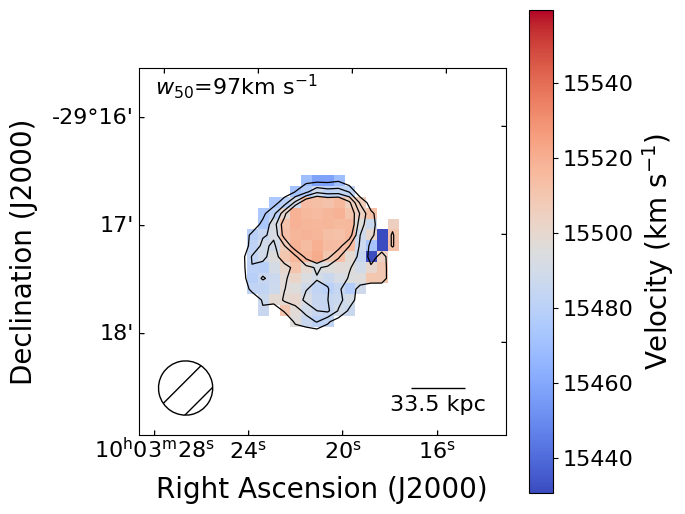}
			\caption{ }
		\end{subfigure}
  \begin{subfigure}[b]{0.24\textwidth}
			\centering
			\includegraphics[width=\textwidth]{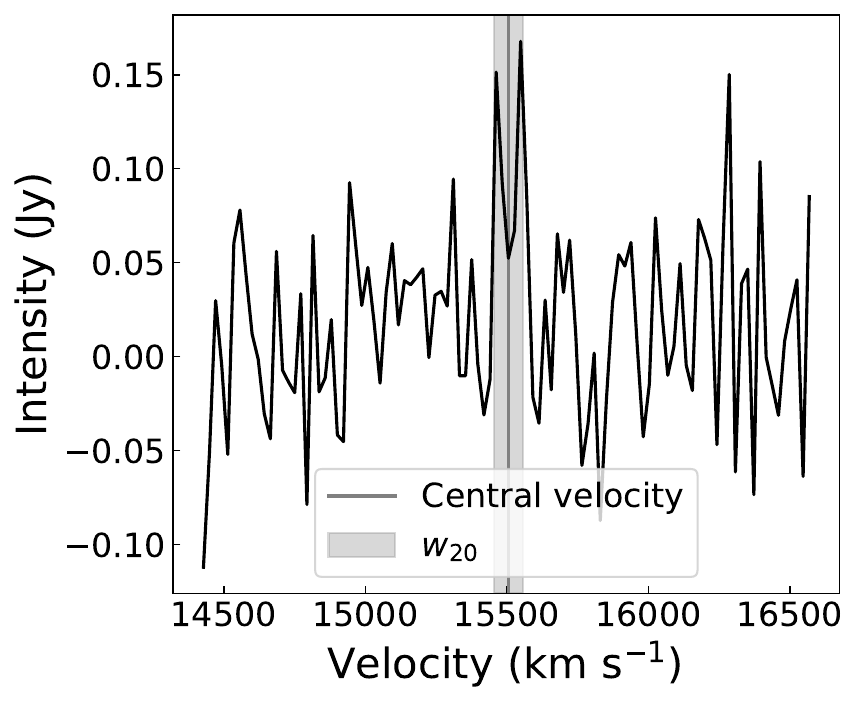}
			\caption{ }
		\end{subfigure}
	\caption{WALLABY J100321-291708 (tidal; Hydra)}
    \label{fig:dgc1}
\end{figure*}

\begin{figure*}
	\centering
		\begin{subfigure}[b]{0.24\textwidth}
			\centering
			\includegraphics[width=\textwidth]{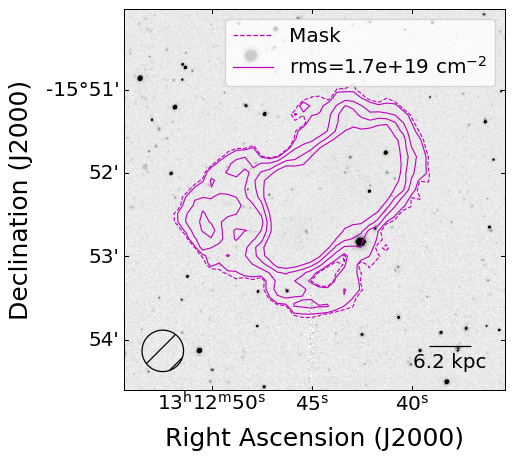}
			\caption{ }
		\end{subfigure}
		\begin{subfigure}[b]{0.24\textwidth}
			\centering
			\includegraphics[width=\textwidth]{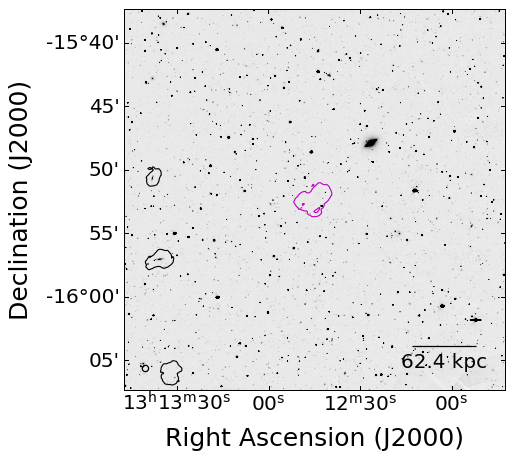}
			\caption{ }
		\end{subfigure}
		\begin{subfigure}[b]{0.26\textwidth}
			\centering
			\includegraphics[width=\textwidth]{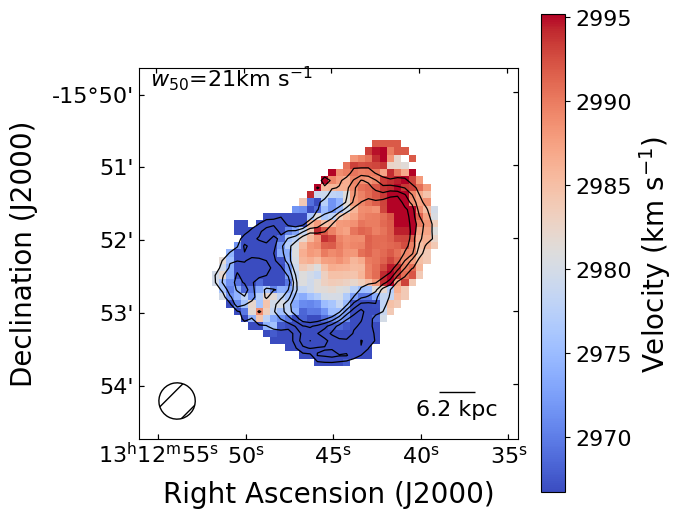}
			\caption{ }
		\end{subfigure}
    \begin{subfigure}[b]{0.24\textwidth}
			\centering
			\includegraphics[width=\textwidth]{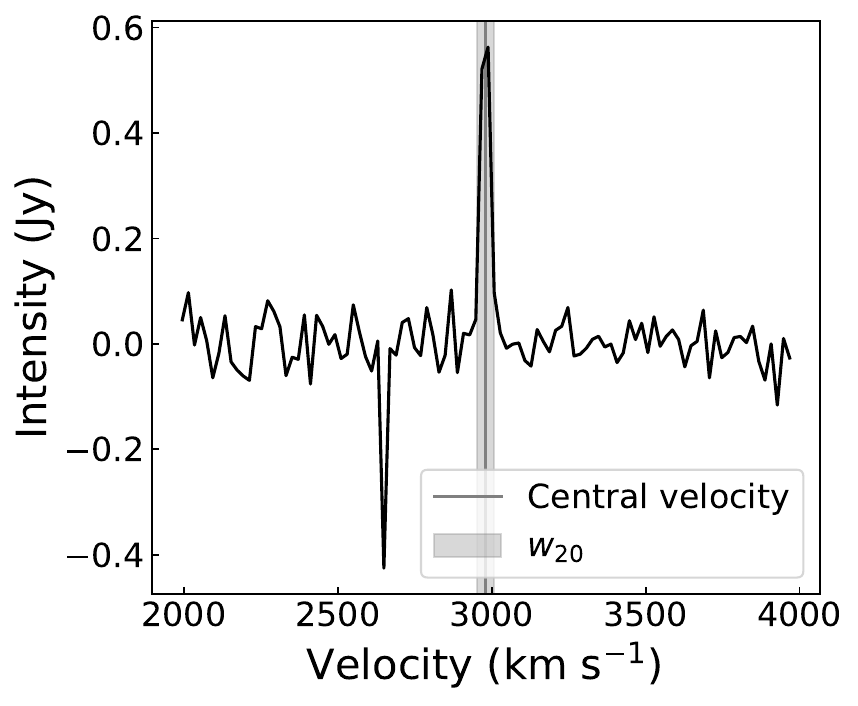}
			\caption{ }
   \end{subfigure}
	\caption{WALLABY J131244-155218 (NGC 5044)}
    \label{fig:faint}
\end{figure*}

\begin{figure*}
	\centering
		\begin{subfigure}[b]{0.24\textwidth}
			\centering
			\includegraphics[width=\textwidth]{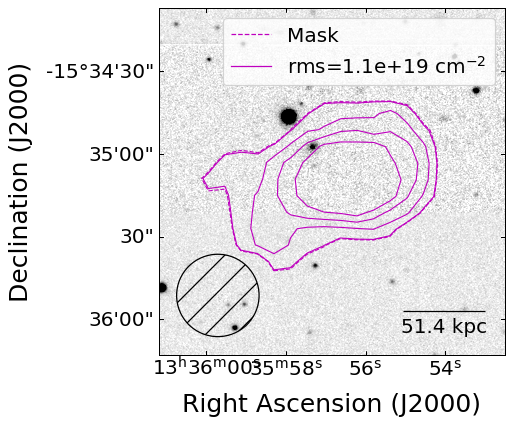}
			\caption{ }
		\end{subfigure}
		\begin{subfigure}[b]{0.24\textwidth}
			\centering
			\includegraphics[width=\textwidth]{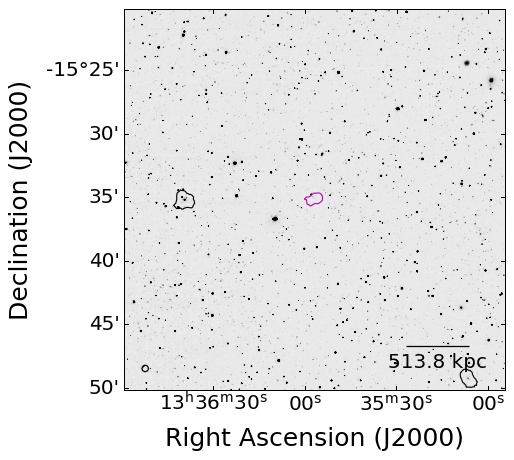}
			\caption{ }
		\end{subfigure}
		\begin{subfigure}[b]{0.26\textwidth}
			\centering
			\includegraphics[width=\textwidth]{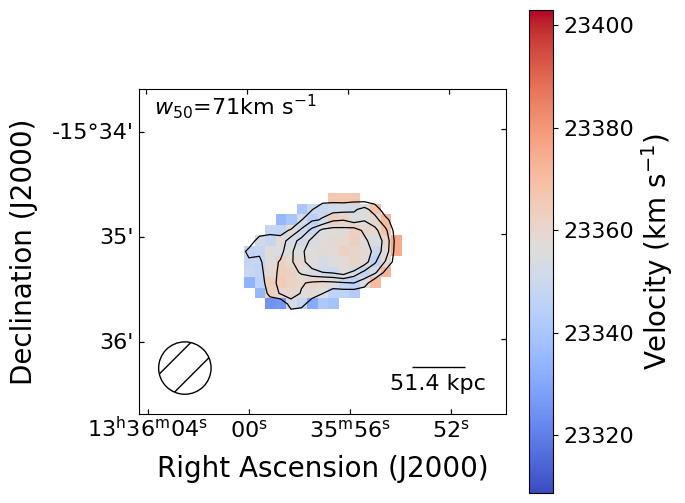}
			\caption{ }
		\end{subfigure}
    \begin{subfigure}[b]{0.24\textwidth}
			\centering
			\includegraphics[width=\textwidth]{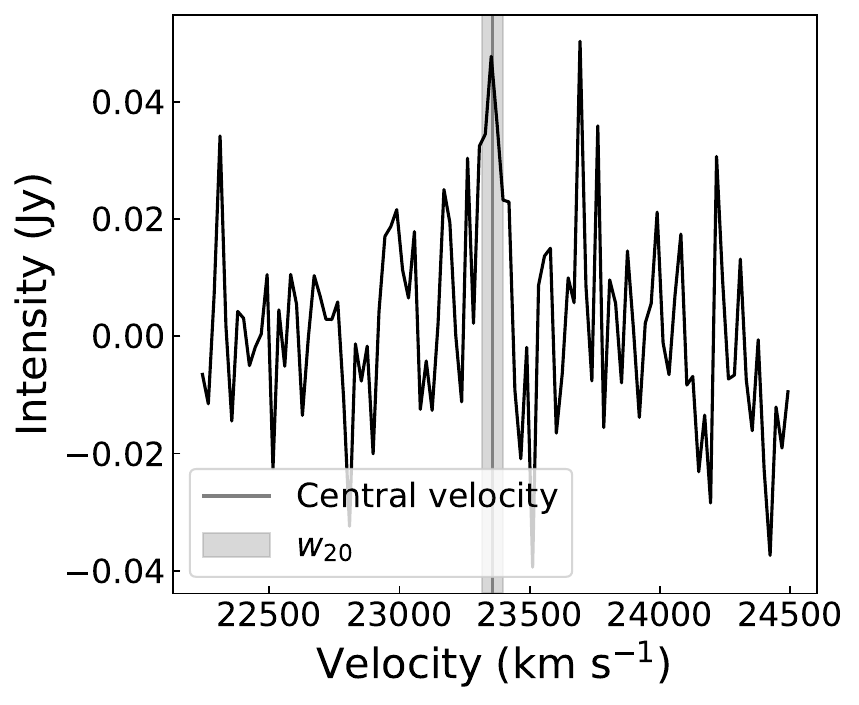}
			\caption{ }
   \end{subfigure}
	\caption{WALLABY J133556-153510 (NGC 5044)}
 \label{fig:lastb}
\end{figure*}

In this section we present the images of the strong dark source detections.  The dark sources that may be tidal remnants from Table \ref{tab:tidal} are noted as `tidal' in the captions. In Figures~\ref{fig:dgc1} to \ref{fig:lastb}, Figure~(a) is the g-band Legacy Survey image with H\,{\sc i} contours of the dark source overlaid.  {The dashed contour represents the edge of the SoFiA mask. The lowest solid contour corresponds to the column density equal to the local rms of the unmasked moment 0 map. Additional contours equal to 3, 5 and 7 times the local rms column density are also shown where possible.} Figure~(b) is the Legacy Survey g-band image zoomed out to 30 arcmin, with the {mask outline and  the column density contour equal to the local rms of the unmasked moment 0 map} of the dark source overlaid in magenta and other WALLABY sources overlaid in black. Figure~(c) is the moment 1 map (velocity field) of the dark source. The WALLABY beam is shown in the lower left corner and the scale is shown in the lower right corner of each image. Figure~(d) is the unmasked H\,{\sc i} spectrum.

\begin{figure*}
	\centering
		\begin{subfigure}[b]{0.24\textwidth}
			\centering
			\includegraphics[width=\textwidth]{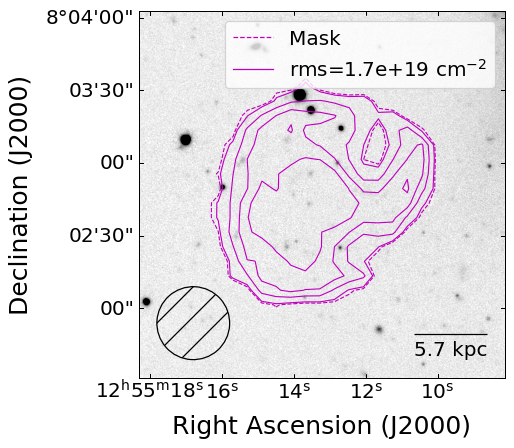}
			\caption{ }
		\end{subfigure}
		\begin{subfigure}[b]{0.24\textwidth}
			\centering
			\includegraphics[width=\textwidth]{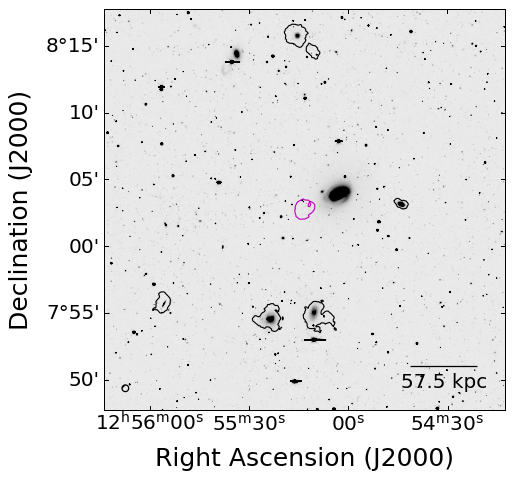}
			\caption{ }
		\end{subfigure}
		\begin{subfigure}[b]{0.26\textwidth}
			\centering
			\includegraphics[width=\textwidth]{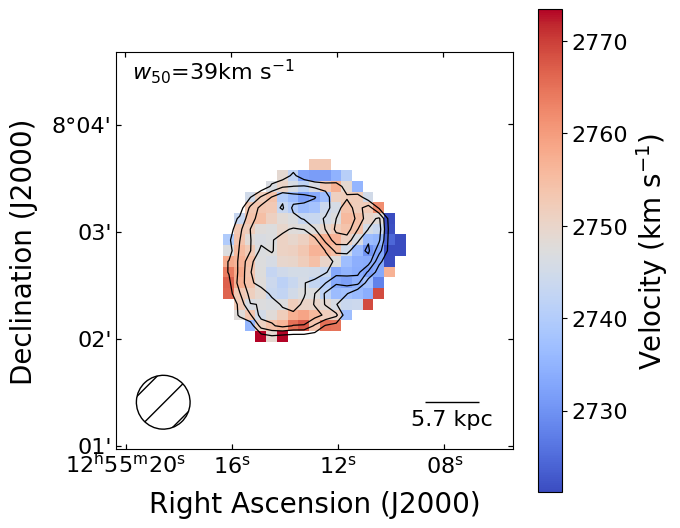}
			\caption{ }
		\end{subfigure}
    \begin{subfigure}[b]{0.24\textwidth}
			\centering
			\includegraphics[width=\textwidth]{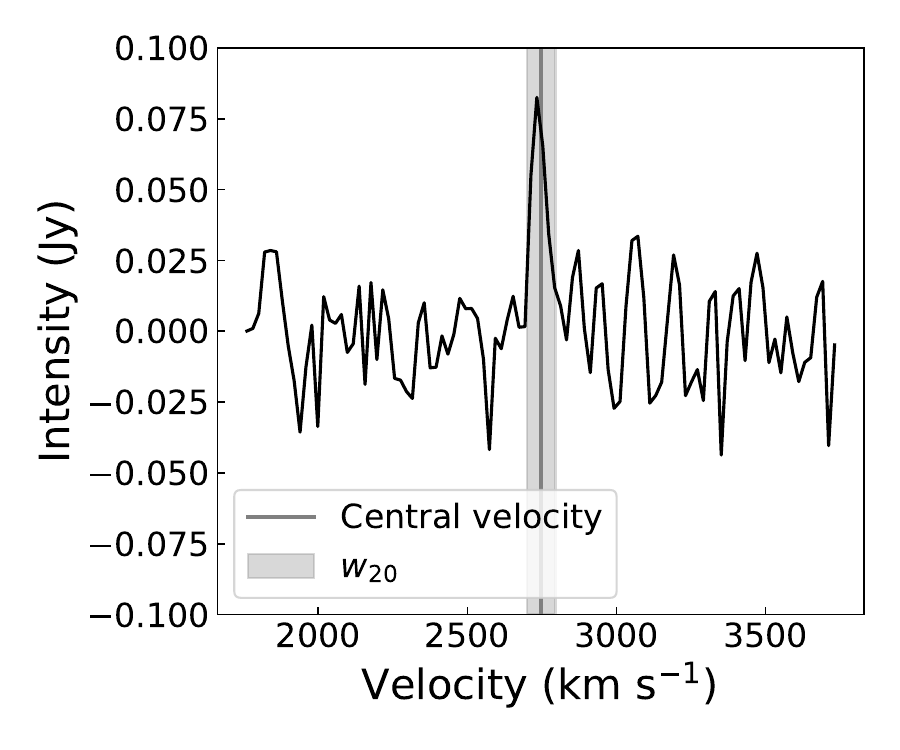}
			\caption{ }
   \end{subfigure}
	\caption{WALLABY J125513+080246 (tidal; NGC 4808)}
\end{figure*}


\begin{figure*}
	\centering
		\begin{subfigure}[b]{0.24\textwidth}
			\centering
			\includegraphics[width=\textwidth]{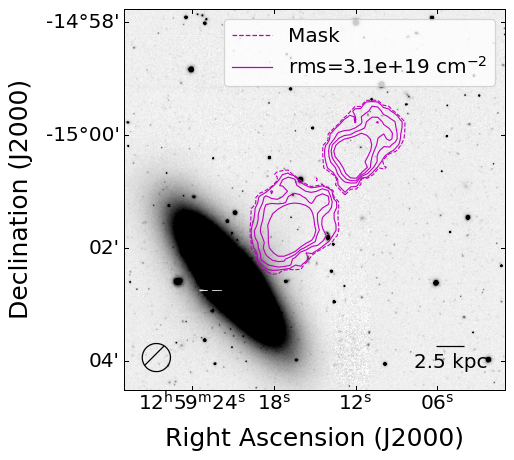}
			\caption{ }
		\end{subfigure}
		\begin{subfigure}[b]{0.24\textwidth}
			\centering
			\includegraphics[width=\textwidth]{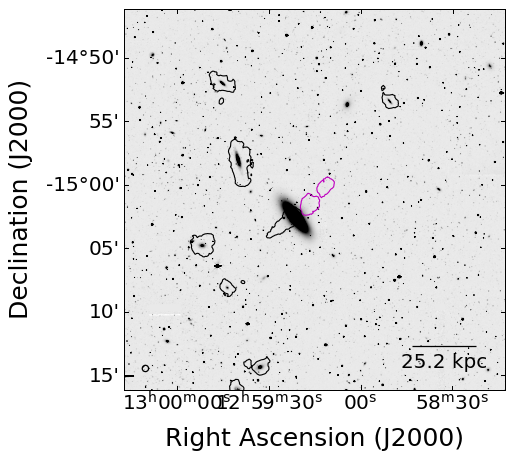}
			\caption{ }
		\end{subfigure}
		\begin{subfigure}[b]{0.26\textwidth}
			\centering
			\includegraphics[width=\textwidth]{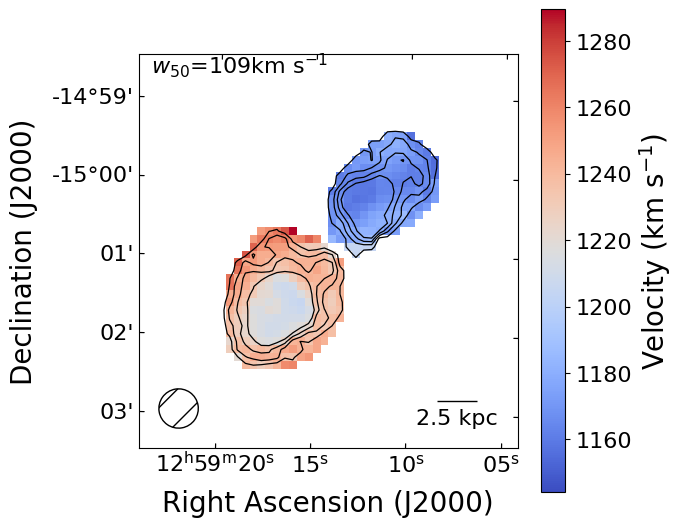}
			\caption{ }
		\end{subfigure}
    \begin{subfigure}[b]{0.24\textwidth}
			\centering
			\includegraphics[width=\textwidth]{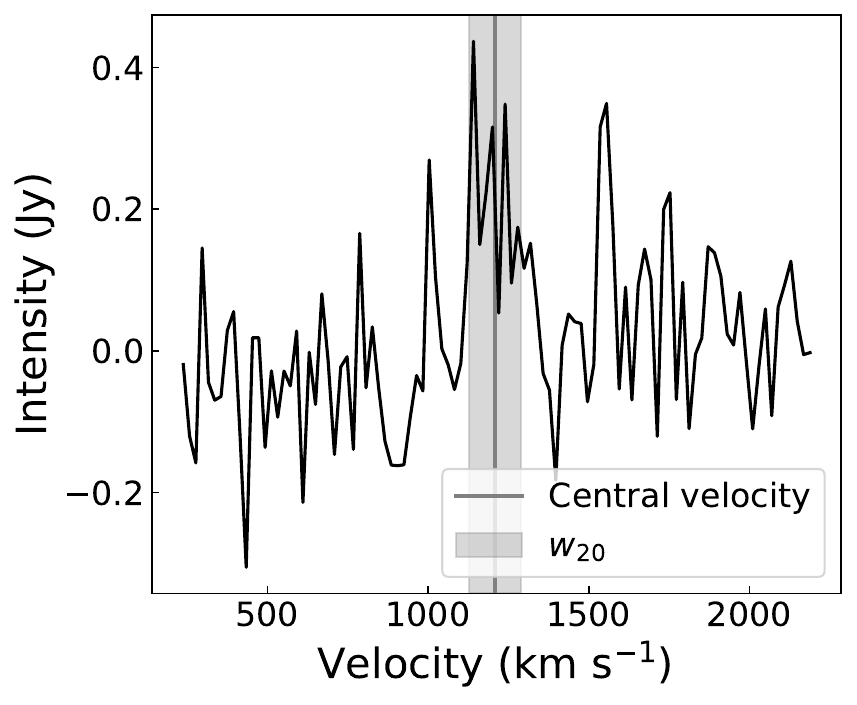}
			\caption{ }
   \end{subfigure}
	\caption{WALLABY J125915-150108 (tidal; NGC 5044 field). Note, H\,{\sc i} is detected on the other side of the galaxy, so this dark tidal cloud candidate may be part of a larger structure such as an outflow or polar ring.}
\end{figure*}

\begin{figure*}
	\centering
		\begin{subfigure}[b]{0.24\textwidth}
			\centering
			\includegraphics[width=\textwidth]{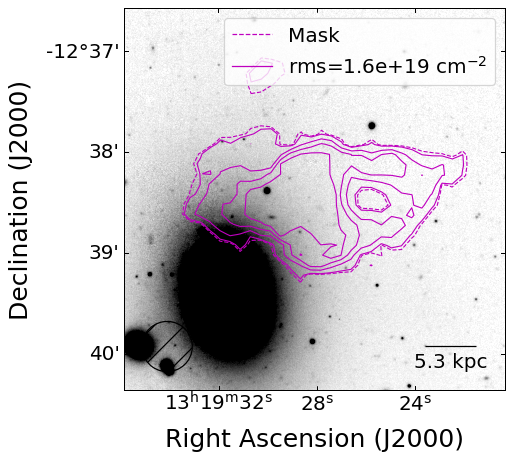}
			\caption{ }
		\end{subfigure}
		\begin{subfigure}[b]{0.24\textwidth}
			\centering
			\includegraphics[width=\textwidth]{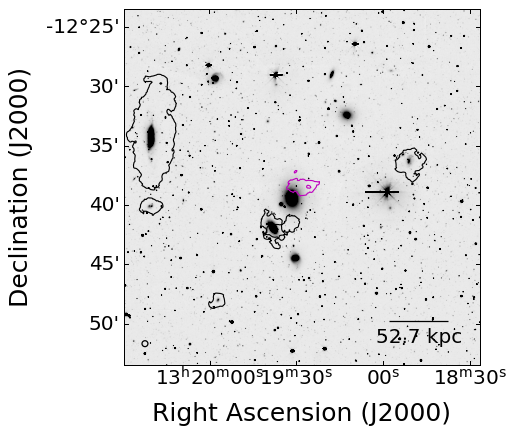}
			\caption{ }
		\end{subfigure}
		\begin{subfigure}[b]{0.26\textwidth}
			\centering
			\includegraphics[width=\textwidth]{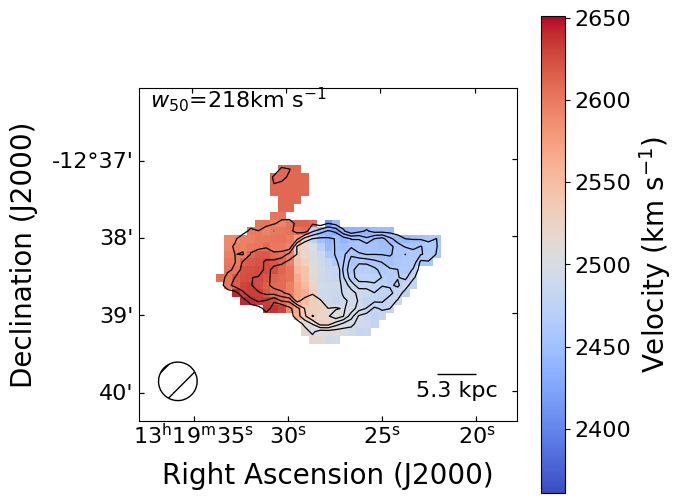}
			\caption{ }
		\end{subfigure}
    \begin{subfigure}[b]{0.24\textwidth}
			\centering
			\includegraphics[width=\textwidth]{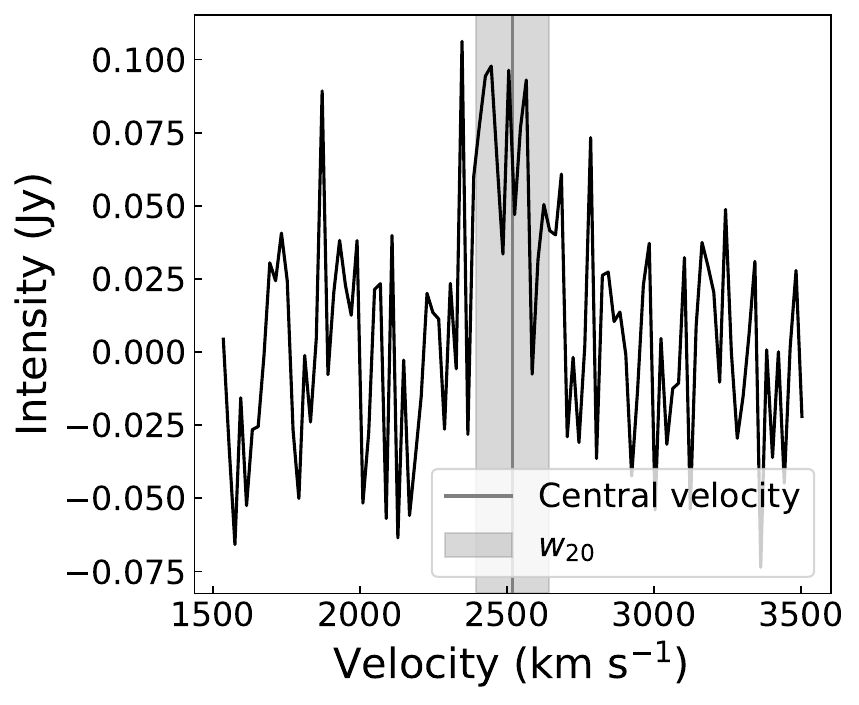}
			\caption{ }
   \end{subfigure}
	\caption{WALLABY J131928-123828 (tidal; NGC 5044 field)}
\end{figure*}


\begin{figure*}
	\centering
		\begin{subfigure}[b]{0.24\textwidth}
			\centering
			\includegraphics[width=\textwidth]{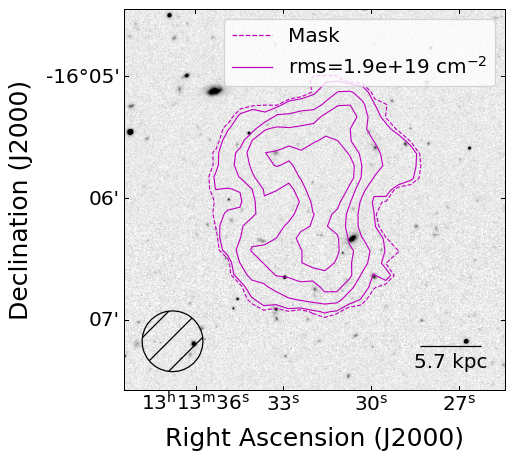}
			\caption{ }
		\end{subfigure}
		\begin{subfigure}[b]{0.24\textwidth}
			\centering
			\includegraphics[width=\textwidth]{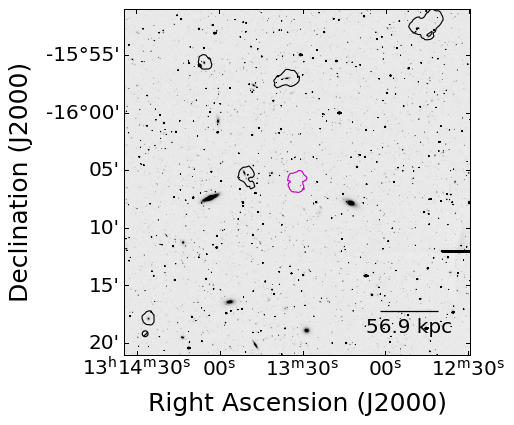}
			\caption{ }
		\end{subfigure}
		\begin{subfigure}[b]{0.26\textwidth}
			\centering
			\includegraphics[width=\textwidth]{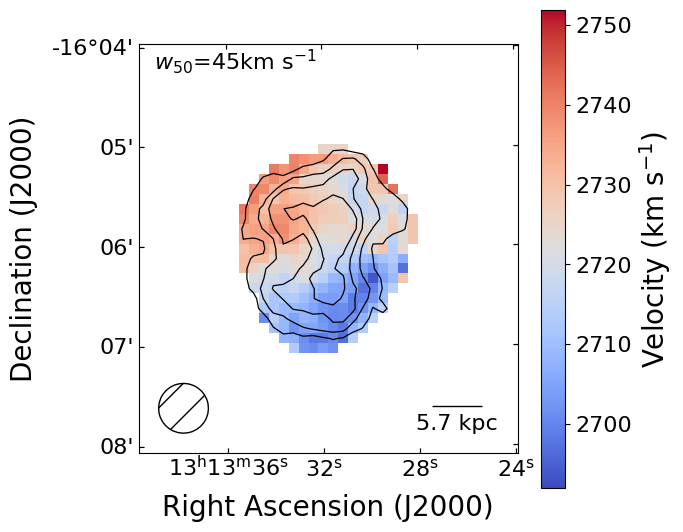}
			\caption{ }
		\end{subfigure}
    \begin{subfigure}[b]{0.24\textwidth}
			\centering
			\includegraphics[width=\textwidth]{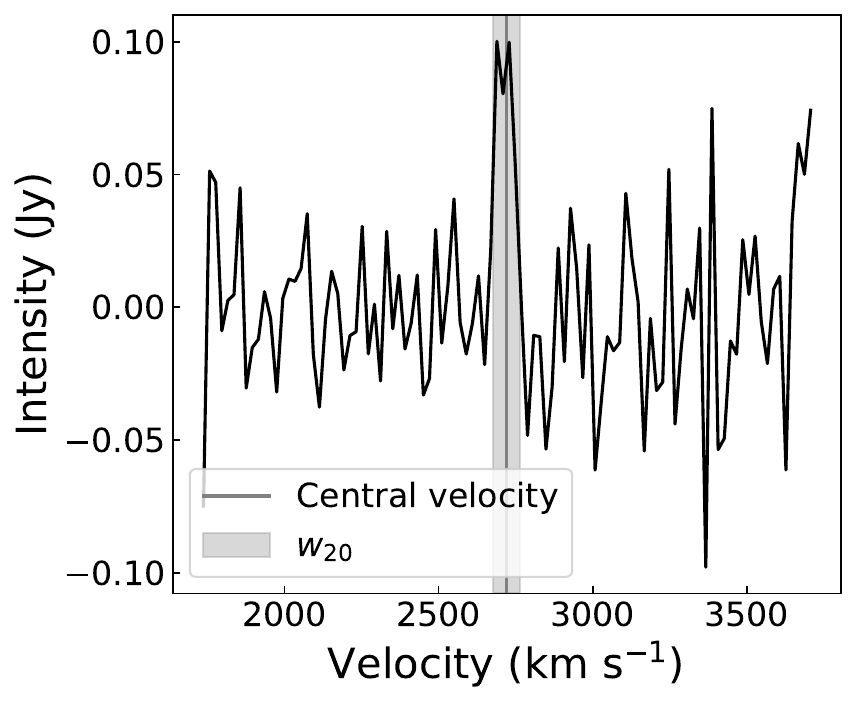}
			\caption{ }
   \end{subfigure}
	\caption{WALLABY J131331-160600 (tidal; NGC 5044)}
\end{figure*}

\begin{figure*}
	\centering
		\begin{subfigure}[b]{0.24\textwidth}
			\centering
			\includegraphics[width=\textwidth]{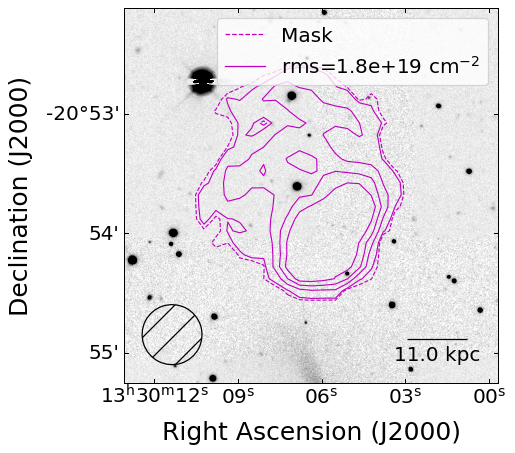}
			\caption{ }
		\end{subfigure}
		\begin{subfigure}[b]{0.24\textwidth}
			\centering
			\includegraphics[width=\textwidth]{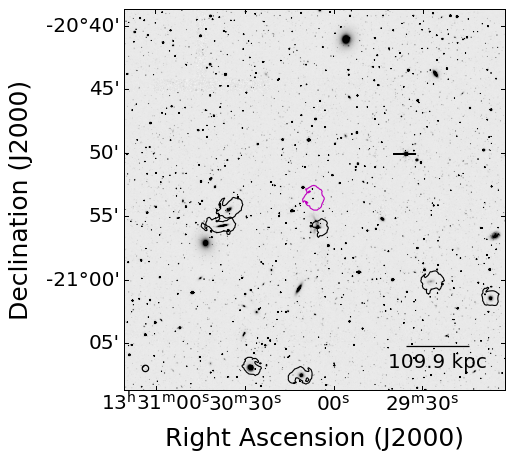}
			\caption{ }
		\end{subfigure}
		\begin{subfigure}[b]{0.26\textwidth}
			\centering
			\includegraphics[width=\textwidth]{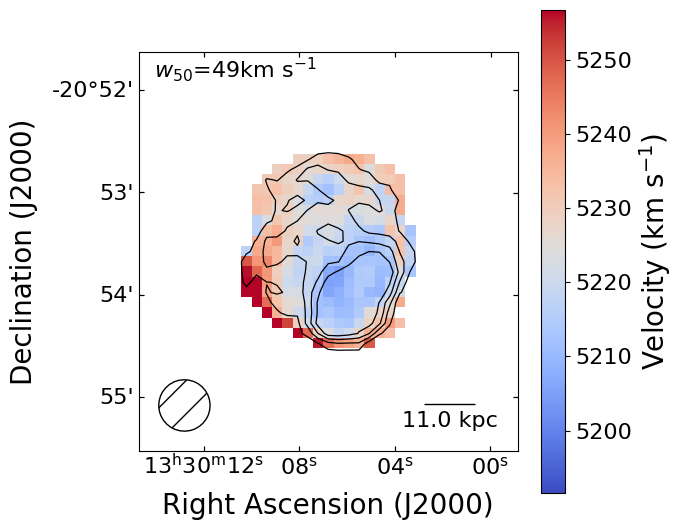}
			\caption{ }
		\end{subfigure}
    \begin{subfigure}[b]{0.24\textwidth}
			\centering
			\includegraphics[width=\textwidth]{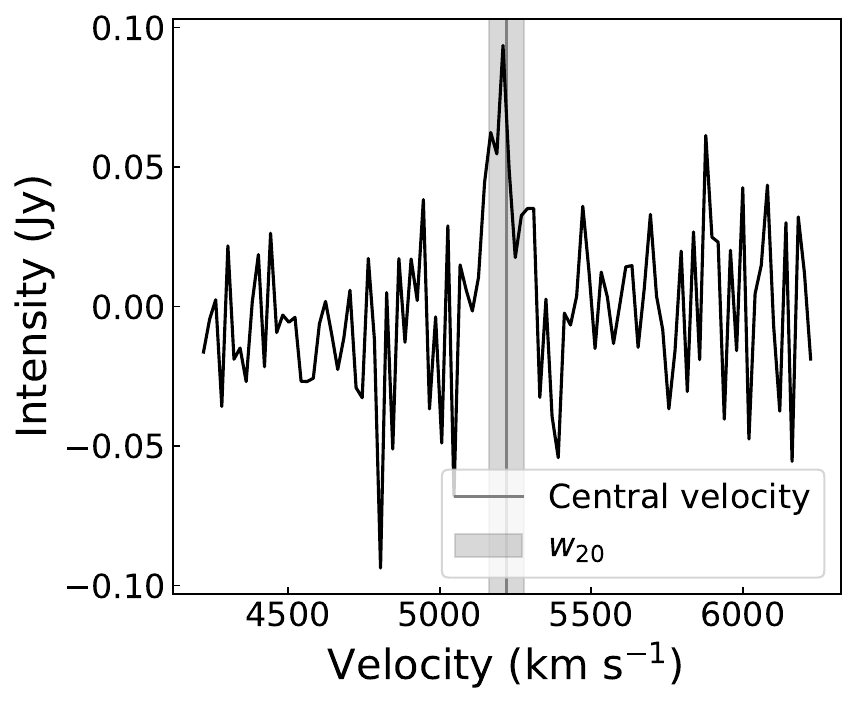}
			\caption{ }
   \end{subfigure}
	\caption{WALLABY J133006-205341 (tidal; NGC 5044)}
\end{figure*}


\begin{figure*}
	\centering
		\begin{subfigure}[b]{0.24\textwidth}
			\centering
			\includegraphics[width=\textwidth]{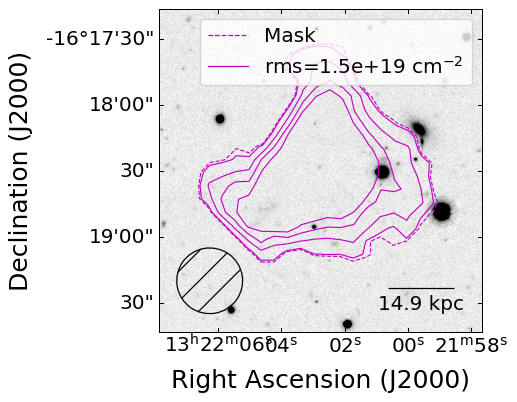}
			\caption{ }
		\end{subfigure}
		\begin{subfigure}[b]{0.24\textwidth}
			\centering
			\includegraphics[width=\textwidth]{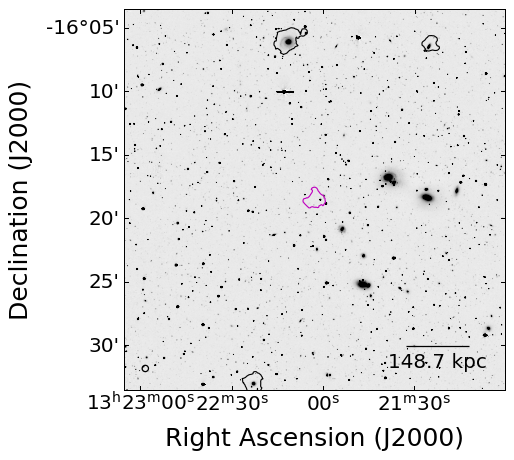}
			\caption{ }
		\end{subfigure}
		\begin{subfigure}[b]{0.26\textwidth}
			\centering
			\includegraphics[width=\textwidth]{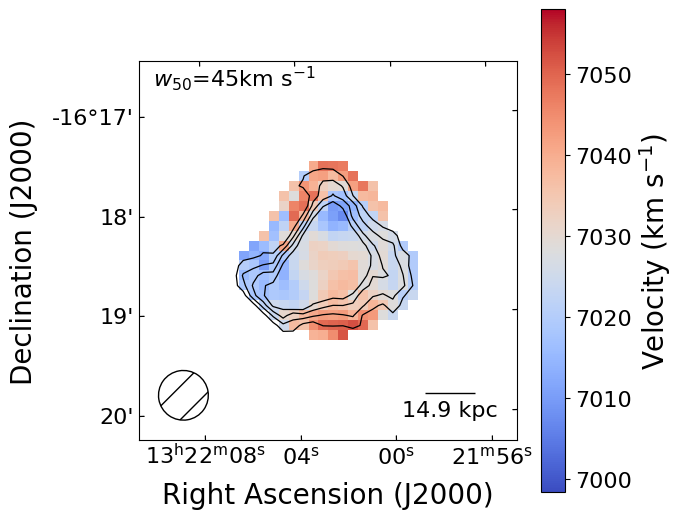}
			\caption{ }
		\end{subfigure}
    \begin{subfigure}[b]{0.24\textwidth}
			\centering
			\includegraphics[width=\textwidth]{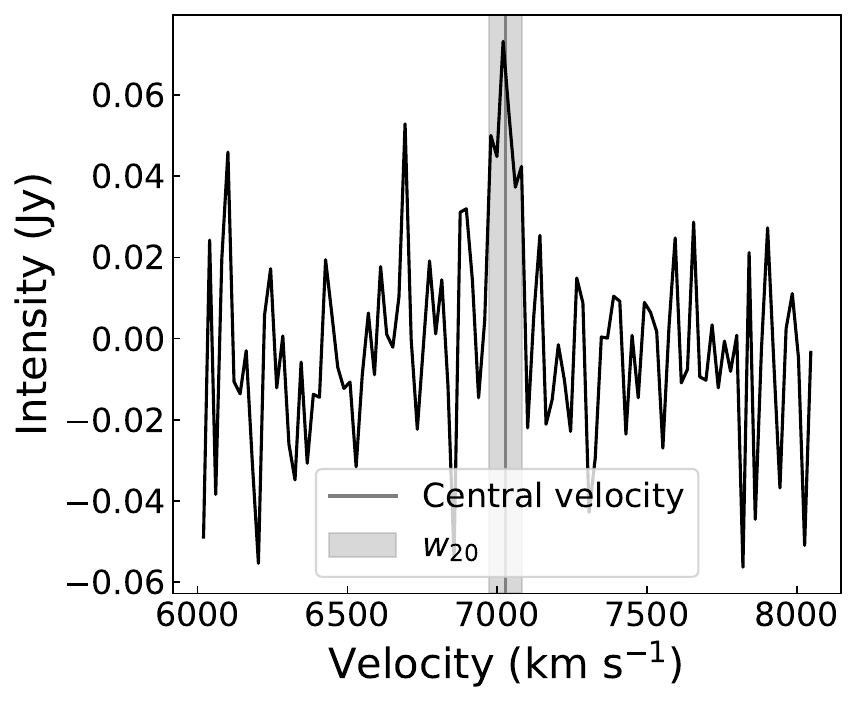}
			\caption{ }
   \end{subfigure}
	\caption{WALLABY J132202-161829 (tidal; NGC 5044)}
\end{figure*}

\begin{figure*}
	\centering
		\begin{subfigure}[b]{0.24\textwidth}
			\centering
			\includegraphics[width=\textwidth]{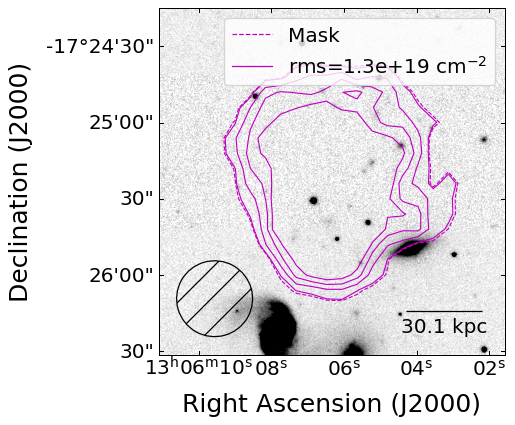}
			\caption{ }
		\end{subfigure}
		\begin{subfigure}[b]{0.24\textwidth}
			\centering
			\includegraphics[width=\textwidth]{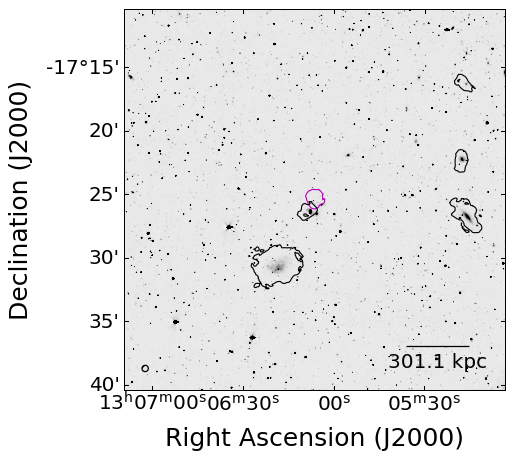}
			\caption{ }
		\end{subfigure}
		\begin{subfigure}[b]{0.26\textwidth}
			\centering
			\includegraphics[width=\textwidth]{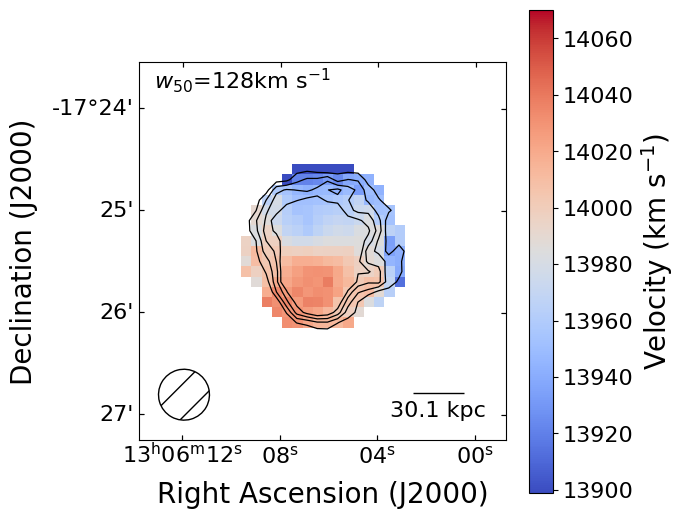}
			\caption{ }
		\end{subfigure}
    \begin{subfigure}[b]{0.24\textwidth}
			\centering
			\includegraphics[width=\textwidth]{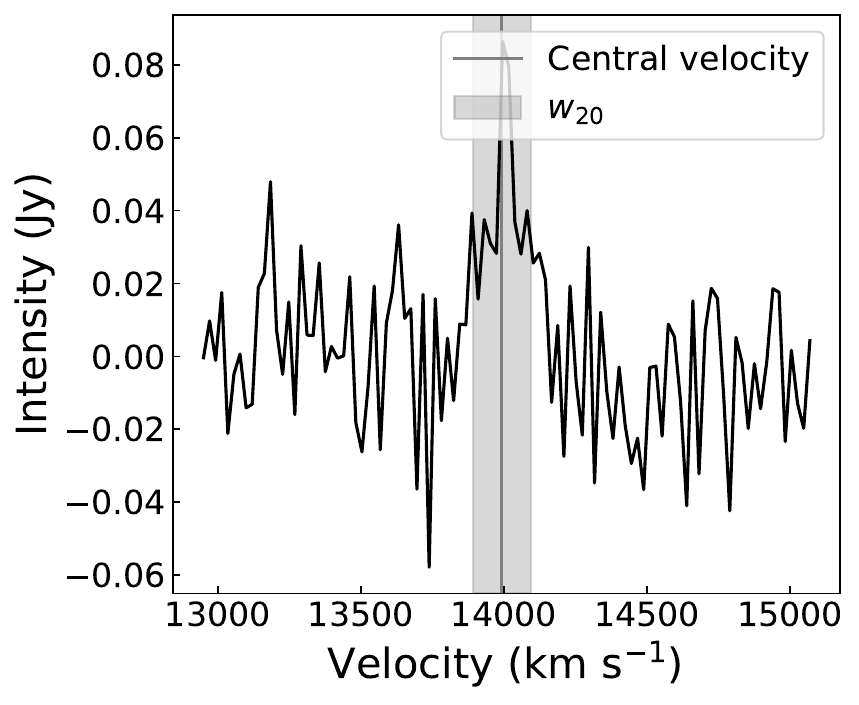}
			\caption{ }
   \end{subfigure}
	\caption{WALLABY J130606-172523 (tidal; NGC 5044)}
\end{figure*}

\begin{figure*}
	\centering
		\begin{subfigure}[b]{0.24\textwidth}
			\centering
			\includegraphics[width=\textwidth]{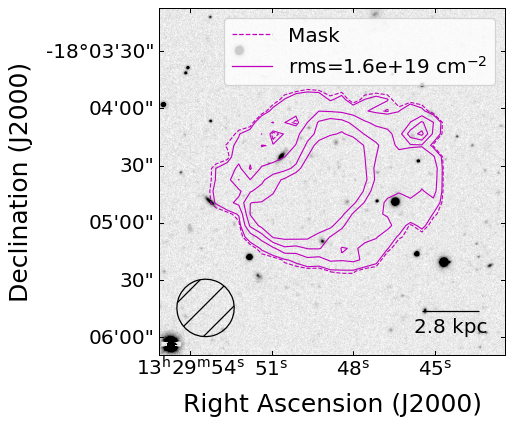}
			\caption{ }
		\end{subfigure}
		\begin{subfigure}[b]{0.24\textwidth}
			\centering
			\includegraphics[width=\textwidth]{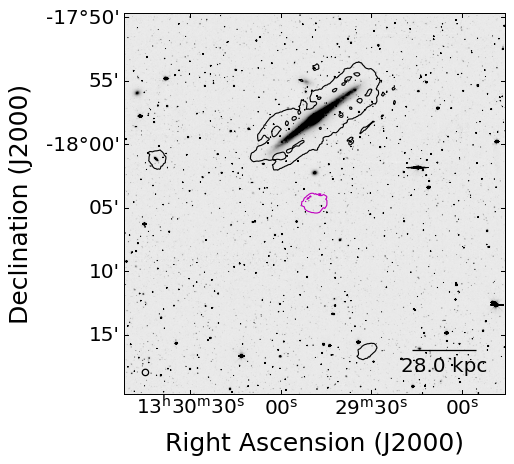}
			\caption{ }
		\end{subfigure}
		\begin{subfigure}[b]{0.26\textwidth}
			\centering
			\includegraphics[width=\textwidth]{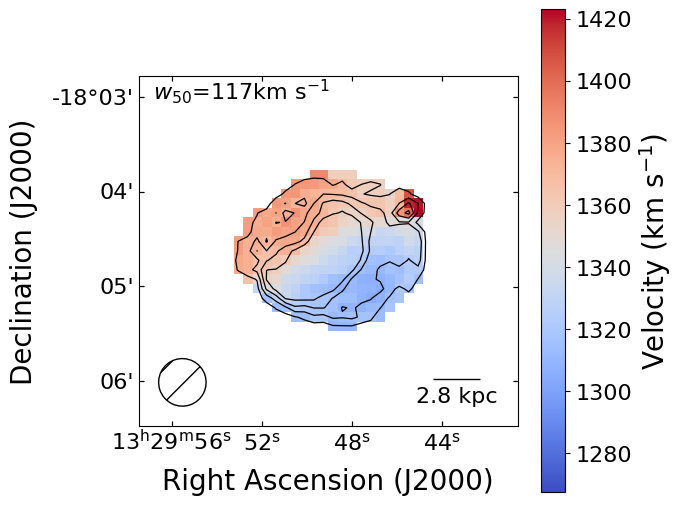}
			\caption{ }
		\end{subfigure}
    \begin{subfigure}[b]{0.24\textwidth}
			\centering
			\includegraphics[width=\textwidth]{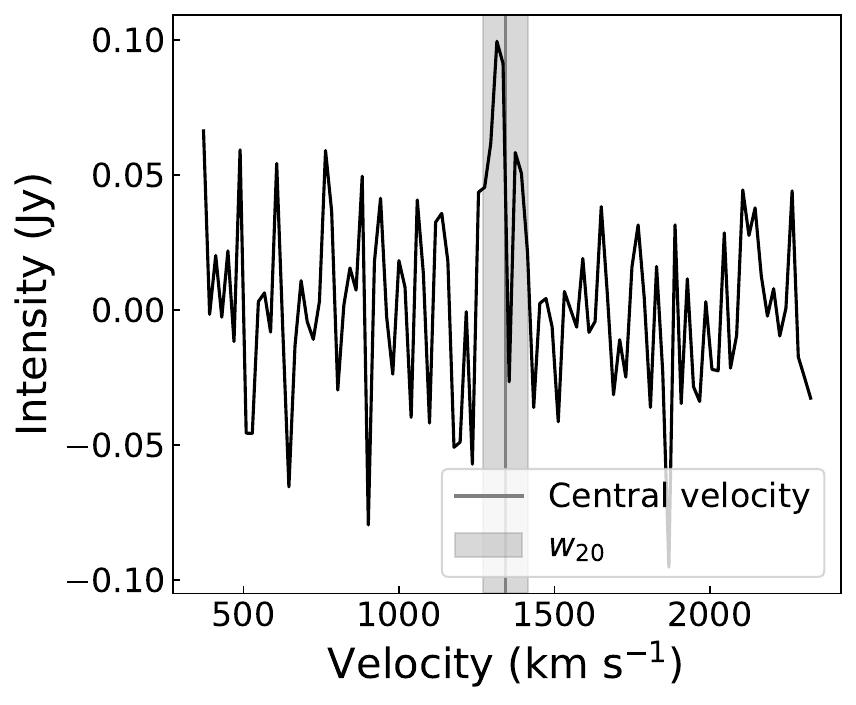}
			\caption{ }
   \end{subfigure}
	\caption{WALLABY J132948-180438 (tidal; NGC 5044)}
\end{figure*}

\begin{figure*}
	\centering
		\begin{subfigure}[b]{0.24\textwidth}
			\centering
			\includegraphics[width=\textwidth]{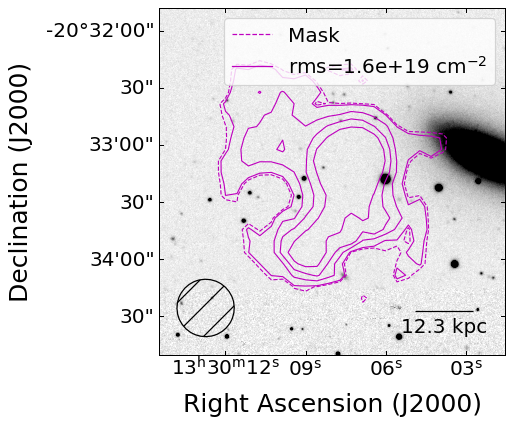}
			\caption{ }
		\end{subfigure}
		\begin{subfigure}[b]{0.24\textwidth}
			\centering
			\includegraphics[width=\textwidth]{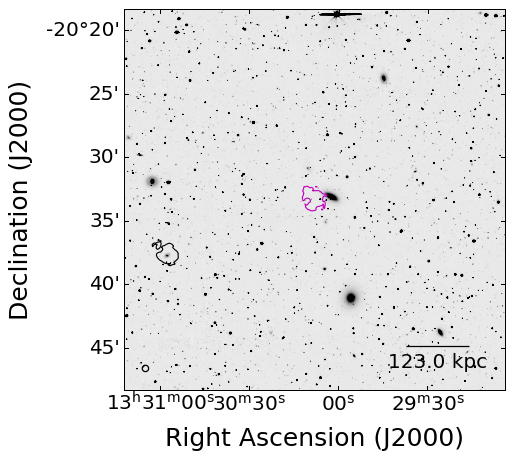}
			\caption{ }
		\end{subfigure}
		\begin{subfigure}[b]{0.26\textwidth}
			\centering
			\includegraphics[width=\textwidth]{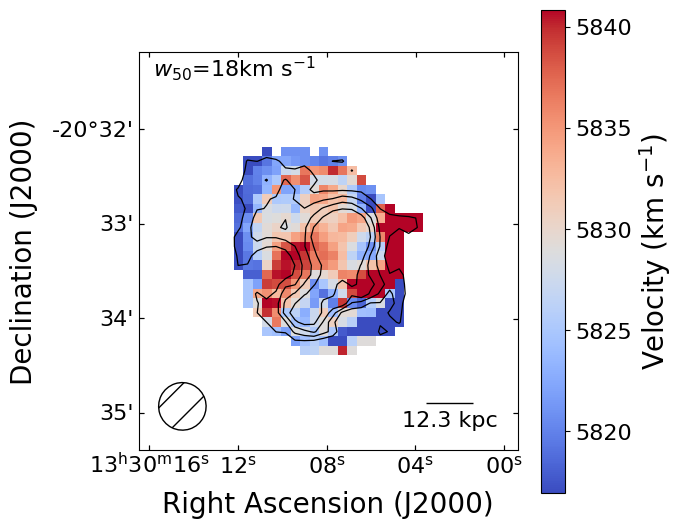}
			\caption{ }
		\end{subfigure}
    \begin{subfigure}[b]{0.24\textwidth}
			\centering
			\includegraphics[width=\textwidth]{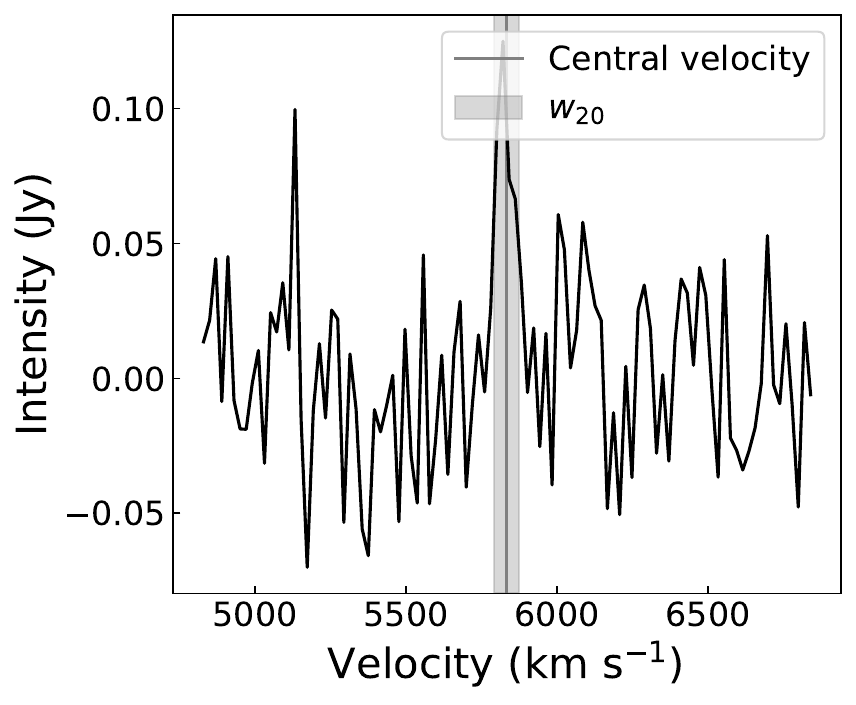}
			\caption{ }
   \end{subfigure}
	\caption{WALLABY J133008-203319 (tidal; NGC 5044)}
\end{figure*}

\begin{figure*}
	\centering
		\begin{subfigure}[b]{0.24\textwidth}
			\centering
			\includegraphics[width=\textwidth]{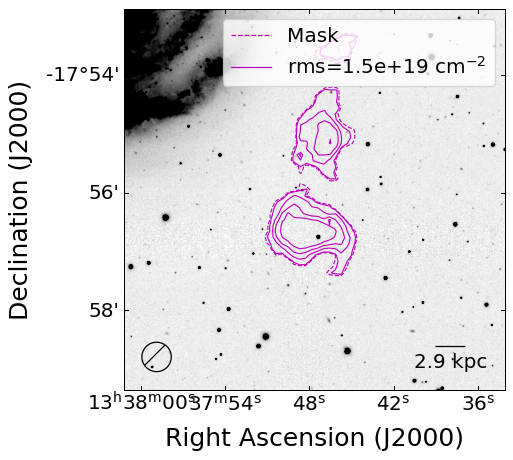}
			\caption{ }
		\end{subfigure}
		\begin{subfigure}[b]{0.24\textwidth}
			\centering
			\includegraphics[width=\textwidth]{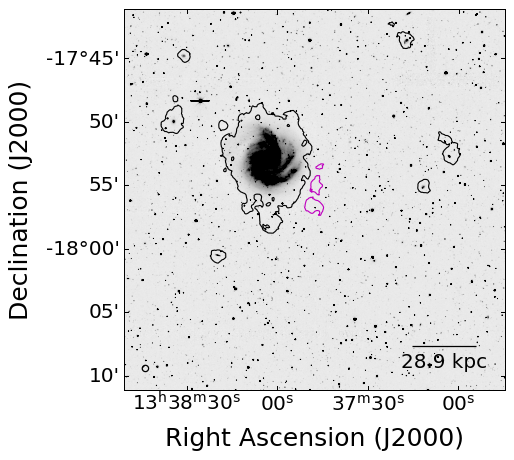}
			\caption{ }
		\end{subfigure}
		\begin{subfigure}[b]{0.26\textwidth}
			\centering
			\includegraphics[width=\textwidth]{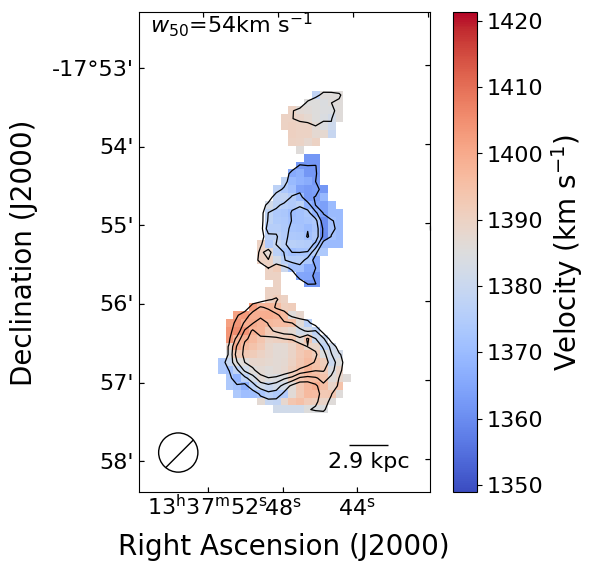}
			\caption{ }
		\end{subfigure}
    \begin{subfigure}[b]{0.24\textwidth}
			\centering
			\includegraphics[width=\textwidth]{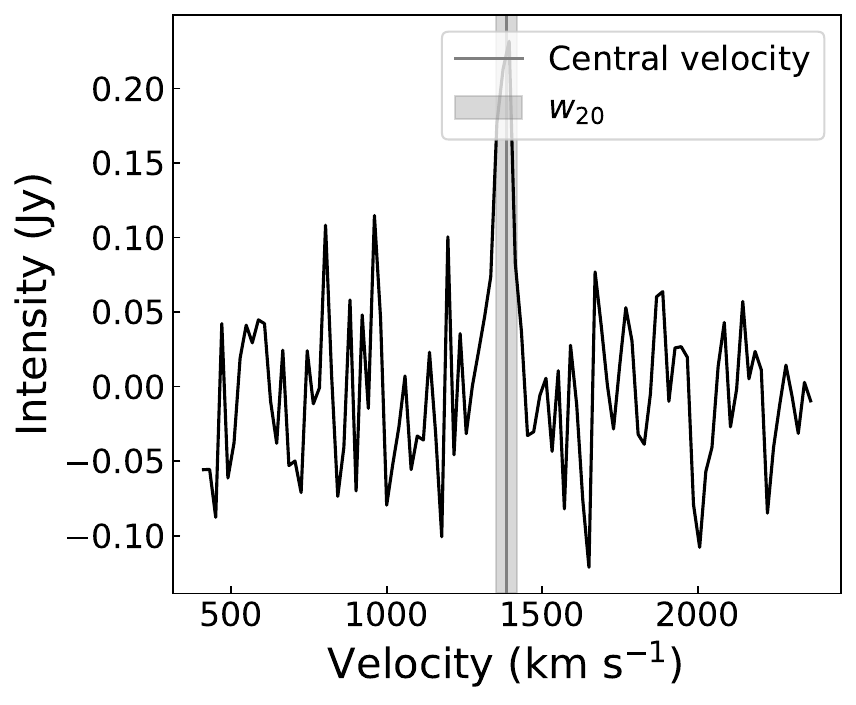}
			\caption{ }
   \end{subfigure}
	\caption{WALLABY J133747-175606 (tidal; NGC 5044)}
\end{figure*}

\begin{figure*}
	\centering
		\begin{subfigure}[b]{0.24\textwidth}
			\centering
			\includegraphics[width=\textwidth]{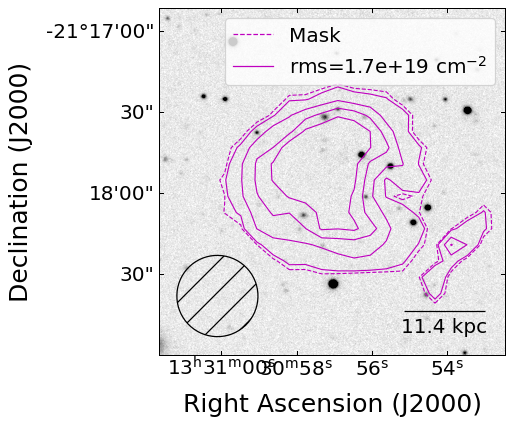}
			\caption{ }
		\end{subfigure}
		\begin{subfigure}[b]{0.24\textwidth}
			\centering
			\includegraphics[width=\textwidth]{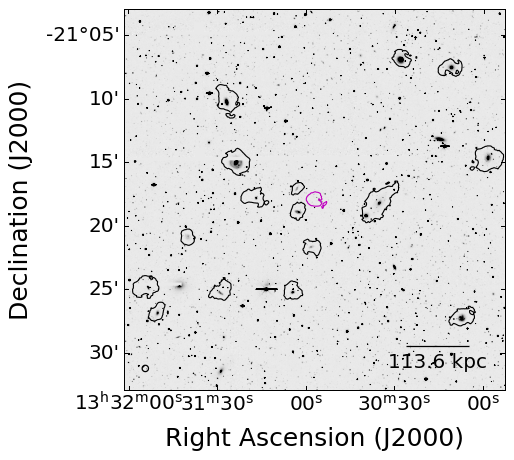}
			\caption{ }
		\end{subfigure}
		\begin{subfigure}[b]{0.26\textwidth}
			\centering
			\includegraphics[width=\textwidth]{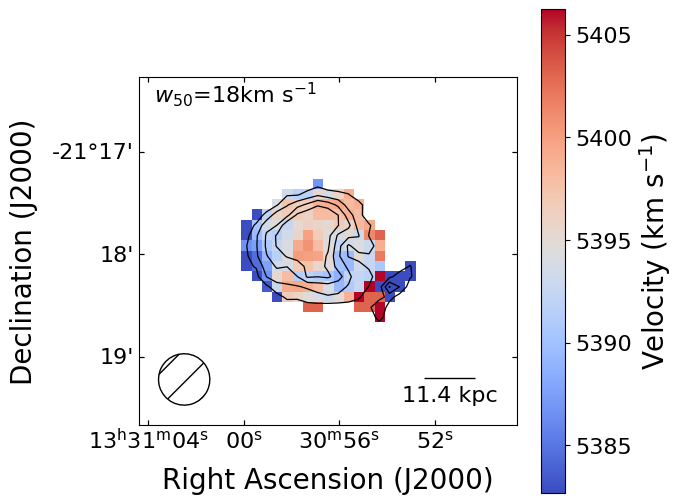}
			\caption{ }
		\end{subfigure}
  \begin{subfigure}[b]{0.24\textwidth}
			\centering
			\includegraphics[width=\textwidth]{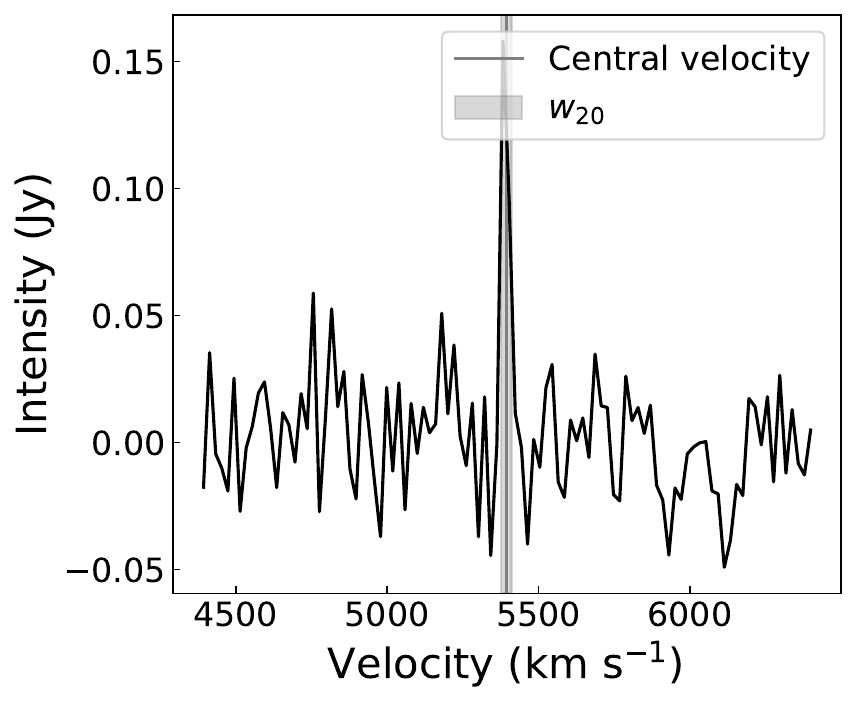}
			\caption{ }
   \end{subfigure}
	\caption{WALLABY J133057-211755 (tidal; NGC 5044)}
\end{figure*}

\begin{figure*}
	\centering
		\begin{subfigure}[b]{0.24\textwidth}
			\centering
			\includegraphics[width=\textwidth]{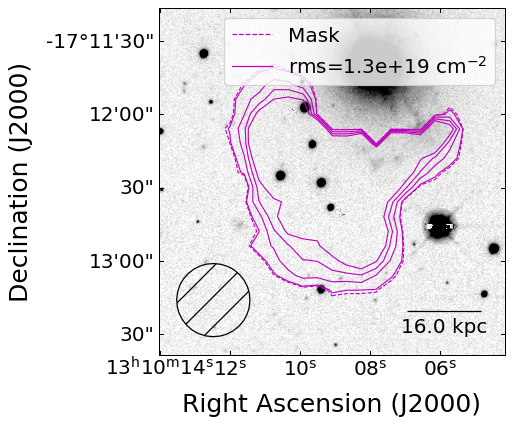}
			\caption{ }
		\end{subfigure}
		\begin{subfigure}[b]{0.24\textwidth}
			\centering
			\includegraphics[width=\textwidth]{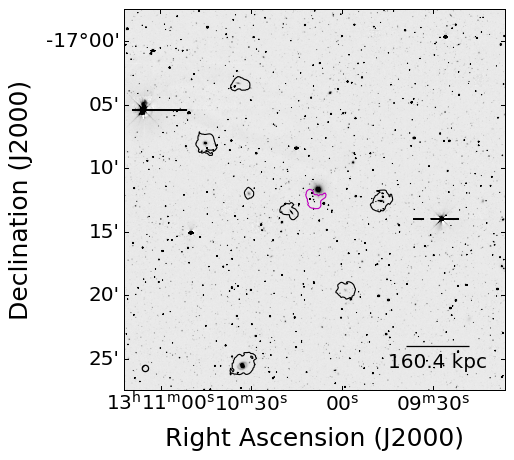}
			\caption{ }
		\end{subfigure}
		\begin{subfigure}[b]{0.26\textwidth}
			\centering
			\includegraphics[width=\textwidth]{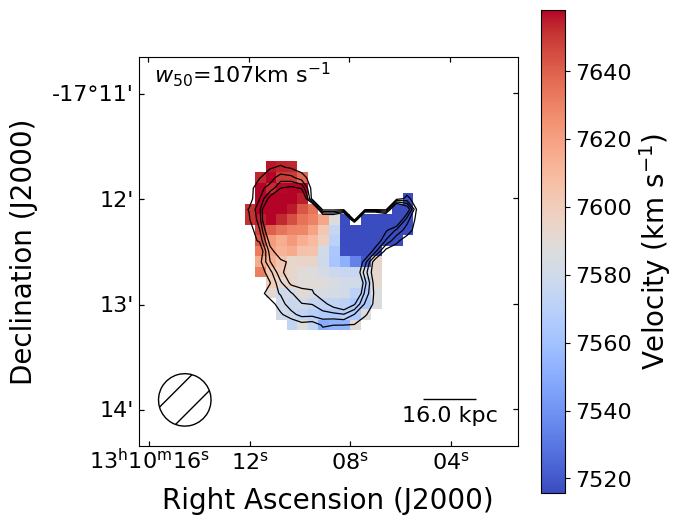}
			\caption{ }
		\end{subfigure}
    \begin{subfigure}[b]{0.24\textwidth}
			\centering
			\includegraphics[width=\textwidth]{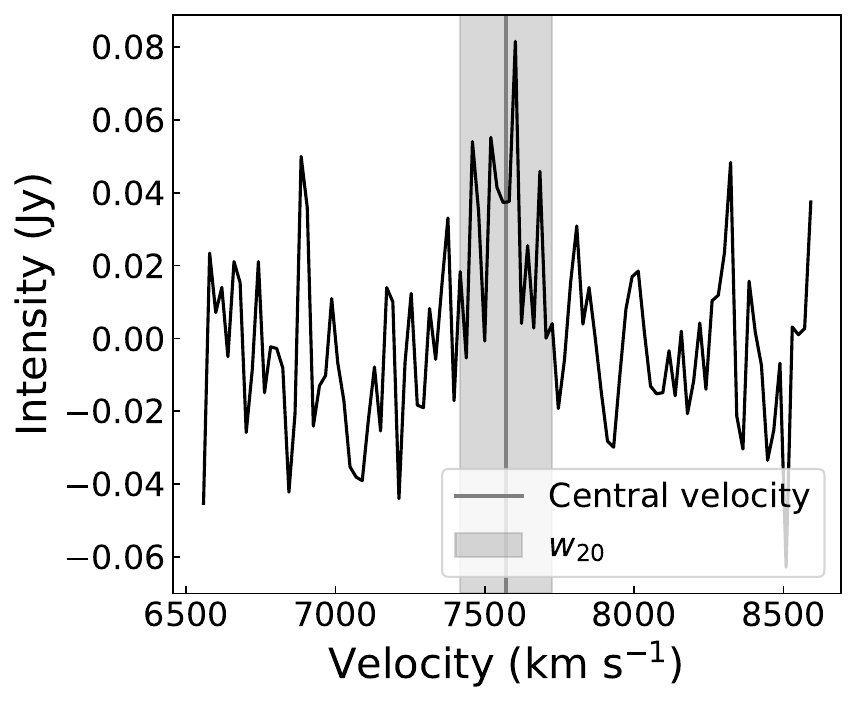}
			\caption{ }
   \end{subfigure}
	\caption{WALLABY J131009-171227 (tidal; NGC 5044)}
\end{figure*}

\begin{figure*}
	\centering
		\begin{subfigure}[b]{0.24\textwidth}
			\centering
			\includegraphics[width=\textwidth]{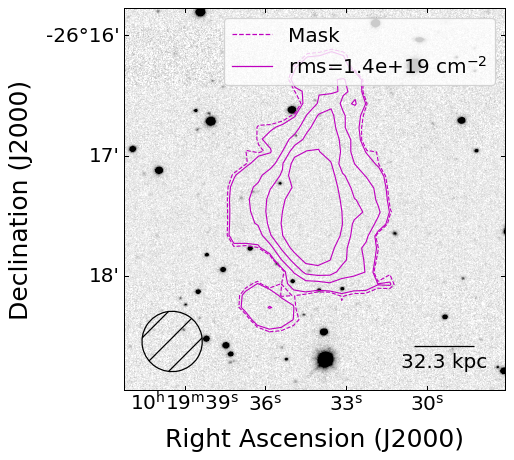}
			\caption{ }
		\end{subfigure}
		\begin{subfigure}[b]{0.24\textwidth}
			\centering
			\includegraphics[width=\textwidth]{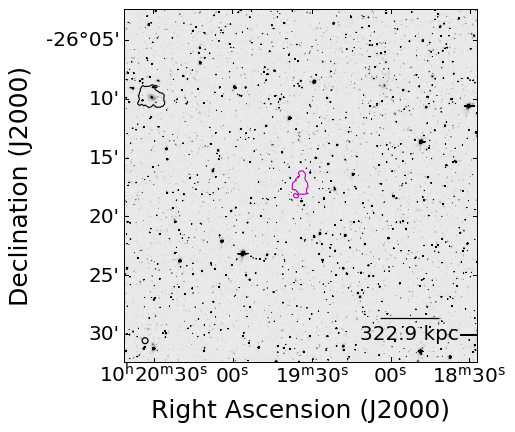}
			\caption{ }
		\end{subfigure}
		\begin{subfigure}[b]{0.26\textwidth}
			\centering
			\includegraphics[width=\textwidth]{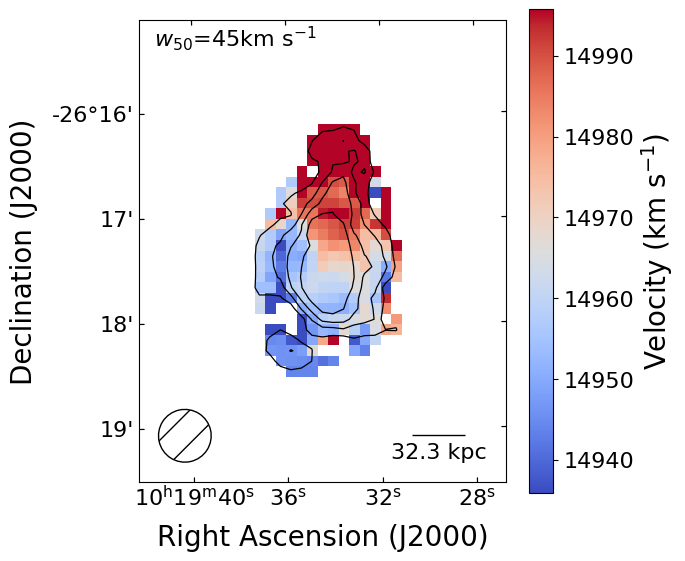}
			\caption{ }
		\end{subfigure}
   \begin{subfigure}[b]{0.24\textwidth}
			\centering
			\includegraphics[width=\textwidth]{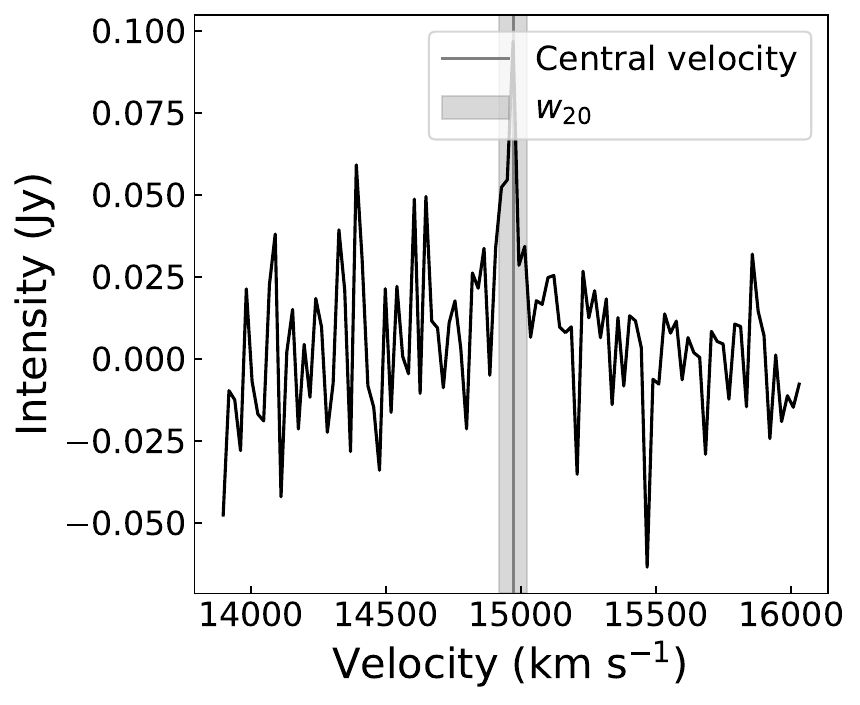}
			\caption{ }
		\end{subfigure}
	\caption{WALLABY J101934-261721 (Hydra)}
\end{figure*}

\begin{figure*}
	\centering
		\begin{subfigure}[b]{0.24\textwidth}
			\centering
			\includegraphics[width=\textwidth]{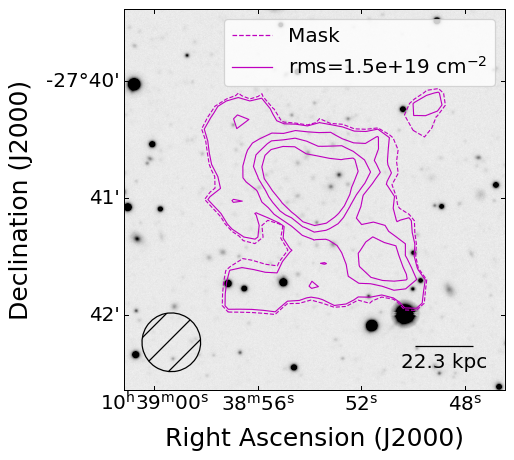}
			\caption{ }
		\end{subfigure}
		\begin{subfigure}[b]{0.24\textwidth}
			\centering
			\includegraphics[width=\textwidth]{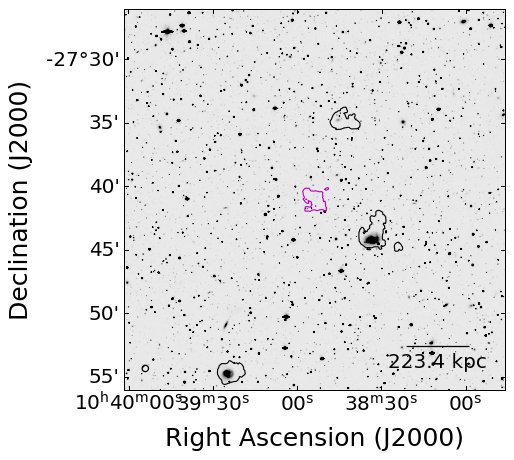}
			\caption{ }
		\end{subfigure}
		\begin{subfigure}[b]{0.26\textwidth}
			\centering
			\includegraphics[width=\textwidth]{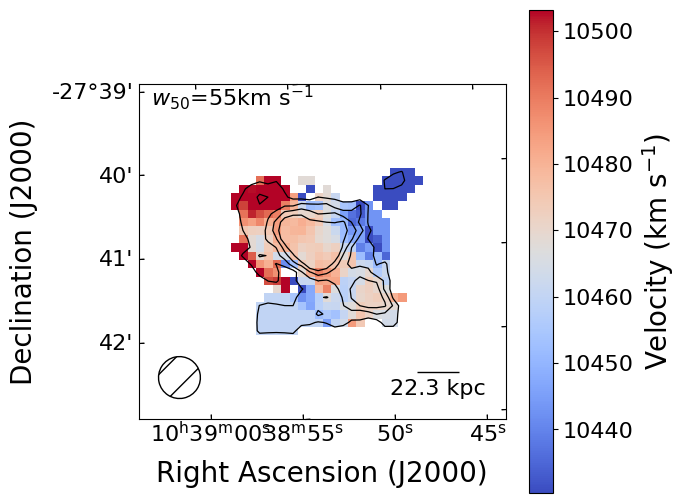}
			\caption{ }
		\end{subfigure}
    \begin{subfigure}[b]{0.24\textwidth}
			\centering
			\includegraphics[width=\textwidth]{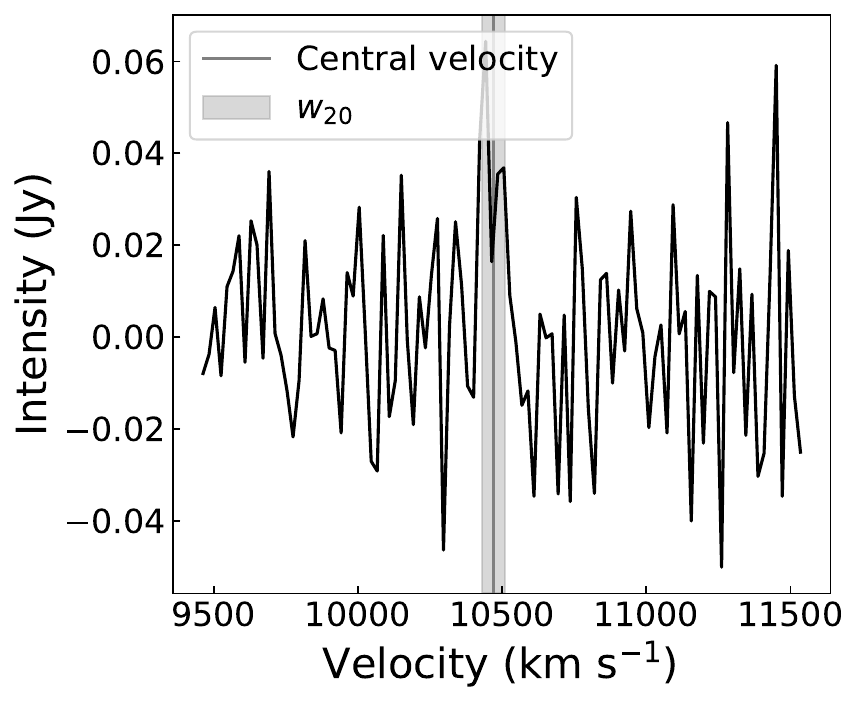}
			\caption{ }
   \end{subfigure}
	\caption{WALLABY J103853-274100 (Hydra)}
\end{figure*}

\newpage

\begin{figure*}
	\centering
		\begin{subfigure}[b]{0.24\textwidth}
			\centering
			\includegraphics[width=\textwidth]{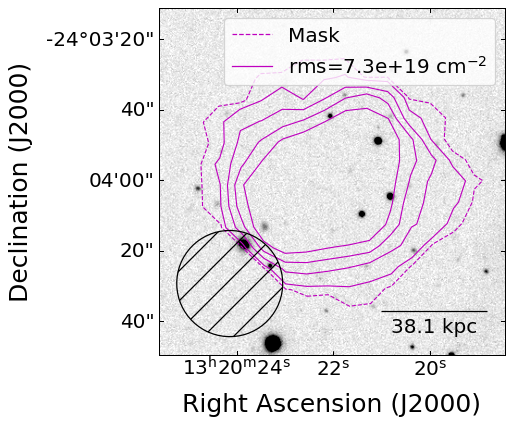}
			\caption{ }
		\end{subfigure}
		\begin{subfigure}[b]{0.24\textwidth}
			\centering
			\includegraphics[width=\textwidth]{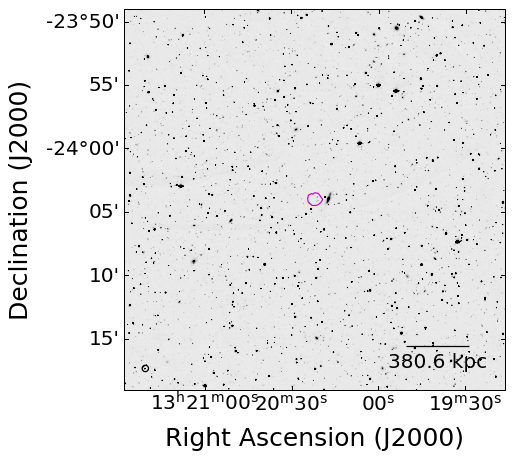}
			\caption{ }
		\end{subfigure}
		\begin{subfigure}[b]{0.26\textwidth}
			\centering
			\includegraphics[width=\textwidth]{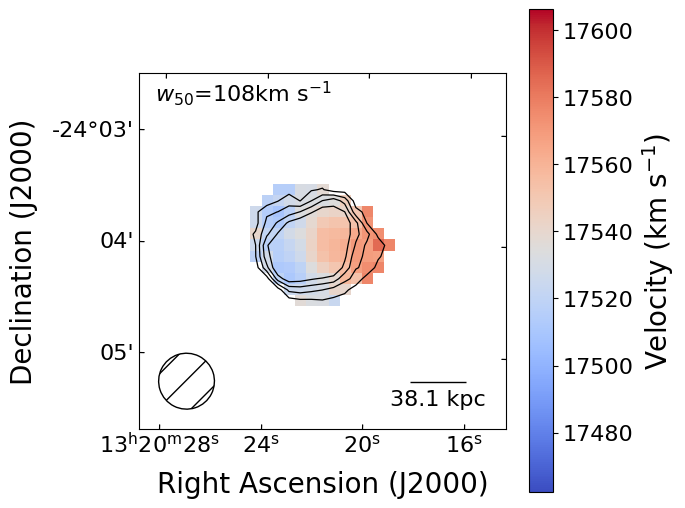}
			\caption{ }
		\end{subfigure}
    \begin{subfigure}[b]{0.24\textwidth}
			\centering
			\includegraphics[width=\textwidth]{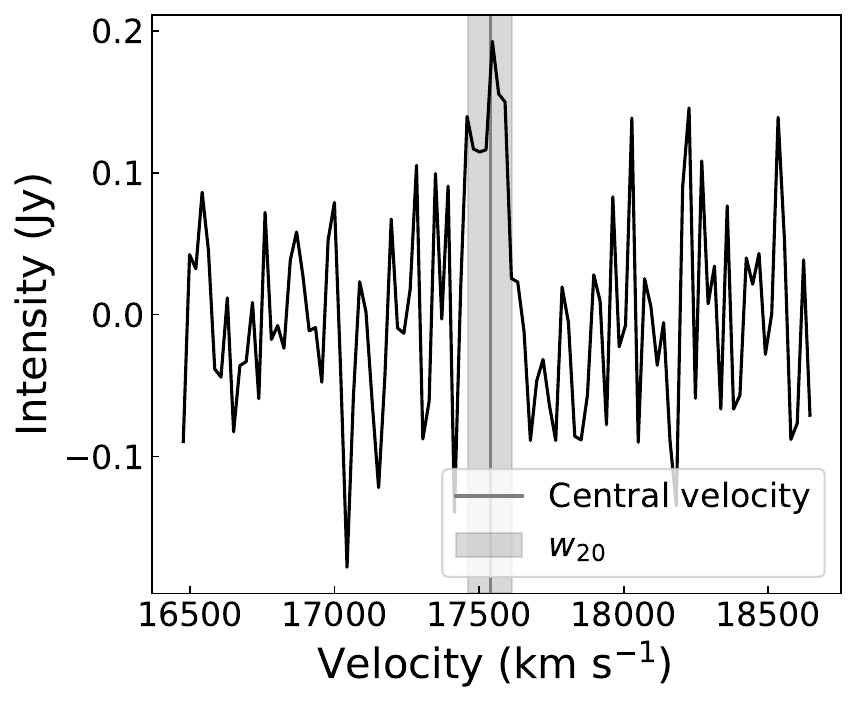}
			\caption{ }
   \end{subfigure}
	\caption{WALLABY J132022-240400 (NGC 5044)}
\end{figure*}

\begin{figure*}
	\centering
		\begin{subfigure}[b]{0.24\textwidth}
			\centering
			\includegraphics[width=\textwidth]{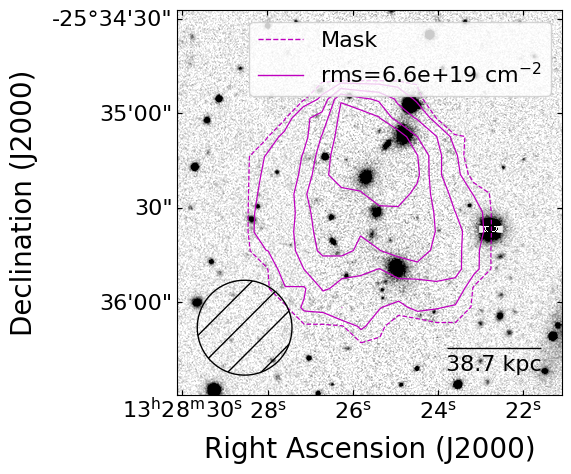}
			\caption{ }
		\end{subfigure}
		\begin{subfigure}[b]{0.24\textwidth}
			\centering
			\includegraphics[width=\textwidth]{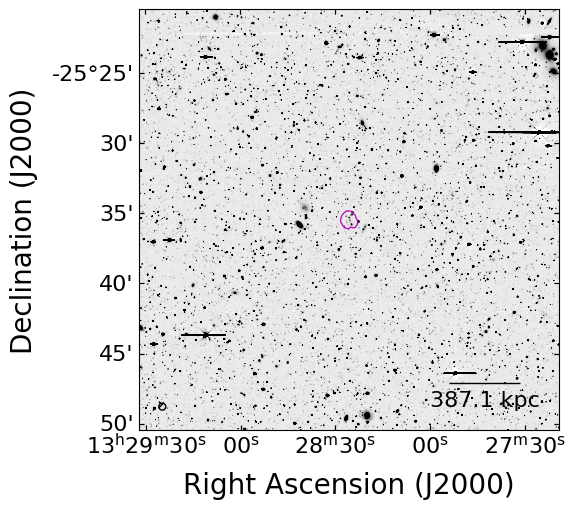}
			\caption{ }
		\end{subfigure}
		\begin{subfigure}[b]{0.26\textwidth}
			\centering
			\includegraphics[width=\textwidth]{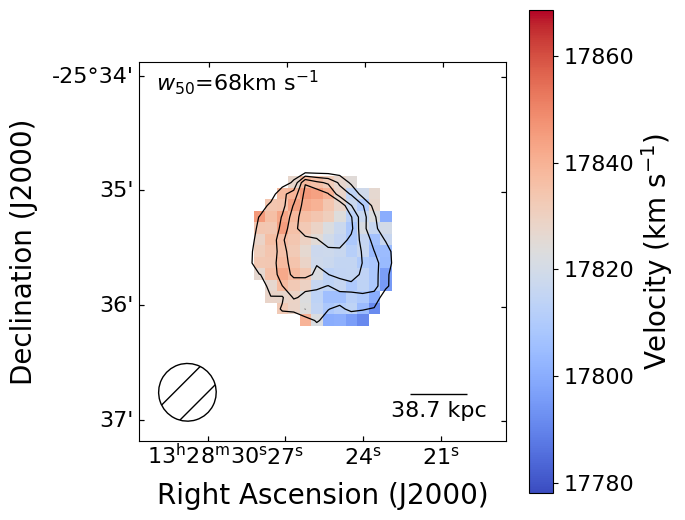}
			\caption{ }
		\end{subfigure}
    \begin{subfigure}[b]{0.24\textwidth}
			\centering
			\includegraphics[width=\textwidth]{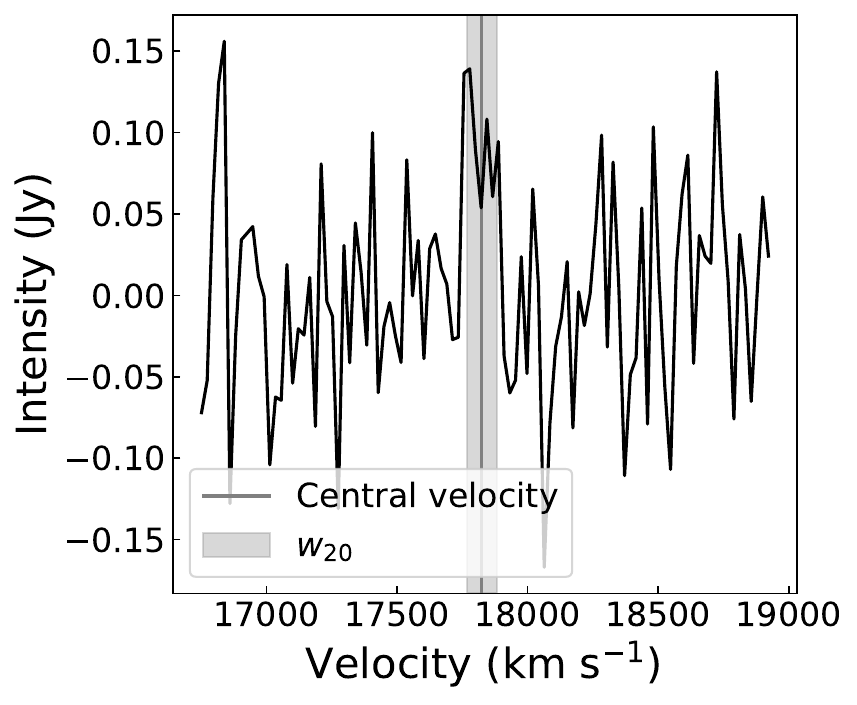}
			\caption{ }
   \end{subfigure}
	\caption{WALLABY J132825-253528 (NGC 5044). $i$-band image shown as $g$-band image is incomplete. While there is no obvious optical counterpart, we note that this source is in a crowded field.}
\end{figure*}

\begin{figure*}
	\centering
		\begin{subfigure}[b]{0.24\textwidth}
			\centering
			\includegraphics[width=\textwidth]{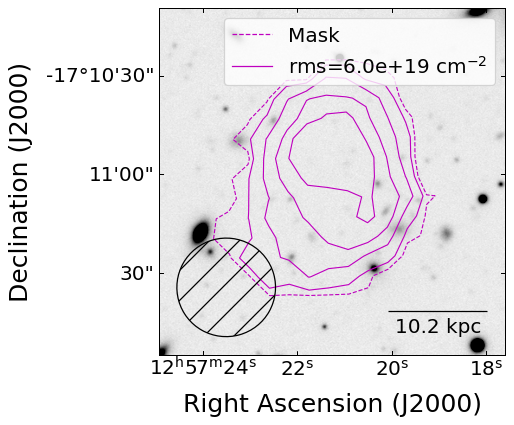}
			\caption{ }
		\end{subfigure}
		\begin{subfigure}[b]{0.24\textwidth}
			\centering
			\includegraphics[width=\textwidth]{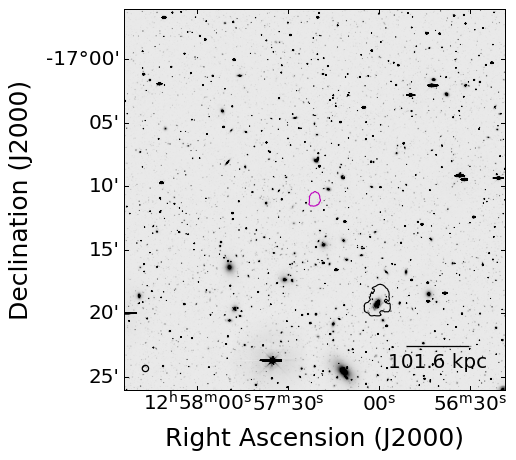}
			\caption{ }
		\end{subfigure}
		\begin{subfigure}[b]{0.26\textwidth}
			\centering
			\includegraphics[width=\textwidth]{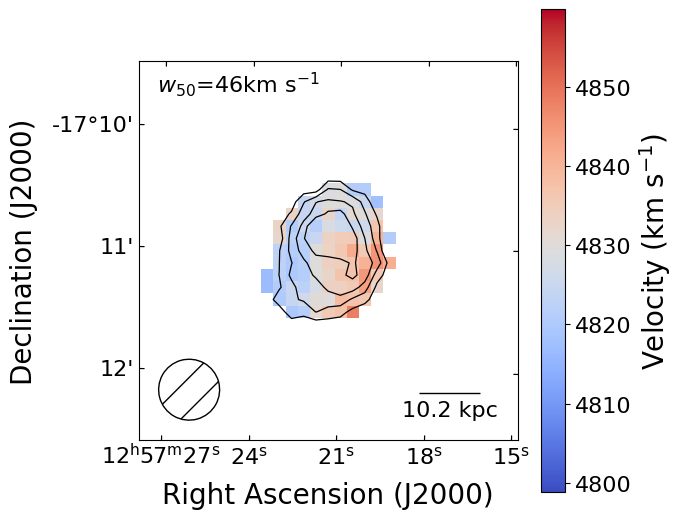}
			\caption{ }
		\end{subfigure}
    \begin{subfigure}[b]{0.24\textwidth}
			\centering
			\includegraphics[width=\textwidth]{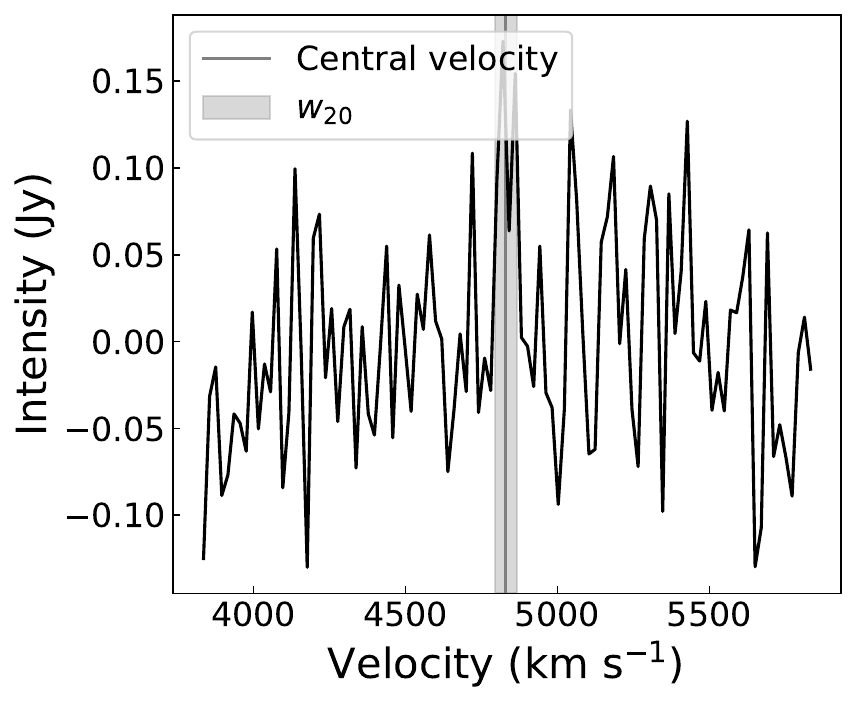}
			\caption{ }
   \end{subfigure}
	\caption{WALLABY J125721-171102 (NGC 5044)}
\end{figure*}

\begin{figure*}
	\centering
		\begin{subfigure}[b]{0.24\textwidth}
			\centering
			\includegraphics[width=\textwidth]{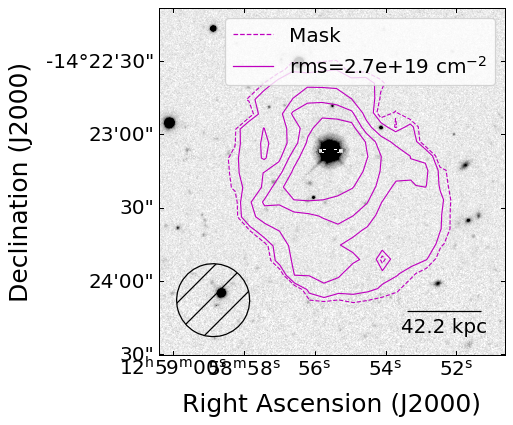}
			\caption{ }
		\end{subfigure}
		\begin{subfigure}[b]{0.24\textwidth}
			\centering
			\includegraphics[width=\textwidth]{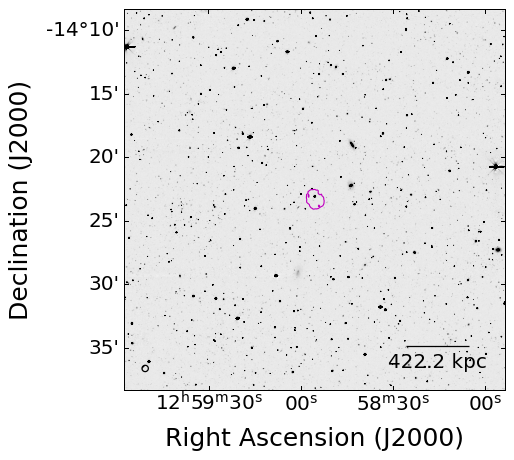}
			\caption{ }
		\end{subfigure}
		\begin{subfigure}[b]{0.26\textwidth}
			\centering
			\includegraphics[width=\textwidth]{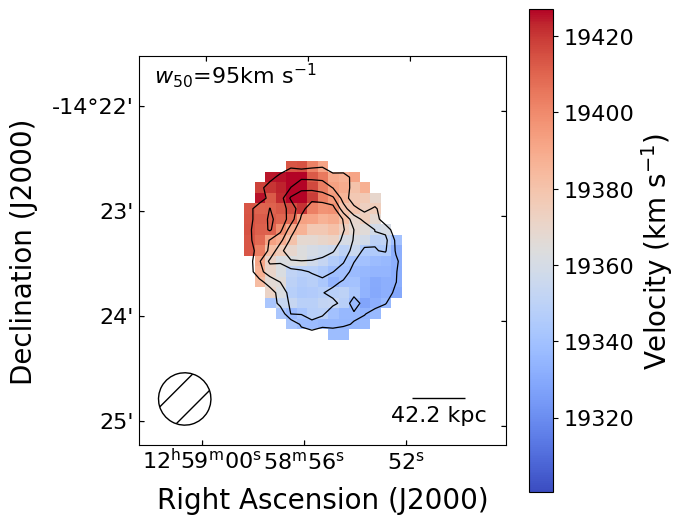}
			\caption{ }
		\end{subfigure}
    \begin{subfigure}[b]{0.24\textwidth}
			\centering
			\includegraphics[width=\textwidth]{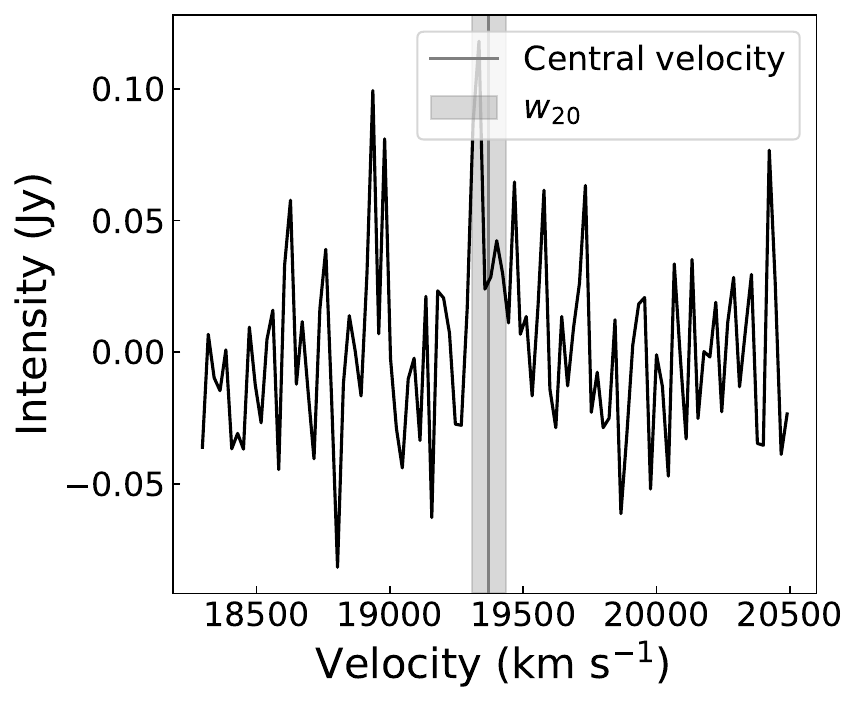}
			\caption{ }
   \end{subfigure}
	\caption{WALLABY J125855-142319 (NGC 5044)}
\end{figure*}

\begin{figure*}
	\centering
		\begin{subfigure}[b]{0.24\textwidth}
			\centering
			\includegraphics[width=\textwidth]{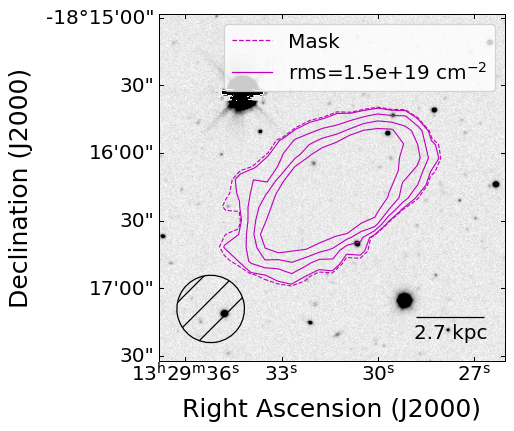}
			\caption{ }
		\end{subfigure}
		\begin{subfigure}[b]{0.24\textwidth}
			\centering
			\includegraphics[width=\textwidth]{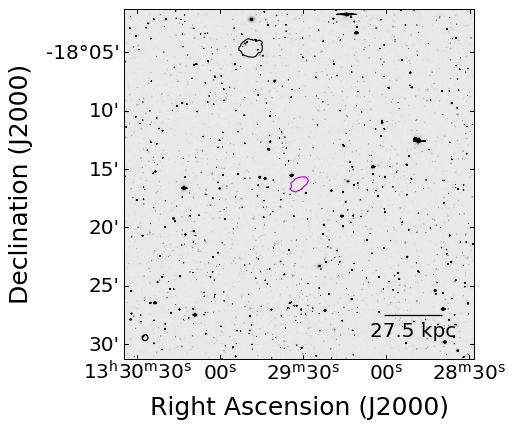}
			\caption{ }
		\end{subfigure}
		\begin{subfigure}[b]{0.26\textwidth}
			\centering
			\includegraphics[width=\textwidth]{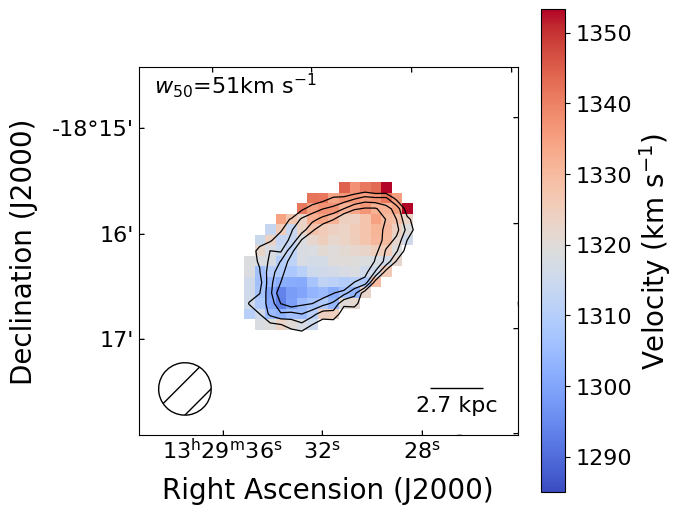}
			\caption{ }
		\end{subfigure}
    \begin{subfigure}[b]{0.24\textwidth}
			\centering
			\includegraphics[width=\textwidth]{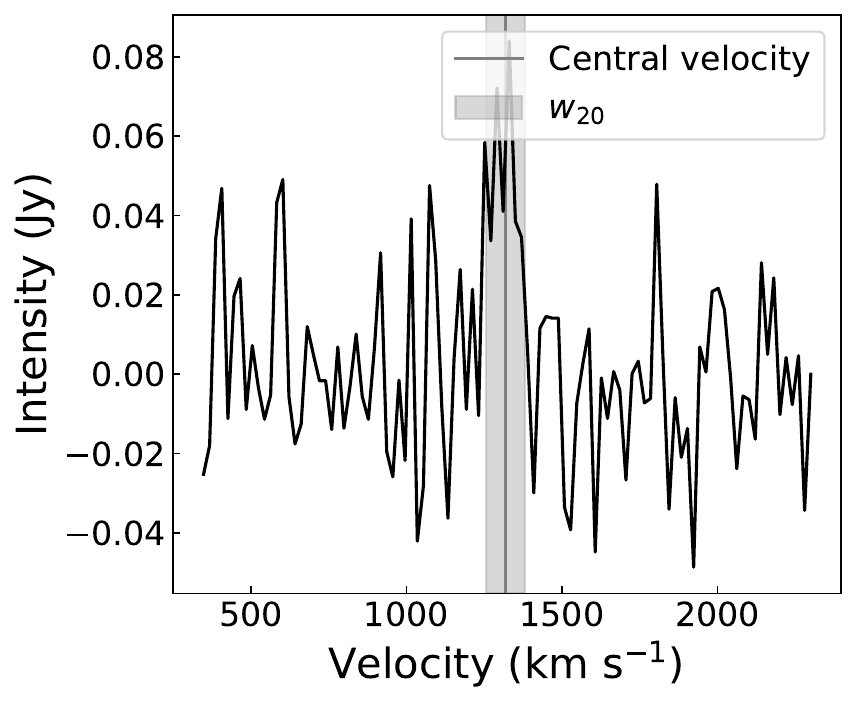}
			\caption{ }
   \end{subfigure}
	\caption{WALLABY J132931-181615 (NGC 5044)}
\end{figure*}


\begin{figure*}
	\centering
		\begin{subfigure}[b]{0.24\textwidth}
			\centering
			\includegraphics[width=\textwidth]{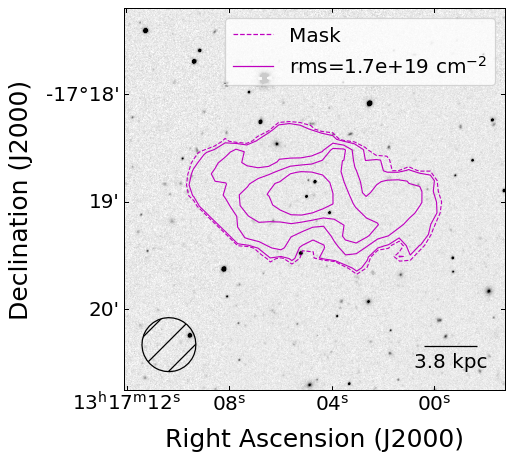}
			\caption{ }
		\end{subfigure}
		\begin{subfigure}[b]{0.24\textwidth}
			\centering
			\includegraphics[width=\textwidth]{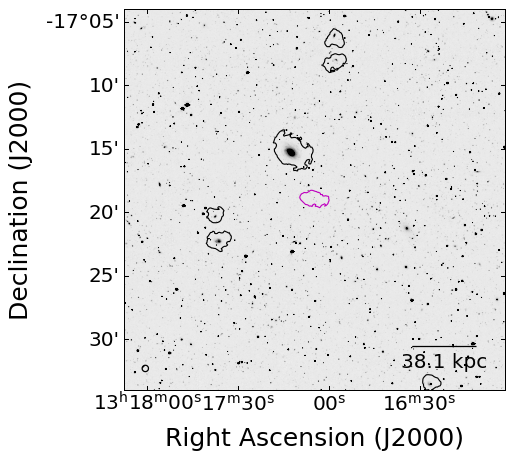}
			\caption{ }
		\end{subfigure}
		\begin{subfigure}[b]{0.26\textwidth}
			\centering
			\includegraphics[width=\textwidth]{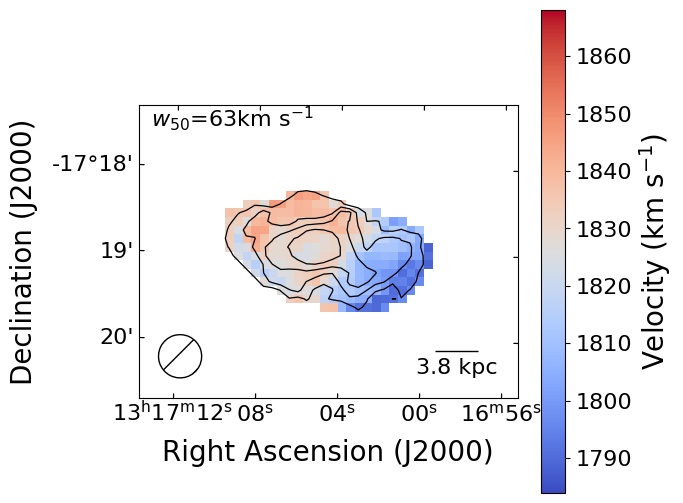}
			\caption{ }
		\end{subfigure}
    \begin{subfigure}[b]{0.24\textwidth}
			\centering
			\includegraphics[width=\textwidth]{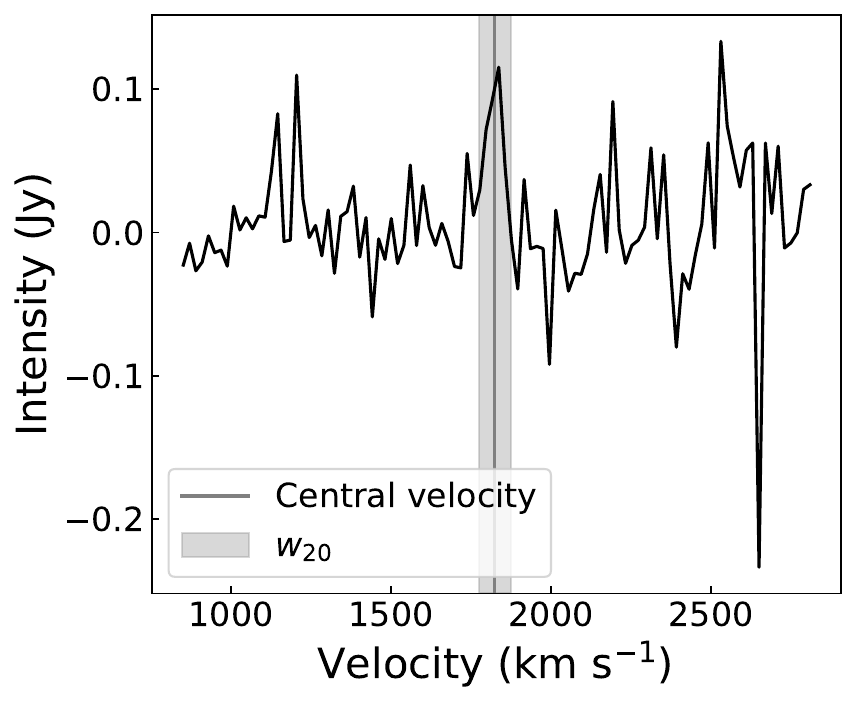}
			\caption{ }
   \end{subfigure}
	\caption{WALLABY J131704-171858 (NGC 5044)}
\end{figure*}

\begin{figure*}
	\centering
		\begin{subfigure}[b]{0.24\textwidth}
			\centering
			\includegraphics[width=\textwidth]{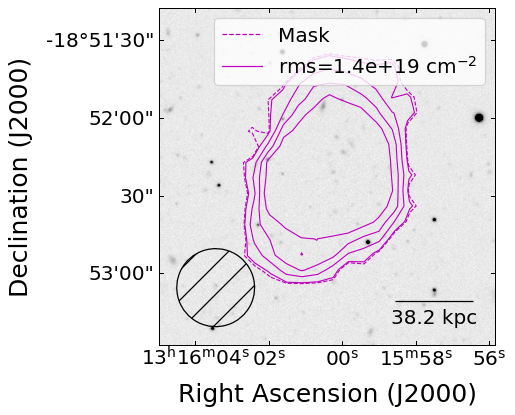}
			\caption{ }
		\end{subfigure}
		\begin{subfigure}[b]{0.24\textwidth}
			\centering
			\includegraphics[width=\textwidth]{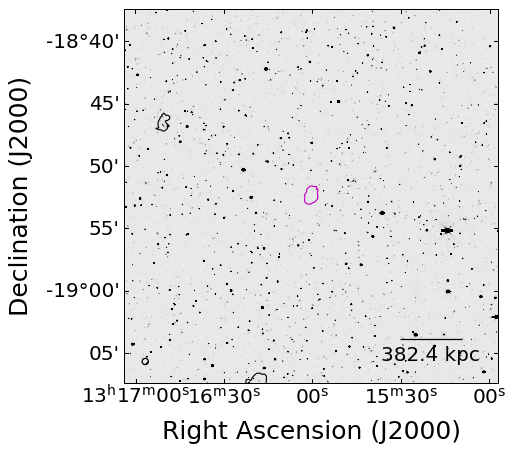}
			\caption{ }
		\end{subfigure}
		\begin{subfigure}[b]{0.26\textwidth}
			\centering
			\includegraphics[width=\textwidth]{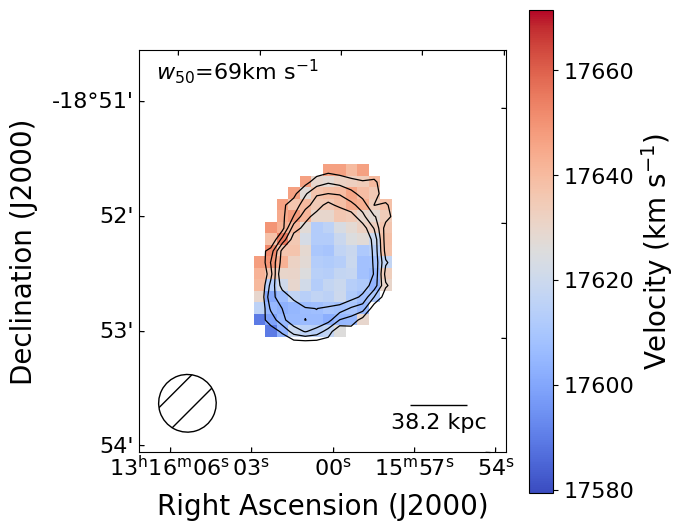}
			\caption{ }
		\end{subfigure}
  \begin{subfigure}[b]{0.24\textwidth}
			\centering
			\includegraphics[width=\textwidth]{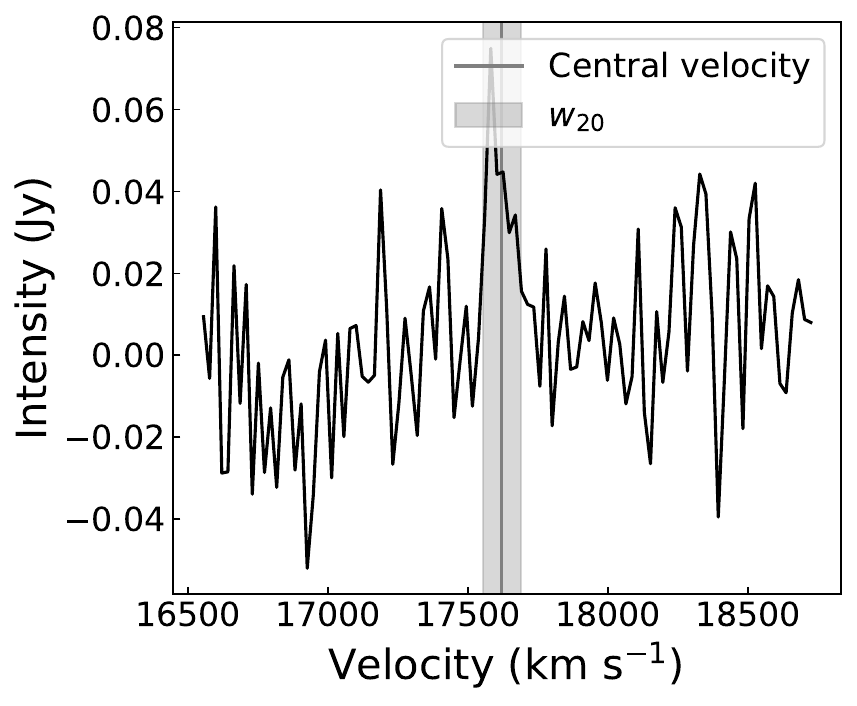}
			\caption{ }
   \end{subfigure}
	\caption{WALLABY J131600-185222 (NGC 5044)}
\end{figure*}
 
\begin{figure*}
	\centering
		\begin{subfigure}[b]{0.24\textwidth}
			\centering
			\includegraphics[width=\textwidth]{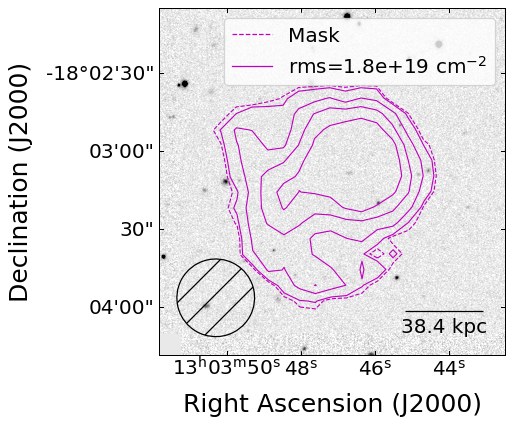}
			\caption{ }
		\end{subfigure}
		\begin{subfigure}[b]{0.24\textwidth}
			\centering
			\includegraphics[width=\textwidth]{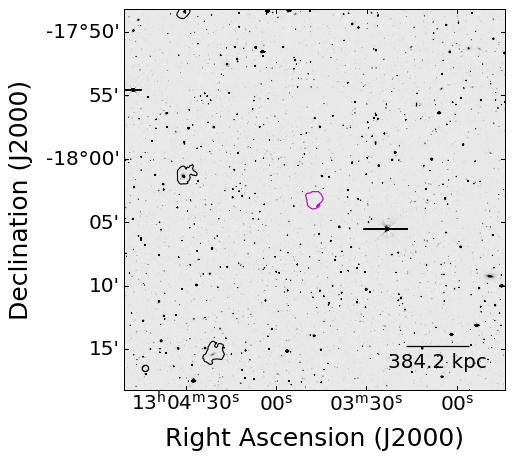}
			\caption{ }
		\end{subfigure}
		\begin{subfigure}[b]{0.26\textwidth}
			\centering
			\includegraphics[width=\textwidth]{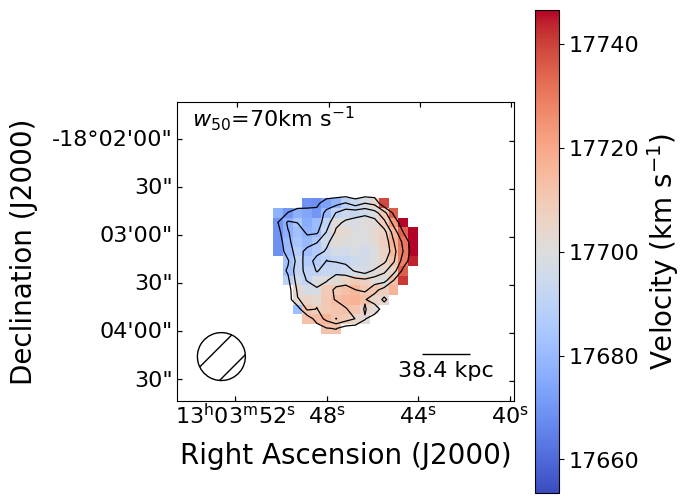}
			\caption{ }
		\end{subfigure}
  \begin{subfigure}[b]{0.24\textwidth}
			\centering
			\includegraphics[width=\textwidth]{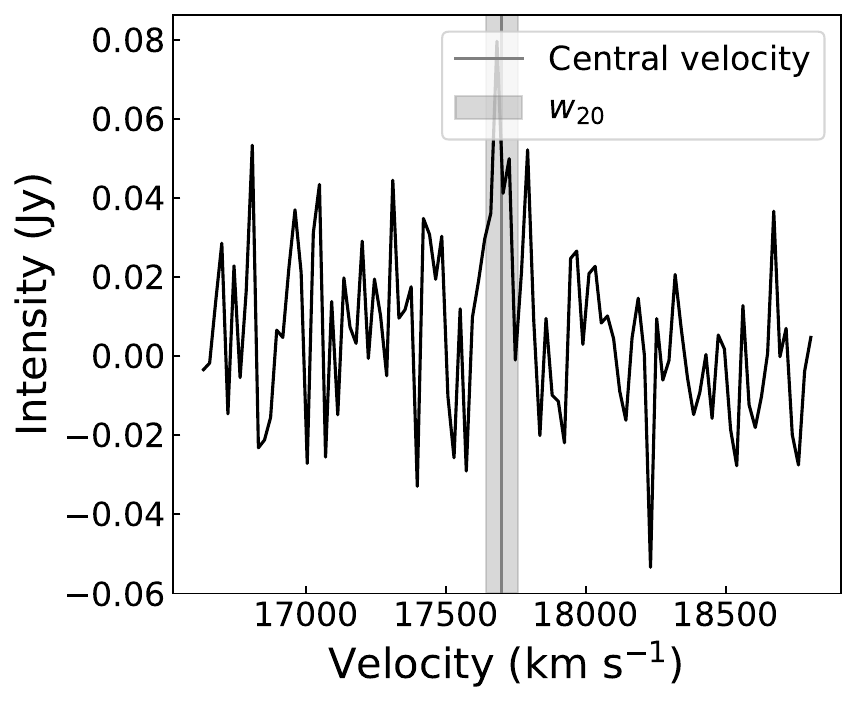}
			\caption{ }
   \end{subfigure}
	\caption{WALLABY J130347-180311 (NGC 5044)}
\end{figure*}
 
\begin{figure*}
	\centering
		\begin{subfigure}[b]{0.24\textwidth}
			\centering
			\includegraphics[width=\textwidth]{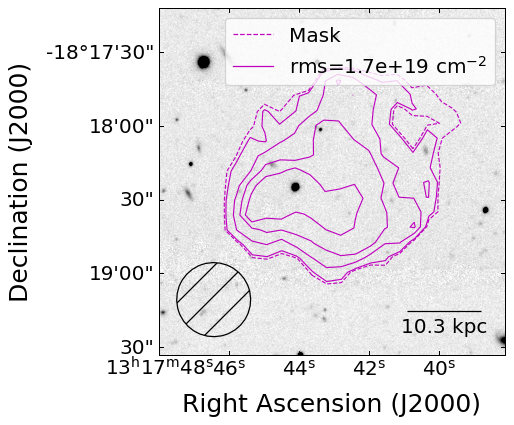}
			\caption{ }
		\end{subfigure}
		\begin{subfigure}[b]{0.24\textwidth}
			\centering
			\includegraphics[width=\textwidth]{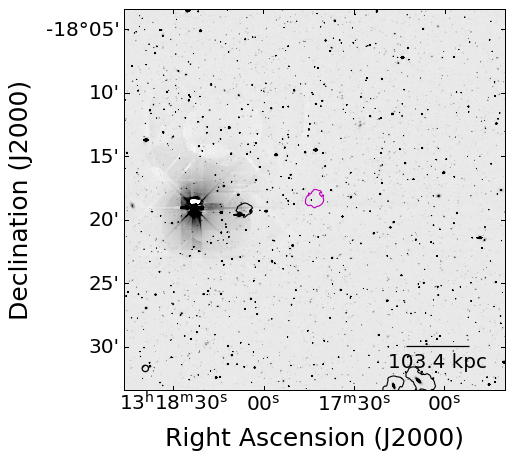}
			\caption{ }
		\end{subfigure}
		\begin{subfigure}[b]{0.26\textwidth}
			\centering
			\includegraphics[width=\textwidth]{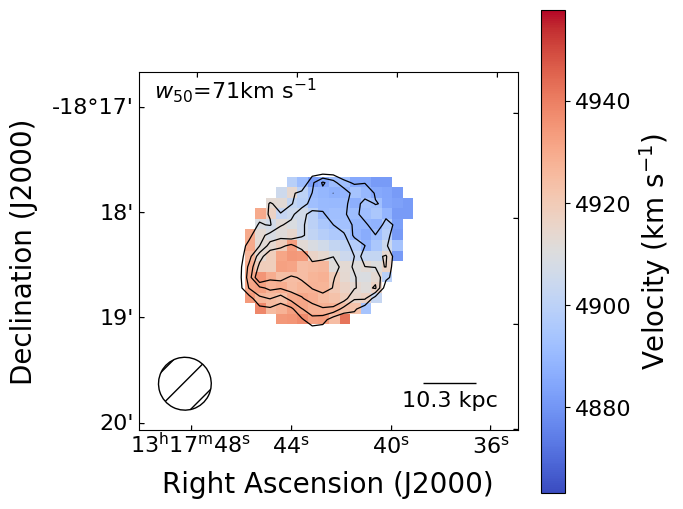}
			\caption{ }
		\end{subfigure}
  \begin{subfigure}[b]{0.24\textwidth}
			\centering
			\includegraphics[width=\textwidth]{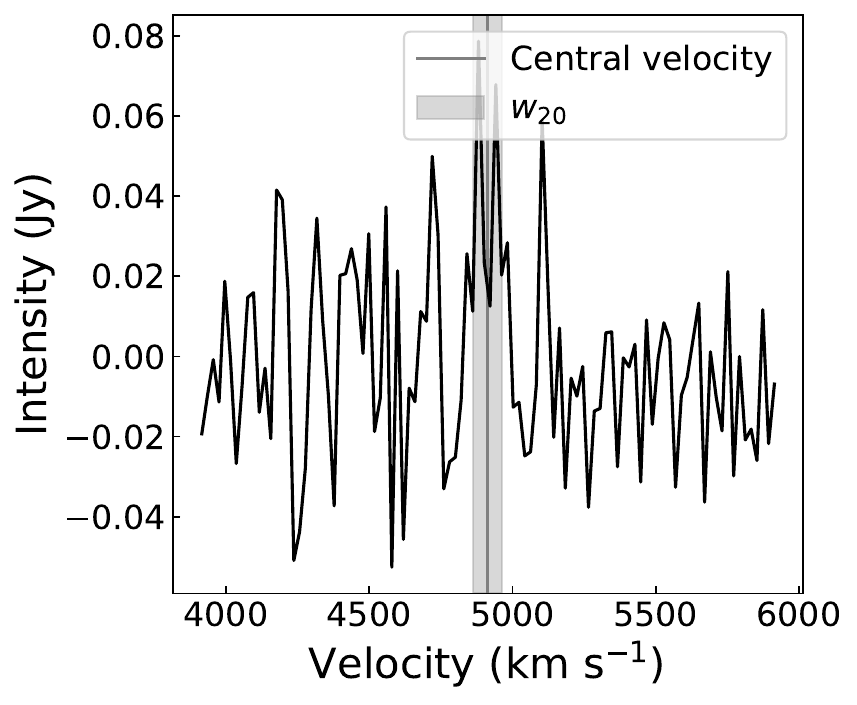}
			\caption{ }
   \end{subfigure}
	\caption{WALLABY J131743-181822  (NGC 5044)}
\end{figure*}
 
\begin{figure*}
	\centering
		\begin{subfigure}[b]{0.24\textwidth}
			\centering
			\includegraphics[width=\textwidth]{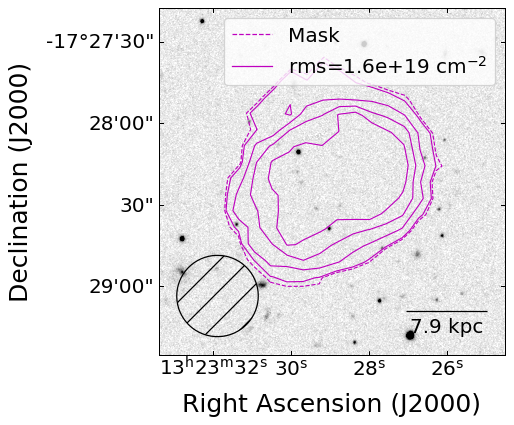}
			\caption{ }
		\end{subfigure}
		\begin{subfigure}[b]{0.24\textwidth}
			\centering
			\includegraphics[width=\textwidth]{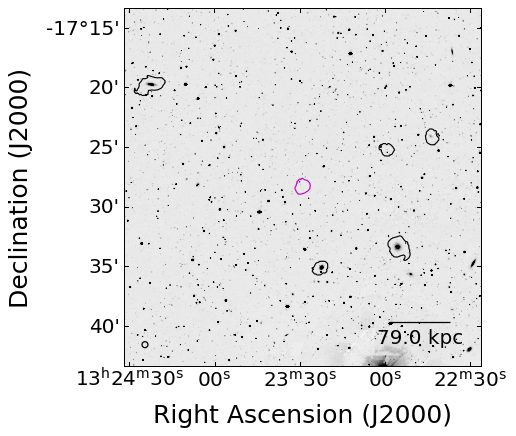}
			\caption{ }
		\end{subfigure}
		\begin{subfigure}[b]{0.26\textwidth}
			\centering
			\includegraphics[width=\textwidth]{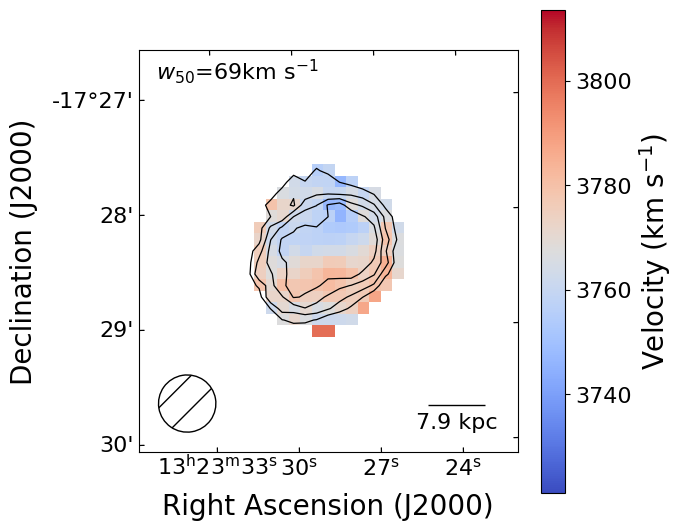}
			\caption{ }
		\end{subfigure}
  \begin{subfigure}[b]{0.24\textwidth}
			\centering
			\includegraphics[width=\textwidth]{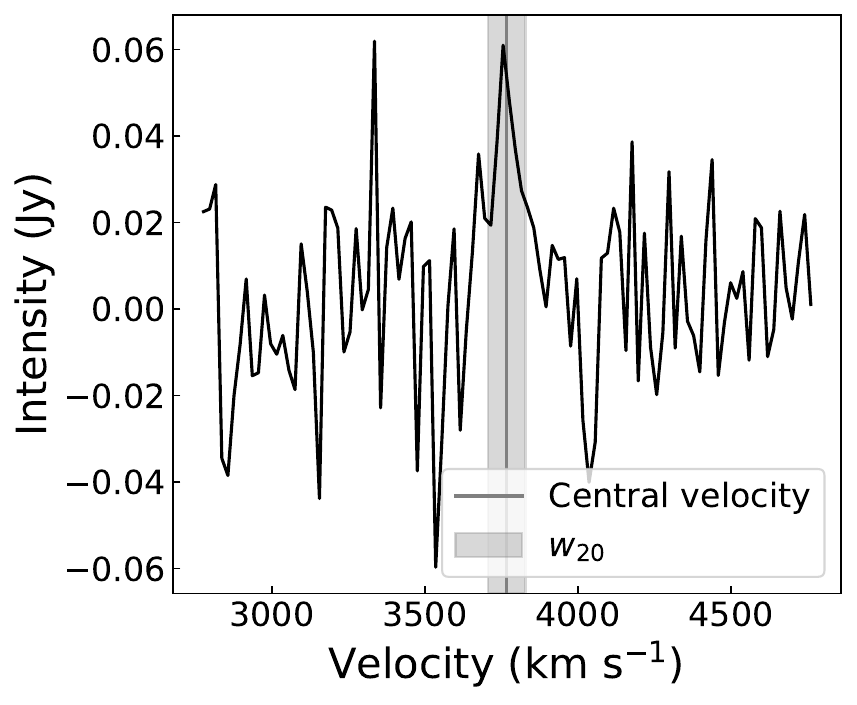}
			\caption{ }
   \end{subfigure}
	\caption{WALLABY J132328-172821 (NGC 5044)}
\end{figure*}
 
\begin{figure*}
	\centering
		\begin{subfigure}[b]{0.24\textwidth}
			\centering
			\includegraphics[width=\textwidth]{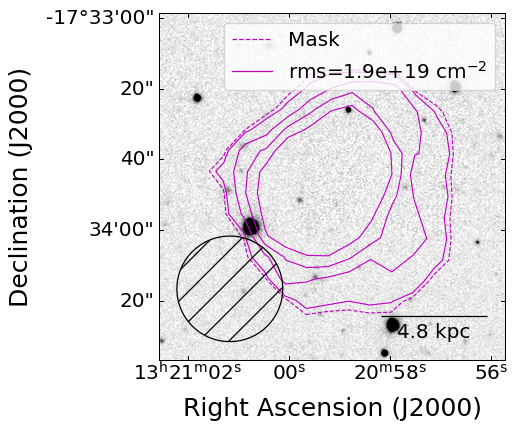}
			\caption{ }
		\end{subfigure}
		\begin{subfigure}[b]{0.24\textwidth}
			\centering
			\includegraphics[width=\textwidth]{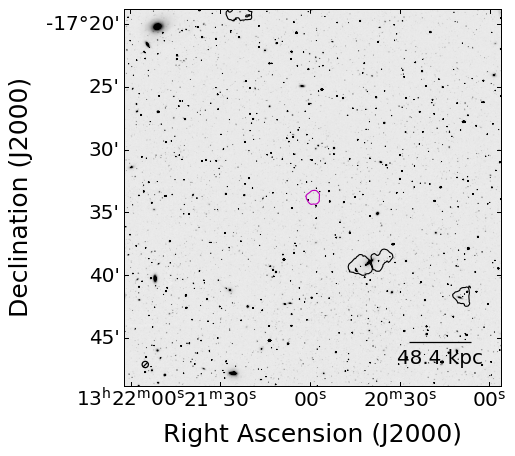}
			\caption{ }
		\end{subfigure}
		\begin{subfigure}[b]{0.26\textwidth}
			\centering
			\includegraphics[width=\textwidth]{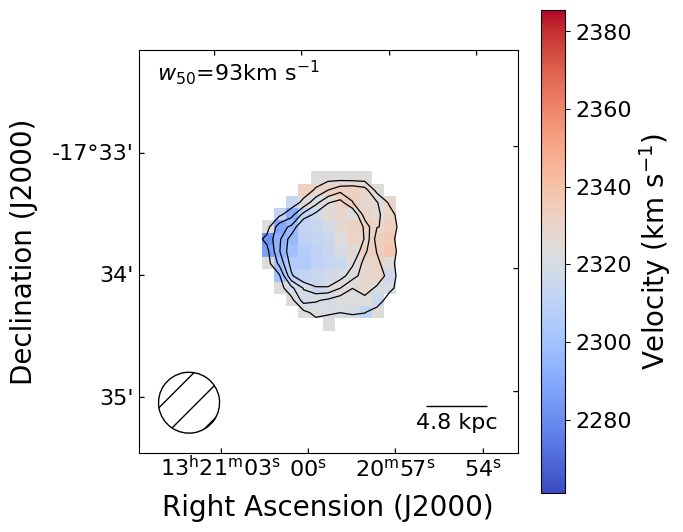}
			\caption{ }
		\end{subfigure}
  \begin{subfigure}[b]{0.24\textwidth}
			\centering
			\includegraphics[width=\textwidth]{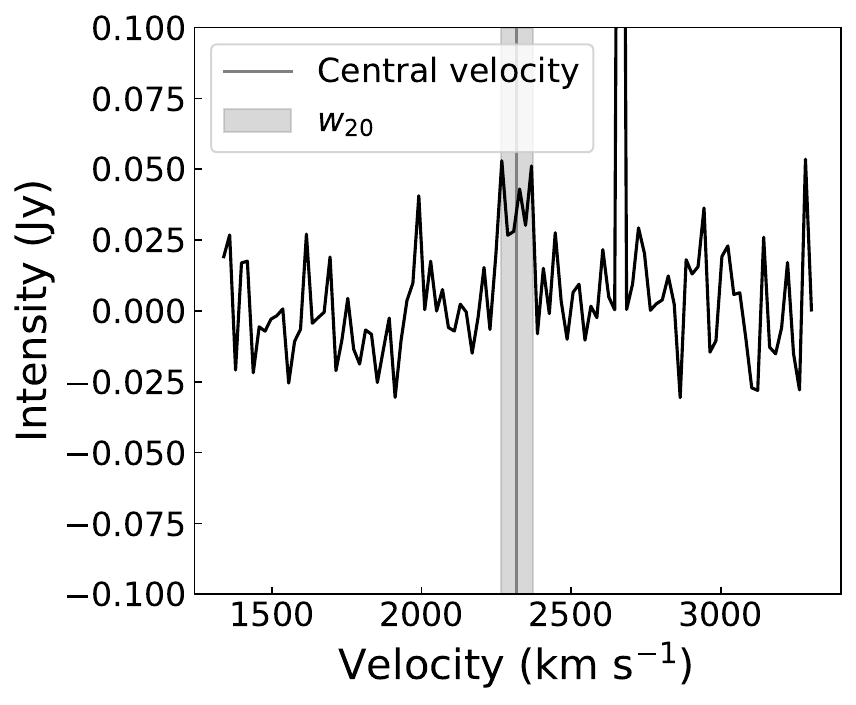}
			\caption{ }
   \end{subfigure}
	\caption{WALLABY J132059-173347 (NGC 5044)}
\end{figure*}

\begin{figure*}
	\centering
		\begin{subfigure}[b]{0.24\textwidth}
			\centering
			\includegraphics[width=\textwidth]{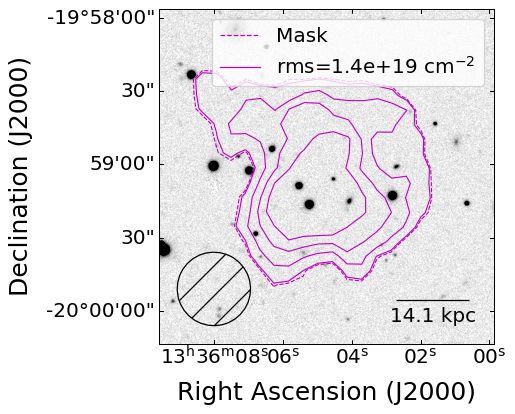}
			\caption{ }
		\end{subfigure}
		\begin{subfigure}[b]{0.24\textwidth}
			\centering
			\includegraphics[width=\textwidth]{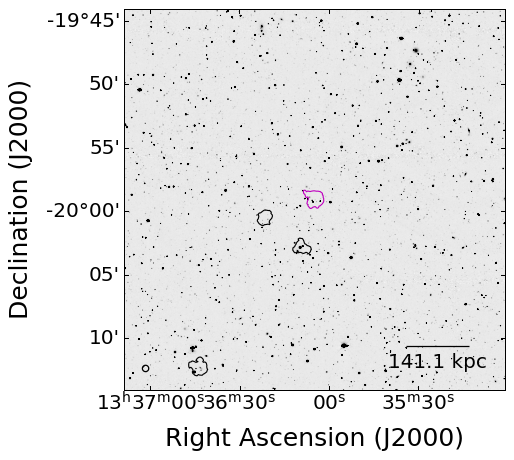}
			\caption{ }
		\end{subfigure}
		\begin{subfigure}[b]{0.26\textwidth}
			\centering
			\includegraphics[width=\textwidth]{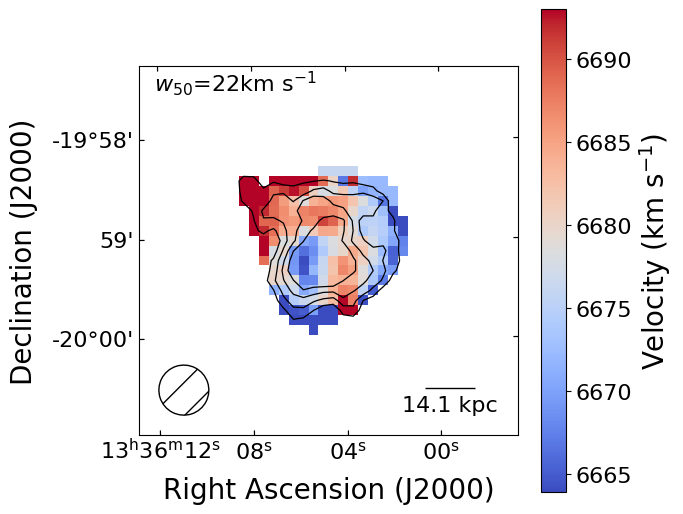}
			\caption{ }
		\end{subfigure}
  \begin{subfigure}[b]{0.24\textwidth}
			\centering
			\includegraphics[width=\textwidth]{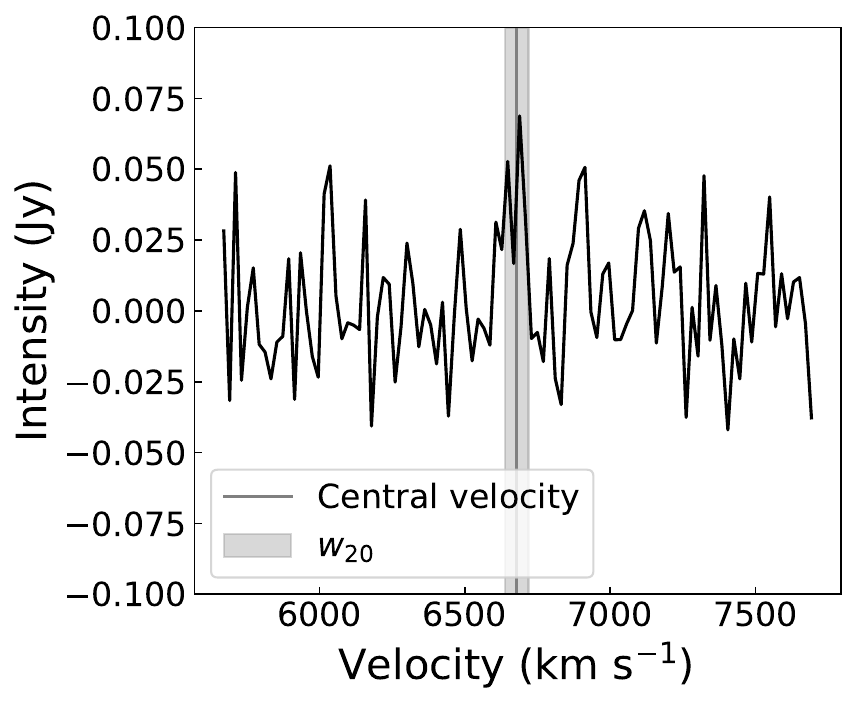}
			\caption{ }
   \end{subfigure}
	\caption{WALLABY J133604-195904 (NGC 5044)}
\end{figure*}
 
\begin{figure*}
	\centering
		\begin{subfigure}[b]{0.24\textwidth}
			\centering
			\includegraphics[width=\textwidth]{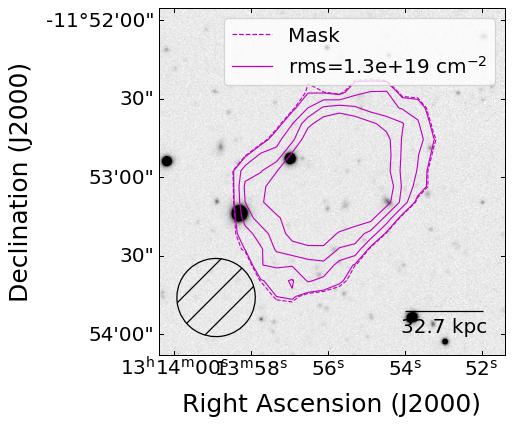}
			\caption{ }
		\end{subfigure}
		\begin{subfigure}[b]{0.24\textwidth}
			\centering
			\includegraphics[width=\textwidth]{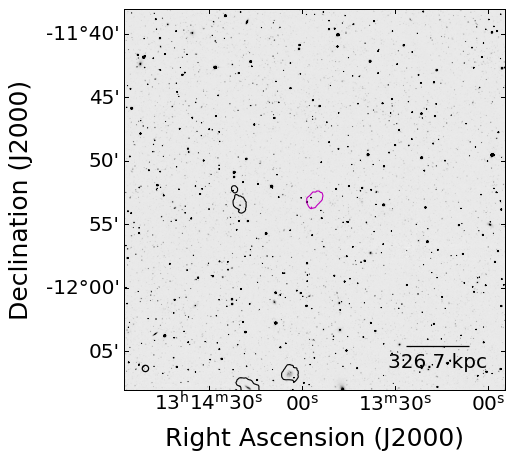}
			\caption{ }
		\end{subfigure}
		\begin{subfigure}[b]{0.26\textwidth}
			\centering
			\includegraphics[width=\textwidth]{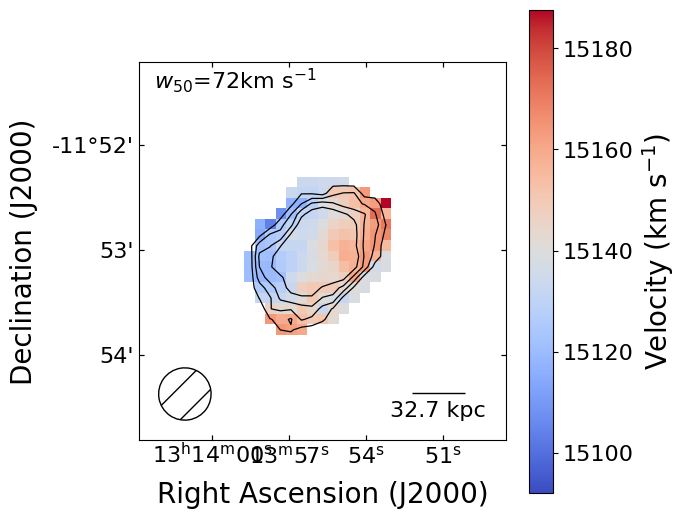}
			\caption{ }
		\end{subfigure}
  \begin{subfigure}[b]{0.24\textwidth}
			\centering
			\includegraphics[width=\textwidth]{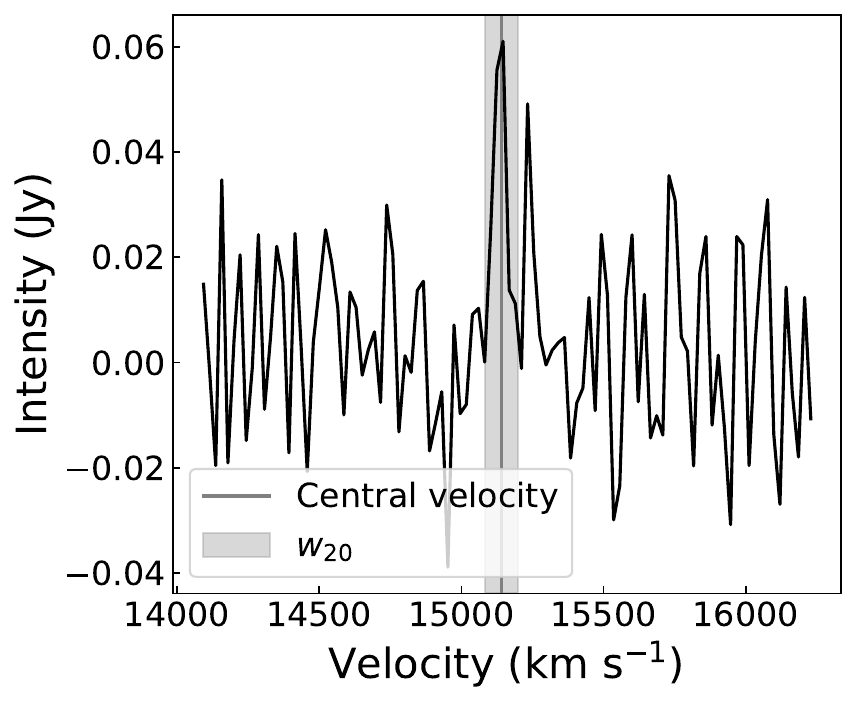}
			\caption{ }
   \end{subfigure}
	\caption{WALLABY J131355-115301 (NGC 5044)}
\end{figure*}
 
\begin{figure*}
	\centering
		\begin{subfigure}[b]{0.24\textwidth}
			\centering
			\includegraphics[width=\textwidth]{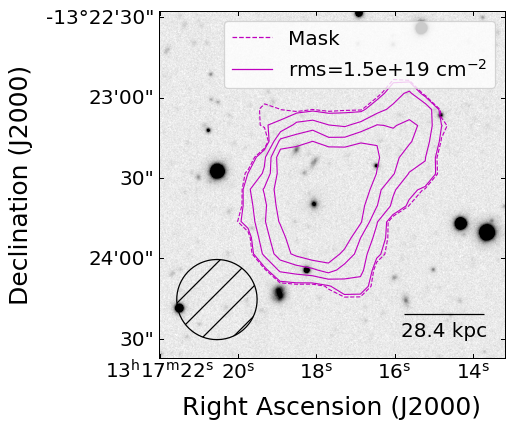}
			\caption{ }
		\end{subfigure}
		\begin{subfigure}[b]{0.24\textwidth}
			\centering
			\includegraphics[width=\textwidth]{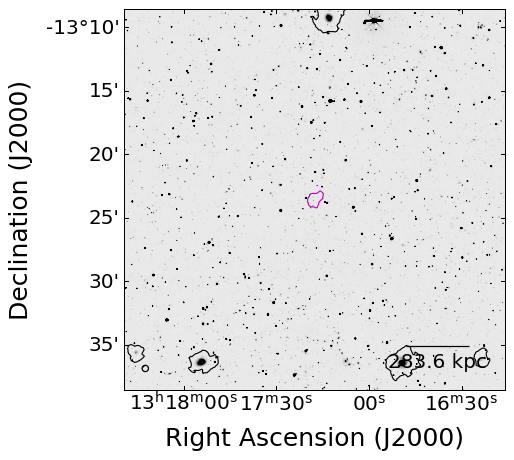}
			\caption{ }
		\end{subfigure}
		\begin{subfigure}[b]{0.26\textwidth}
			\centering
			\includegraphics[width=\textwidth]{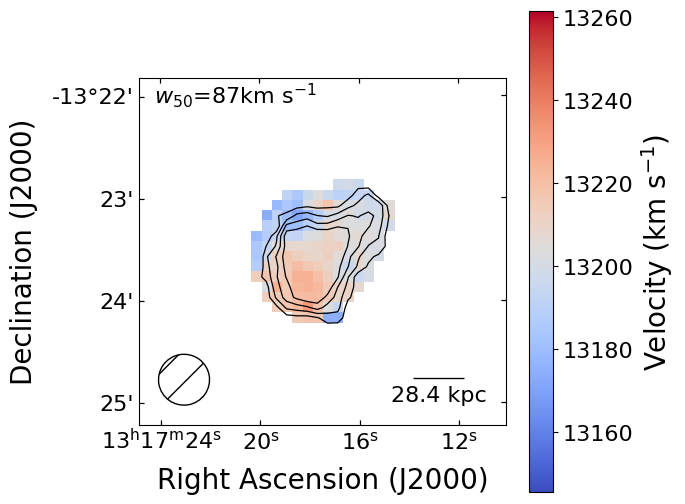}
			\caption{ }
		\end{subfigure}
  \begin{subfigure}[b]{0.24\textwidth}
			\centering
			\includegraphics[width=\textwidth]{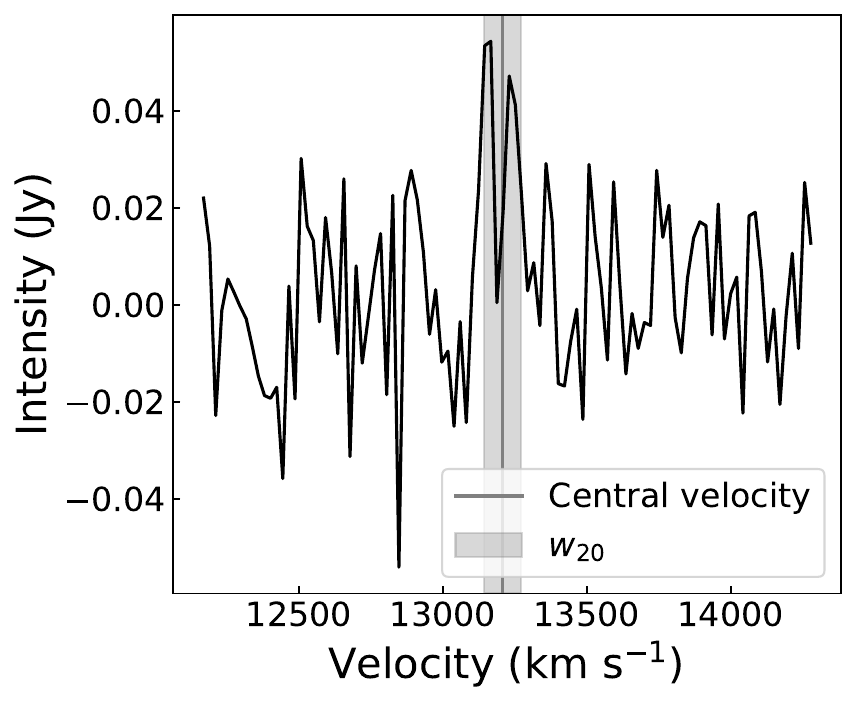}
			\caption{ }
   \end{subfigure}
	\caption{WALLABY J131717-132332 (NGC 5044)}
\end{figure*}

\begin{figure*}
	\centering
		\begin{subfigure}[b]{0.24\textwidth}
			\centering
			\includegraphics[width=\textwidth]{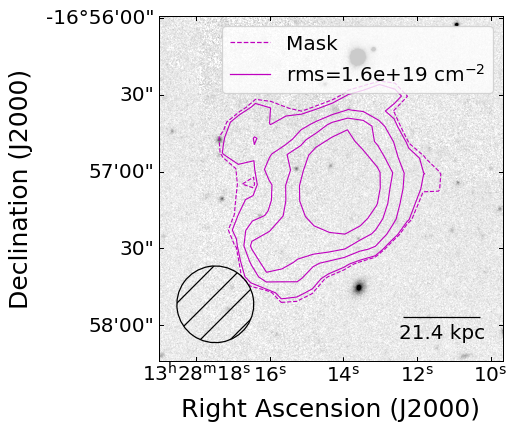}
			\caption{ }
		\end{subfigure}
		\begin{subfigure}[b]{0.24\textwidth}
			\centering
			\includegraphics[width=\textwidth]{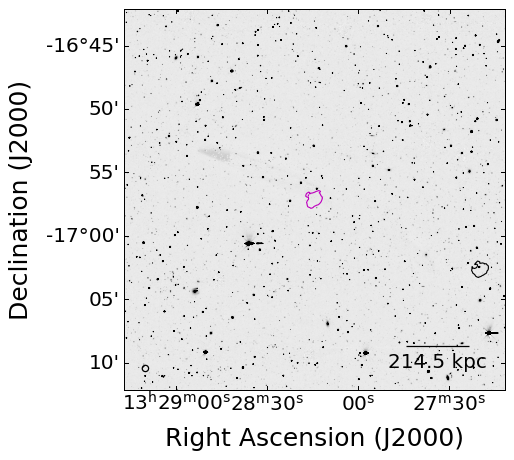}
			\caption{ }
		\end{subfigure}
		\begin{subfigure}[b]{0.26\textwidth}
			\centering
			\includegraphics[width=\textwidth]{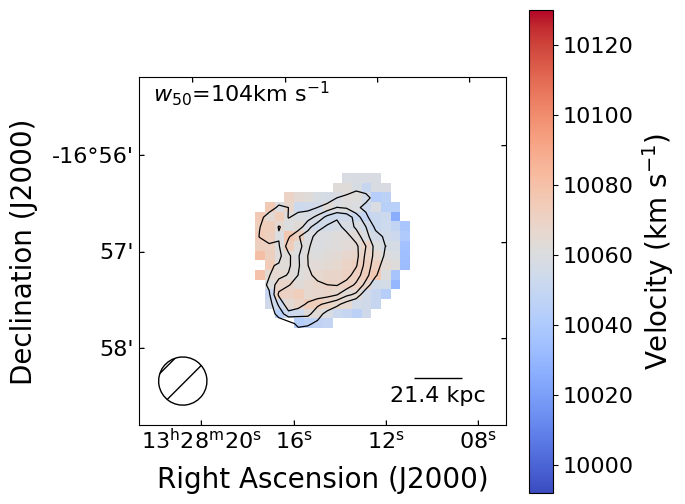}
			\caption{ }
		\end{subfigure}
  \begin{subfigure}[b]{0.24\textwidth}
			\centering
			\includegraphics[width=\textwidth]{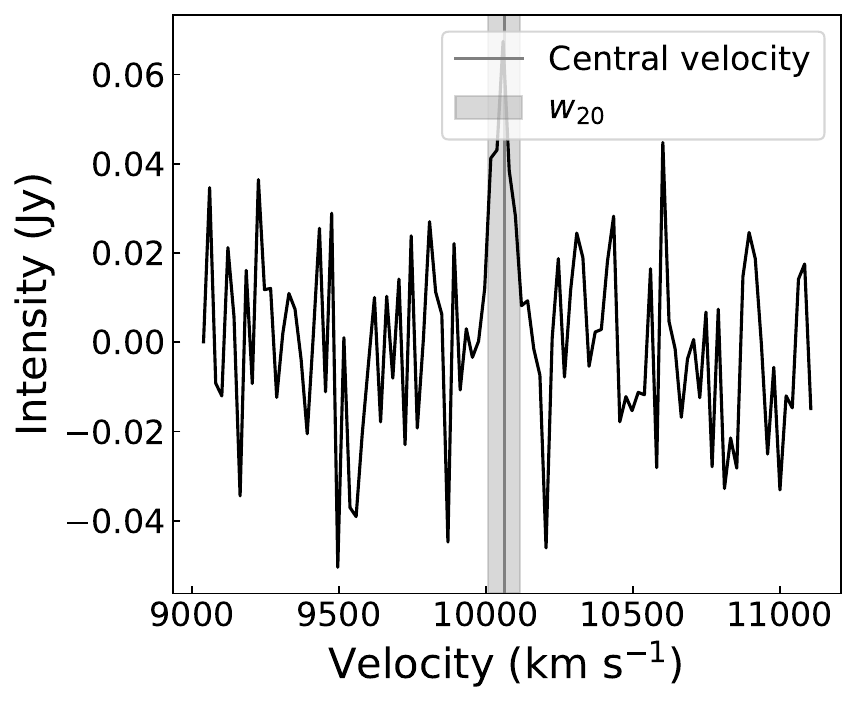}
			\caption{ }
   \end{subfigure}
	\caption{WALLABY J132814-165706 (NGC 5044)}
\end{figure*}

\begin{figure*}
	\centering
		\begin{subfigure}[b]{0.24\textwidth}
			\centering
			\includegraphics[width=\textwidth]{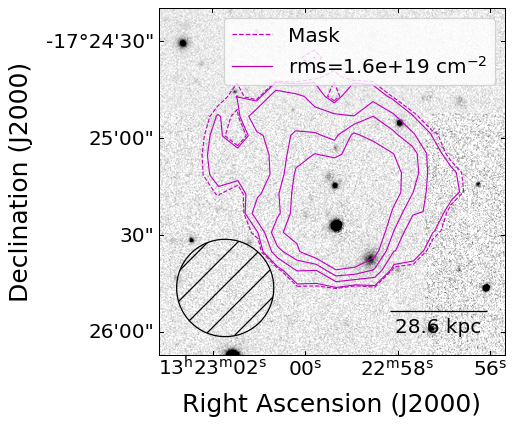}
			\caption{ }
		\end{subfigure}
		\begin{subfigure}[b]{0.24\textwidth}
			\centering
			\includegraphics[width=\textwidth]{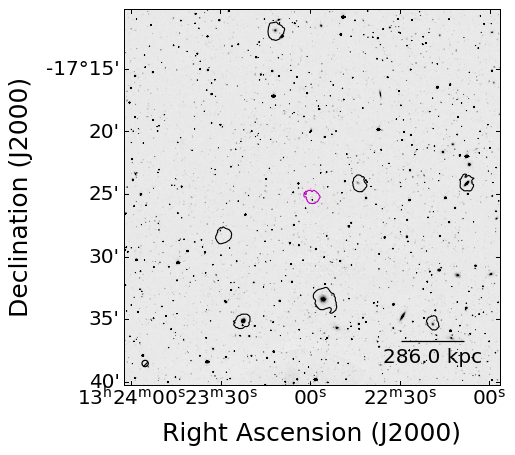}
			\caption{ }
		\end{subfigure}
		\begin{subfigure}[b]{0.26\textwidth}
			\centering
			\includegraphics[width=\textwidth]{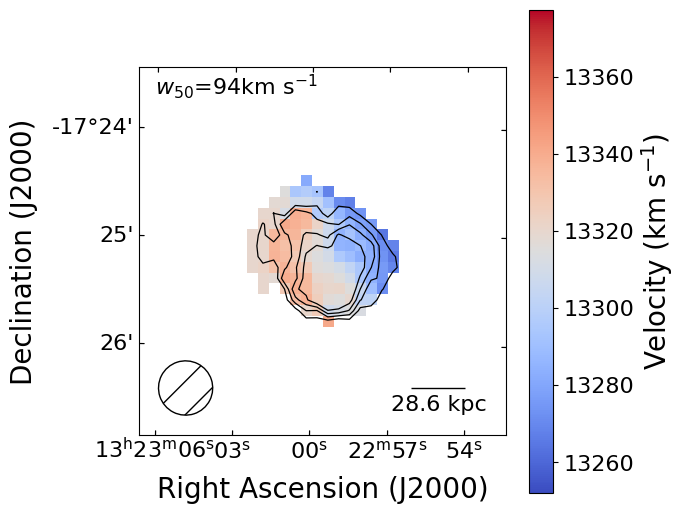}
			\caption{ }
		\end{subfigure}
  \begin{subfigure}[b]{0.24\textwidth}
			\centering
			\includegraphics[width=\textwidth]{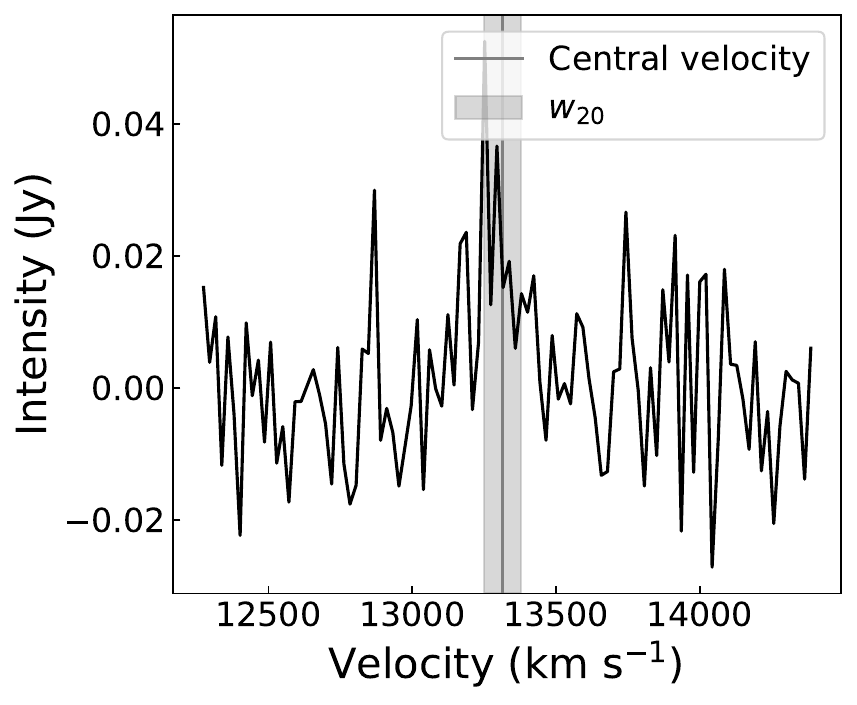}
			\caption{ }
   \end{subfigure}
	\caption{WALLABY J132259-172513 (NGC 5044)}
\end{figure*}
 
\begin{figure*}
	\centering
		\begin{subfigure}[b]{0.24\textwidth}
			\centering
			\includegraphics[width=\textwidth]{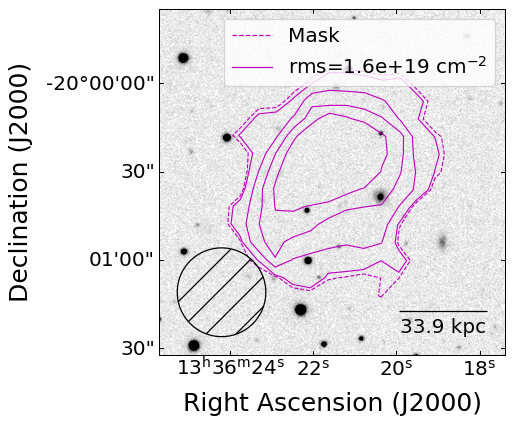}
			\caption{ }
		\end{subfigure}
		\begin{subfigure}[b]{0.24\textwidth}
			\centering
			\includegraphics[width=\textwidth]{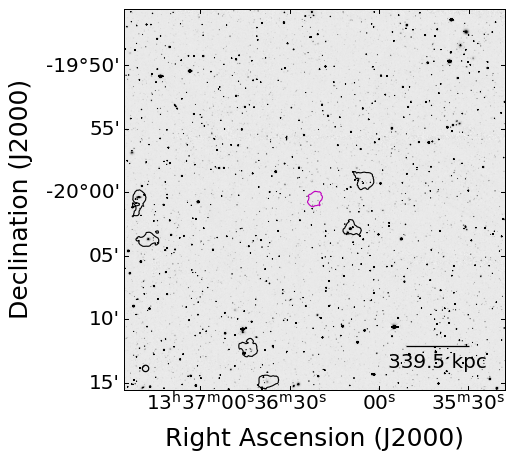}
			\caption{ }
		\end{subfigure}
		\begin{subfigure}[b]{0.26\textwidth}
			\centering
			\includegraphics[width=\textwidth]{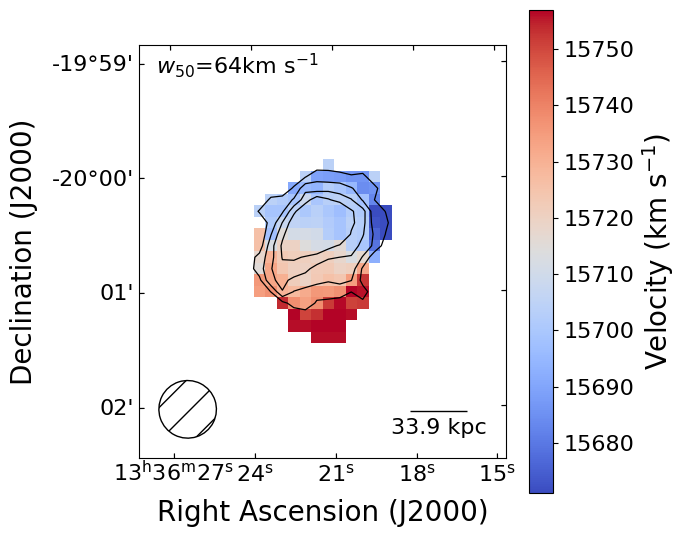}
			\caption{ }
		\end{subfigure}
  \begin{subfigure}[b]{0.24\textwidth}
			\centering
			\includegraphics[width=\textwidth]{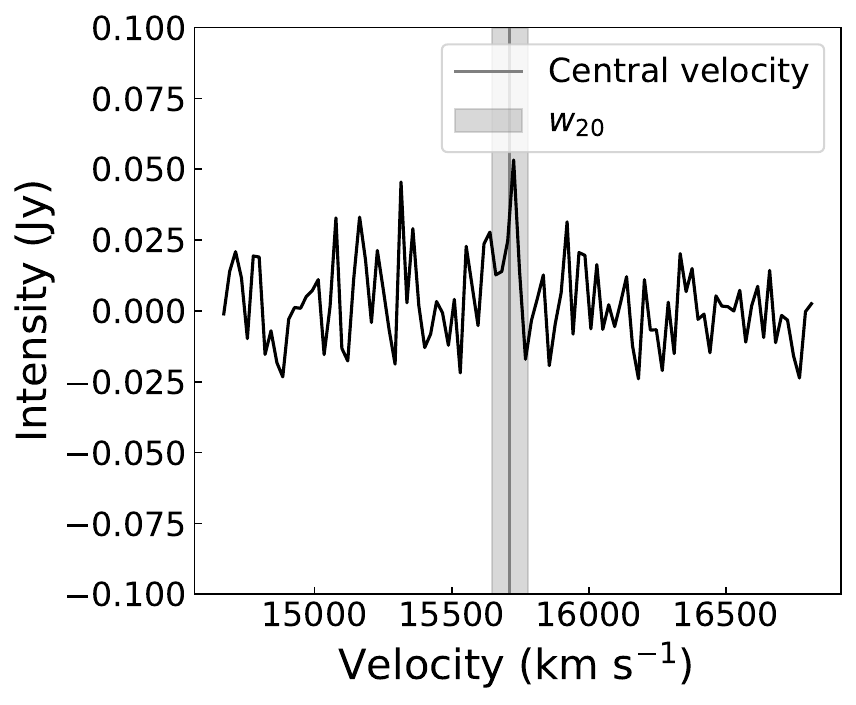}
			\caption{ }
   \end{subfigure}
	\caption{WALLABY J133621-200033 (NGC 5044)}
\end{figure*}
 
\begin{figure*}
	\centering
		\begin{subfigure}[b]{0.24\textwidth}
			\centering
			\includegraphics[width=\textwidth]{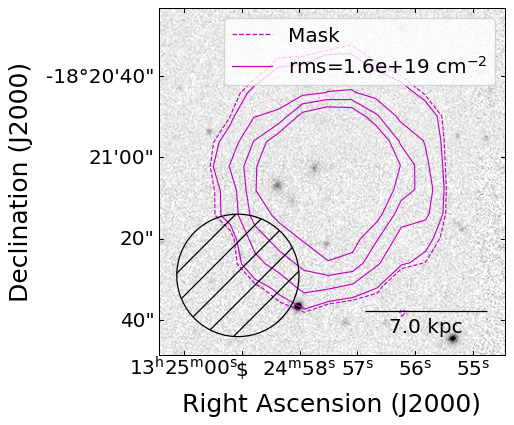}
			\caption{ }
		\end{subfigure}
		\begin{subfigure}[b]{0.24\textwidth}
			\centering
			\includegraphics[width=\textwidth]{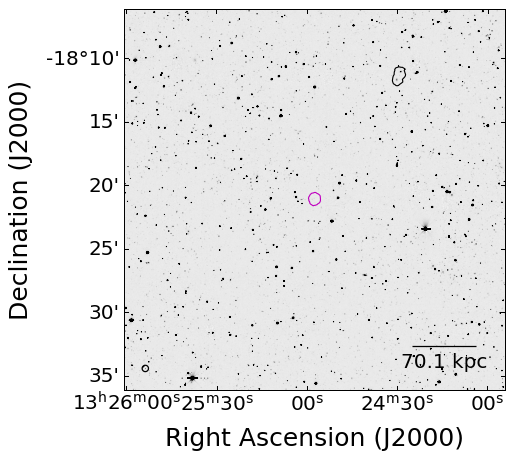}
			\caption{ }
		\end{subfigure}
		\begin{subfigure}[b]{0.26\textwidth}
			\centering
			\includegraphics[width=\textwidth]{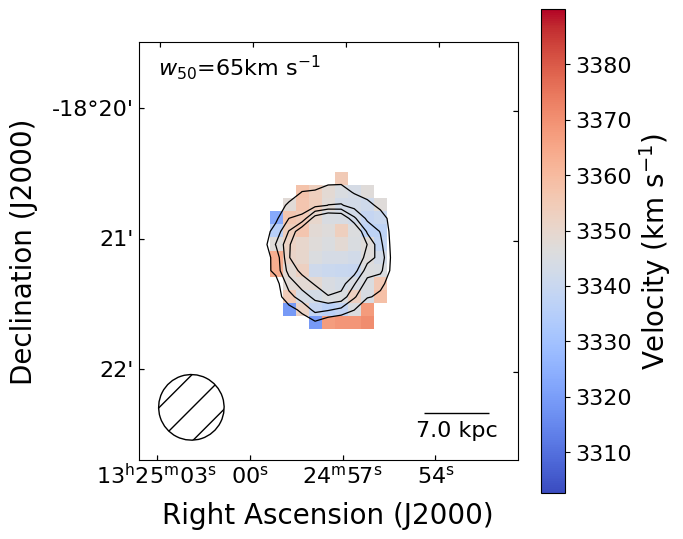}
			\caption{ }
		\end{subfigure}
  \begin{subfigure}[b]{0.24\textwidth}
			\centering
			\includegraphics[width=\textwidth]{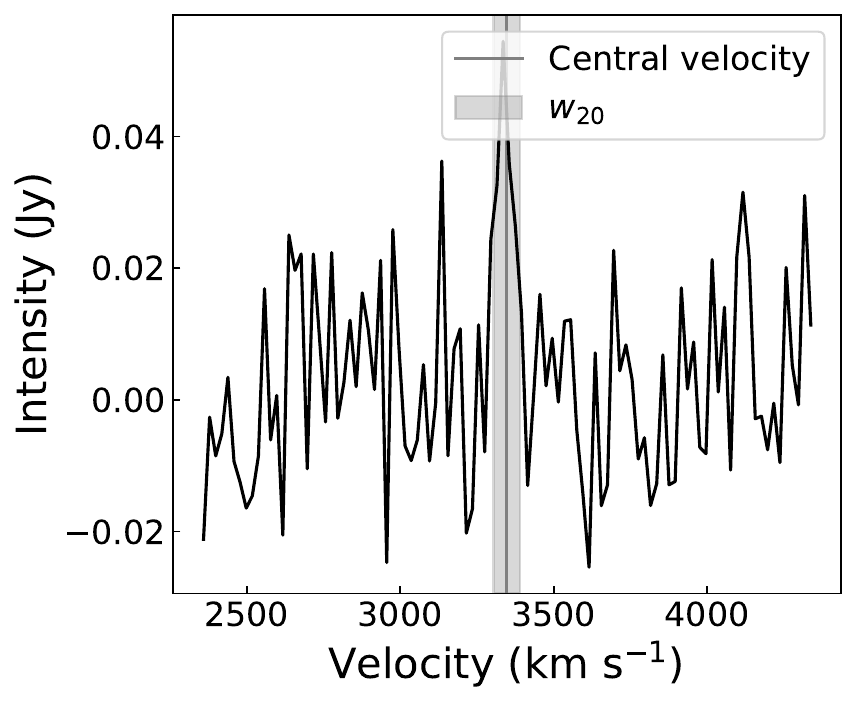}
			\caption{ }
   \end{subfigure}
	\caption{WALLABY J132457-182105 (NGC 5044)}
\end{figure*}
 
\begin{figure*}
	\centering
		\begin{subfigure}[b]{0.24\textwidth}
			\centering
			\includegraphics[width=\textwidth]{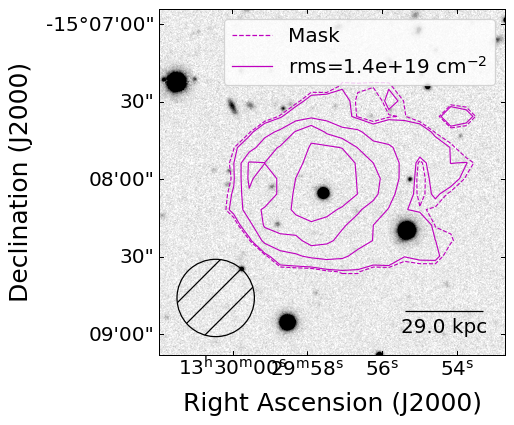}
			\caption{ }
		\end{subfigure}
		\begin{subfigure}[b]{0.24\textwidth}
			\centering
			\includegraphics[width=\textwidth]{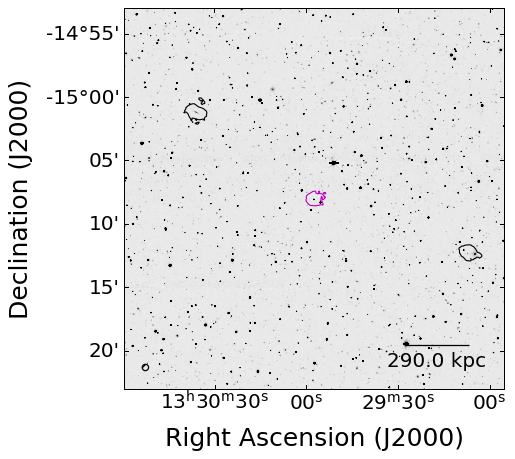}
			\caption{ }
		\end{subfigure}
		\begin{subfigure}[b]{0.26\textwidth}
			\centering
			\includegraphics[width=\textwidth]{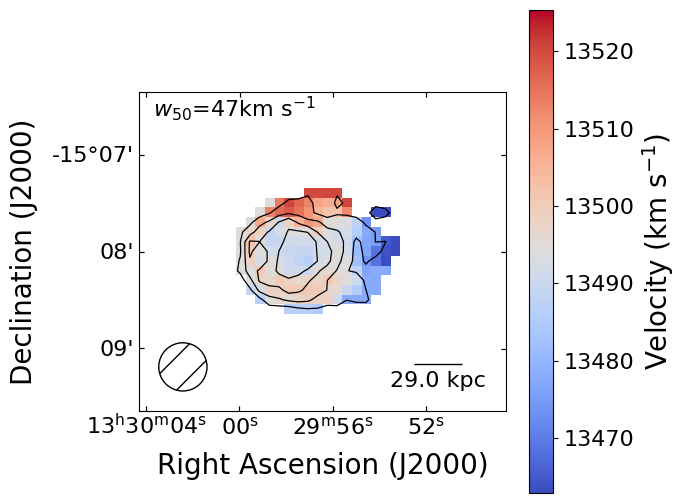}
			\caption{ }
		\end{subfigure}
  \begin{subfigure}[b]{0.24\textwidth}
			\centering
			\includegraphics[width=\textwidth]{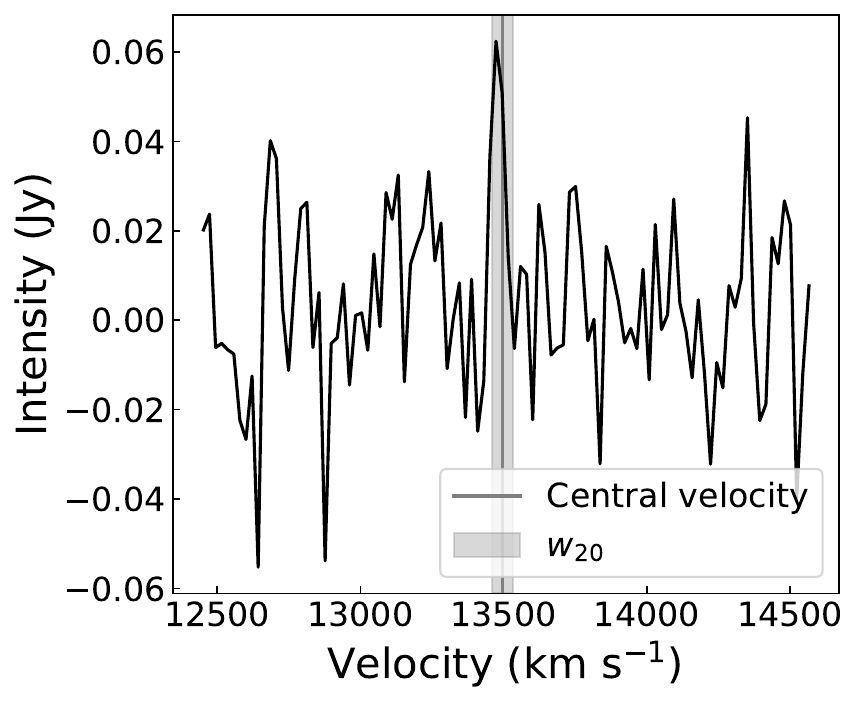}
			\caption{ }
   \end{subfigure}
	\caption{WALLABY J132957-150800 (NGC 5044)}
\end{figure*}


\begin{figure*}
	\centering
		\begin{subfigure}[b]{0.24\textwidth}
			\centering
			\includegraphics[width=\textwidth]{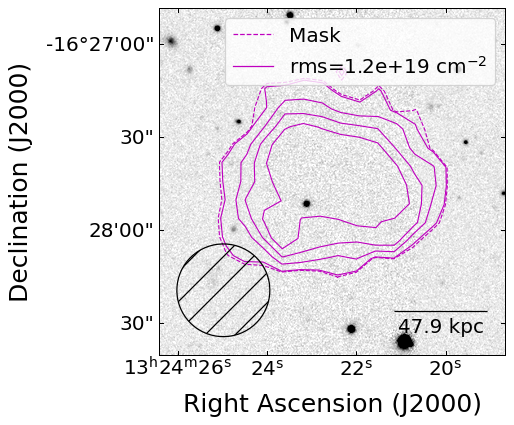}
			\caption{ }
		\end{subfigure}
		\begin{subfigure}[b]{0.24\textwidth}
			\centering
			\includegraphics[width=\textwidth]{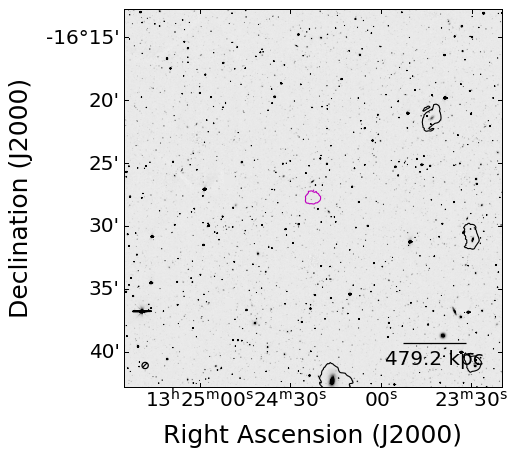}
			\caption{ }
		\end{subfigure}
		\begin{subfigure}[b]{0.26\textwidth}
			\centering
			\includegraphics[width=\textwidth]{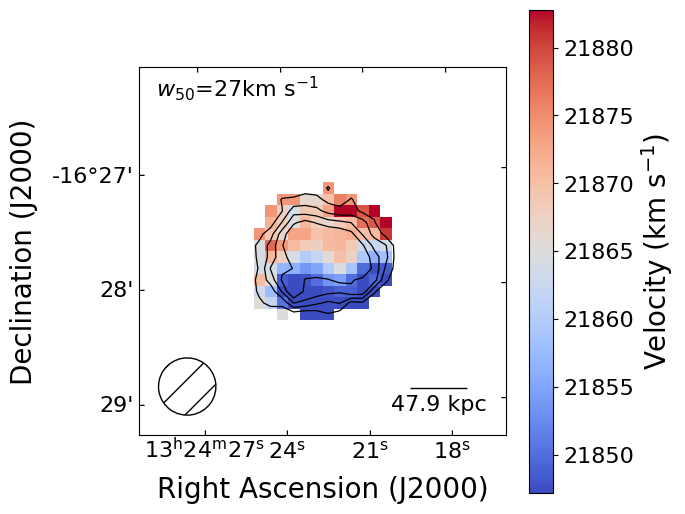}
			\caption{ }
		\end{subfigure}
    \begin{subfigure}[b]{0.24\textwidth}
			\centering
			\includegraphics[width=\textwidth]{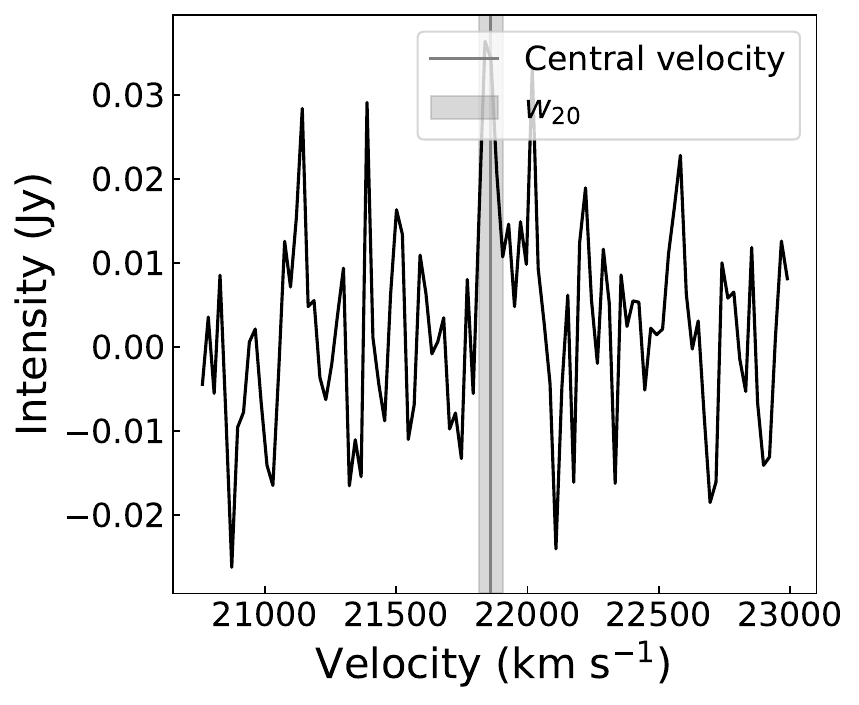}
			\caption{ }
   \end{subfigure}
	\caption{WALLABY J132422-162744 (NGC 5044). This source may be a partial detection.}
\end{figure*}
 
\begin{figure*}
	\centering
		\begin{subfigure}[b]{0.24\textwidth}
			\centering
			\includegraphics[width=\textwidth]{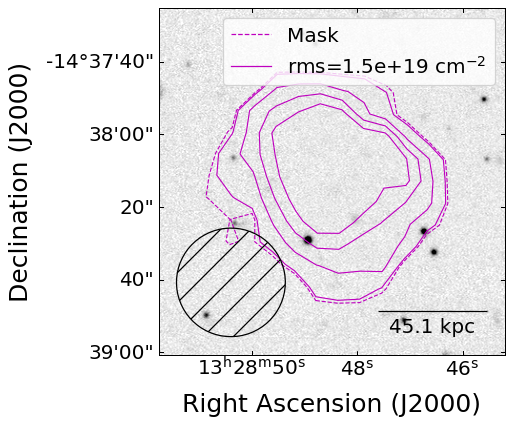}
			\caption{ }
		\end{subfigure}
		\begin{subfigure}[b]{0.24\textwidth}
			\centering
			\includegraphics[width=\textwidth]{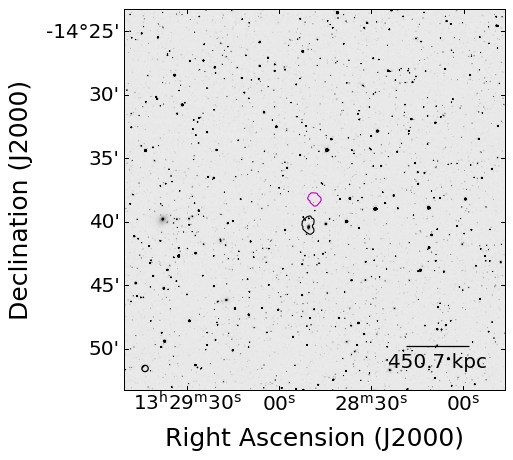}
			\caption{ }
		\end{subfigure}
		\begin{subfigure}[b]{0.26\textwidth}
			\centering
			\includegraphics[width=\textwidth]{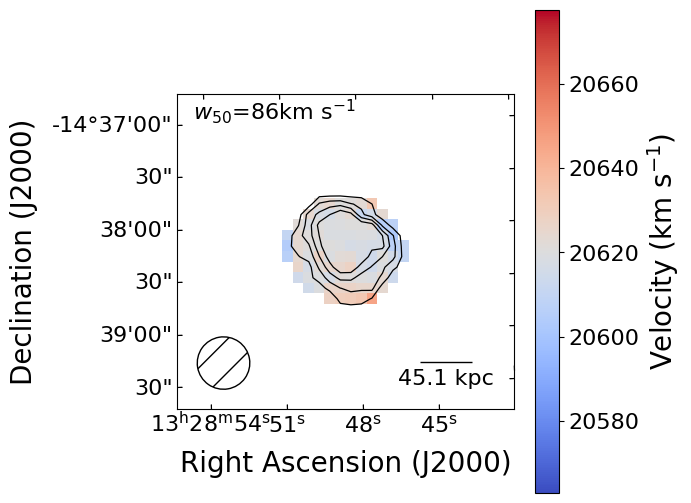}
			\caption{ }
		\end{subfigure}
    \begin{subfigure}[b]{0.24\textwidth}
			\centering
			\includegraphics[width=\textwidth]{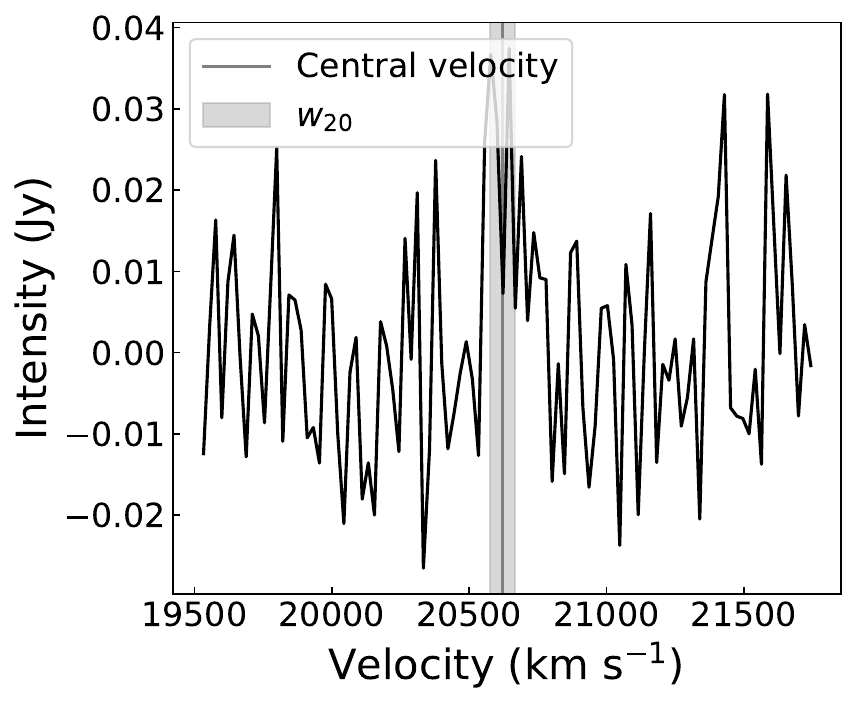}
			\caption{ }
   \end{subfigure}
	\caption{WALLABY J132848-143813 (NGC 5044)}
\end{figure*}

\section{Uncertain Dark Sources}
\label{append:weak}

In this section we present the images of the uncertain dark H\,{\sc i} sources. Table \ref{tab:unreliable} presents the uncertain dark sources.  {Each figure in this appendix contains the same set of images as outlined in Appendix \ref{append:strong}}. To emphasise that these dark candidates are  {uncertain} detections, they are marked with $^{*}$ next to the source name.

\begin{table*}
    \centering
    \caption{Properties of the  {uncertain} dark sources. From left to right the columns are: WALLABY field, WALLABY name, right ascension, declination, central velocity, luminosity distance, emission line width at half maximum, $w_{20}$ emission line width,  H\,{\sc i} size (major axis at 1 M$_{\odot}$ pc$^{-2}$ contour level), H\,{\sc i} mass,  stellar mass $3\sigma$ upper limit, SFR $3\sigma$ upper limit and the signal-to-noise ratio. }
    \label{tab:unreliable}
    \setlength{\tabcolsep}{3pt}
    \begin{tabular}{llccccccccccc}
    \hline
       Field &  Name  & RA (J2000) & DEC (J2000) & $\frac{cz}{\rm km s^{-1}}$ & $\frac{d_L}{\rm Mpc}$ & $\frac{w_{50}}{\rm km~s^{-1}}$ & $\frac{w_{20}}{\rm km~s^{-1}}$ & $\frac{D_{\rm HI}}{\rm kpc}$ & $\log(\frac{M_{HI}}{\rm M_{\odot}})$ & $\log(\frac{M_{*}}{\rm M_{\odot}})$ & $\log(\frac{SFR}{\rm M_{\odot} yr^{-1}})$ & SNR \\
       \hline
Hydra & J103543-255954 & 158.933 & -25.998 & 12936 & 190.8 & 141.9 & 180.9 &    & 9.88$\pm$0.04 & $<5.52$ & $<-1.78$ & 11.9 \\
& J103818-285023 & 159.578 & -28.840 & 13500 & 199.3 & 303.7 & 341.5 & 88.5$\pm8.8$ & 10.19$\pm$0.03 & $<6.71$ & $<-2.82$ & 14.9 \\
\hline
NGC 4808& J130011+065105 & 195.050 & 6.851 & 9793 & 143.3 & 27.3 & 64.5 &    & 9.09$\pm$0.07 &  & & 6.9 \\
 & J130119+053553 & 195.330 & 5.598 & 25748 & 391.3 & 146.3 & 153.7 &    & 10.49$\pm$0.06 & $<6.30$ & $<-2.87$ & 8.0 \\
\hline
& J125656-202606 & 194.234 & -20.435 & 3678 & 53.0 & 9.6 & 19.4 &    & 8.87$\pm$0.11 & $<4.15$ & $<-2.56$ & 4.5 \\
 & J130514-203447 & 196.310 & -20.580 & 13503 & 199.4 & 119.1 & 166.3 &    & 9.63$\pm$0.08 & $<5.53$ & $<-1.58$ & 6.0 \\
& J130835-143159 & 197.148 & -14.533 & 23969 & 362.8 & 108.7 & 122.1 &    & 9.88$\pm$0.06 & $<6.32$ & $<-1.65$ & 7.4 \\
& J131247-132906 & 198.198 & -13.485 & 3936 & 56.8 & 63.1 & 188.9 &    & 8.49$\pm$0.07 & $<4.00$ & $<-3.74$ & 6.6 \\
& J131805-200055 & 199.522 & -20.015 & 12873 & 189.8 & 27.8 & 32.1 &    & 9.17$\pm$0.07 & $<5.92$ & $<-1.84$ & 6.5 \\
& J131844-113805 & 199.684 & -11.635 & 8944 & 130.6 & 71.8 & 113.3 & 23.3$\pm6.0$ & 9.43$\pm$0.06 & $<5.54$ &   & 7.5 \\
NGC 5044 & J131847-210939 & 199.699 & -21.161 & 20013 & 300.1 & 197.7 & 282.6 &    & 10.50$\pm$0.03 & $<5.99$ & $<-1.38$ & 13.6 \\
& J132238-204726 & 200.660 & -20.791 & 26365 & 401.2 & 78.1 & 115.9 &    & 10.19$\pm$0.08 & $<6.17$ & $<-2.07$ & 5.9 \\
& J132359-235510 & 200.998 & -23.920 & 18769 & 280.6 & 77.6 & 112.6 &    & 9.99$\pm$0.08 & $<5.42$ & $<-1.43$ & 6.0 \\
& J132709-163509 & 201.790 & -16.586 & 8405 & 122.6 & 95.8 & 124.6 &    & 9.21$\pm$0.07 &  &   & 6.6 \\
& J132719-170237 & 201.833 & -17.044 & 19130 & 286.3 & 92.3 & 127.2 &    & 9.79$\pm$0.08 & $<5.30$ & $<-2.29$ & 5.7 \\
& J132810-151352 & 202.042 & -15.231 & 16440 & 244.5 & 93.0 & 108.0 &    & 9.51$\pm$0.10 &  &  & 4.8 \\
& J132935-153750 & 202.397 & -15.631 & 14951 & 221.5 & 78.5 & 103.2 &    & 9.45$\pm$0.07 & $<5.46$ & $<-1.61$ & 7.1 \\
        \hline
    \end{tabular}
\end{table*}

\begin{figure*}
	\centering
		\begin{subfigure}[b]{0.24\textwidth}
			\centering
			\includegraphics[width=\textwidth]{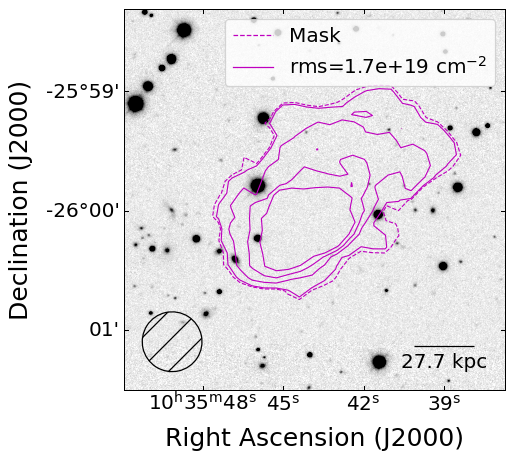}
			\caption{ }
		\end{subfigure}
		\begin{subfigure}[b]{0.24\textwidth}
			\centering
			\includegraphics[width=\textwidth]{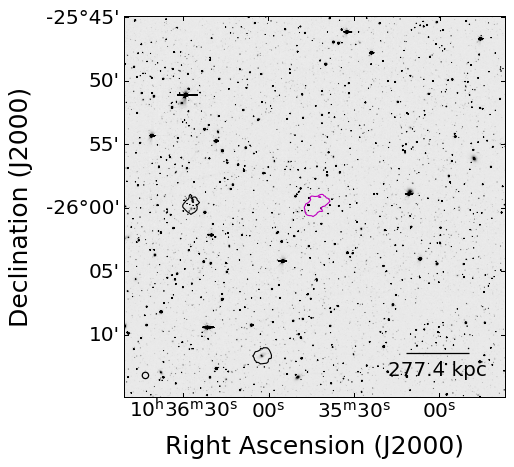}
			\caption{ }
		\end{subfigure}
		\begin{subfigure}[b]{0.26\textwidth}
			\centering
			\includegraphics[width=\textwidth]{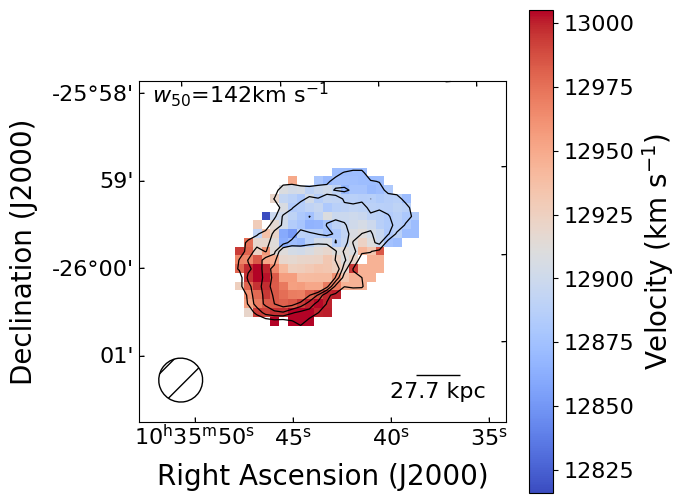}
			\caption{ }
		\end{subfigure}
   \begin{subfigure}[b]{0.24\textwidth}
			\centering
			\includegraphics[width=\textwidth]{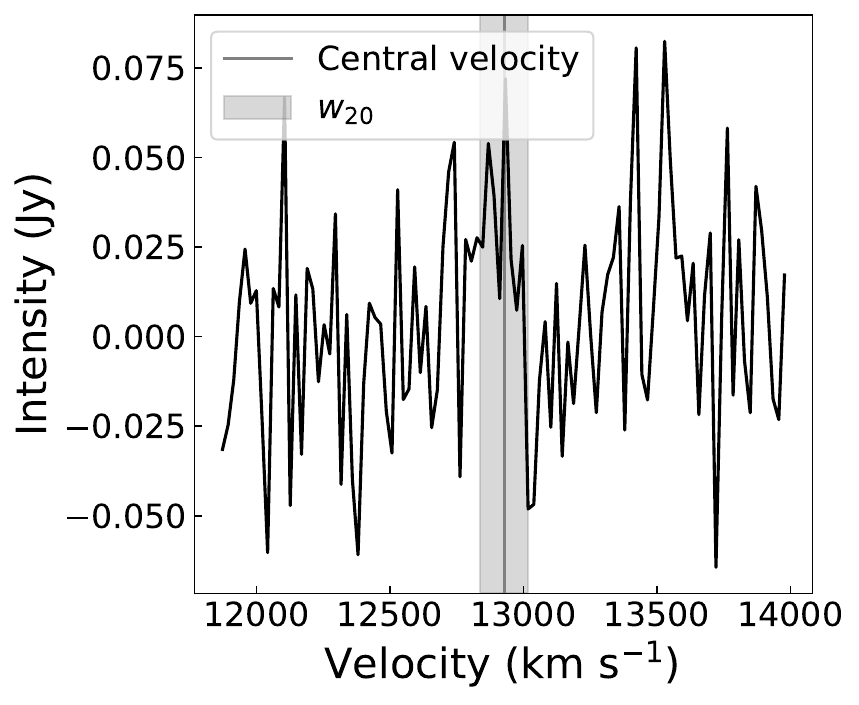}
			\caption{ }
		\end{subfigure}
	\caption{WALLABY J103543-255954$^{*}$ (Hydra)}
 \label{fig:firstc}
\end{figure*}


\begin{figure*}
	\centering
		\begin{subfigure}[b]{0.24\textwidth}
			\centering
			\includegraphics[width=\textwidth]{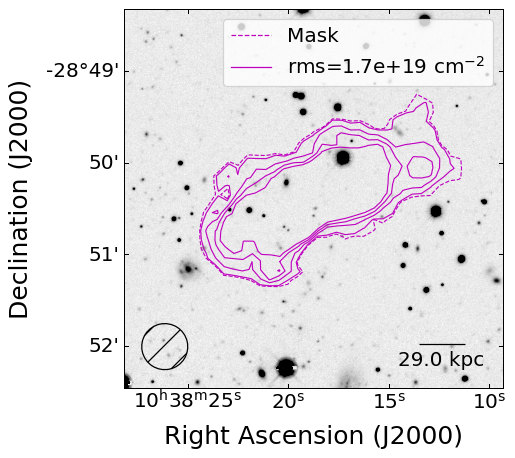}
			\caption{ }
		\end{subfigure}
		\begin{subfigure}[b]{0.24\textwidth}
			\centering
			\includegraphics[width=\textwidth]{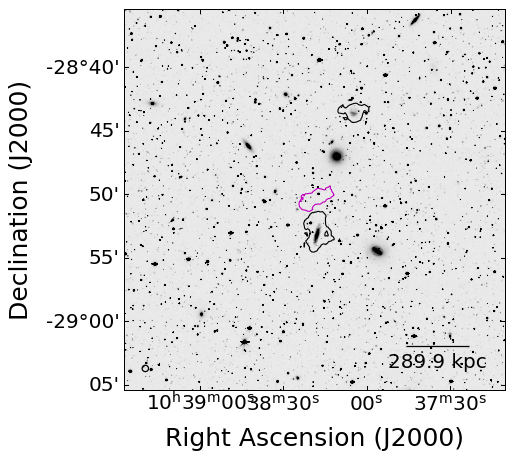}
			\caption{ }
		\end{subfigure}
		\begin{subfigure}[b]{0.26\textwidth}
			\centering
			\includegraphics[width=\textwidth]{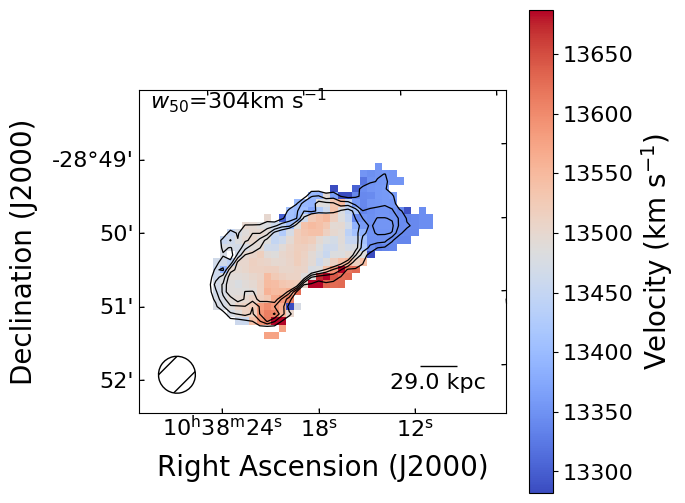}
			\caption{ }
		\end{subfigure}
  \begin{subfigure}[b]{0.24\textwidth}
			\centering
			\includegraphics[width=\textwidth]{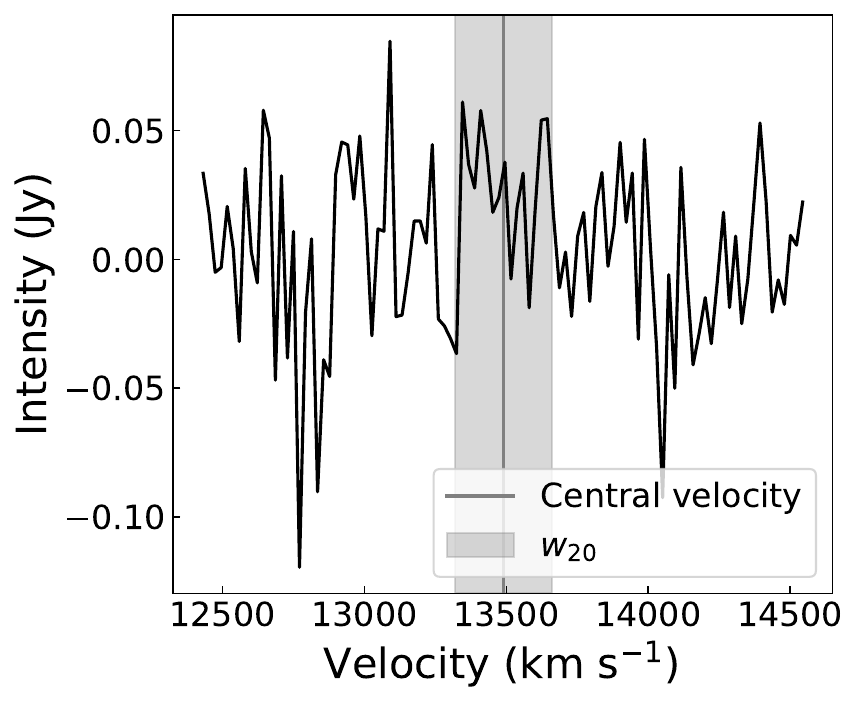}
			\caption{ }
   \end{subfigure}
	\caption{WALLABY J103818-285023$^{*}$ (Hydra)}
\end{figure*}

\begin{figure*}
	\centering
		\begin{subfigure}[b]{0.24\textwidth}
			\centering
			\includegraphics[width=\textwidth]{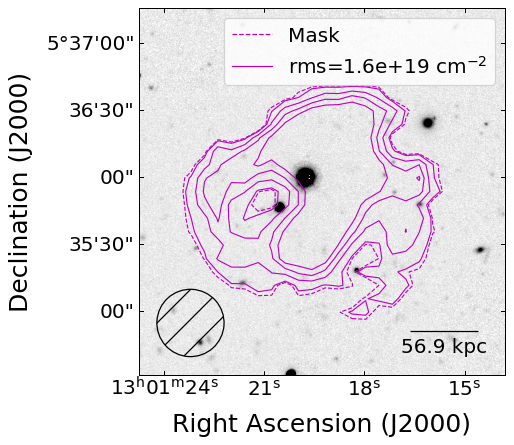}
			\caption{ }
		\end{subfigure}
		\begin{subfigure}[b]{0.24\textwidth}
			\centering
			\includegraphics[width=\textwidth]{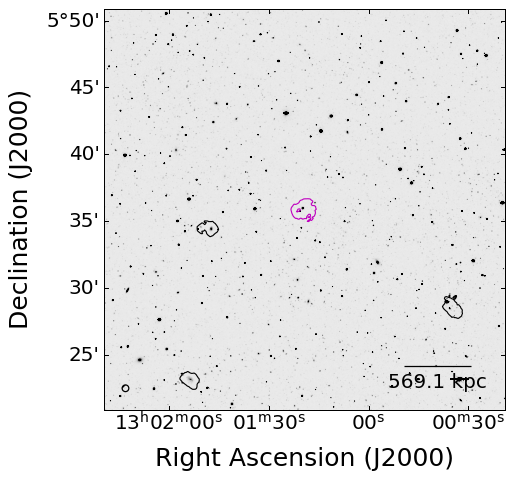}
			\caption{ }
		\end{subfigure}
		\begin{subfigure}[b]{0.26\textwidth}
			\centering
			\includegraphics[width=\textwidth]{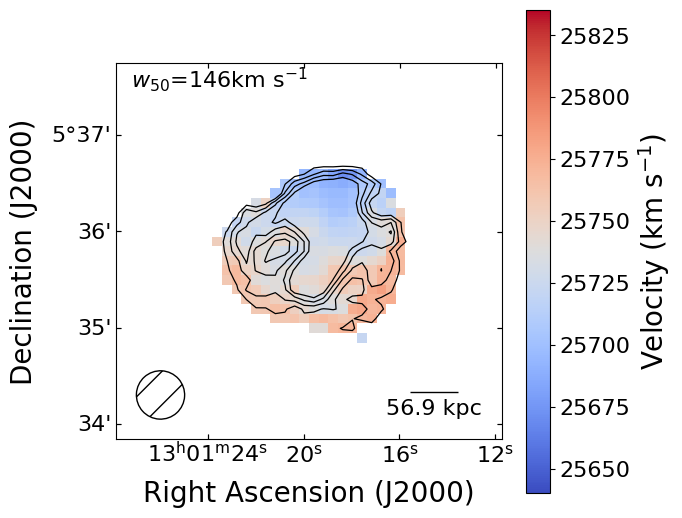}
			\caption{ }
		\end{subfigure}
    \begin{subfigure}[b]{0.24\textwidth}
			\centering
			\includegraphics[width=\textwidth]{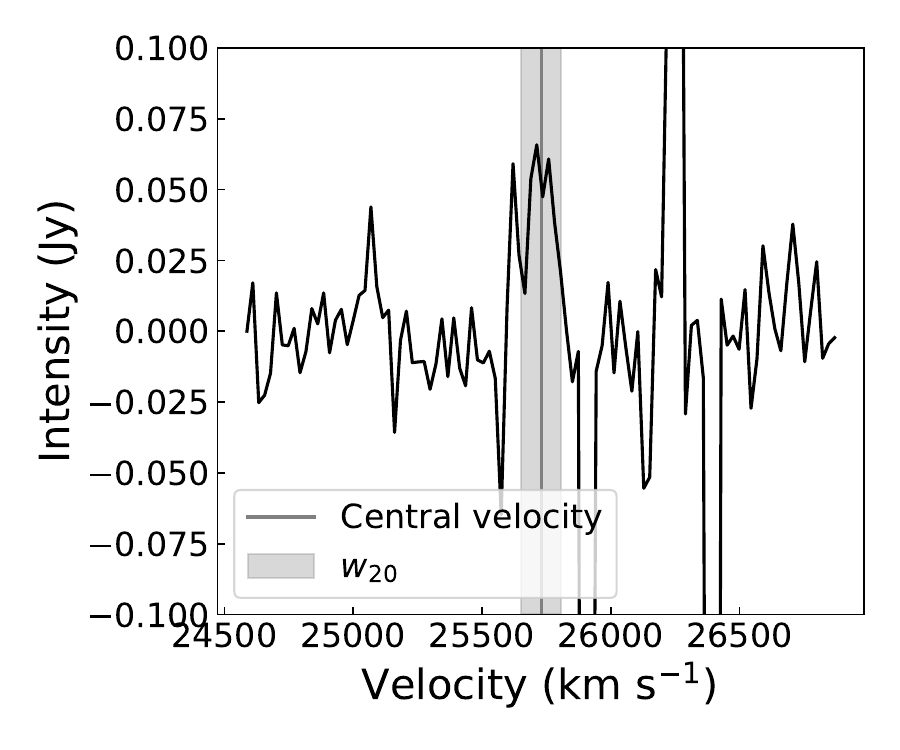}
			\caption{ }
   \end{subfigure}
	\caption{WALLABY J130119+053553$^{*}$ (NGC 4808). Possible RFI from GPS satellite at 1308 MHz.}
\end{figure*}


\begin{figure*}
	\centering
		\begin{subfigure}[b]{0.24\textwidth}
			\centering
			\includegraphics[width=\textwidth]{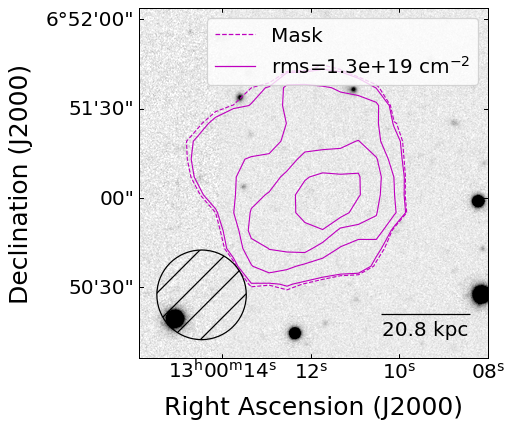}
			\caption{ }
		\end{subfigure}
		\begin{subfigure}[b]{0.24\textwidth}
			\centering
			\includegraphics[width=\textwidth]{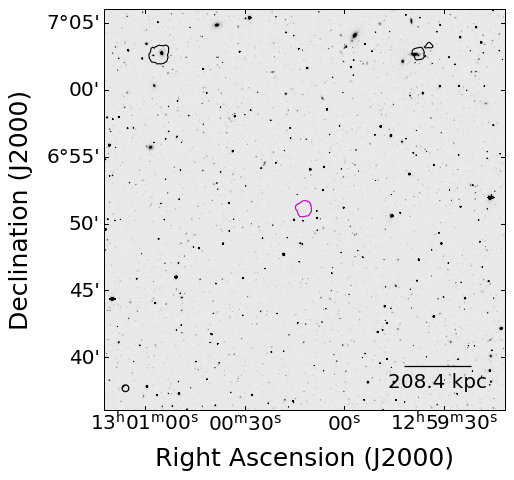}
			\caption{ }
		\end{subfigure}
		\begin{subfigure}[b]{0.26\textwidth}
			\centering
			\includegraphics[width=\textwidth]{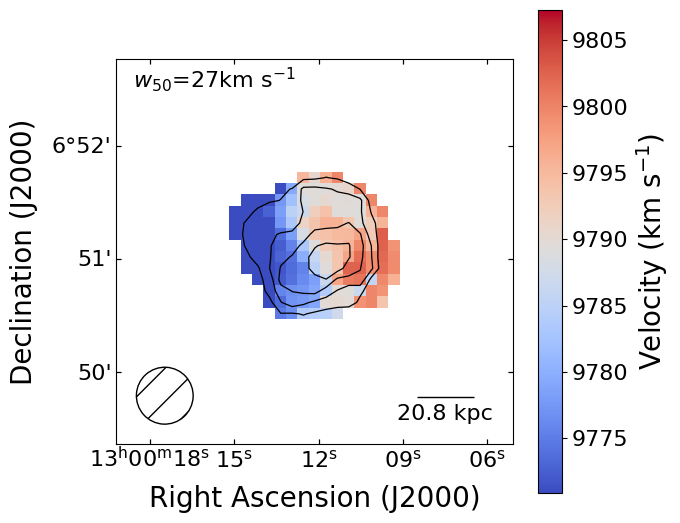}
			\caption{ }
		\end{subfigure}
    \begin{subfigure}[b]{0.24\textwidth}
			\centering
			\includegraphics[width=\textwidth]{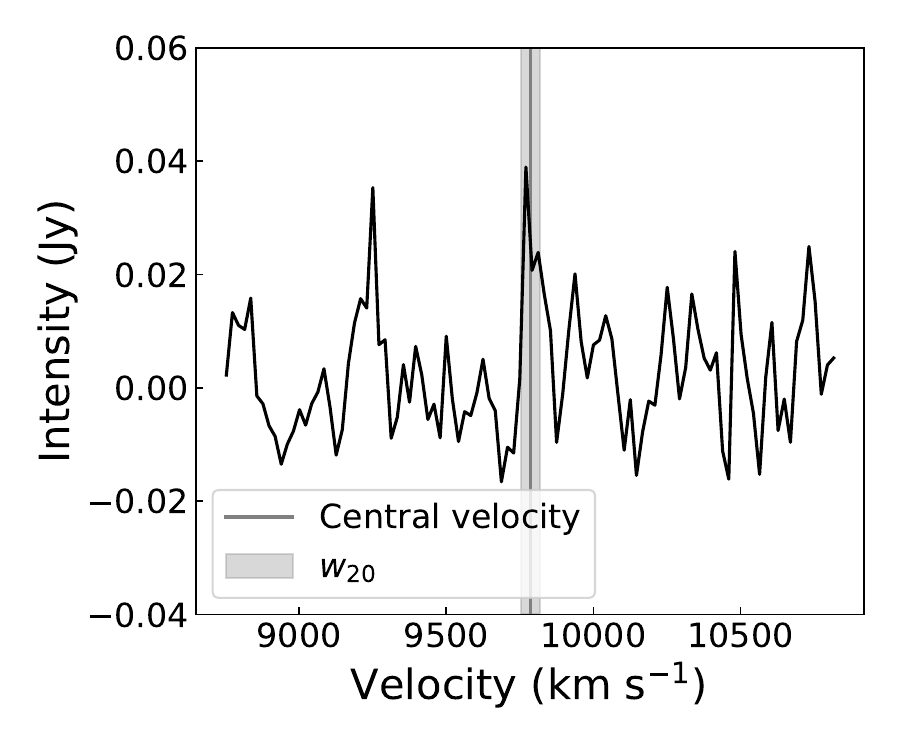}
			\caption{ }
   \end{subfigure}
	\caption{WALLABY J130011+065105$^{*}$ (NGC 4808)}
\end{figure*}

\begin{figure*}
	\centering
		\begin{subfigure}[b]{0.24\textwidth}
			\centering
			\includegraphics[width=\textwidth]{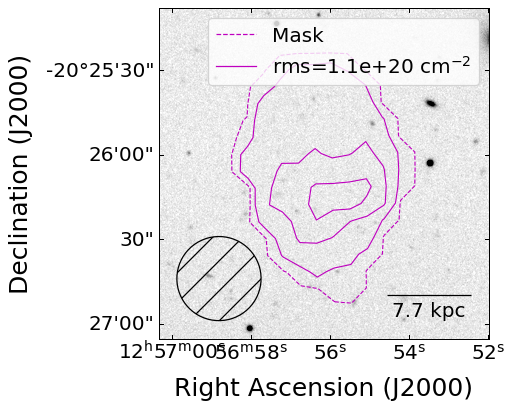}
			\caption{ }
		\end{subfigure}
		\begin{subfigure}[b]{0.24\textwidth}
			\centering
			\includegraphics[width=\textwidth]{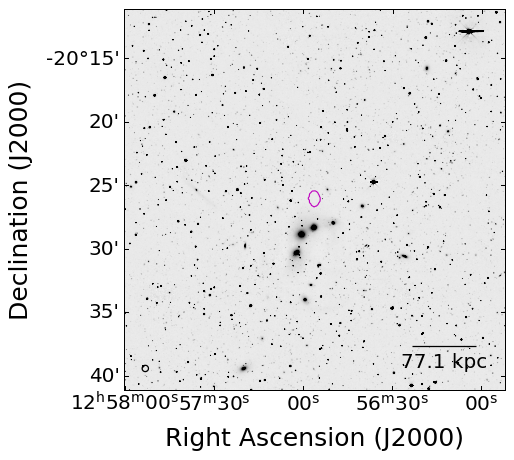}
			\caption{ }
		\end{subfigure}
		\begin{subfigure}[b]{0.26\textwidth}
			\centering
			\includegraphics[width=\textwidth]{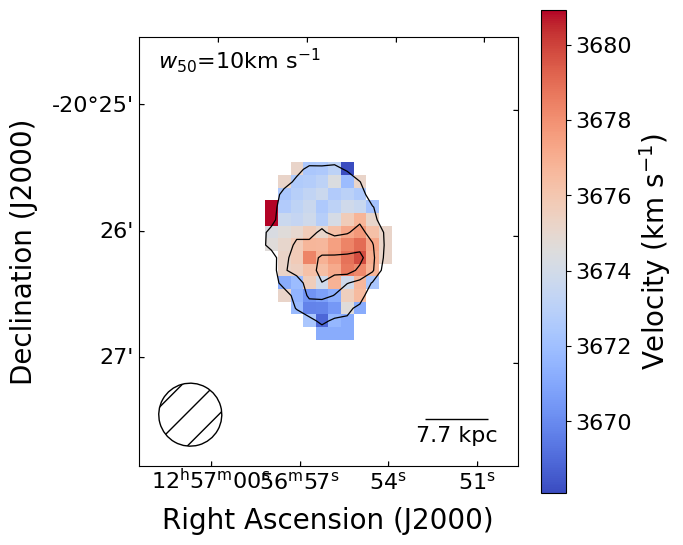}
			\caption{ }
		\end{subfigure}
    \begin{subfigure}[b]{0.24\textwidth}
			\centering
			\includegraphics[width=\textwidth]{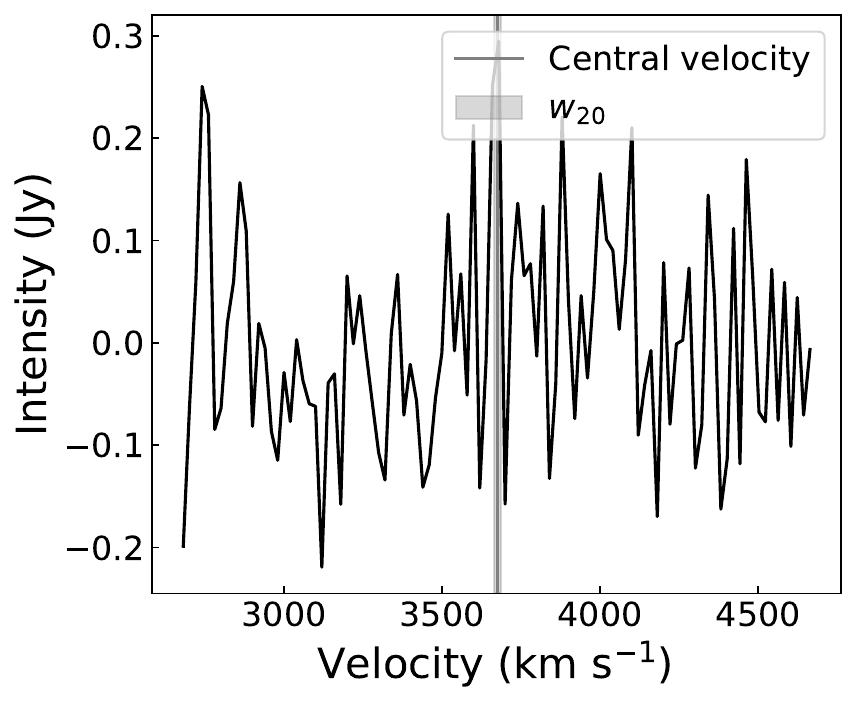}
			\caption{ }
   \end{subfigure}
	\caption{WALLABY J125656-202606$^{*}$ (NGC 5044)}
\end{figure*}

\begin{figure*}
	\centering
		\begin{subfigure}[b]{0.24\textwidth}
			\centering
			\includegraphics[width=\textwidth]{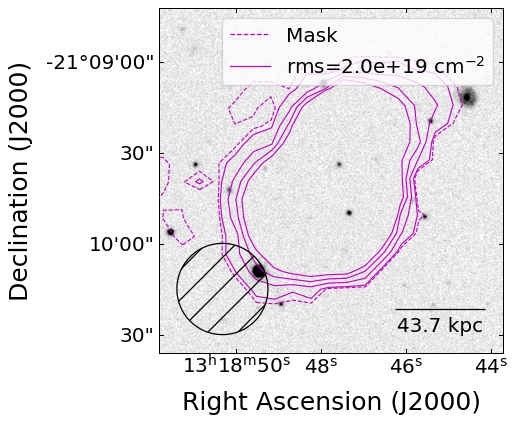}
			\caption{ }
		\end{subfigure}
		\begin{subfigure}[b]{0.24\textwidth}
			\centering
			\includegraphics[width=\textwidth]{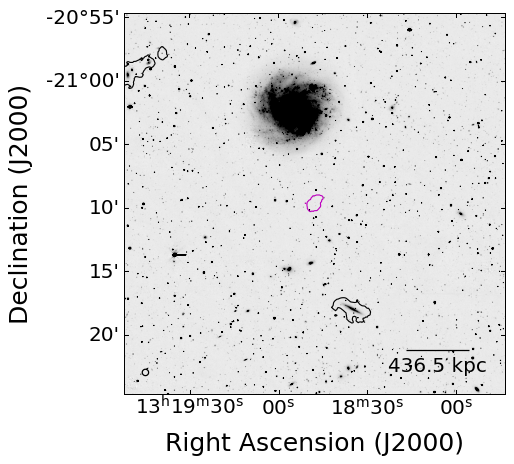}
			\caption{ }
		\end{subfigure}
		\begin{subfigure}[b]{0.26\textwidth}
			\centering
			\includegraphics[width=\textwidth]{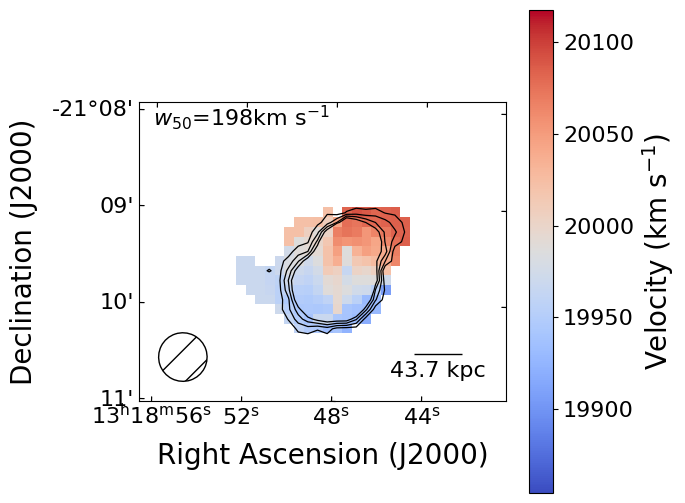}
			\caption{ }
		\end{subfigure}
    \begin{subfigure}[b]{0.24\textwidth}
			\centering
			\includegraphics[width=\textwidth]{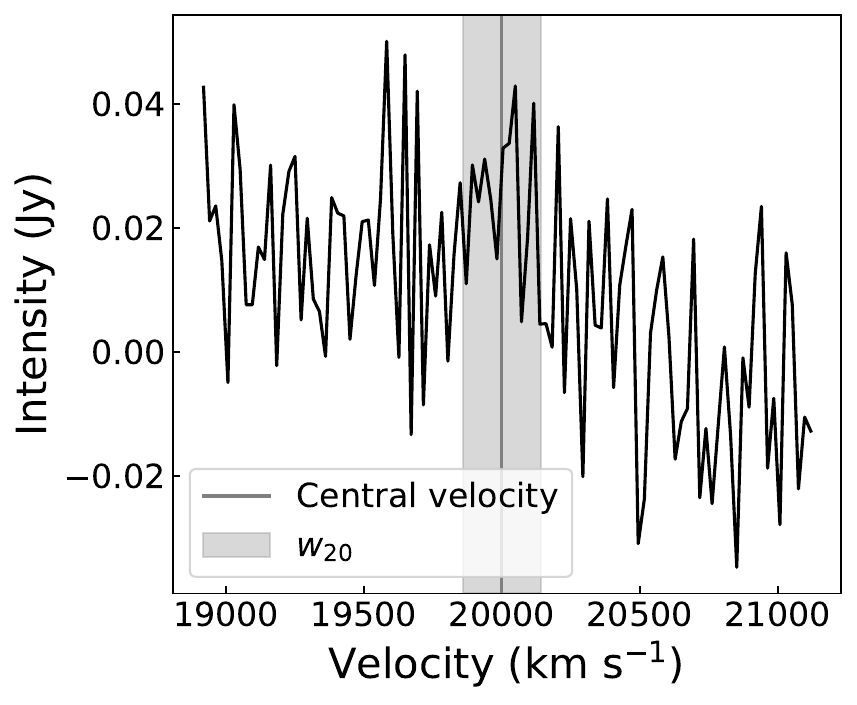}
			\caption{ }
   \end{subfigure}
	\caption{WALLABY J131847-210939$^{*}$ (NGC 5044)}
\end{figure*}


\begin{figure*}
	\centering
		\begin{subfigure}[b]{0.24\textwidth}
			\centering
			\includegraphics[width=\textwidth]{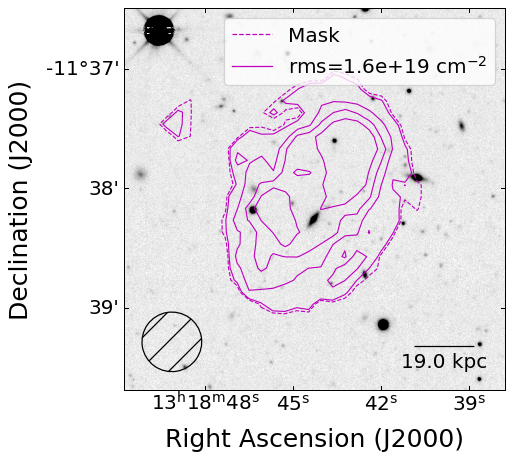}
			\caption{ }
		\end{subfigure}
		\begin{subfigure}[b]{0.24\textwidth}
			\centering
			\includegraphics[width=\textwidth]{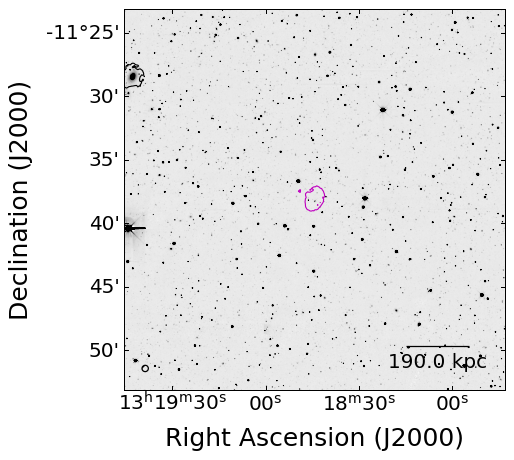}
			\caption{ }
		\end{subfigure}
		\begin{subfigure}[b]{0.26\textwidth}
			\centering
			\includegraphics[width=\textwidth]{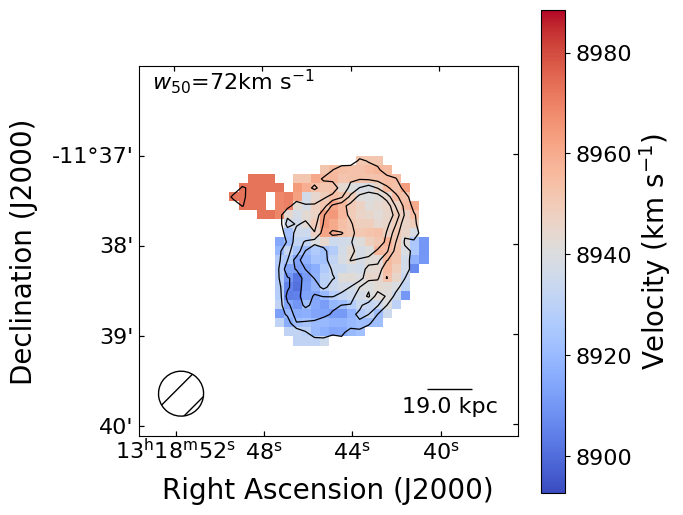}
			\caption{ }
		\end{subfigure}
    \begin{subfigure}[b]{0.24\textwidth}
			\centering
			\includegraphics[width=\textwidth]{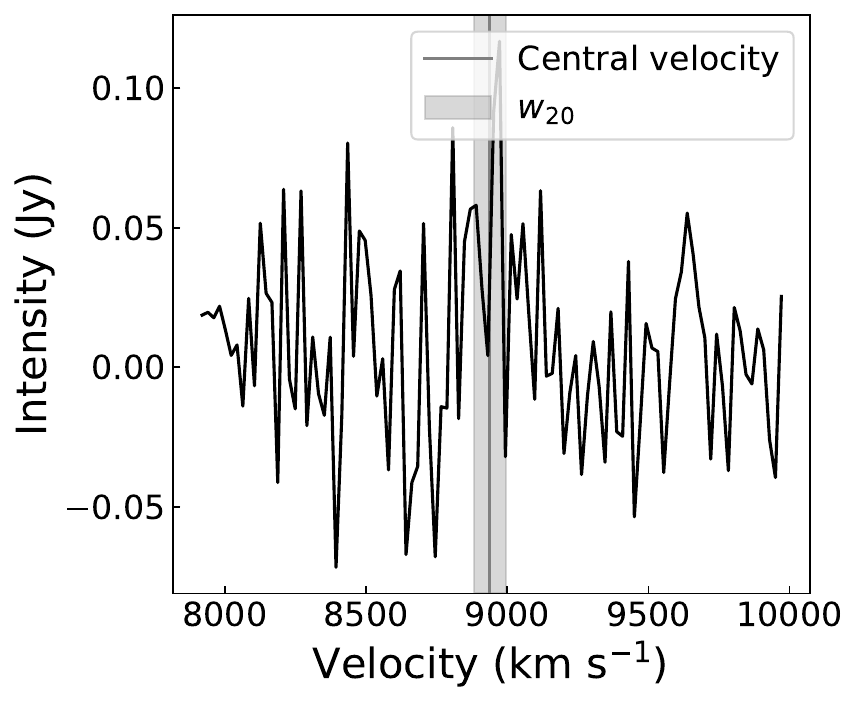}
			\caption{ }
   \end{subfigure}
	\caption{WALLABY J131844-113805$^{*}$ (NGC 5044)}
\end{figure*}

\begin{figure*}
	\centering
		\begin{subfigure}[b]{0.24\textwidth}
			\centering
			\includegraphics[width=\textwidth]{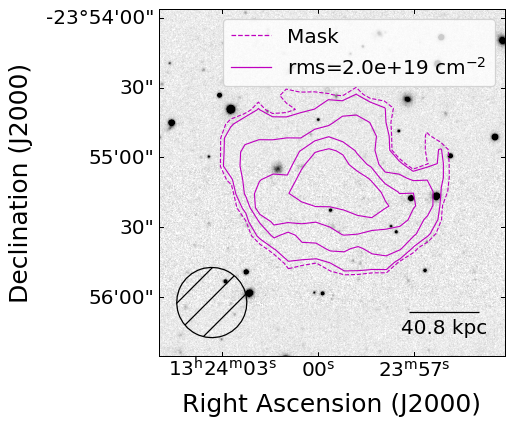}
			\caption{ }
		\end{subfigure}
		\begin{subfigure}[b]{0.24\textwidth}
			\centering
			\includegraphics[width=\textwidth]{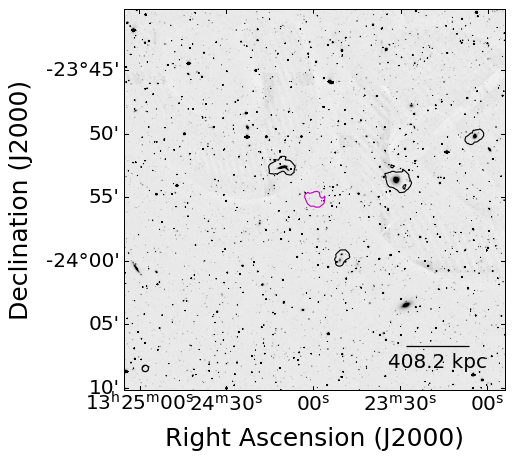}
			\caption{ }
		\end{subfigure}
		\begin{subfigure}[b]{0.26\textwidth}
			\centering
			\includegraphics[width=\textwidth]{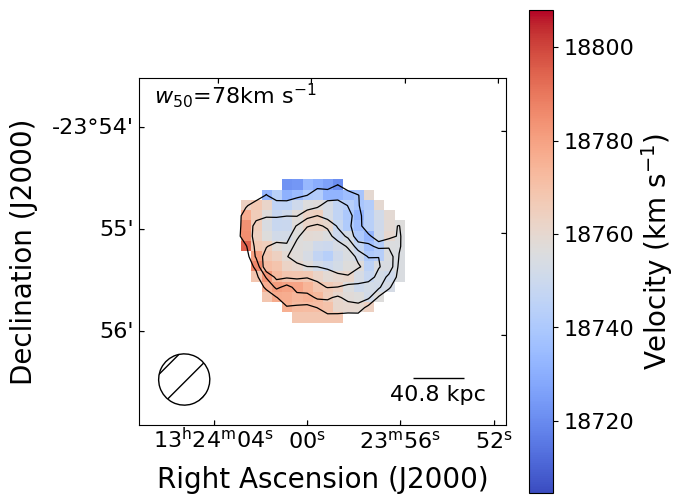}
			\caption{ }
		\end{subfigure}
  \begin{subfigure}[b]{0.24\textwidth}
			\centering
			\includegraphics[width=\textwidth]{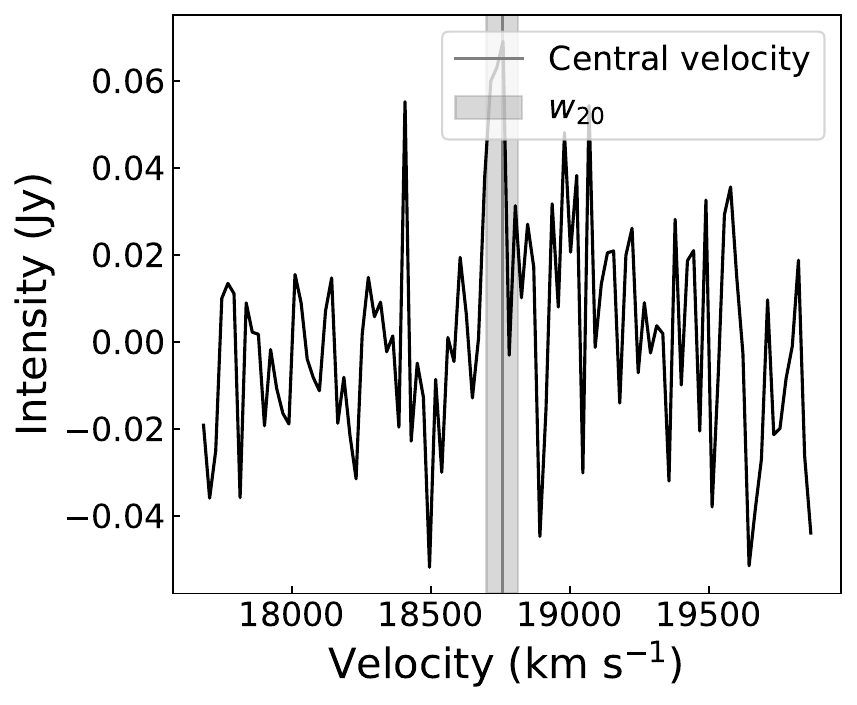}
			\caption{ }
   \end{subfigure}
	\caption{WALLABY J132359-235510$^{*}$ (NGC 5044)}
\end{figure*}

\begin{figure*}
	\centering
		\begin{subfigure}[b]{0.24\textwidth}
			\centering
			\includegraphics[width=\textwidth]{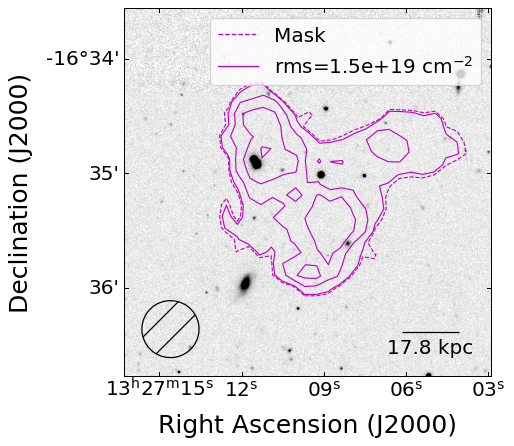}
			\caption{ }
		\end{subfigure}
		\begin{subfigure}[b]{0.24\textwidth}
			\centering
			\includegraphics[width=\textwidth]{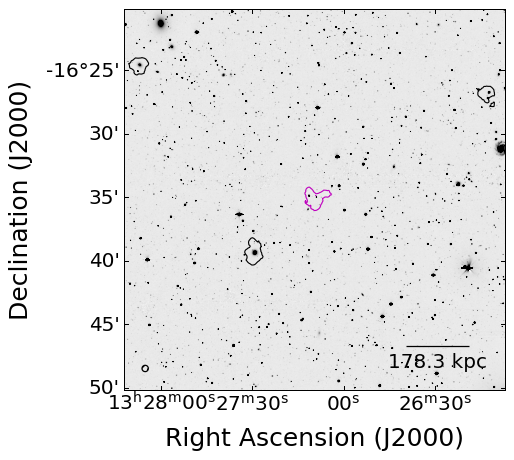}
			\caption{ }
		\end{subfigure}
		\begin{subfigure}[b]{0.26\textwidth}
			\centering
			\includegraphics[width=\textwidth]{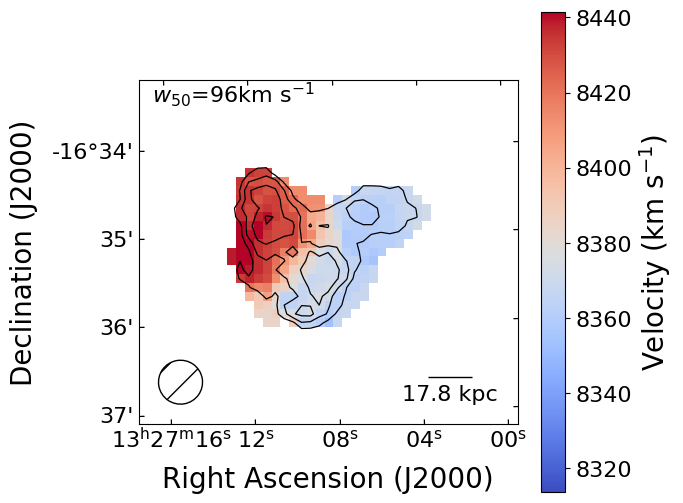}
			\caption{ }
		\end{subfigure}
  \begin{subfigure}[b]{0.24\textwidth}
			\centering
			\includegraphics[width=\textwidth]{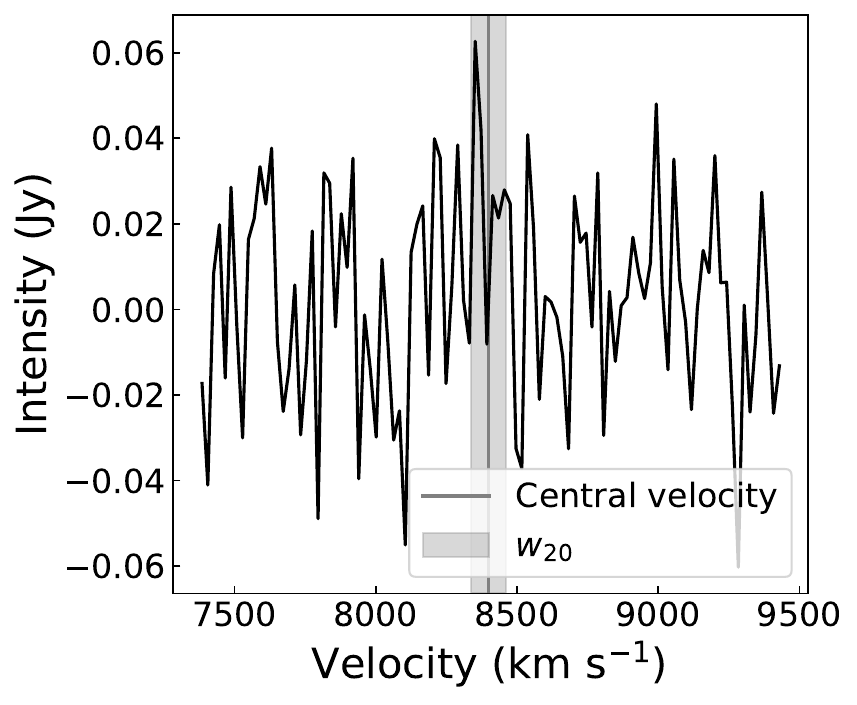}
			\caption{ }
   \end{subfigure}
	\caption{WALLABY J132709-163509$^{*}$ (NGC 5044). Possible RFI from GPS satellite at 1381 MHz.}
\end{figure*}
 
\begin{figure*}
	\centering
		\begin{subfigure}[b]{0.24\textwidth}
			\centering
			\includegraphics[width=\textwidth]{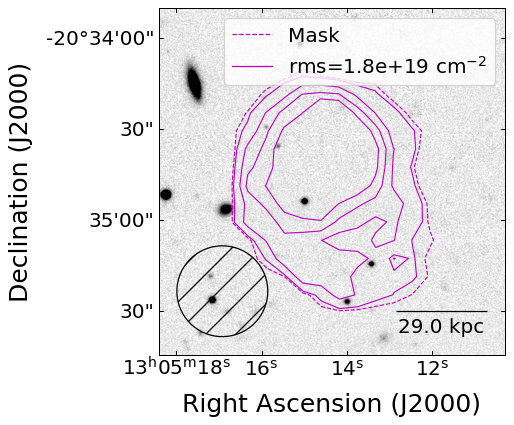}
			\caption{ }
		\end{subfigure}
		\begin{subfigure}[b]{0.24\textwidth}
			\centering
			\includegraphics[width=\textwidth]{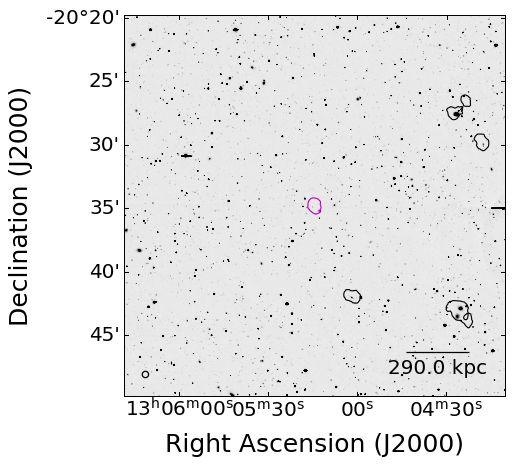}
			\caption{ }
		\end{subfigure}
		\begin{subfigure}[b]{0.26\textwidth}
			\centering
			\includegraphics[width=\textwidth]{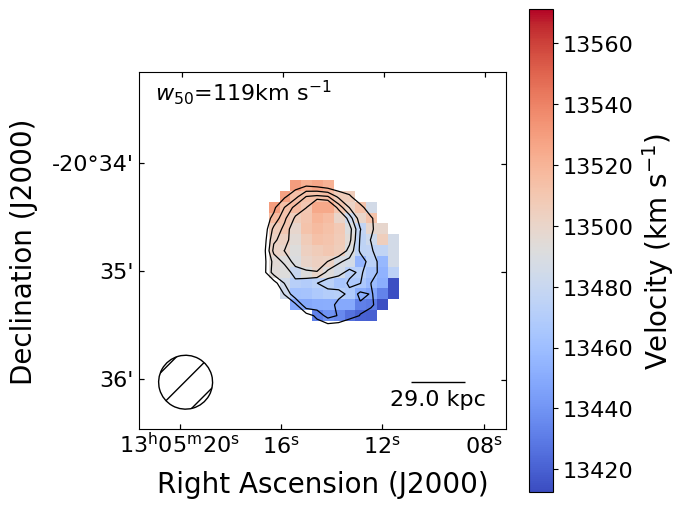}
			\caption{ }
		\end{subfigure}
  \begin{subfigure}[b]{0.24\textwidth}
			\centering
			\includegraphics[width=\textwidth]{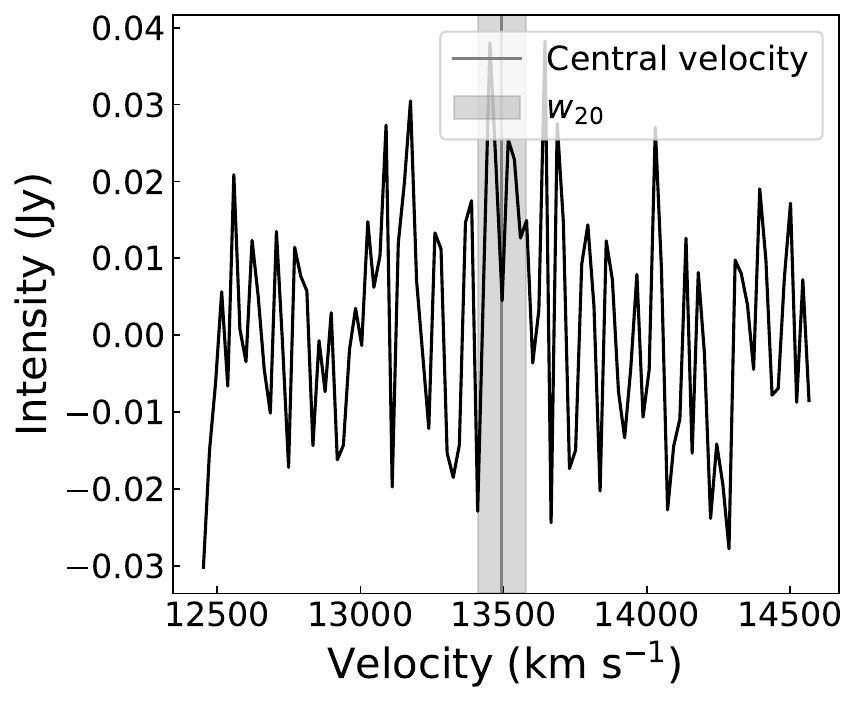}
			\caption{ }
   \end{subfigure}
	\caption{WALLABY J130514-203447$^{*}$ (NGC 5044)}
\end{figure*}

\begin{figure*}
	\centering
		\begin{subfigure}[b]{0.24\textwidth}
			\centering
			\includegraphics[width=\textwidth]{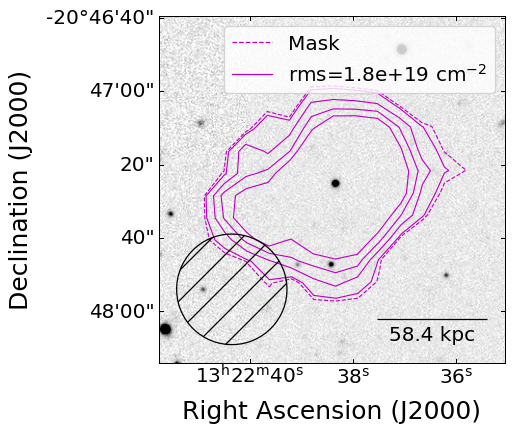}
			\caption{ }
		\end{subfigure}
		\begin{subfigure}[b]{0.24\textwidth}
			\centering
			\includegraphics[width=\textwidth]{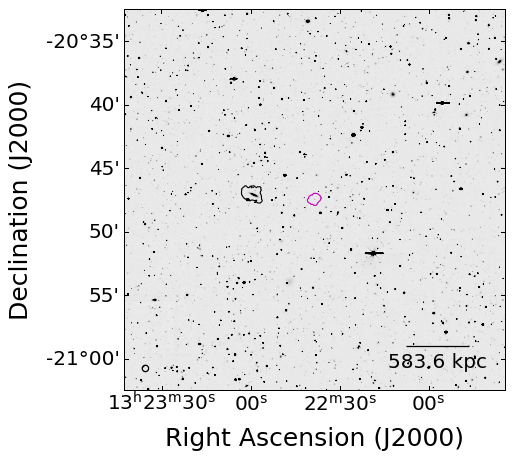}
			\caption{ }
		\end{subfigure}
		\begin{subfigure}[b]{0.26\textwidth}
			\centering
			\includegraphics[width=\textwidth]{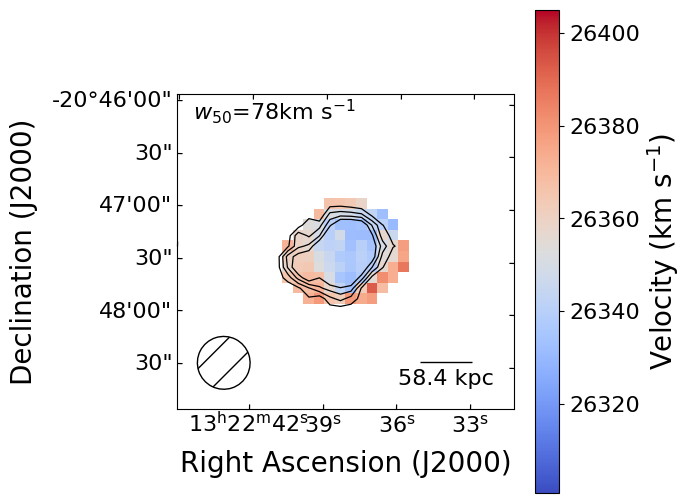}
			\caption{ }
		\end{subfigure}
  \begin{subfigure}[b]{0.24\textwidth}
			\centering
			\includegraphics[width=\textwidth]{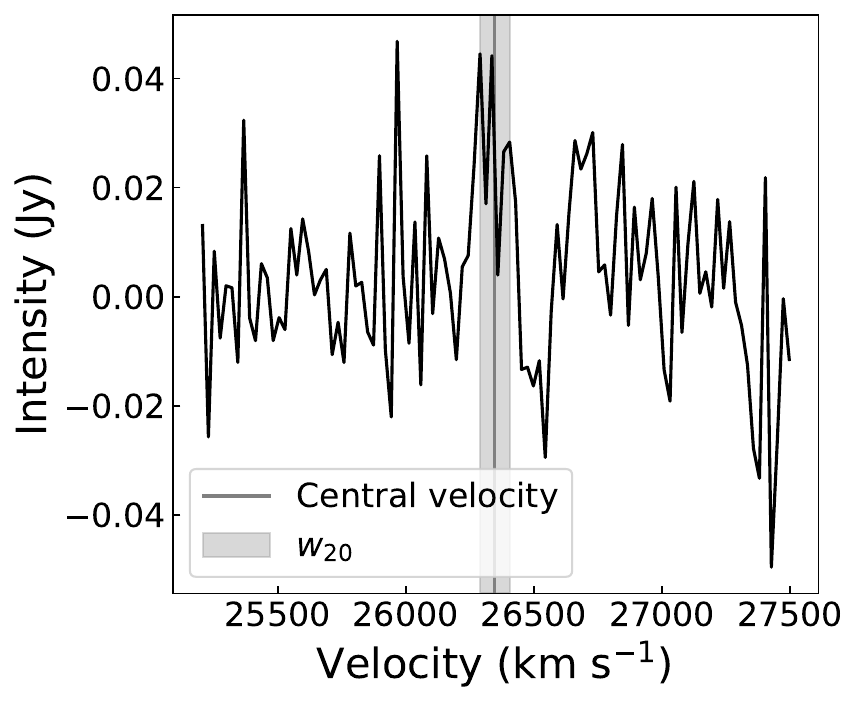}
			\caption{ }
   \end{subfigure}
	\caption{WALLABY J132238-204726$^{*}$ (NGC 5044). Possible RFI from GPS satellite at 1305 MHz.}
\end{figure*}
 
\begin{figure*}
	\centering
		\begin{subfigure}[b]{0.24\textwidth}
			\centering
			\includegraphics[width=\textwidth]{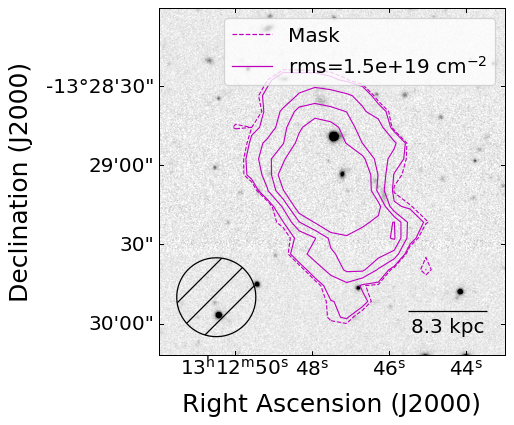}
			\caption{ }
		\end{subfigure}
		\begin{subfigure}[b]{0.24\textwidth}
			\centering
			\includegraphics[width=\textwidth]{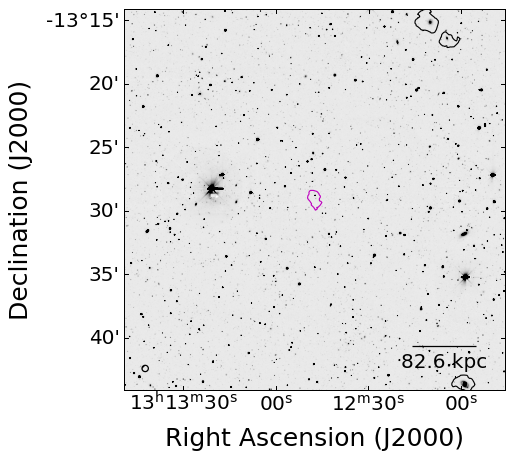}
			\caption{ }
		\end{subfigure}
		\begin{subfigure}[b]{0.26\textwidth}
			\centering
			\includegraphics[width=\textwidth]{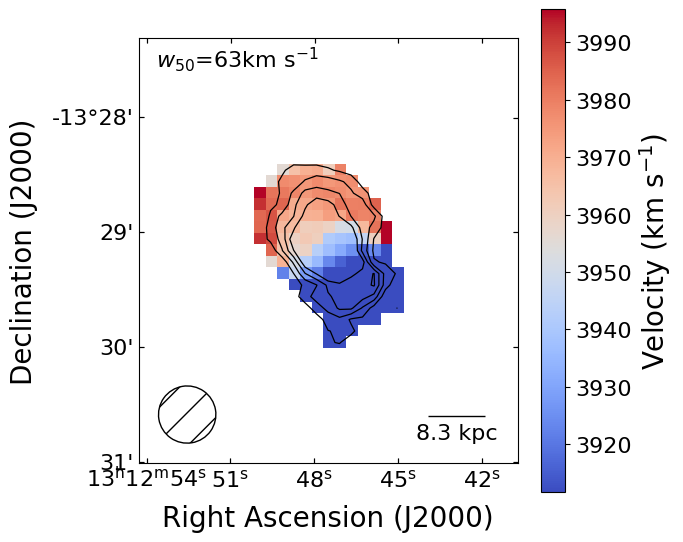}
			\caption{ }
		\end{subfigure}
  \begin{subfigure}[b]{0.24\textwidth}
			\centering
			\includegraphics[width=\textwidth]{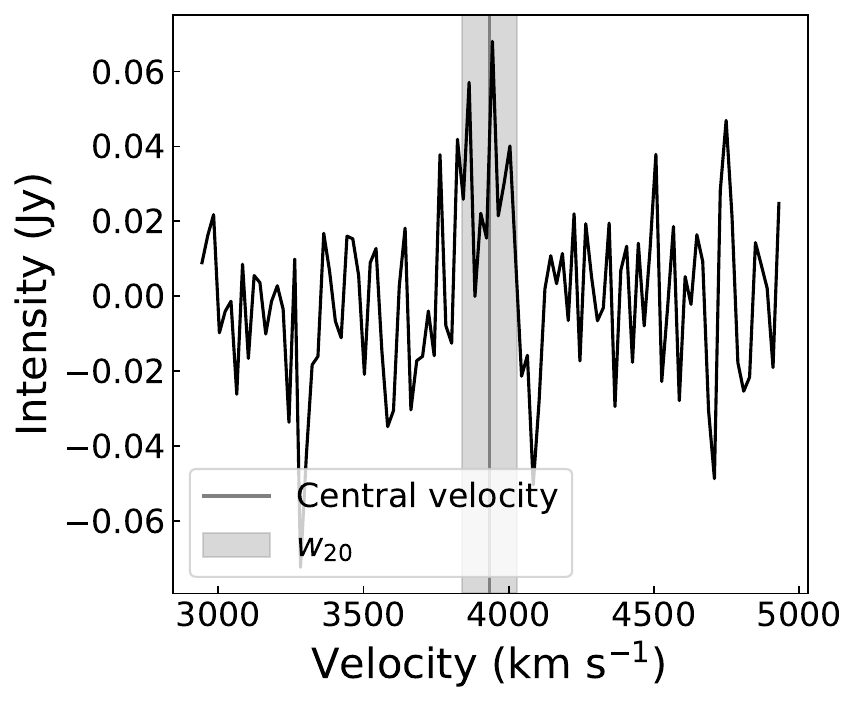}
			\caption{ }
   \end{subfigure}
	\caption{WALLABY J131247-132906$^{*}$ (NGC 5044)}
\end{figure*}

\begin{figure*}
	\centering
		\begin{subfigure}[b]{0.24\textwidth}
			\centering
			\includegraphics[width=\textwidth]{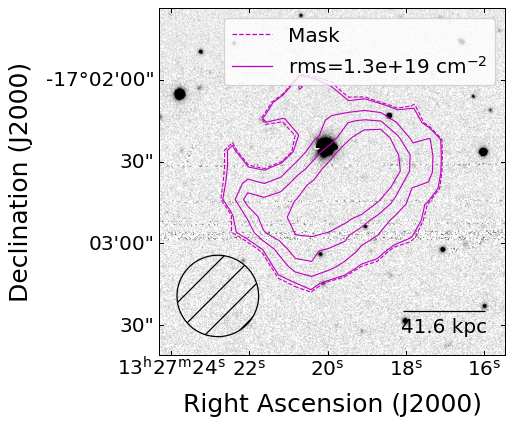}
			\caption{ }
		\end{subfigure}
		\begin{subfigure}[b]{0.24\textwidth}
			\centering
			\includegraphics[width=\textwidth]{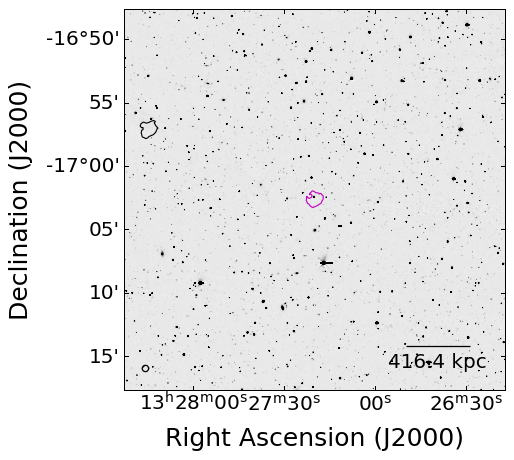}
			\caption{ }
		\end{subfigure}
		\begin{subfigure}[b]{0.26\textwidth}
			\centering
			\includegraphics[width=\textwidth]{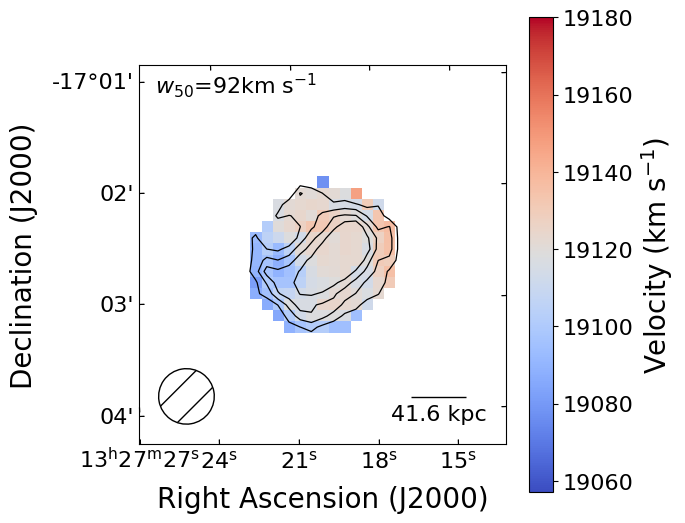}
			\caption{ }
		\end{subfigure}
  \begin{subfigure}[b]{0.24\textwidth}
			\centering
			\includegraphics[width=\textwidth]{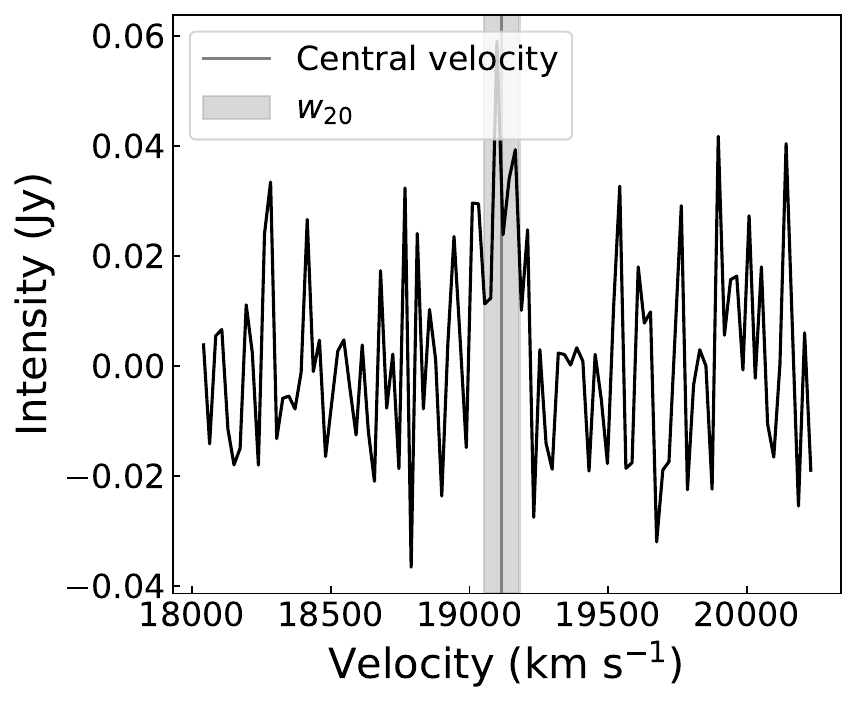}
			\caption{ }
   \end{subfigure}
	\caption{WALLABY J132719-170237$^{*}$ (NGC 5044)}
\end{figure*}

\begin{figure*}
	\centering
		\begin{subfigure}[b]{0.24\textwidth}
			\centering
			\includegraphics[width=\textwidth]{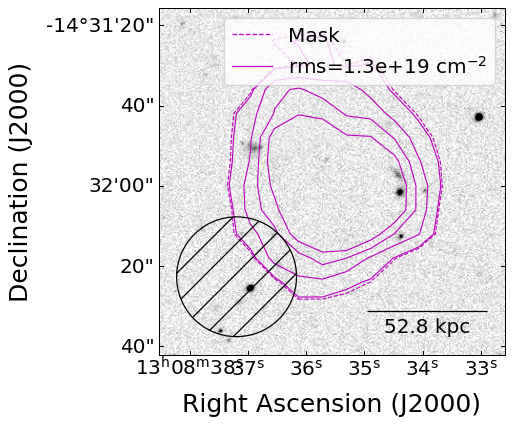}
			\caption{ }
		\end{subfigure}
		\begin{subfigure}[b]{0.24\textwidth}
			\centering
			\includegraphics[width=\textwidth]{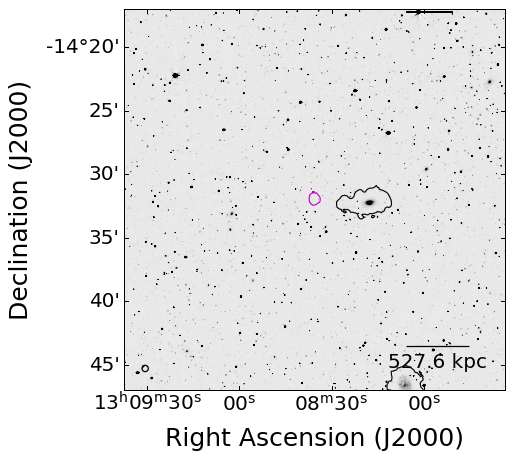}
			\caption{ }
		\end{subfigure}
		\begin{subfigure}[b]{0.26\textwidth}
			\centering
			\includegraphics[width=\textwidth]{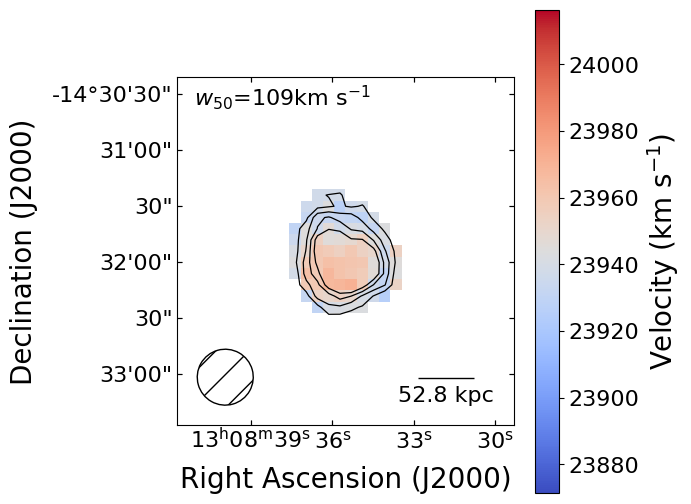}
			\caption{ }
		\end{subfigure}
    \begin{subfigure}[b]{0.24\textwidth}
			\centering
			\includegraphics[width=\textwidth]{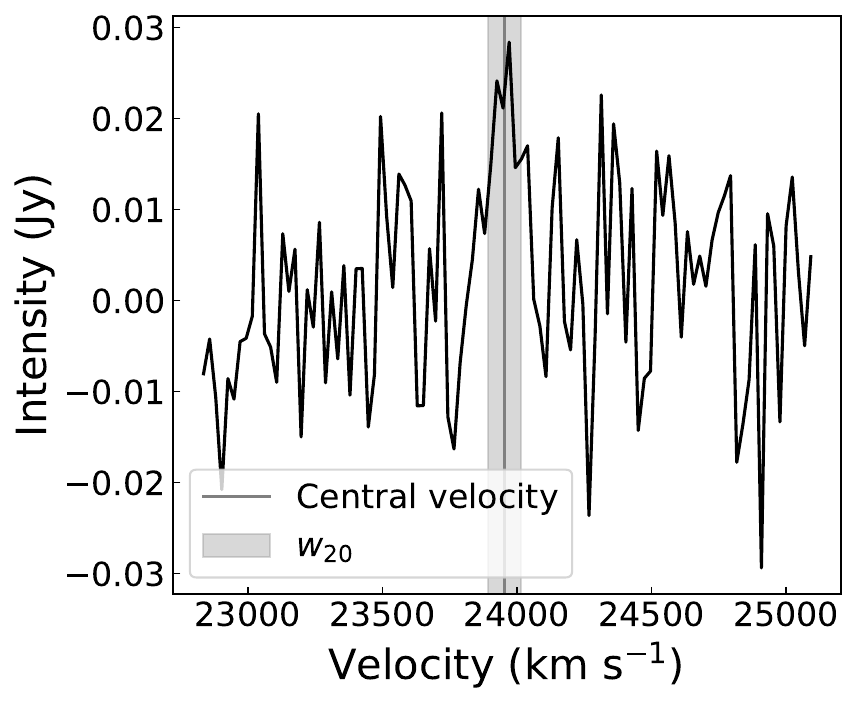}
			\caption{ }
   \end{subfigure}
	\caption{WALLABY J130835-143159$^{*}$ (NGC 5044)}
\end{figure*}
 
\begin{figure*}
	\centering
		\begin{subfigure}[b]{0.24\textwidth}
			\centering
			\includegraphics[width=\textwidth]{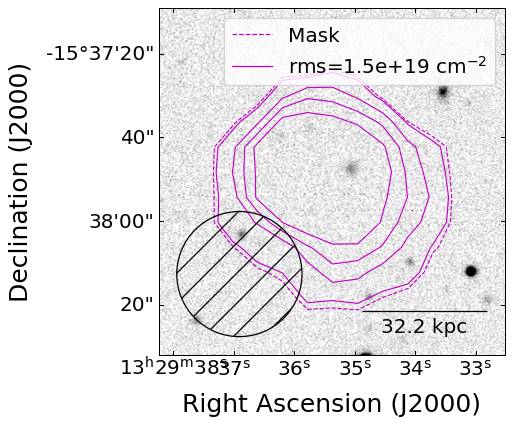}
			\caption{ }
		\end{subfigure}
		\begin{subfigure}[b]{0.24\textwidth}
			\centering
			\includegraphics[width=\textwidth]{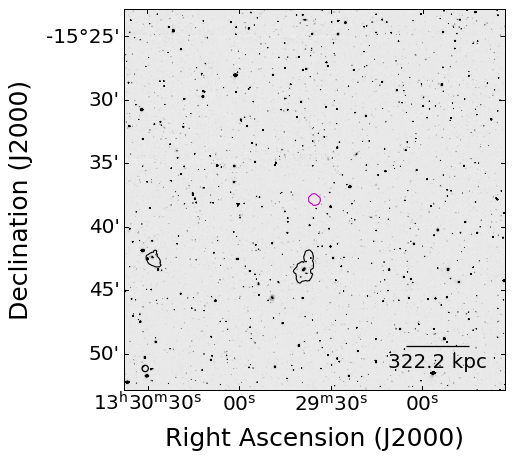}
			\caption{ }
		\end{subfigure}
		\begin{subfigure}[b]{0.26\textwidth}
			\centering
			\includegraphics[width=\textwidth]{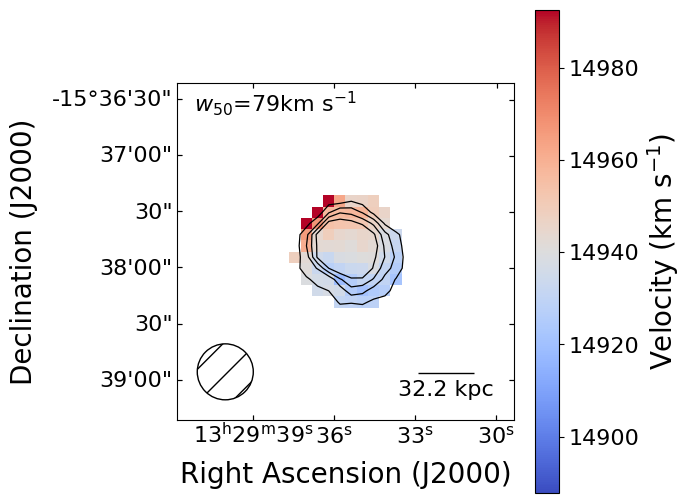}
			\caption{ }
		\end{subfigure}
    \begin{subfigure}[b]{0.24\textwidth}
			\centering
			\includegraphics[width=\textwidth]{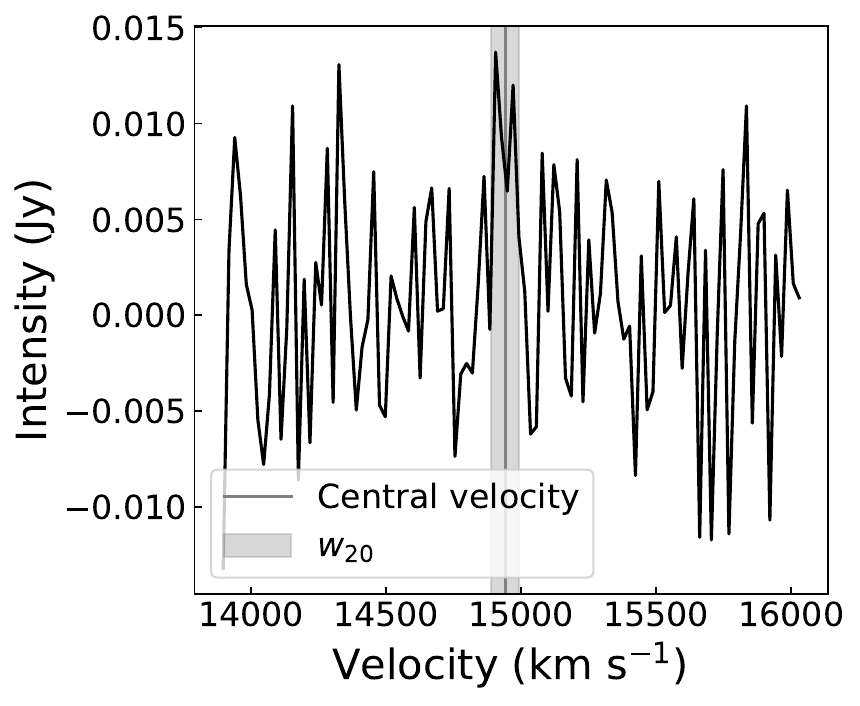}
			\caption{ }
   \end{subfigure}
	\caption{WALLABY J132935-153750$^{*}$ (NGC 5044)}
\end{figure*}
 
\begin{figure*}
	\centering
		\begin{subfigure}[b]{0.24\textwidth}
			\centering
			\includegraphics[width=\textwidth]{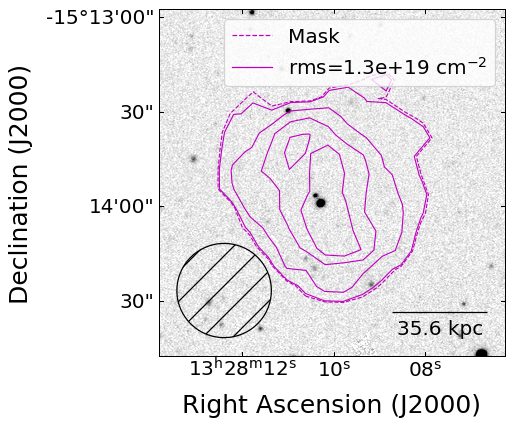}
			\caption{ }
		\end{subfigure}
		\begin{subfigure}[b]{0.24\textwidth}
			\centering
			\includegraphics[width=\textwidth]{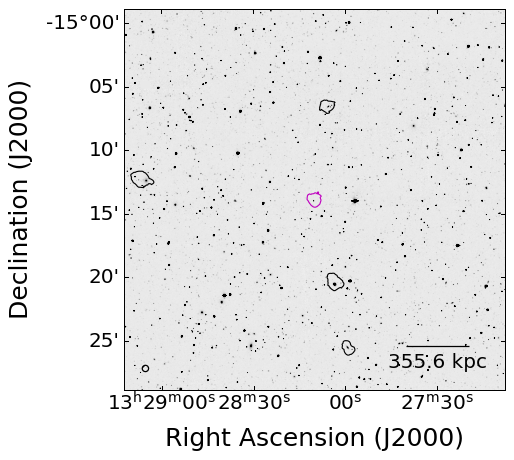}
			\caption{ }
		\end{subfigure}
		\begin{subfigure}[b]{0.26\textwidth}
			\centering
			\includegraphics[width=\textwidth]{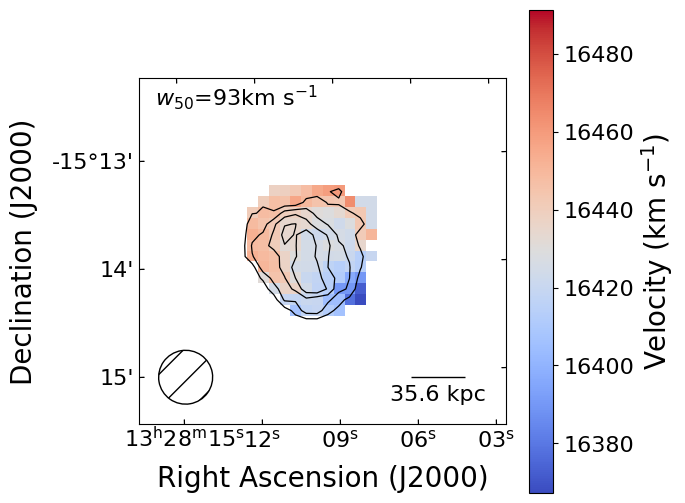}
			\caption{ }
		\end{subfigure}
    \begin{subfigure}[b]{0.24\textwidth}
			\centering
			\includegraphics[width=\textwidth]{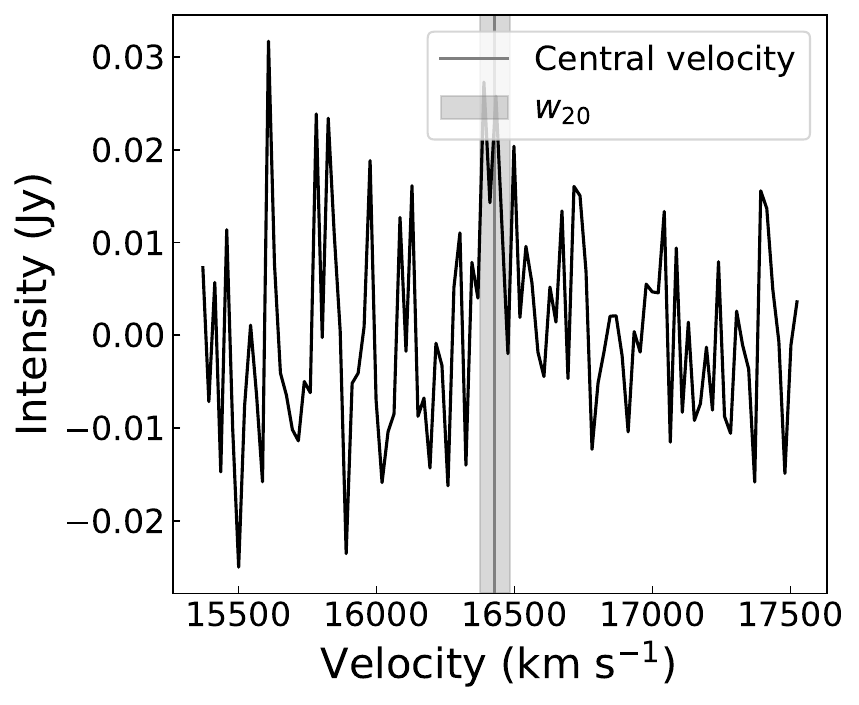}
			\caption{ }
   \end{subfigure}
	\caption{WALLABY J132810-151352$^{*}$ (NGC 5044)}
\end{figure*}
 
\begin{figure*}
	\centering
		\begin{subfigure}[b]{0.24\textwidth}
			\centering
			\includegraphics[width=\textwidth]{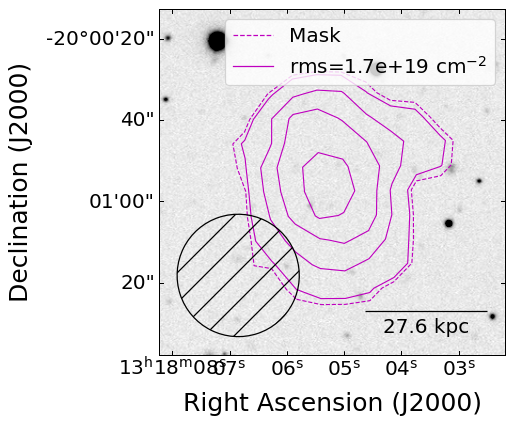}
			\caption{ }
		\end{subfigure}
		\begin{subfigure}[b]{0.24\textwidth}
			\centering
			\includegraphics[width=\textwidth]{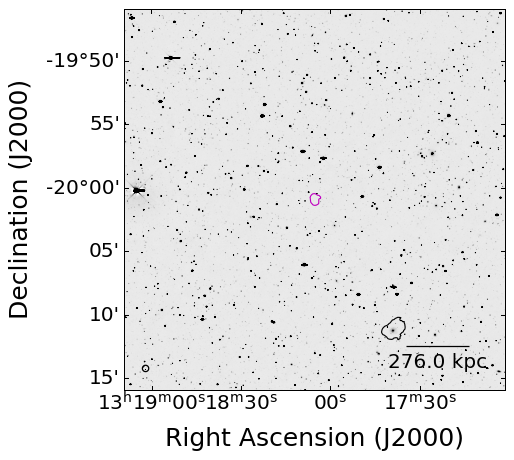}
			\caption{ }
		\end{subfigure}
		\begin{subfigure}[b]{0.26\textwidth}
			\centering
			\includegraphics[width=\textwidth]{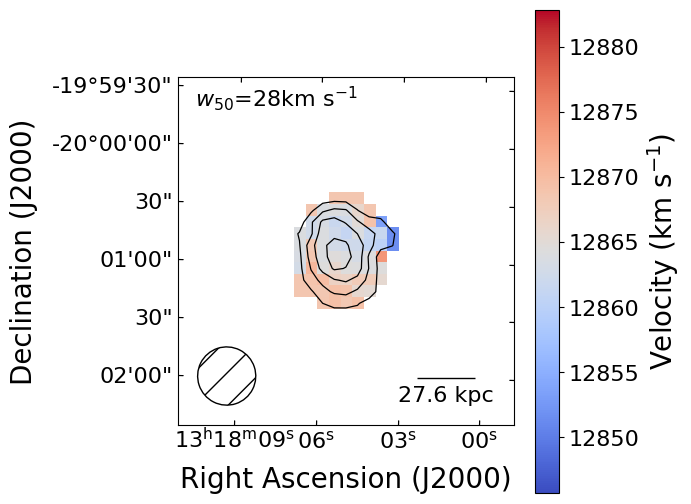}
			\caption{ }
		\end{subfigure}
    \begin{subfigure}[b]{0.24\textwidth}
			\centering
			\includegraphics[width=\textwidth]{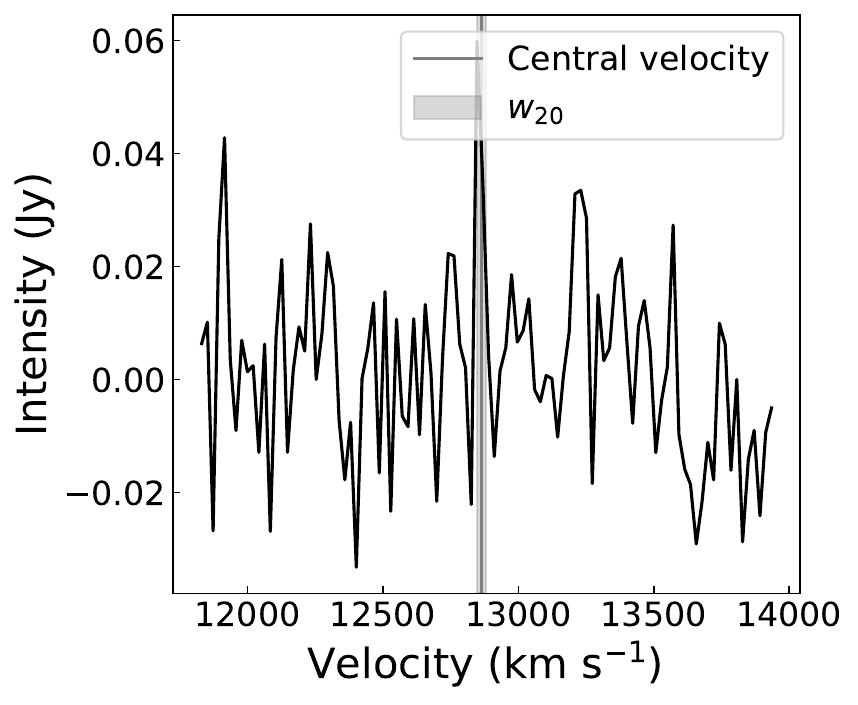}
			\caption{ }
   \end{subfigure}
	\caption{WALLABY J131805-200055$^{*}$ (NGC 5044)}
 \label{dgc_last}
\end{figure*}

\section{Co-added Images}
\label{append:coadd}

 {Here we present all the co-added images of the dark sources. For each source, we co-add the $g$, $r$, $i$ and $z$-band images, convolving with a boxcar kernel with a size of 2.6 arcsec by 2.6 arcsec to degrade the resolution and enhance the surface brightness sensitivity, greater enabling the detection of diffuse emission. The same H{\sc i} contours shown are the same as those presented in Appendices \ref{append:strong} and \ref{append:weak}. WALLABY J131244-155218 (also presented in Figure \ref{fig:faint} in Section \ref{sec:disc-dgc}) is the only source that shows evidence of an optical counterpart.}

\clearpage

\begin{figure*}
     \centering
    \begin{subfigure}[b]{0.24\textwidth}
         \centering
         \includegraphics[width=\textwidth]{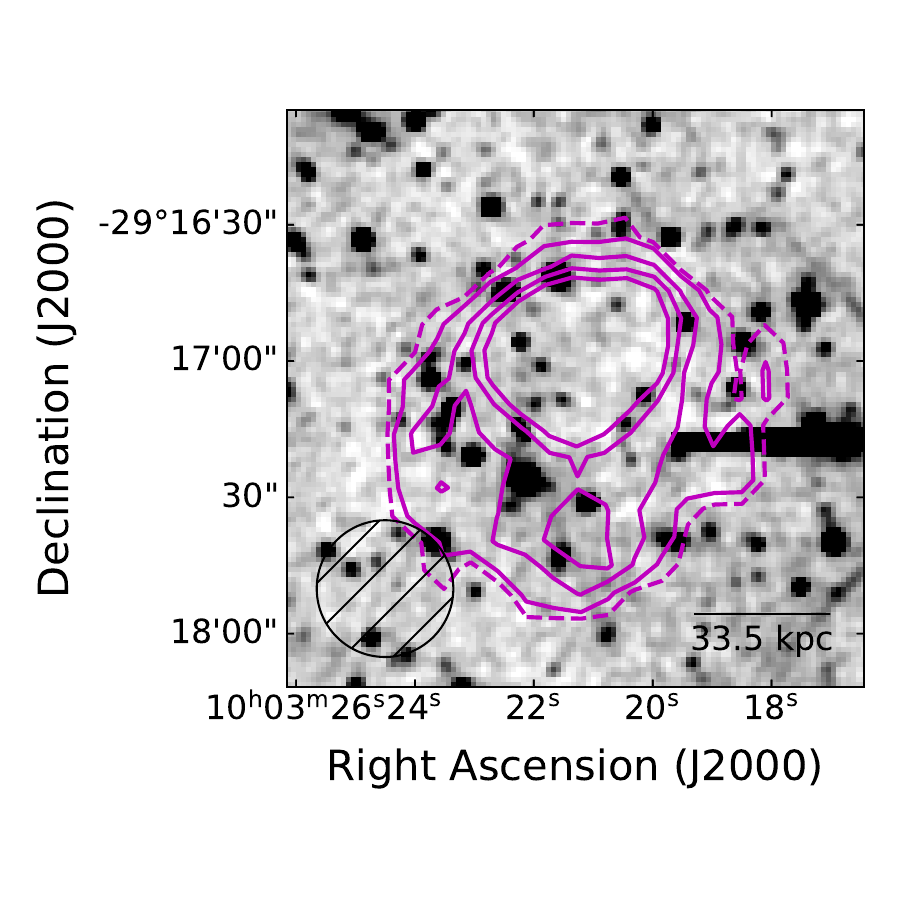}
         \caption{WALLABY J100321-291708}
     \end{subfigure}
     \begin{subfigure}[b]{0.24\textwidth}
         \centering
         \includegraphics[width=\textwidth]{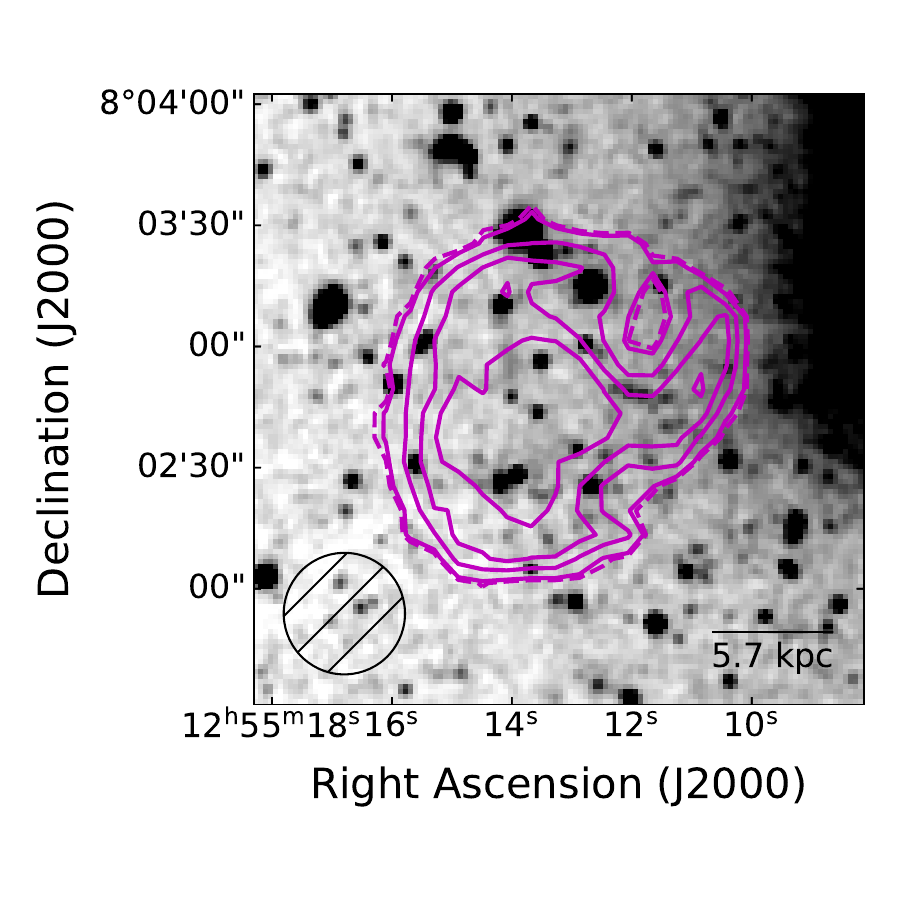}
         \caption{WALLABY J125513+080246}
     \end{subfigure}
     \begin{subfigure}[b]{0.24\textwidth}
         \centering
         \includegraphics[width=\textwidth]{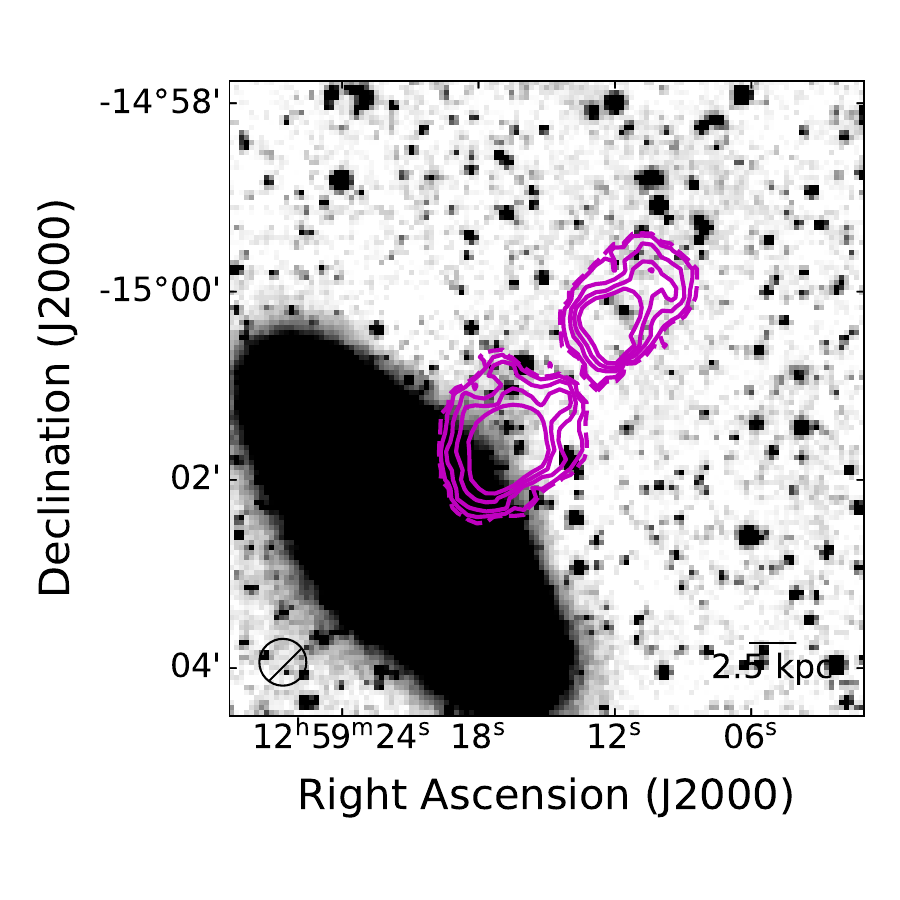}
         \caption{WALLABY J125915-150108}
     \end{subfigure}
     \begin{subfigure}[b]{0.24\textwidth}
         \centering
         \includegraphics[width=\textwidth]{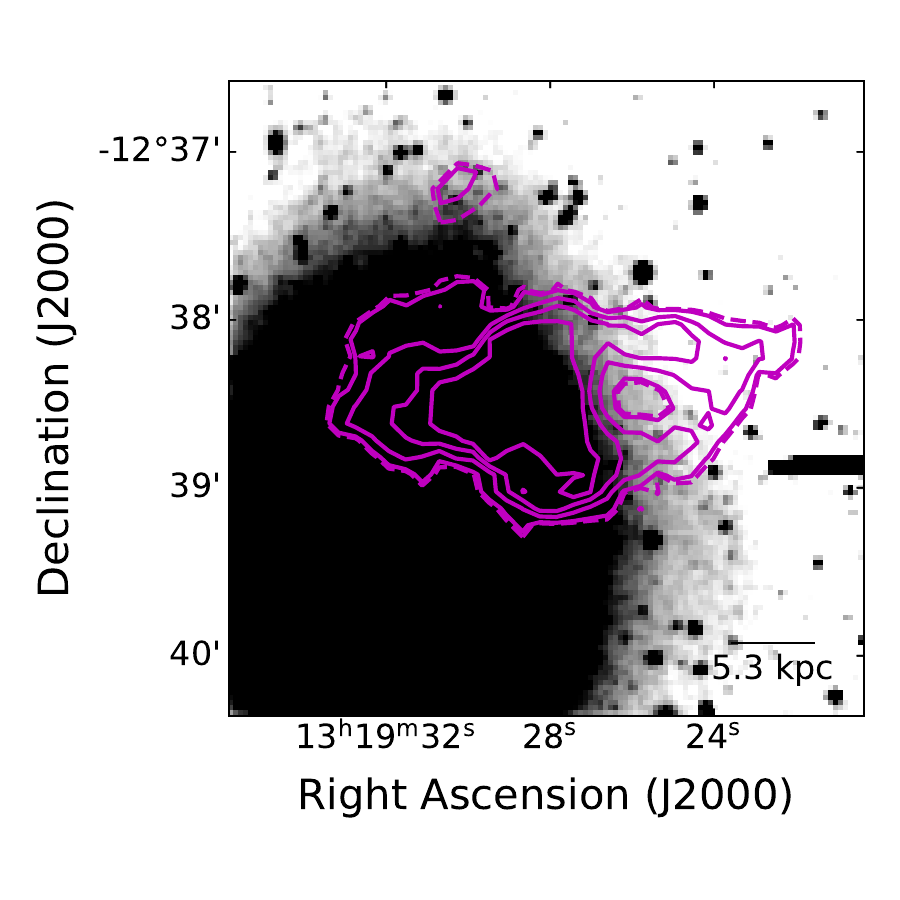}
         \caption{WALLABY J131928-123828}
     \end{subfigure}
     
     \begin{subfigure}[b]{0.24\textwidth}
         \centering
         \includegraphics[width=\textwidth]{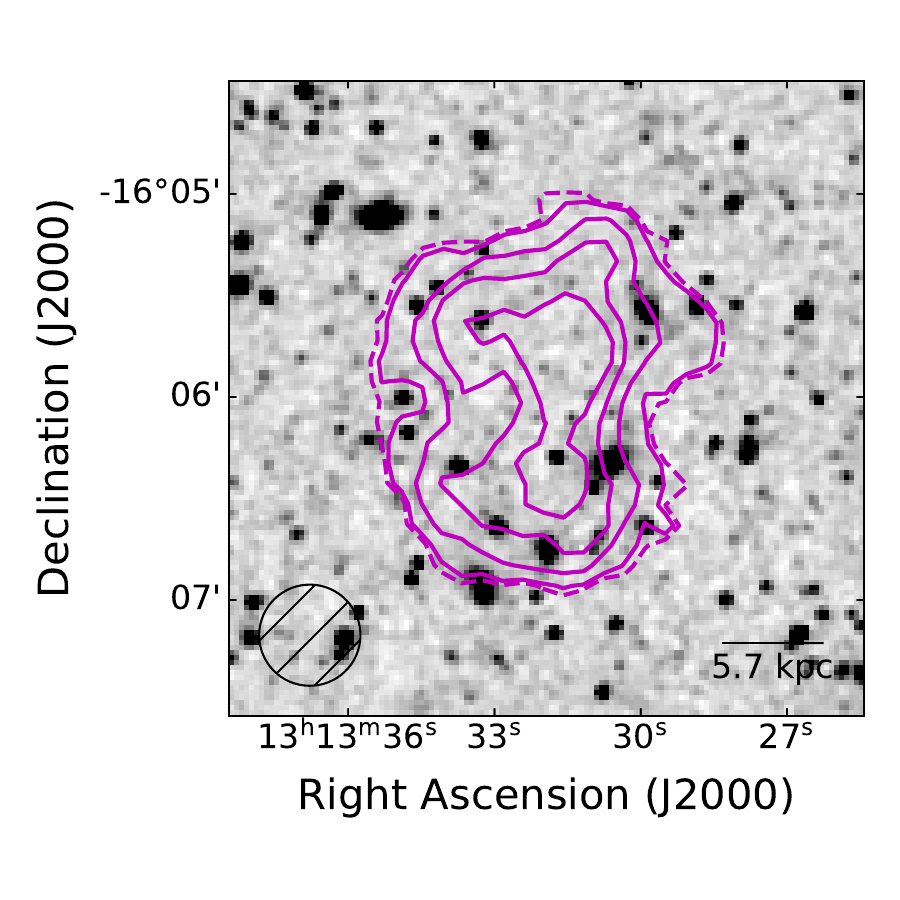}
         \caption{WALLABY J131331-160600}
     \end{subfigure}
     \begin{subfigure}[b]{0.24\textwidth}
         \centering
         \includegraphics[width=\textwidth]{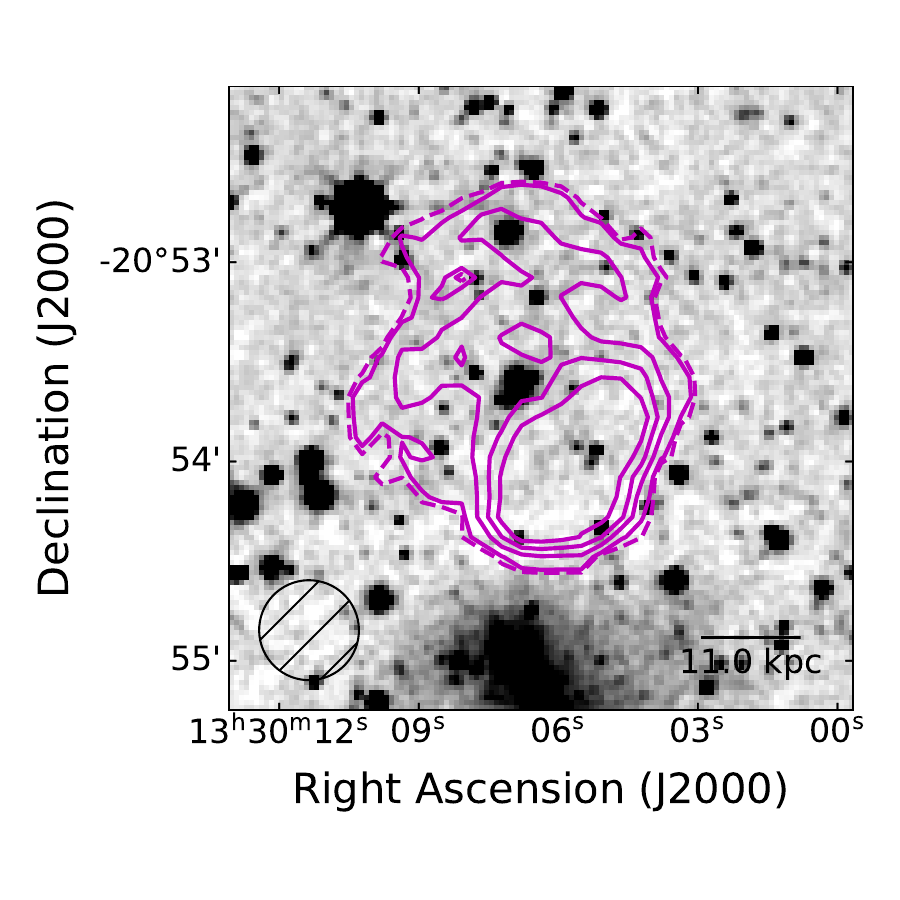}
         \caption{WALLABY J133006-205341}
     \end{subfigure}
     \begin{subfigure}[b]{0.24\textwidth}
         \centering
         \includegraphics[width=\textwidth]{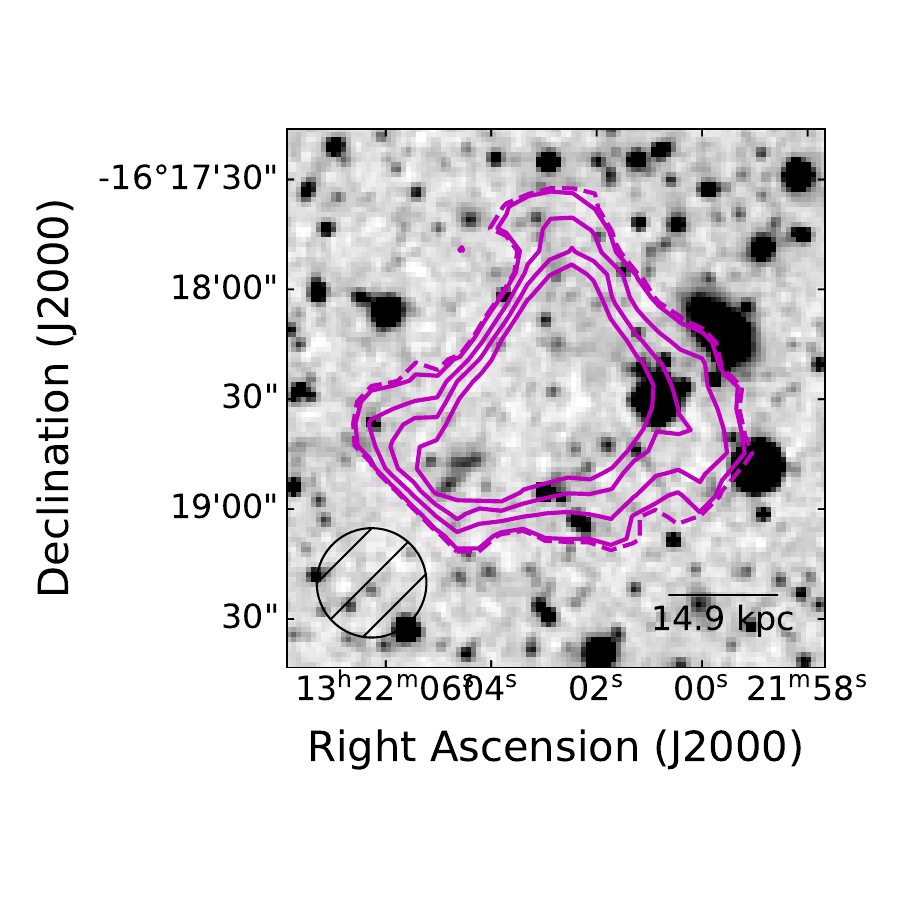}
         \caption{WALLABY J132202-161829}
     \end{subfigure}
     \begin{subfigure}[b]{0.24\textwidth}
         \centering
         \includegraphics[width=\textwidth]{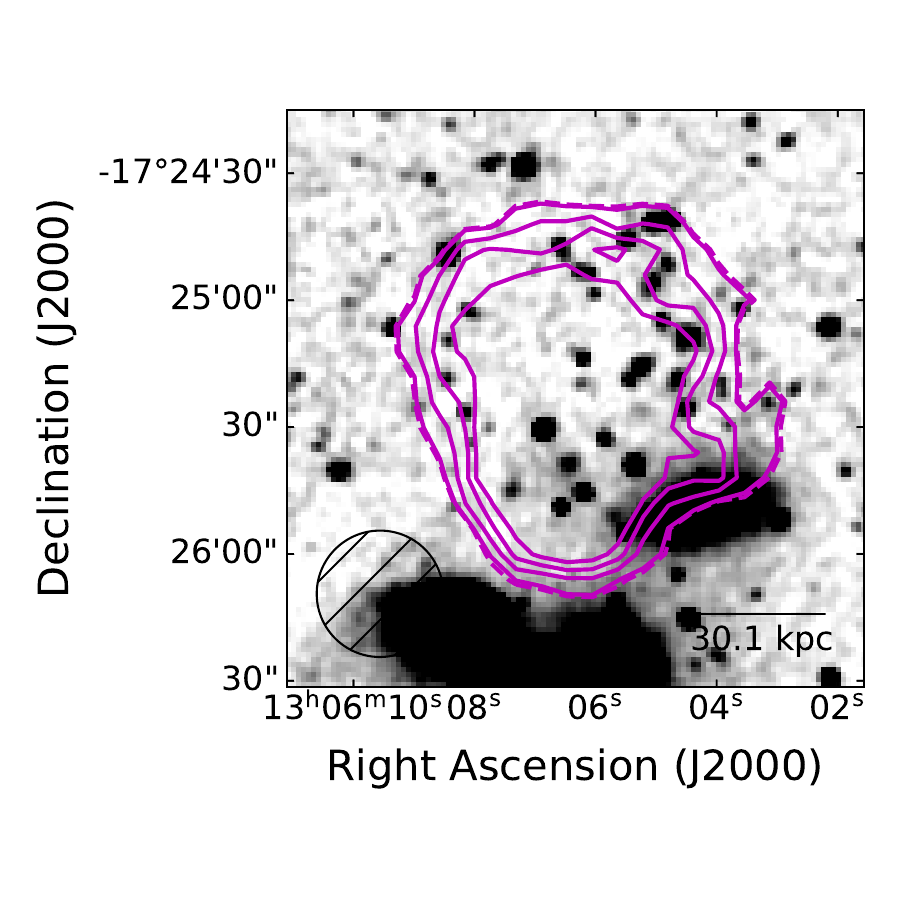}
         \caption{WALLABY J130606-17252}
     \end{subfigure}

     \begin{subfigure}[b]{0.24\textwidth}
         \centering
         \includegraphics[width=\textwidth]{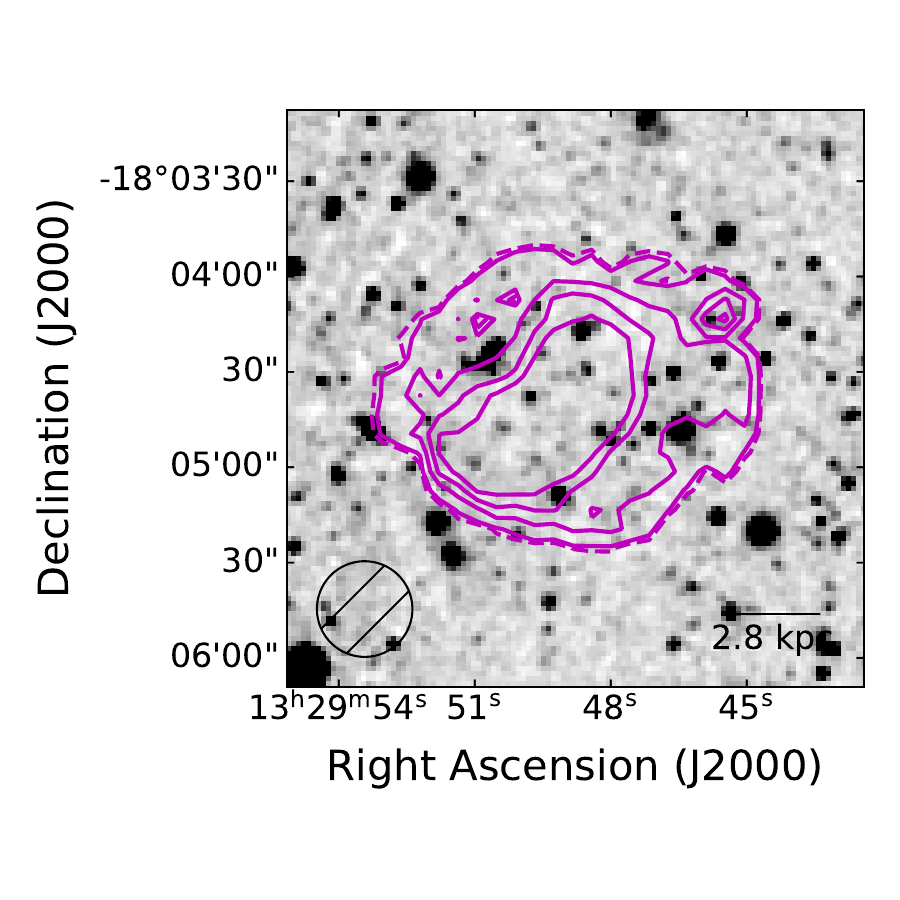}
         \caption{WALLABY J132948-180438}
     \end{subfigure}
     \begin{subfigure}[b]{0.24\textwidth}
         \centering
         \includegraphics[width=\textwidth]{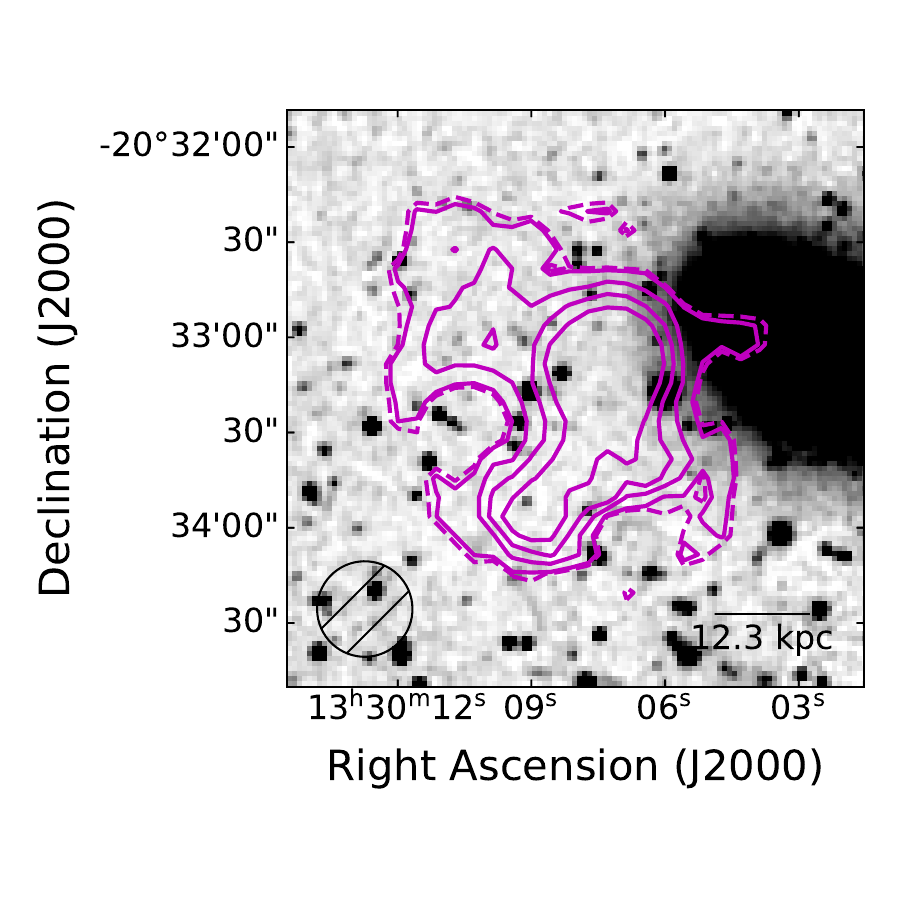}
         \caption{WALLABY J133008-203319}
     \end{subfigure}
     \begin{subfigure}[b]{0.24\textwidth}
         \centering
         \includegraphics[width=\textwidth]{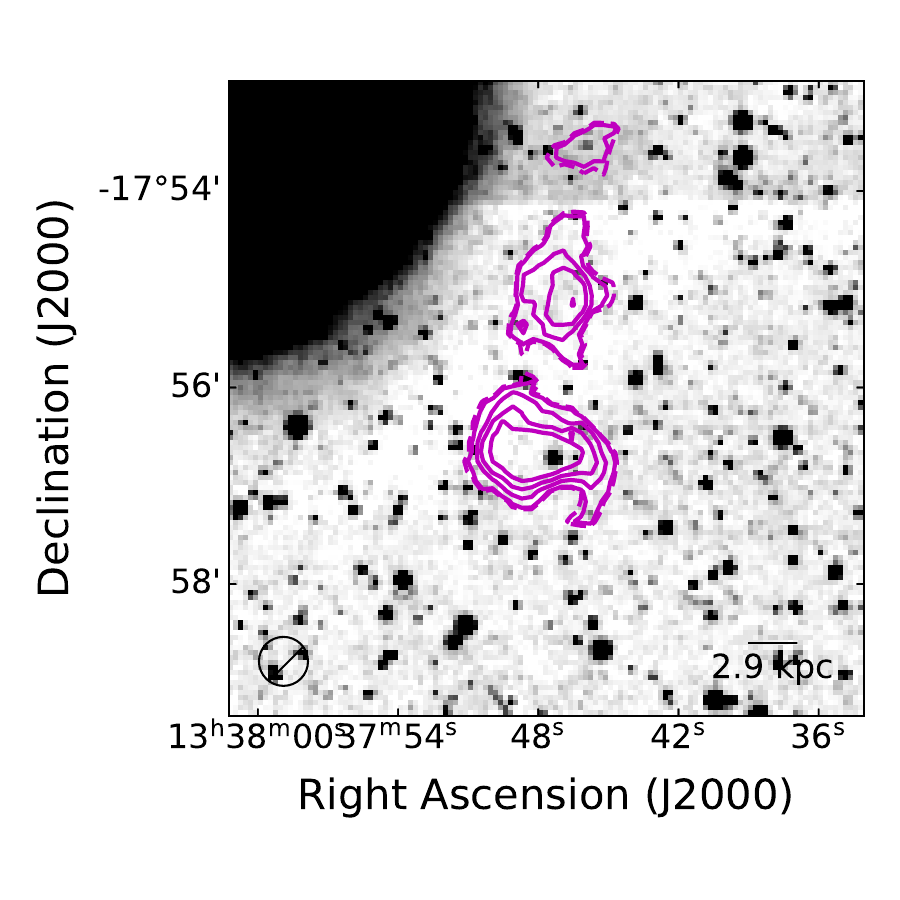}
         \caption{WALLABY J133747-175606}
     \end{subfigure}
     \begin{subfigure}[b]{0.24\textwidth}
         \centering
         \includegraphics[width=\textwidth]{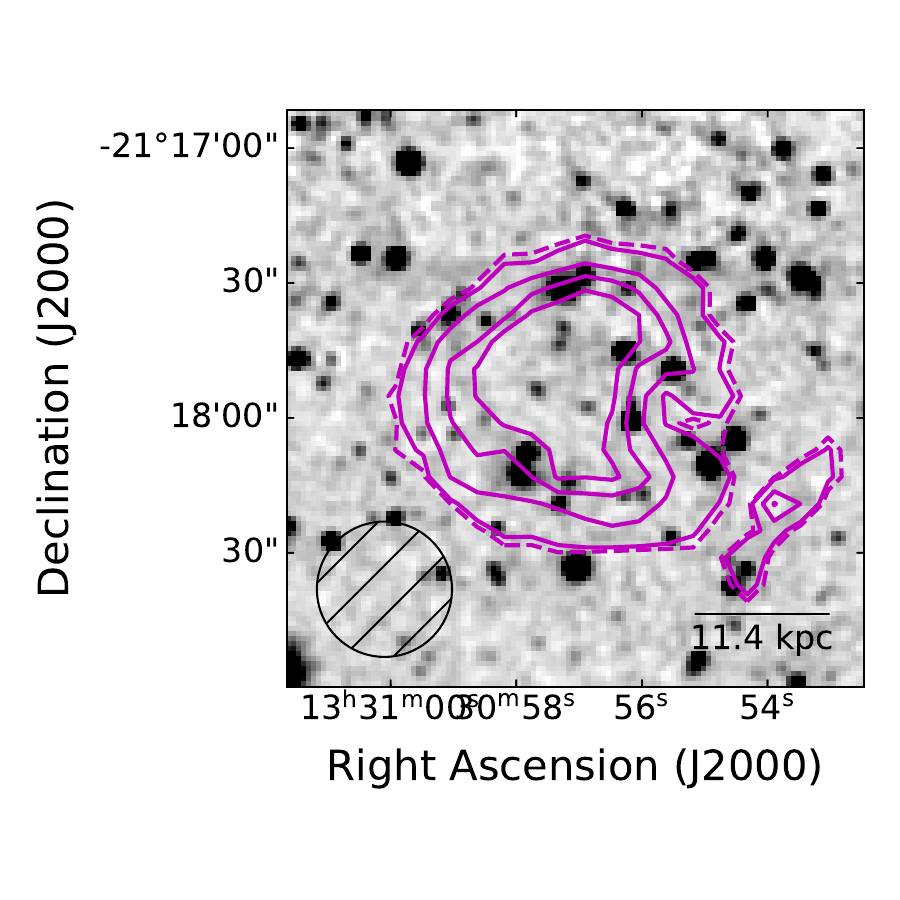}
         \caption{WALLABY J133057-211755}
     \end{subfigure}

     \begin{subfigure}[b]{0.24\textwidth}
         \centering
         \includegraphics[width=\textwidth]{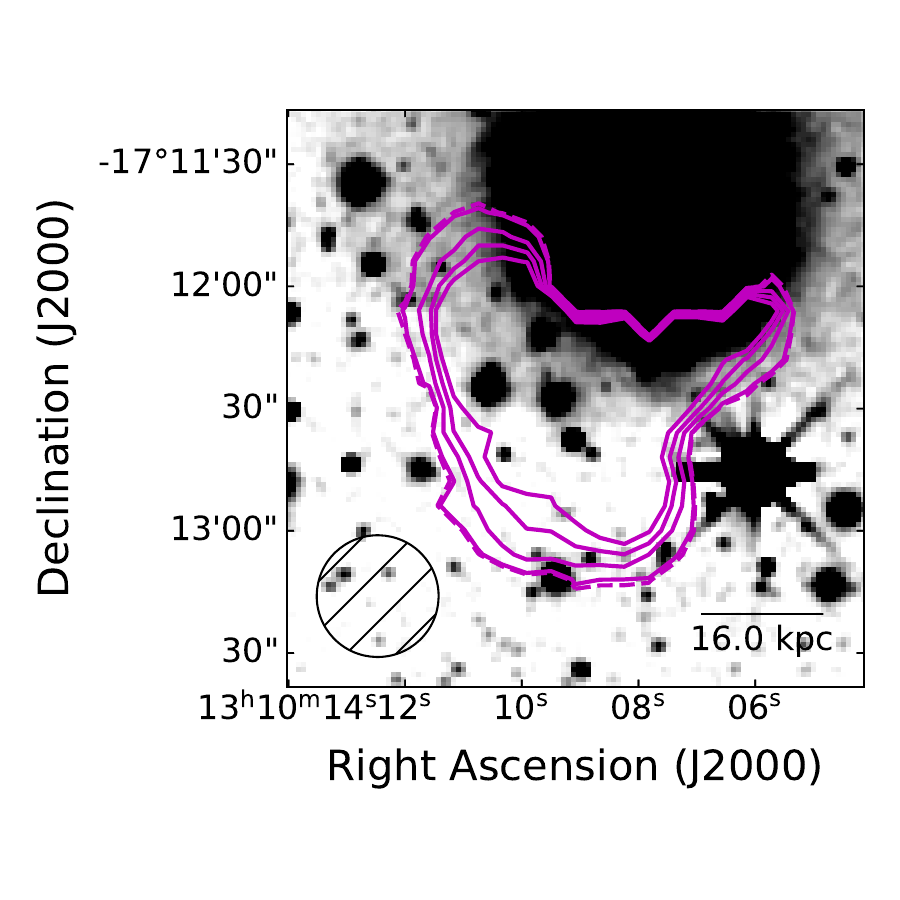}
         \caption{WALLABY J131009-171227}
     \end{subfigure}

    \caption{Co-added images of tidal sources}
    \label{fig:coadd_t}
\end{figure*}

\begin{figure*}
     \centering
    \begin{subfigure}[b]{0.24\textwidth}
         \centering
         \includegraphics[width=\textwidth]{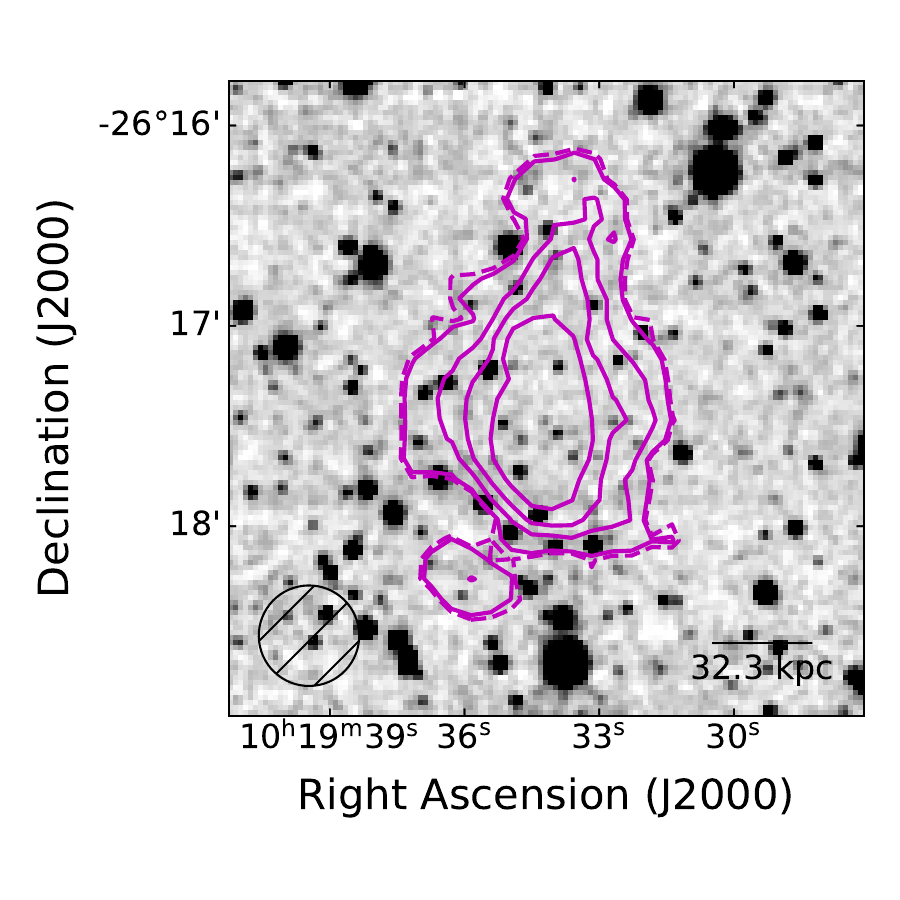}
         \caption{WALLABY J101934-261721}
     \end{subfigure}
     \begin{subfigure}[b]{0.24\textwidth}
         \centering
         \includegraphics[width=\textwidth]{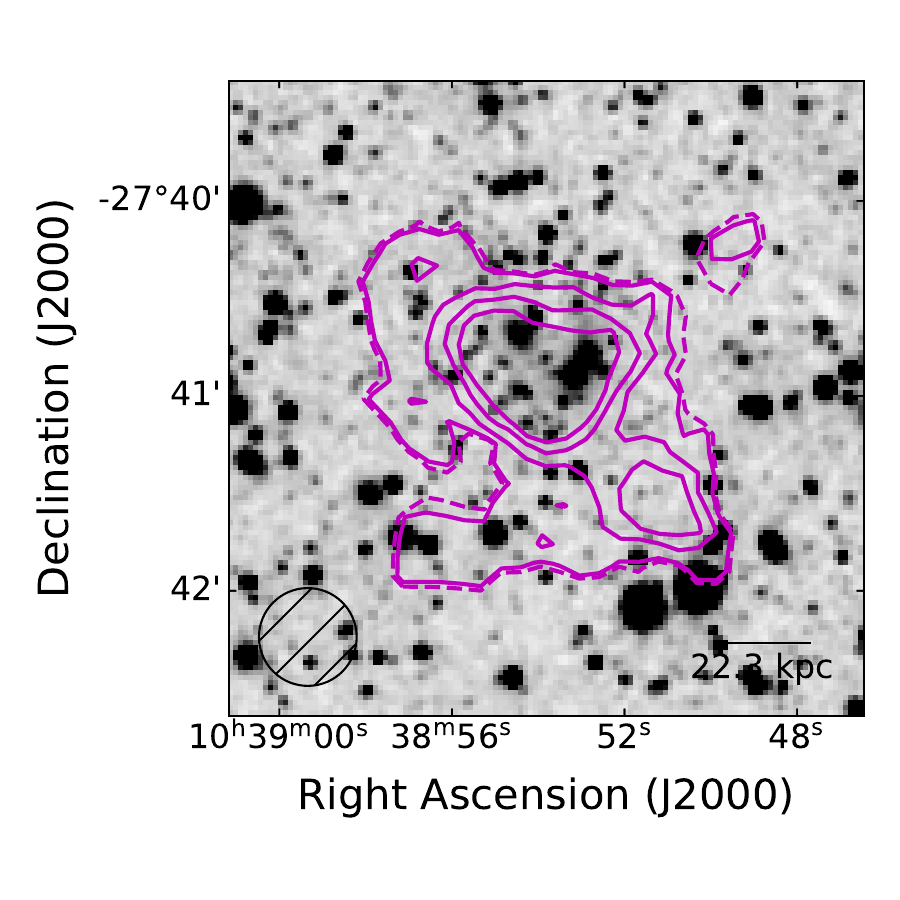}
         \caption{WALLABY J103853-274100}
     \end{subfigure}
     \begin{subfigure}[b]{0.24\textwidth}
         \centering
         \includegraphics[width=\textwidth]{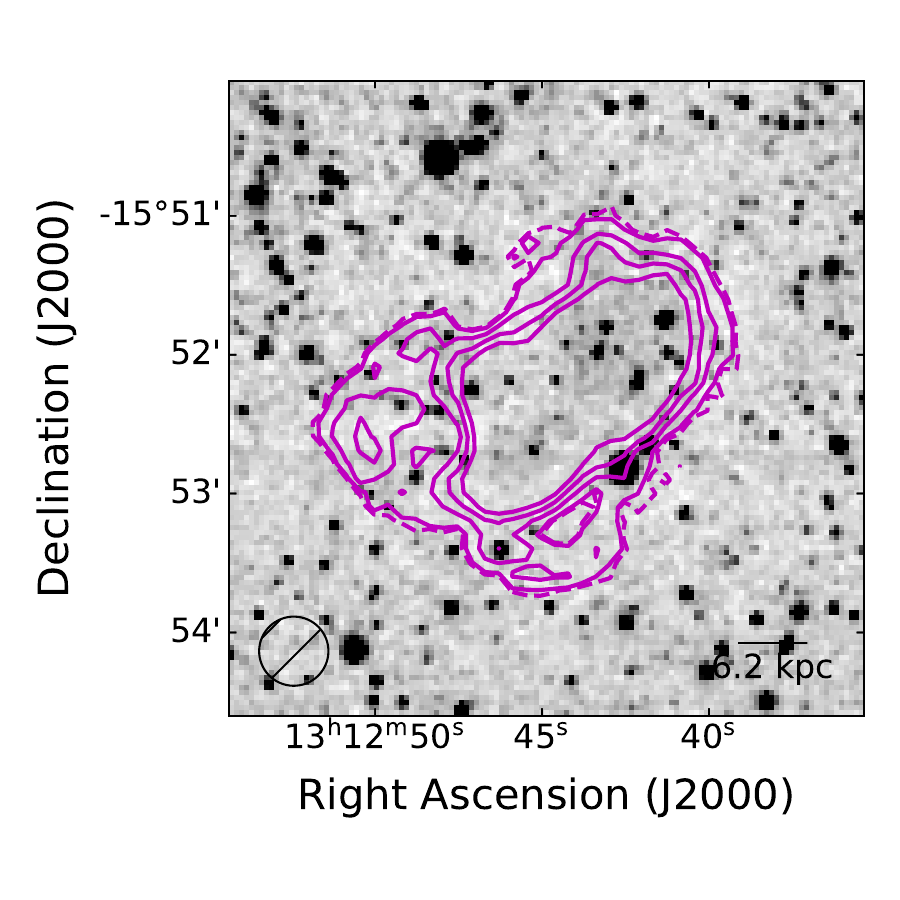}
         \caption{WALLABY J131244-155218}
     \end{subfigure}
     \begin{subfigure}[b]{0.24\textwidth}
         \centering
         \includegraphics[width=\textwidth]{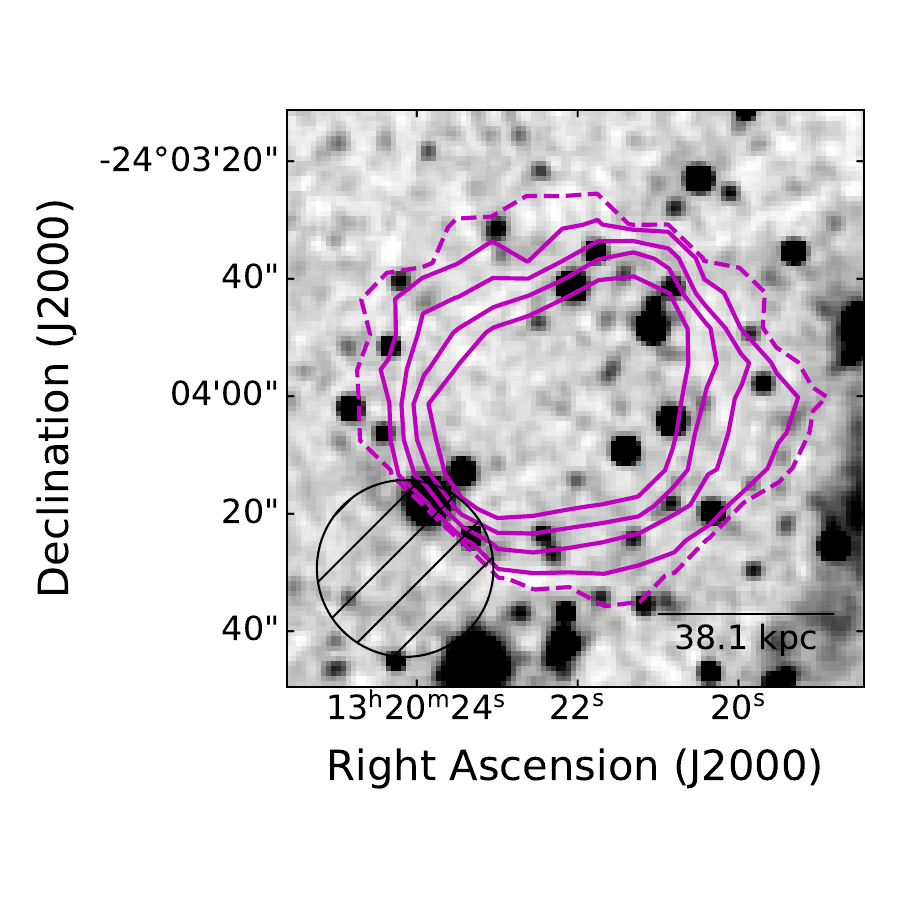}
         \caption{WALLABY J132022-240400}
     \end{subfigure}

    \begin{subfigure}[b]{0.24\textwidth}
         \centering
         \includegraphics[width=\textwidth]{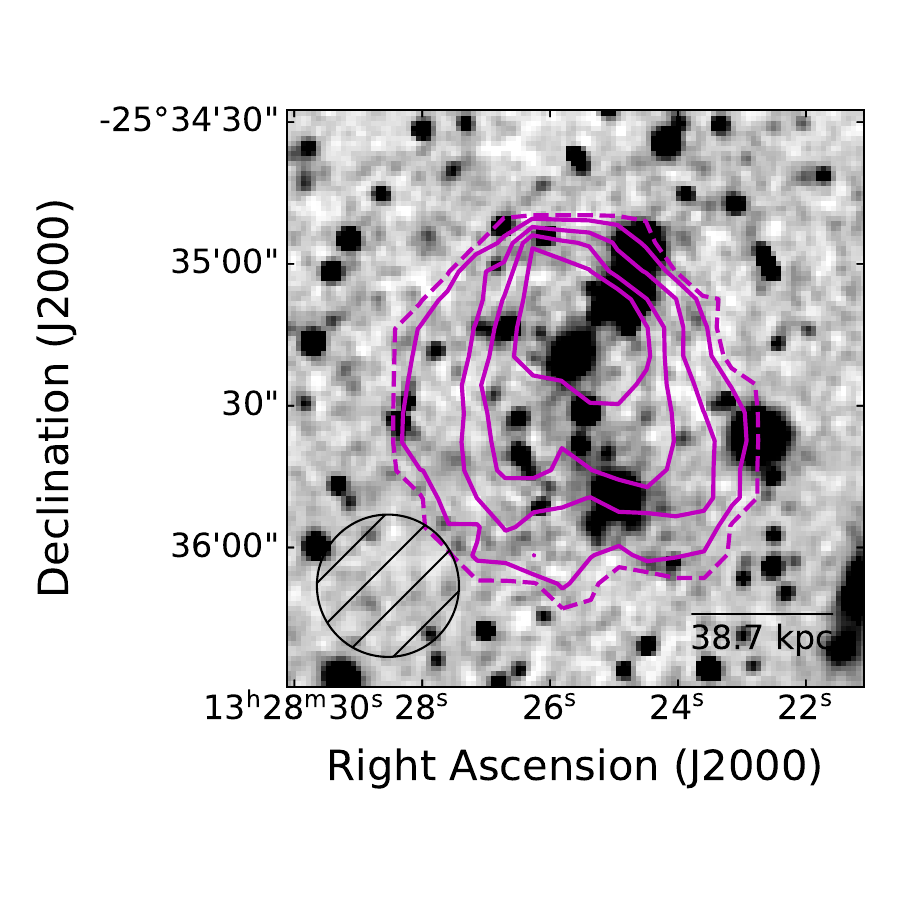}
         \caption{WALLABY J132825-253528}
     \end{subfigure}
     \begin{subfigure}[b]{0.24\textwidth}
         \centering
         \includegraphics[width=\textwidth]{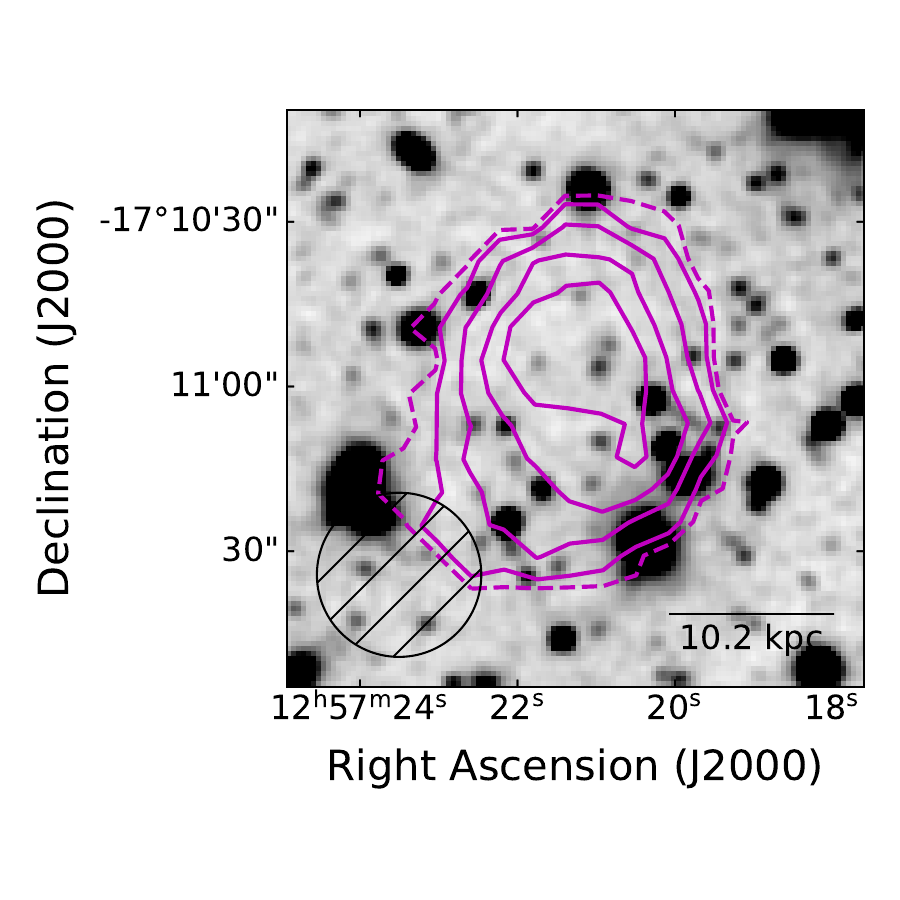}
         \caption{WALLABY J125721-171102}
     \end{subfigure}
     \begin{subfigure}[b]{0.24\textwidth}
         \centering
         \includegraphics[width=\textwidth]{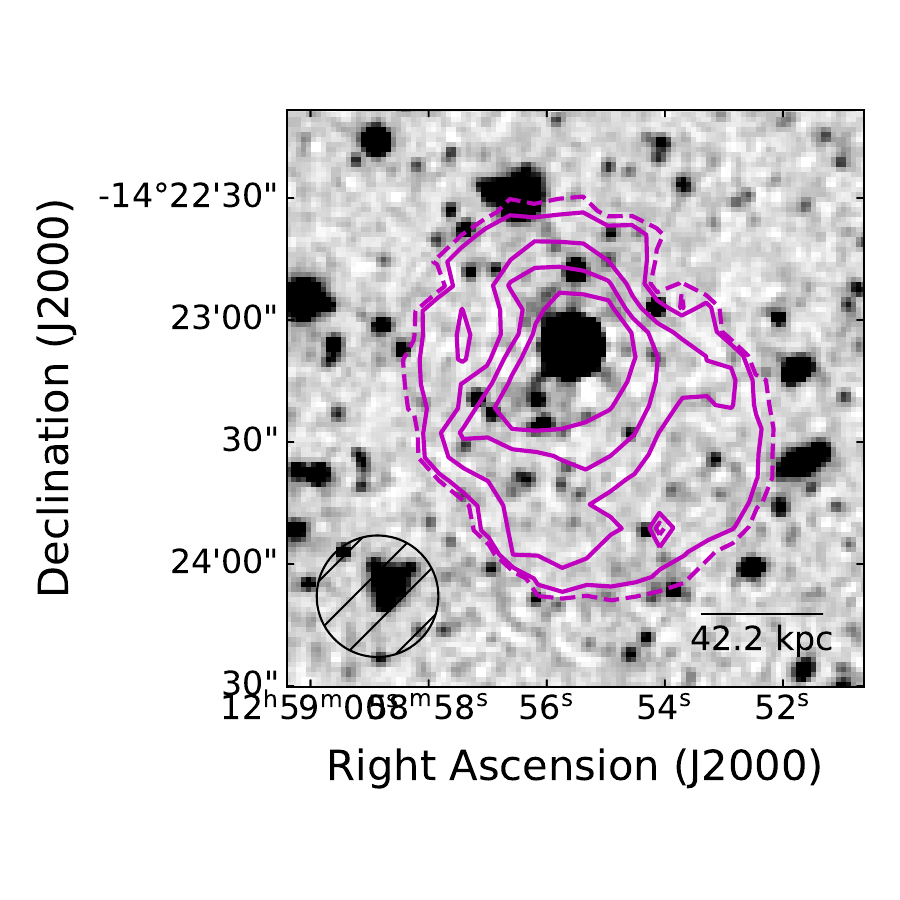}
         \caption{WALLABY J125855-142319}
     \end{subfigure}
     \begin{subfigure}[b]{0.24\textwidth}
         \centering
         \includegraphics[width=\textwidth]{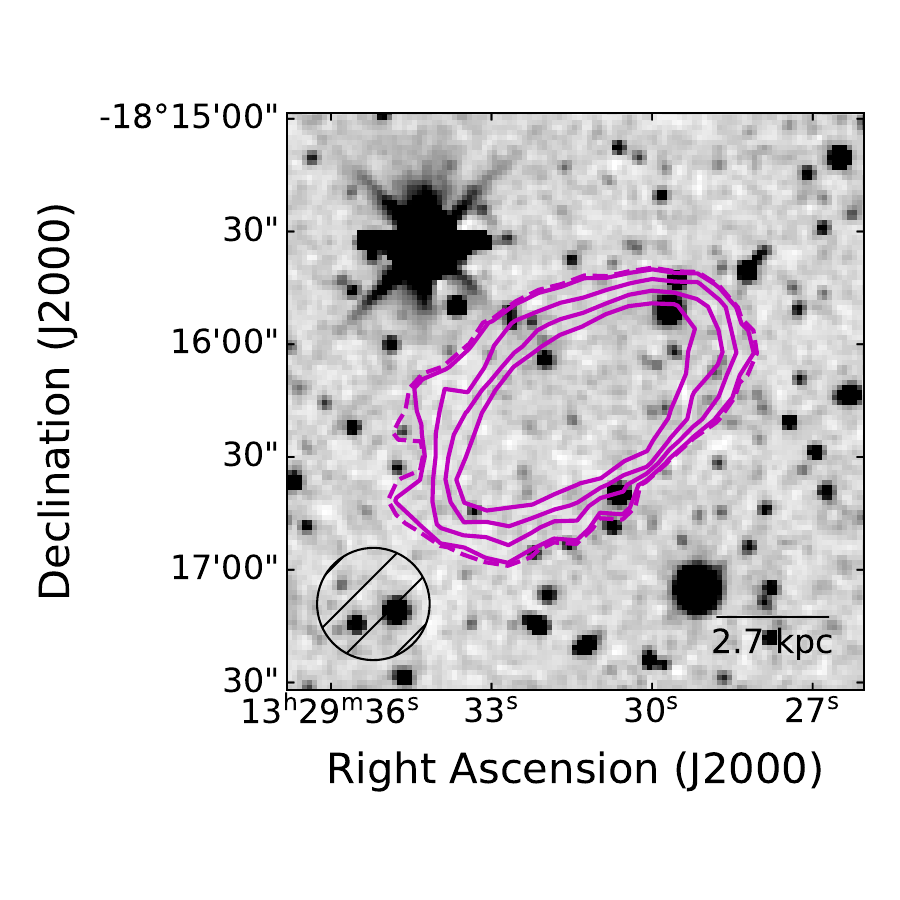}
         \caption{WALLABY J132931-181615}
     \end{subfigure}

     \begin{subfigure}[b]{0.24\textwidth}
         \centering
         \includegraphics[width=\textwidth]{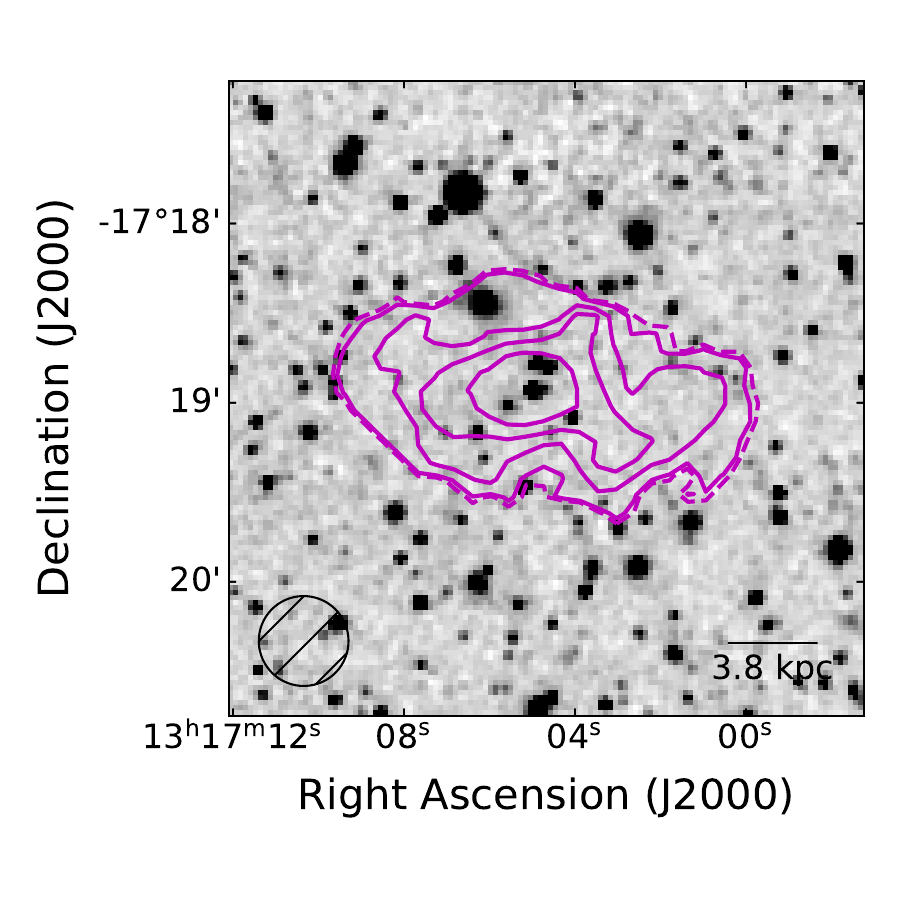}
         \caption{WALLABY J131704-171858}
     \end{subfigure}
     \begin{subfigure}[b]{0.24\textwidth}
         \centering
         \includegraphics[width=\textwidth]{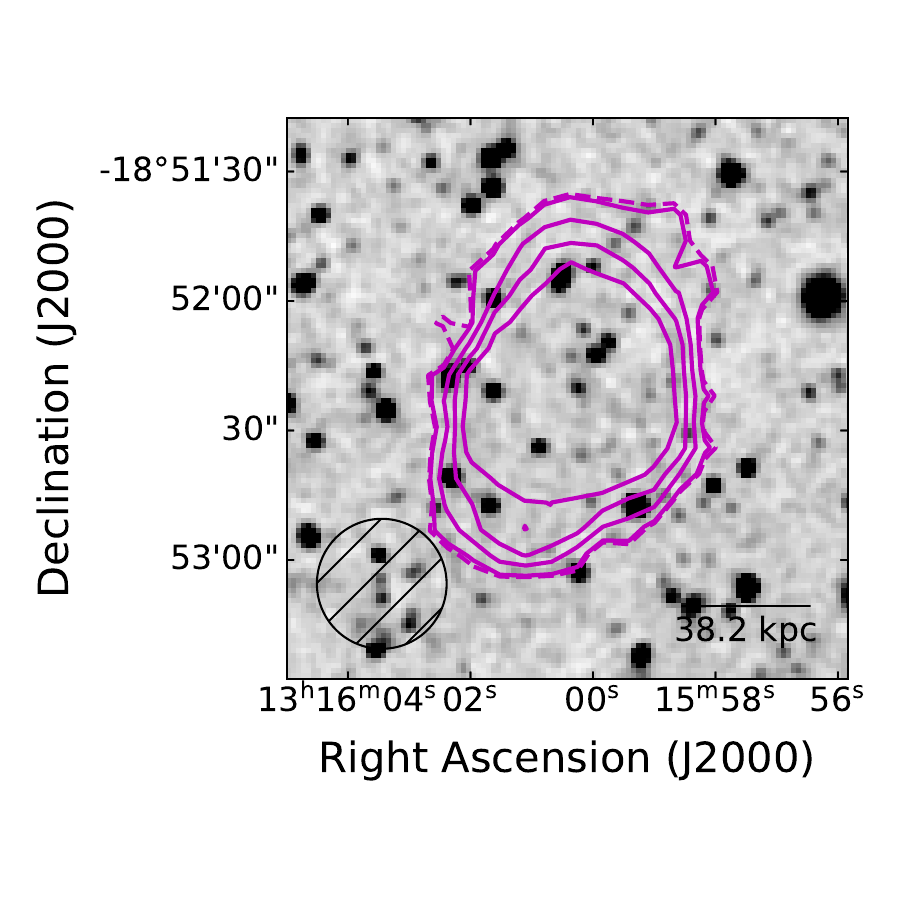}
         \caption{WALLABY J131600-185222}
     \end{subfigure}
     \begin{subfigure}[b]{0.24\textwidth}
         \centering
         \includegraphics[width=\textwidth]{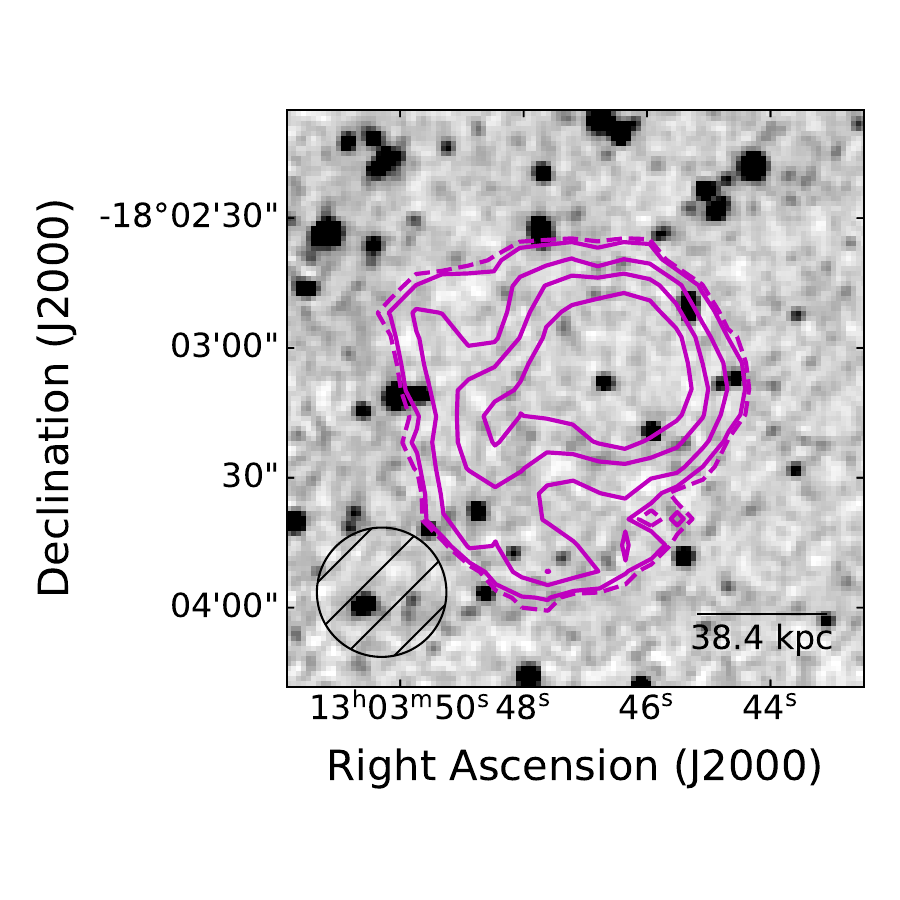}
         \caption{WALLABY J130347-180311}
     \end{subfigure}
     \begin{subfigure}[b]{0.24\textwidth}
         \centering
         \includegraphics[width=\textwidth]{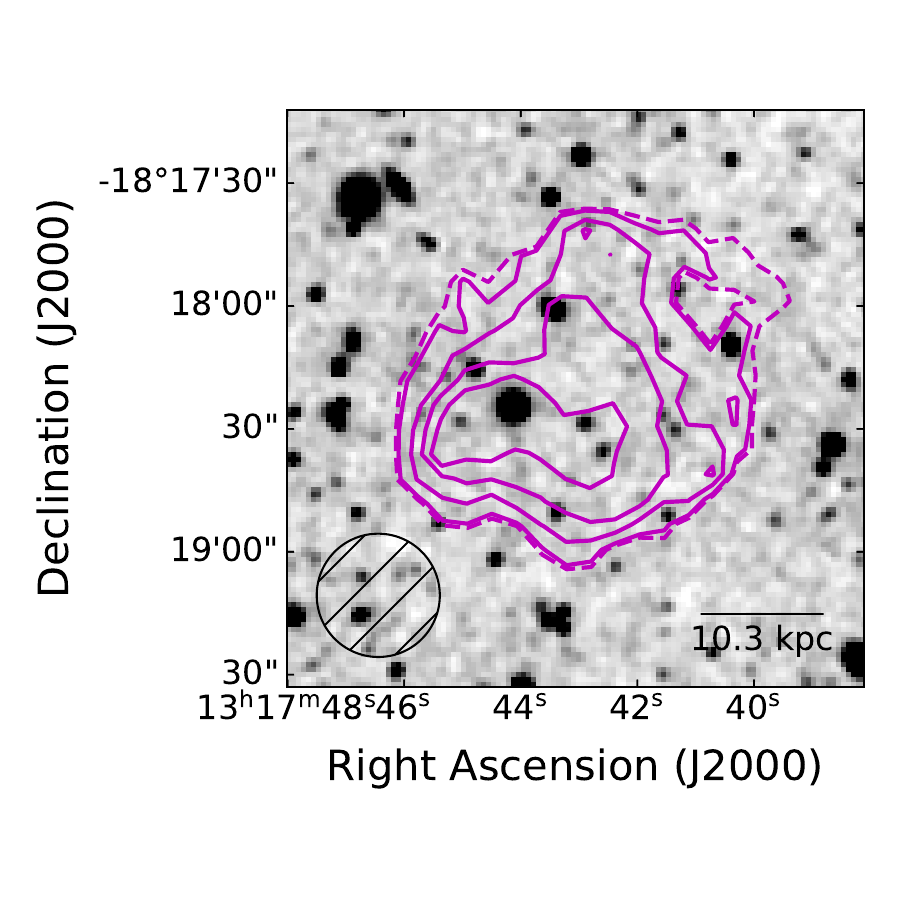}
         \caption{WALLABY J131743-181822}
     \end{subfigure}

     \begin{subfigure}[b]{0.24\textwidth}
         \centering
         \includegraphics[width=\textwidth]{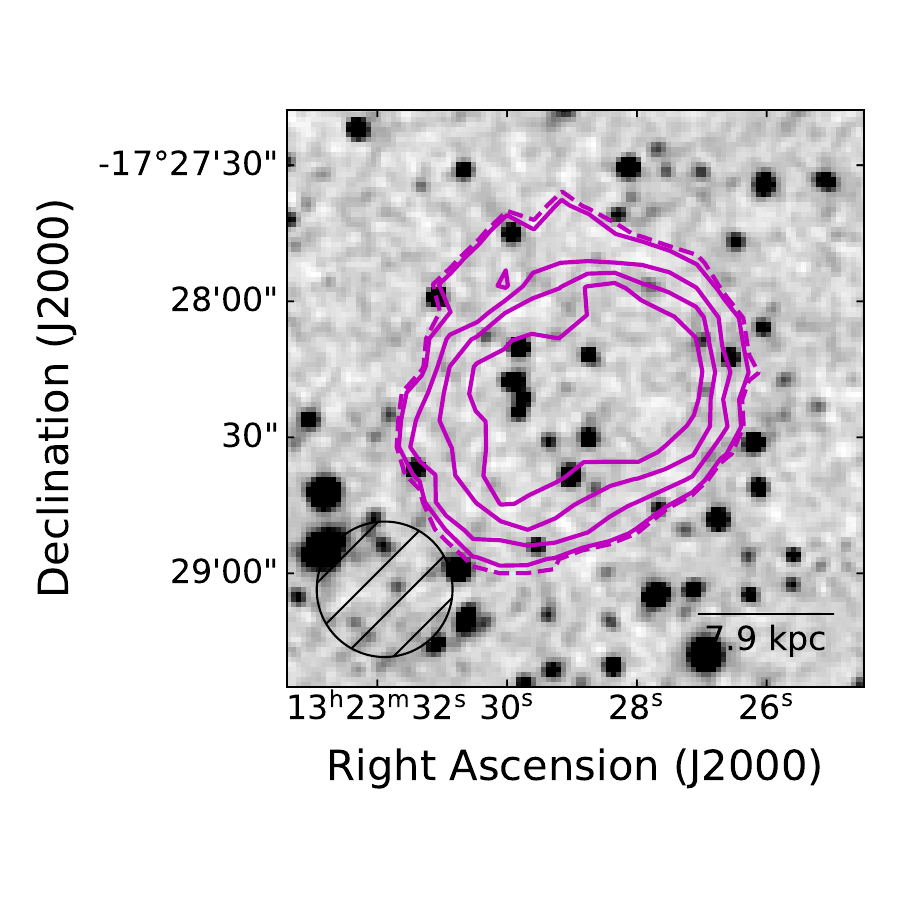}
         \caption{WALLABY J132328-172821}
     \end{subfigure}
     \begin{subfigure}[b]{0.24\textwidth}
         \centering
         \includegraphics[width=\textwidth]{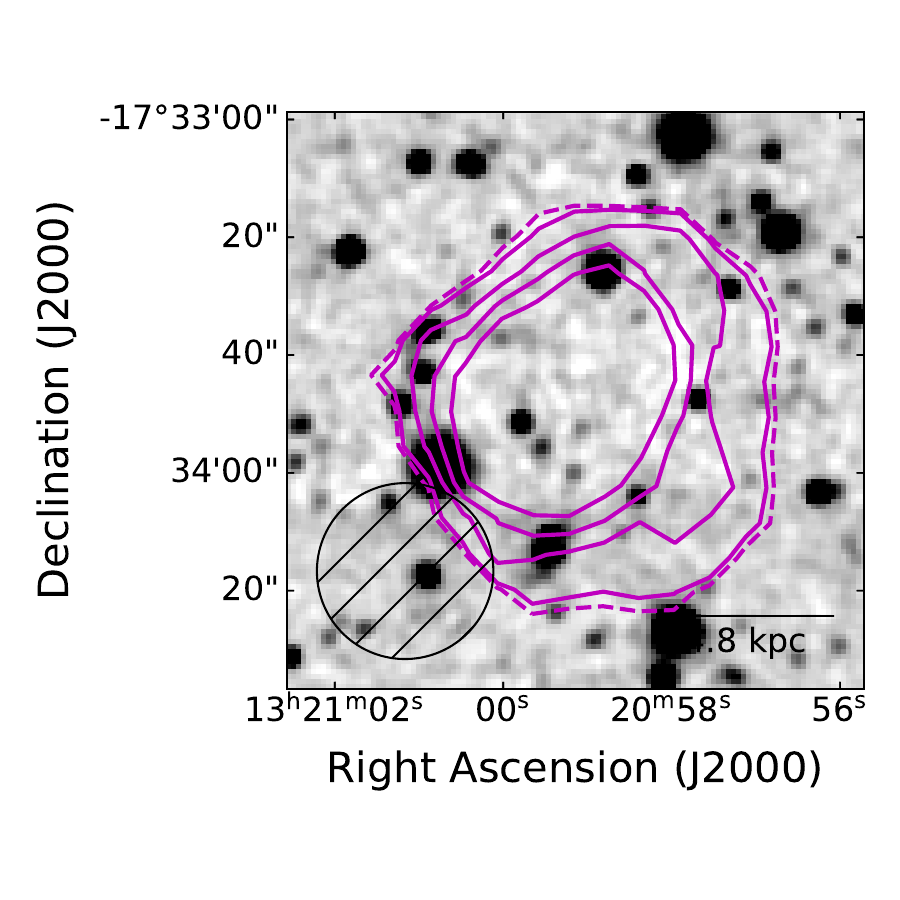}
         \caption{WALLABY J132059-173347}
     \end{subfigure}
     \begin{subfigure}[b]{0.24\textwidth}
         \centering
         \includegraphics[width=\textwidth]{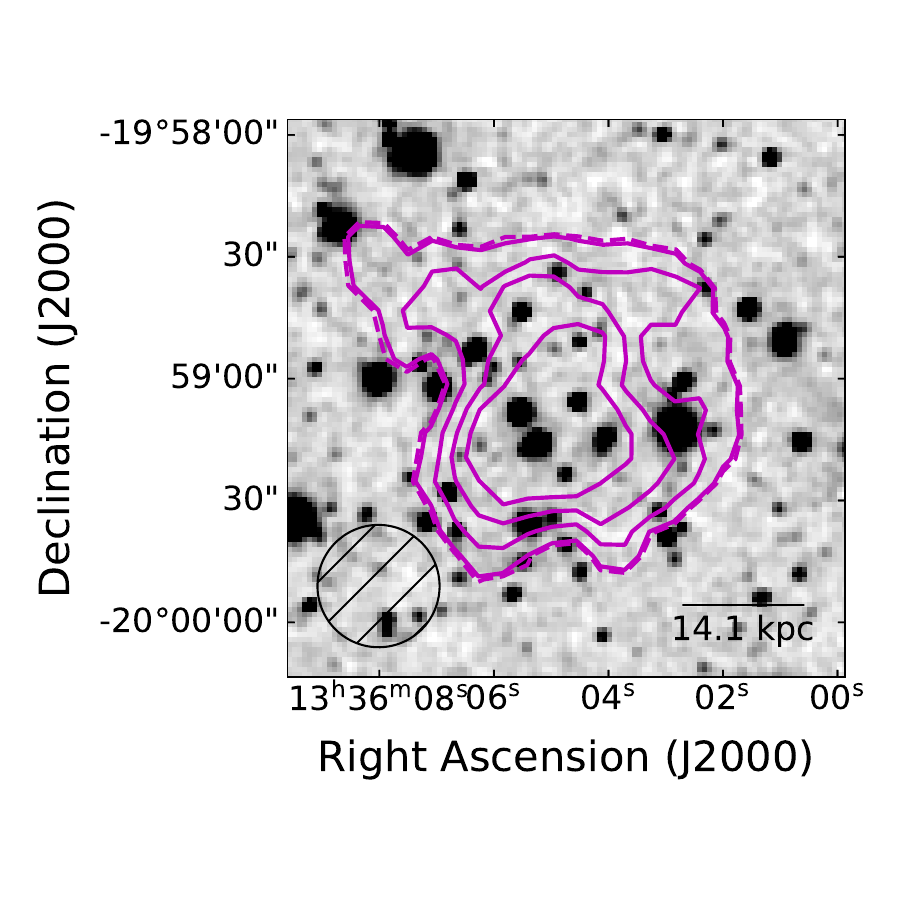}
         \caption{WALLABY J133604-195904}
     \end{subfigure}
     \begin{subfigure}[b]{0.24\textwidth}
         \centering
         \includegraphics[width=\textwidth]{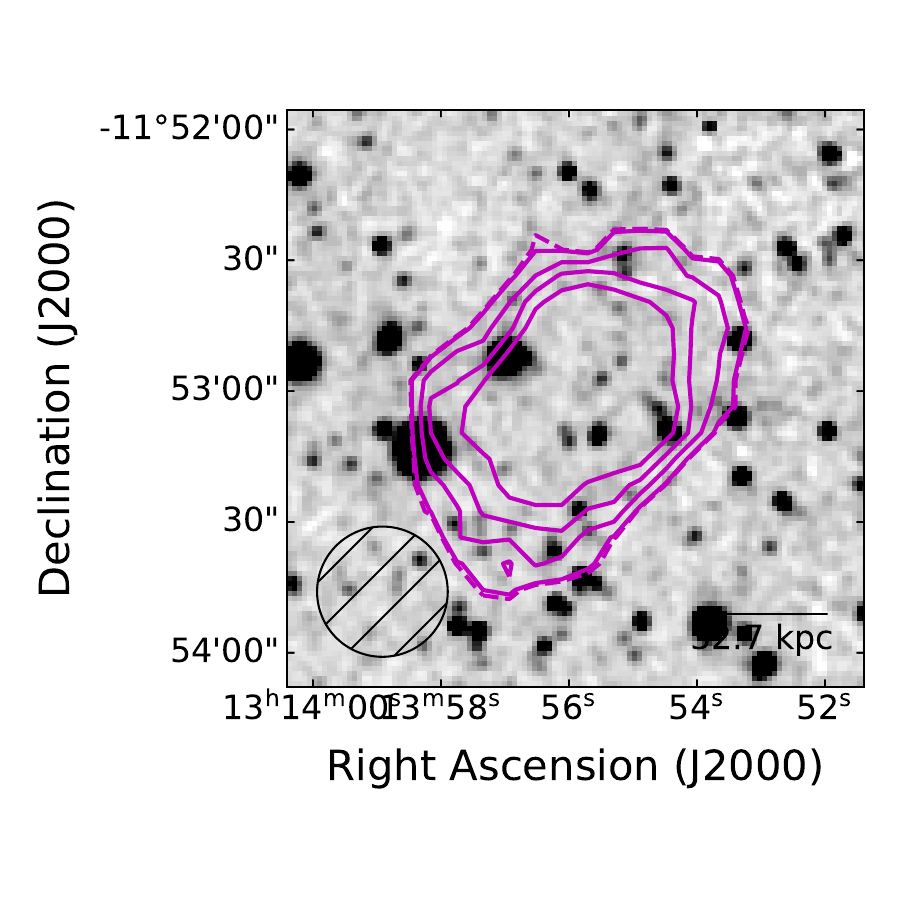}
         \caption{WALLABY J131355-115301}
     \end{subfigure}

    \caption{Co-added images of other strong dark source detections}
    \label{fig:coadd_s}
\end{figure*}

\begin{figure*}\ContinuedFloat
     \centering
     
     \begin{subfigure}[b]{0.24\textwidth}
         \centering
         \includegraphics[width=\textwidth]{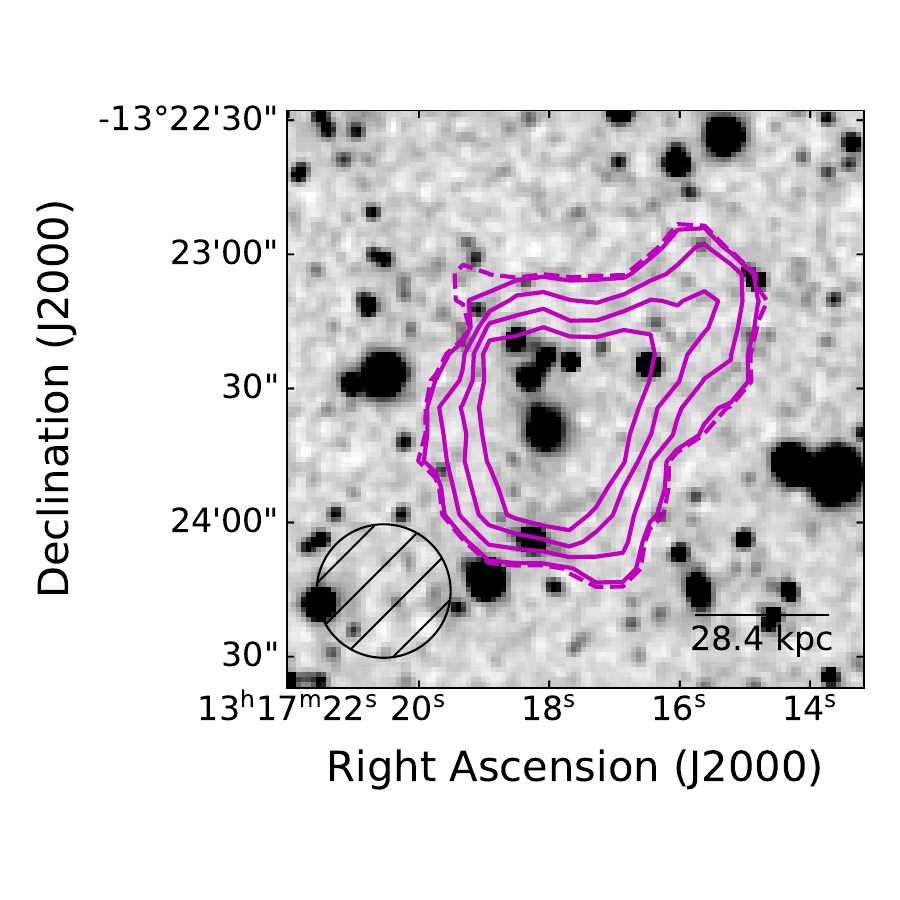}
         \caption{WALLABY J131717-132332}
     \end{subfigure}
     \begin{subfigure}[b]{0.24\textwidth}
         \centering
         \includegraphics[width=\textwidth]{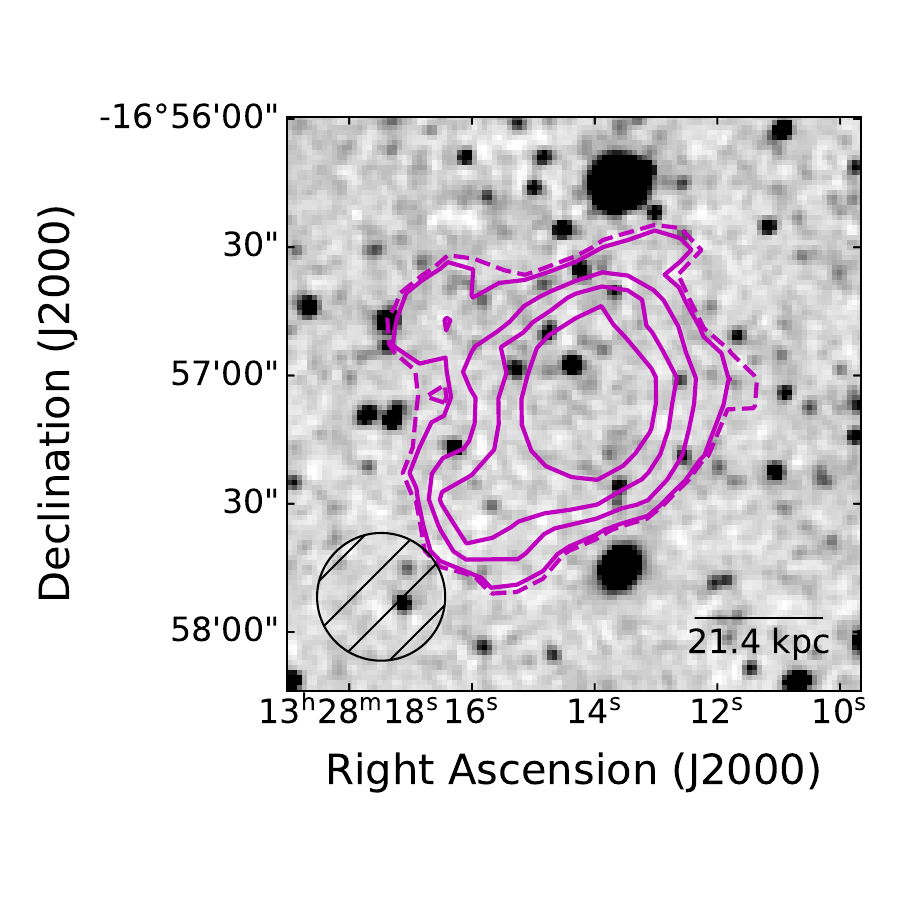}
         \caption{WALLABY J132814-165706}
     \end{subfigure}
     \begin{subfigure}[b]{0.24\textwidth}
         \centering
         \includegraphics[width=\textwidth]{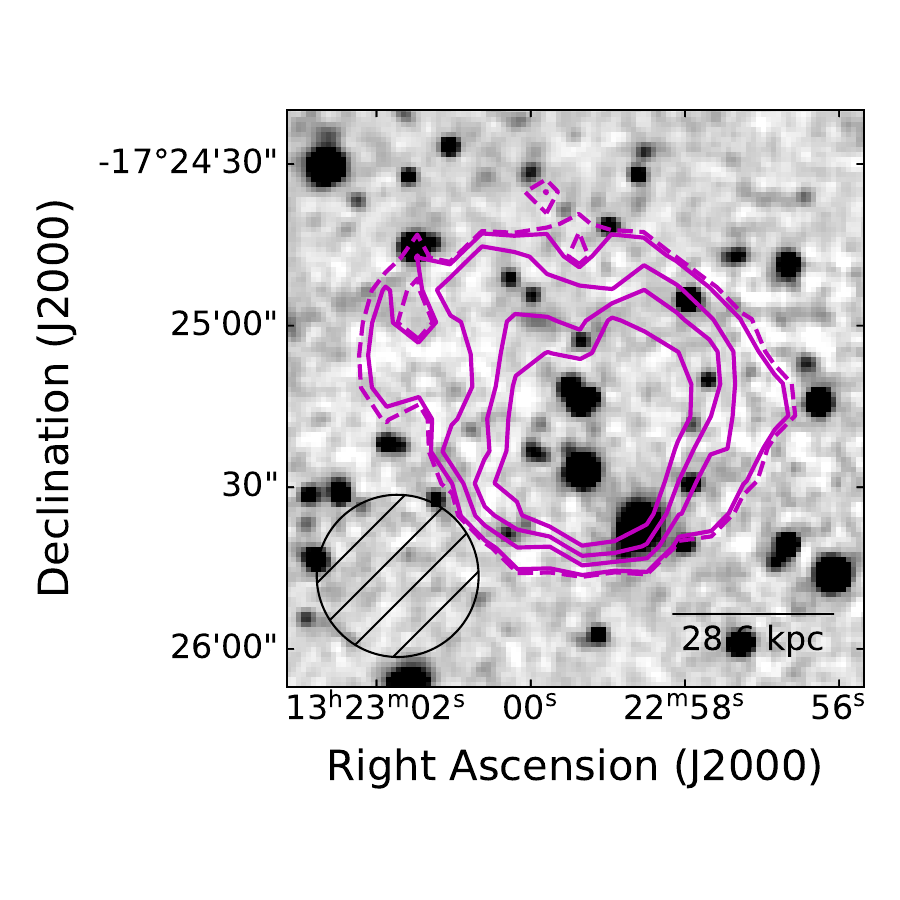}
         \caption{WALLABY J132259-172513}
     \end{subfigure}
     \begin{subfigure}[b]{0.24\textwidth}
         \centering
         \includegraphics[width=\textwidth]{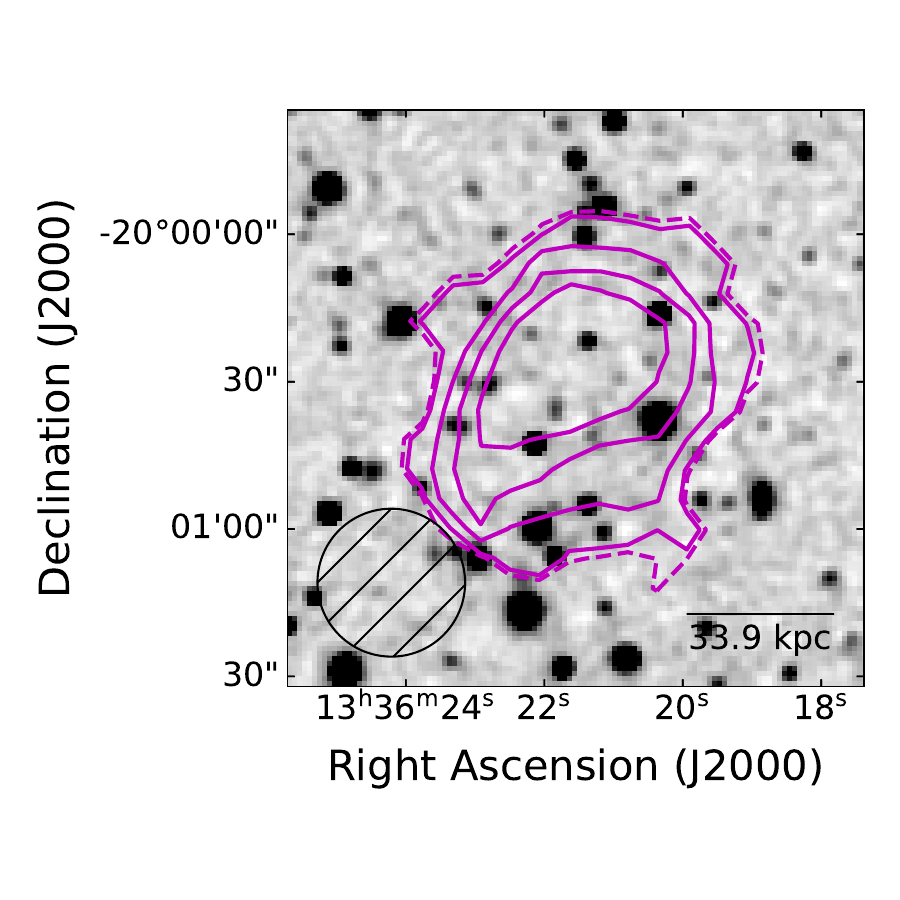}
         \caption{WALLABY J133621-200033}
     \end{subfigure}

    \begin{subfigure}[b]{0.24\textwidth}
         \centering
         \includegraphics[width=\textwidth]{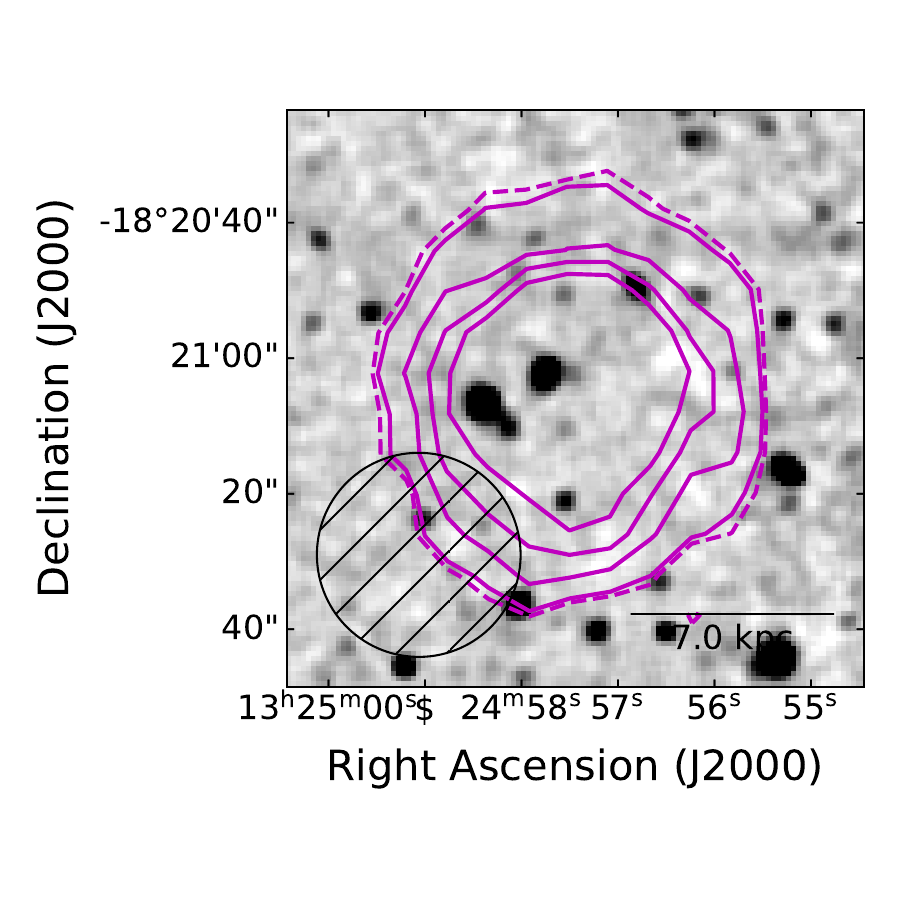}
         \caption{WALLABY J132457-182105}
     \end{subfigure}
     \begin{subfigure}[b]{0.24\textwidth}
         \centering
         \includegraphics[width=\textwidth]{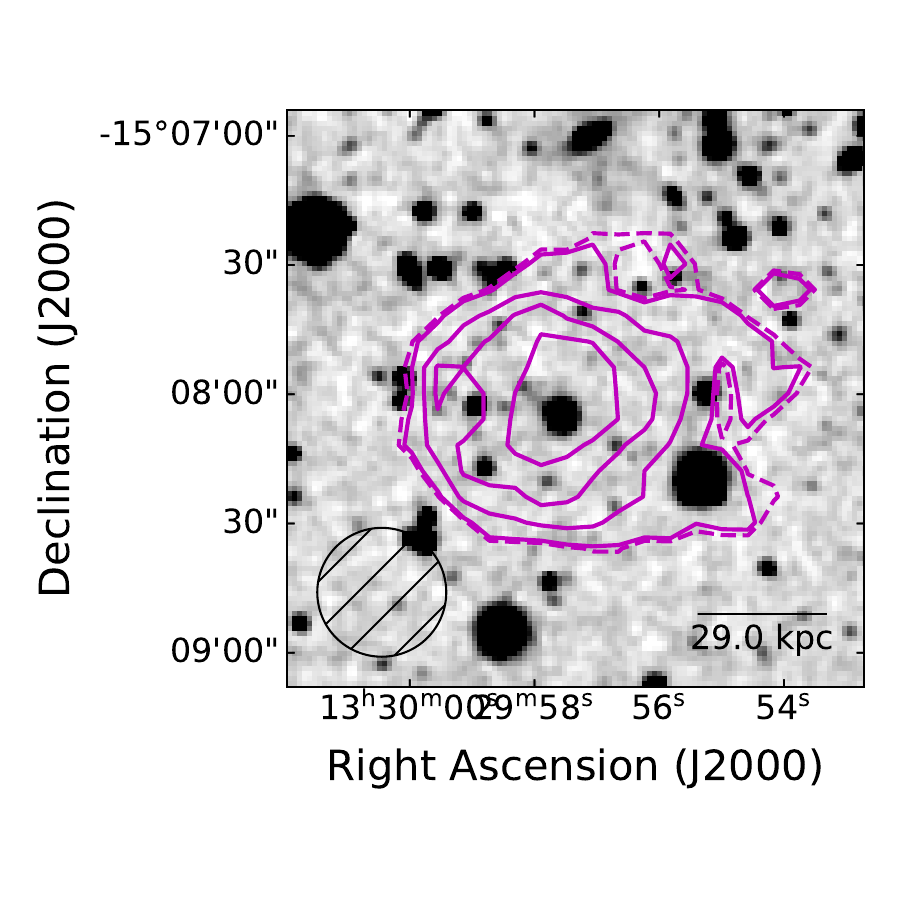}
         \caption{WALLABY J132957-150800}
     \end{subfigure}
     \begin{subfigure}[b]{0.24\textwidth}
         \centering
         \includegraphics[width=\textwidth]{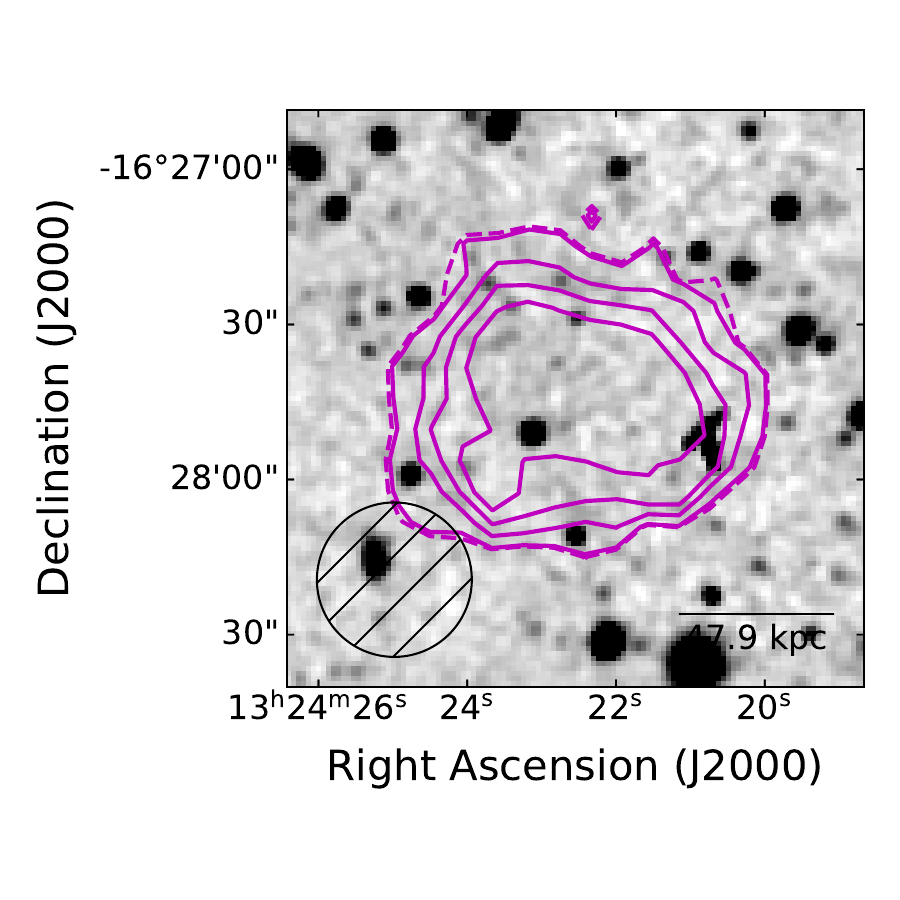}
         \caption{WALLABY J132422-162744}
     \end{subfigure}
     \begin{subfigure}[b]{0.24\textwidth}
         \centering
         \includegraphics[width=\textwidth]{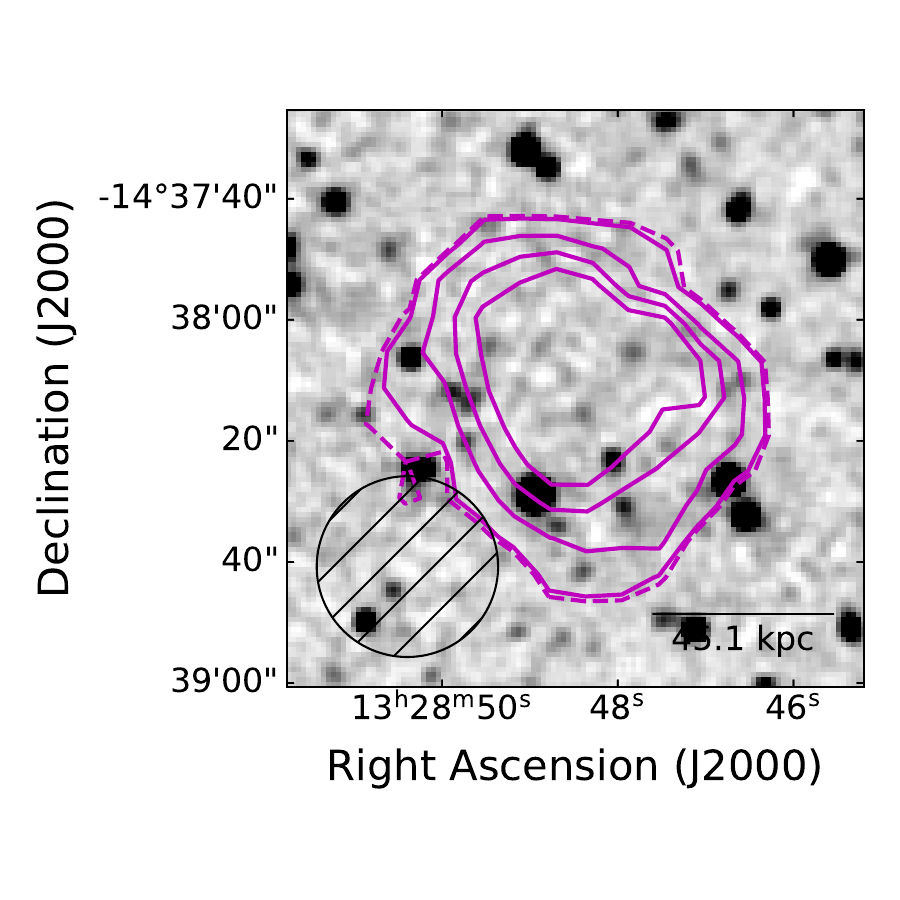}
         \caption{WALLABY J132848-143813}
     \end{subfigure}

     \begin{subfigure}[b]{0.24\textwidth}
         \centering
         \includegraphics[width=\textwidth]{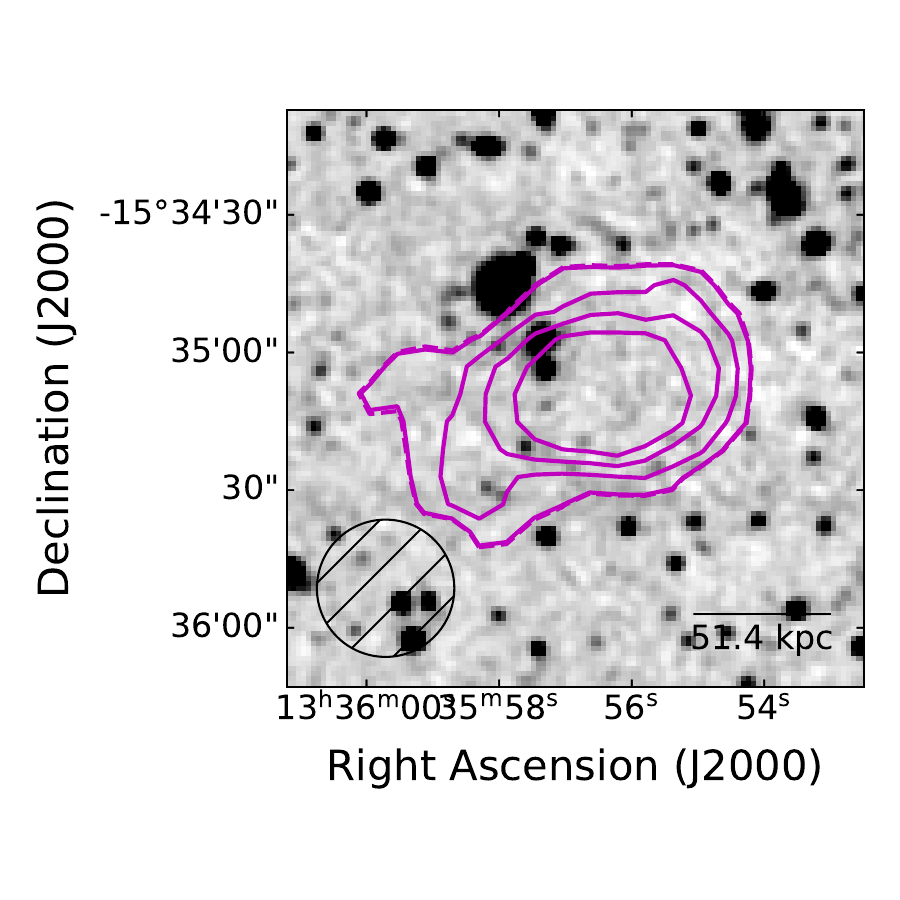}
         \caption{WALLABY J133556-153510}
     \end{subfigure}

    \caption{Co-added images of other strong dark source detections}
    \label{fig:coadd_s}
\end{figure*}

\begin{figure*}
     \centering

     \begin{subfigure}[b]{0.24\textwidth}
         \centering
         \includegraphics[width=\textwidth]{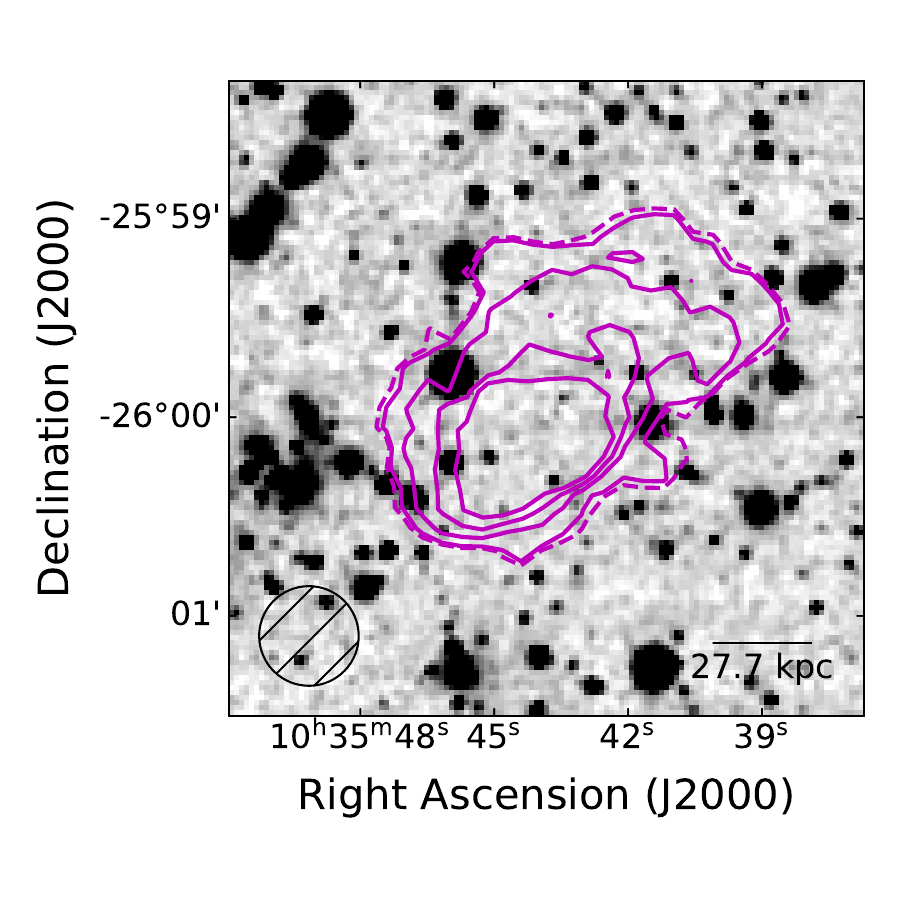}
         \caption{WALLABY J103543-255954}
     \end{subfigure}
     \begin{subfigure}[b]{0.24\textwidth}
         \centering
         \includegraphics[width=\textwidth]{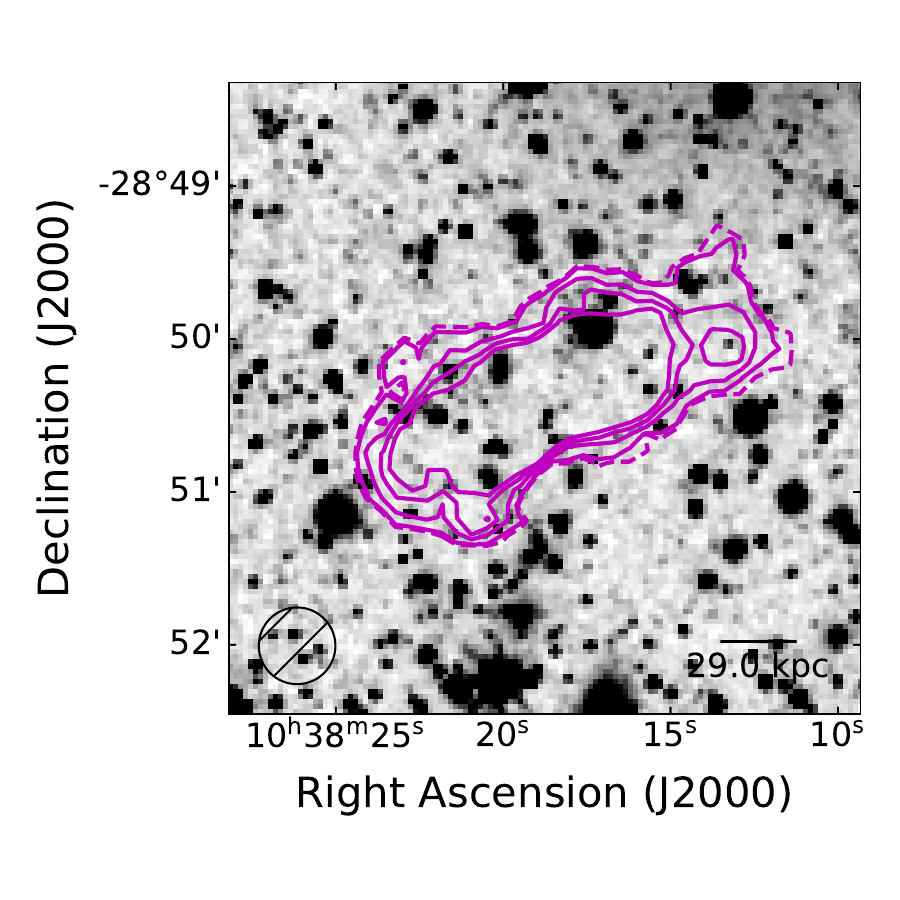}
         \caption{WALLABY J103818-285023}
     \end{subfigure}
     \begin{subfigure}[b]{0.24\textwidth}
         \centering
         \includegraphics[width=\textwidth]{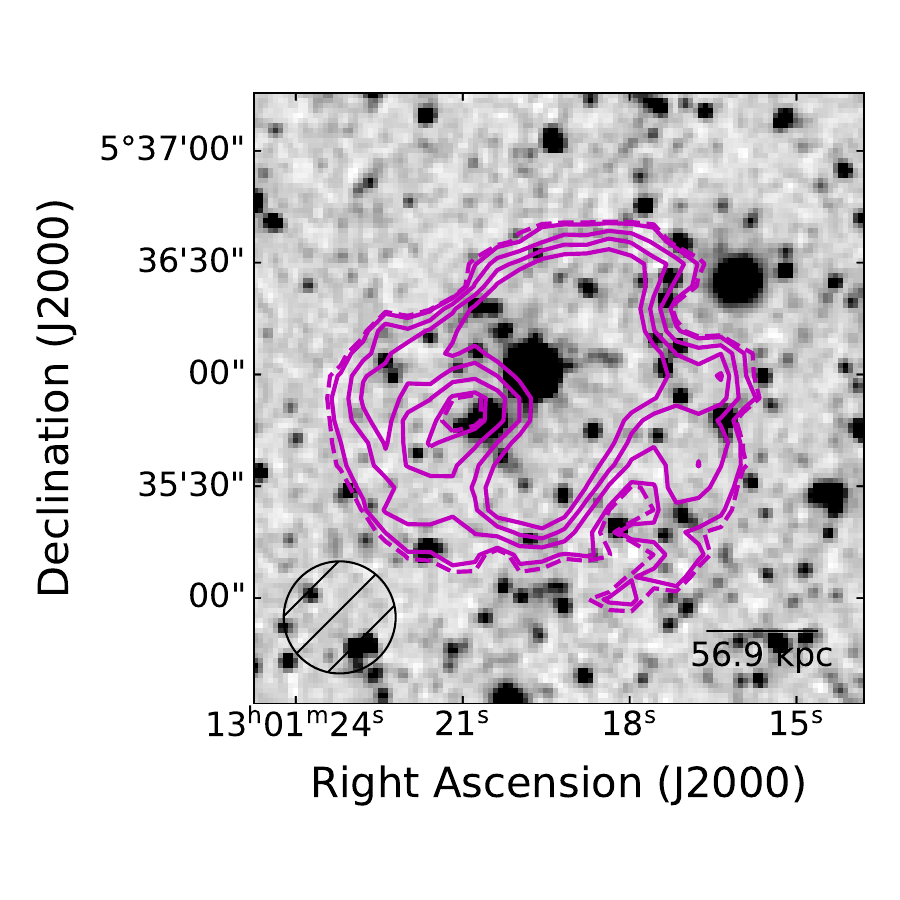}
         \caption{WALLABY J130119+053553}
     \end{subfigure}
     \begin{subfigure}[b]{0.24\textwidth}
         \centering
         \includegraphics[width=\textwidth]{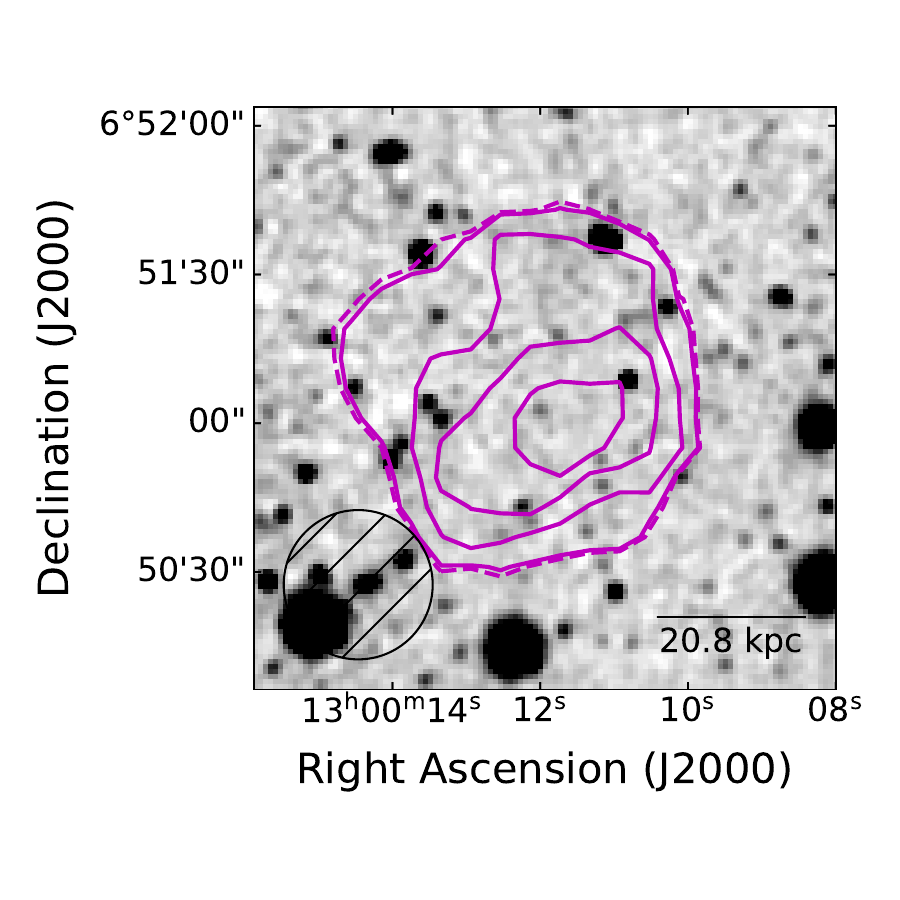}
         \caption{WALLABY J130011+065105}
     \end{subfigure}

    \caption{Co-added images of uncertain dark sources}
    \label{fig:coadd_u}
\end{figure*}

\begin{figure*}\ContinuedFloat
     \centering

     \begin{subfigure}[b]{0.24\textwidth}
         \centering
         \includegraphics[width=\textwidth]{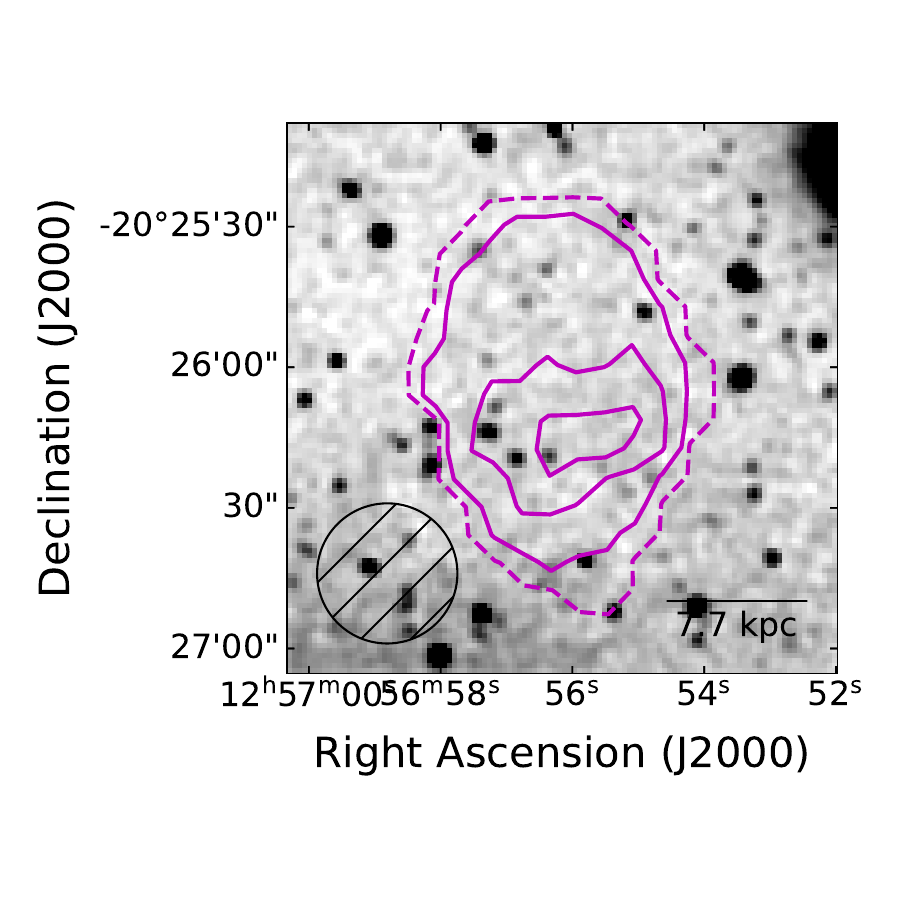}
         \caption{WALLABY J125656-202606}
     \end{subfigure}
     \begin{subfigure}[b]{0.24\textwidth}
         \centering
         \includegraphics[width=\textwidth]{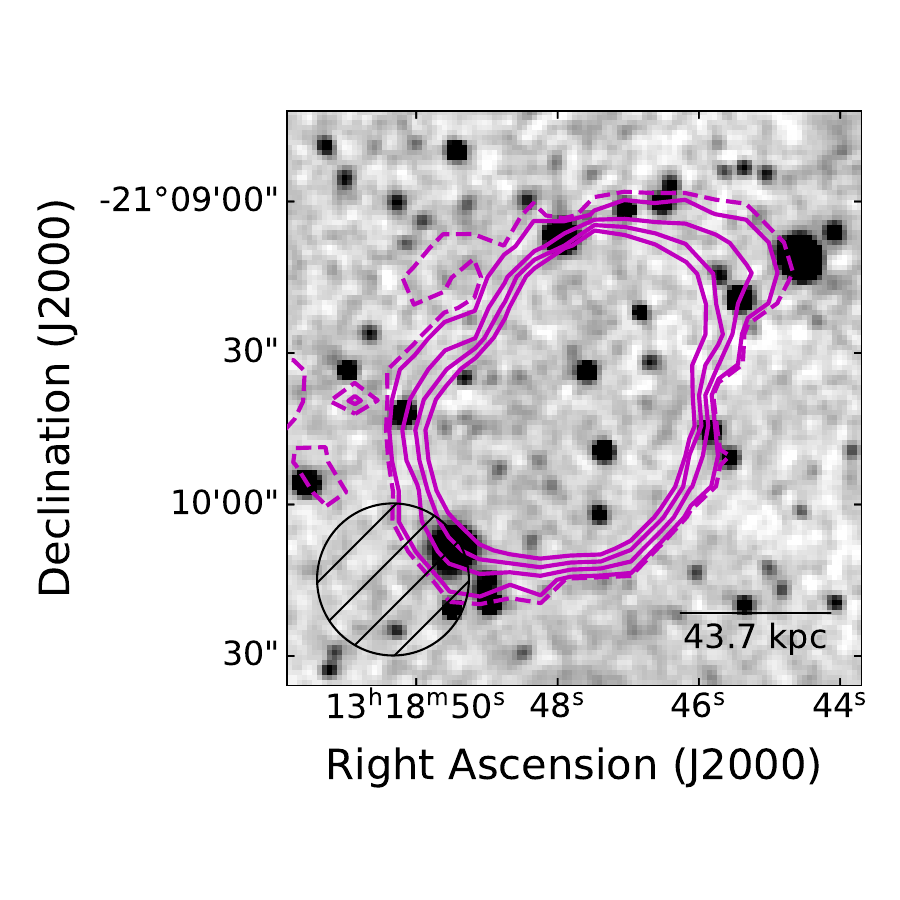}
         \caption{WALLABY J131847-210939}
     \end{subfigure}
    \begin{subfigure}[b]{0.24\textwidth}
         \centering
         \includegraphics[width=\textwidth]{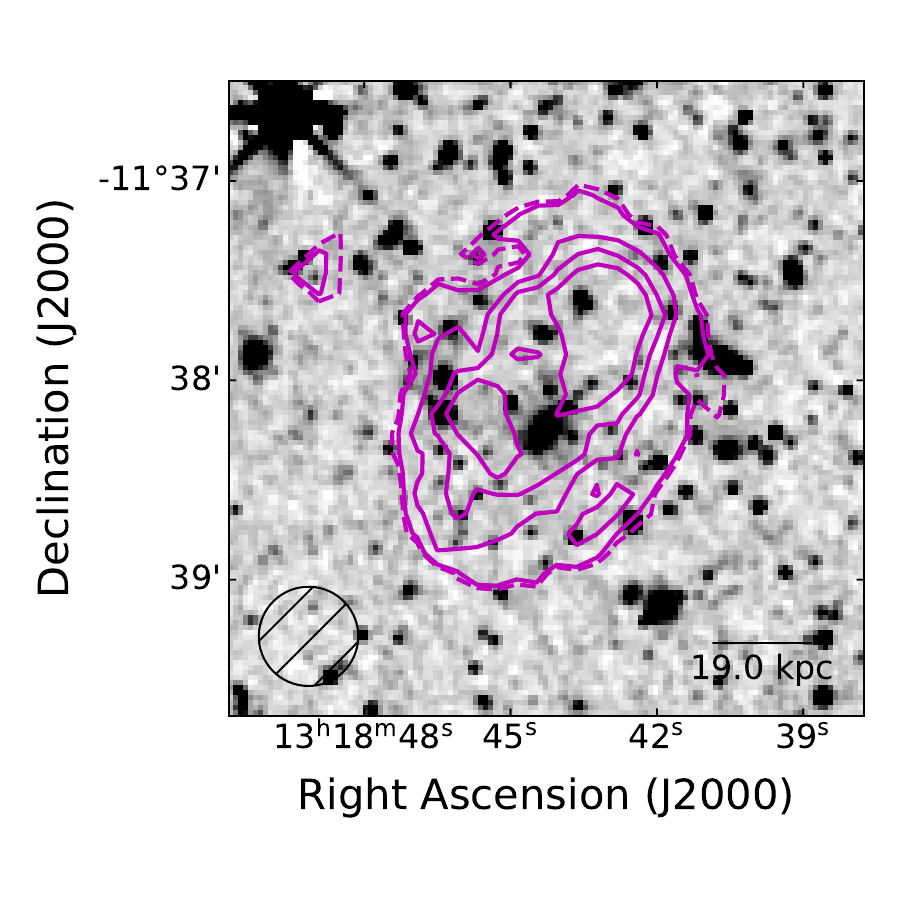}
         \caption{WALLABY J131844-113805}
     \end{subfigure}
     \begin{subfigure}[b]{0.24\textwidth}
         \centering
         \includegraphics[width=\textwidth]{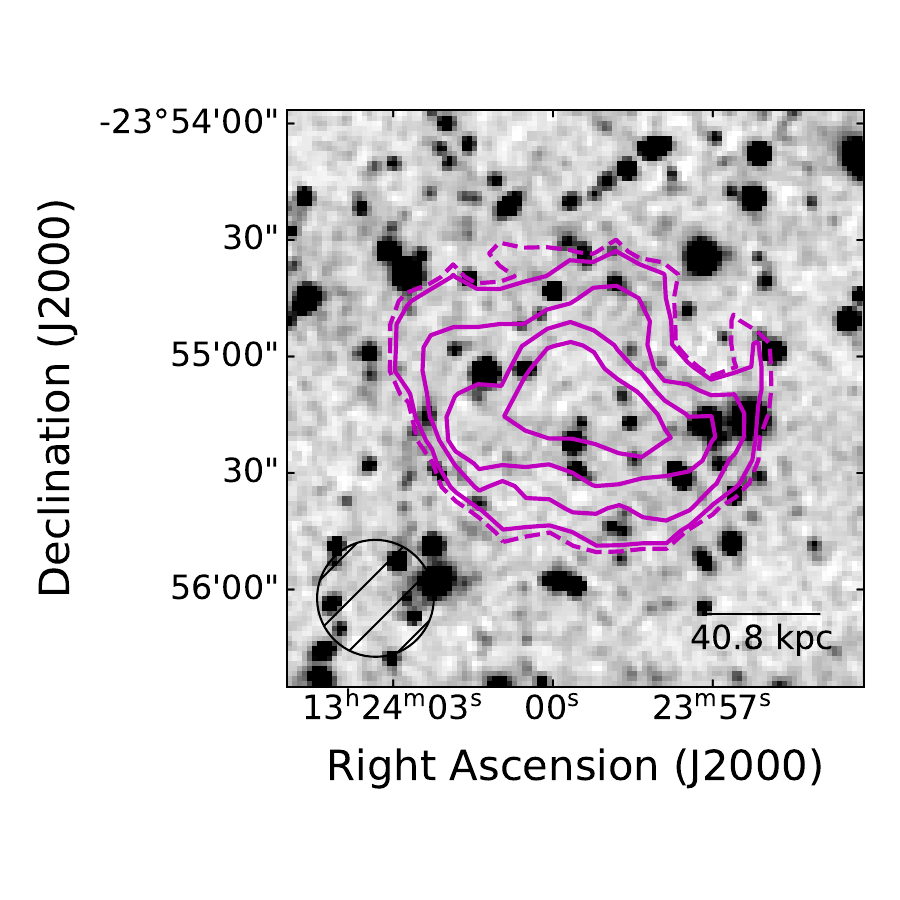}
         \caption{WALLABY J132359-235510}
     \end{subfigure}

     \begin{subfigure}[b]{0.24\textwidth}
         \centering
         \includegraphics[width=\textwidth]{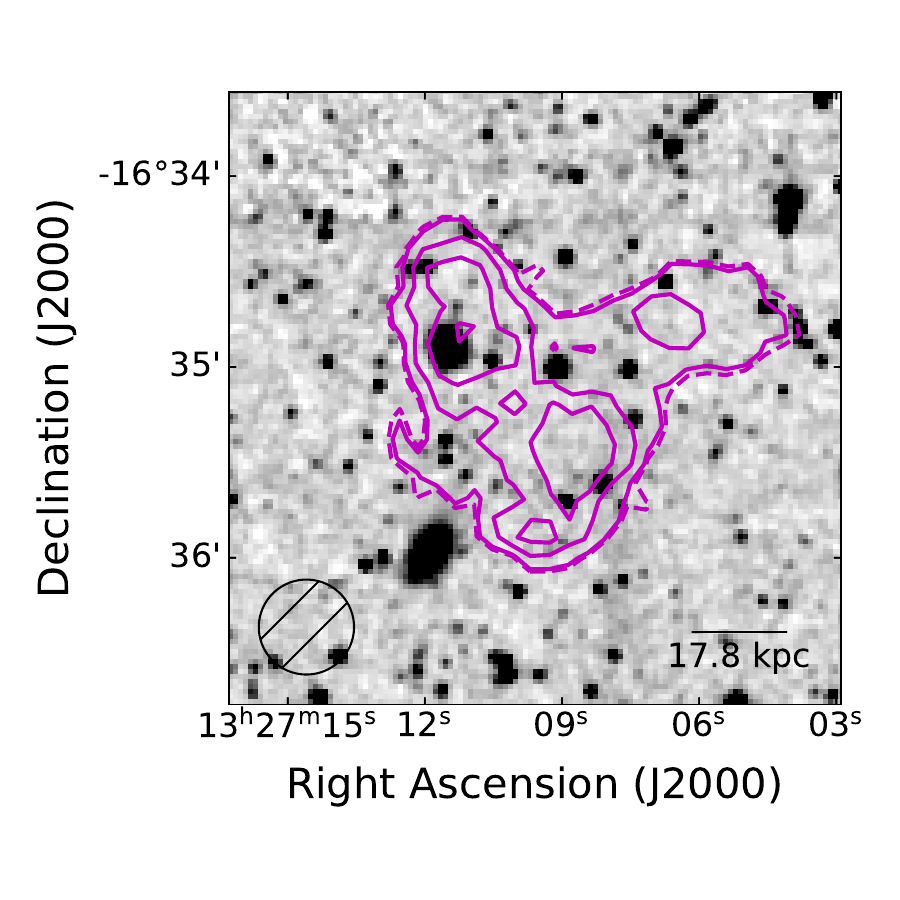}
         \caption{WALLABY J132709-163509}
     \end{subfigure}
     \begin{subfigure}[b]{0.24\textwidth}
         \centering
         \includegraphics[width=\textwidth]{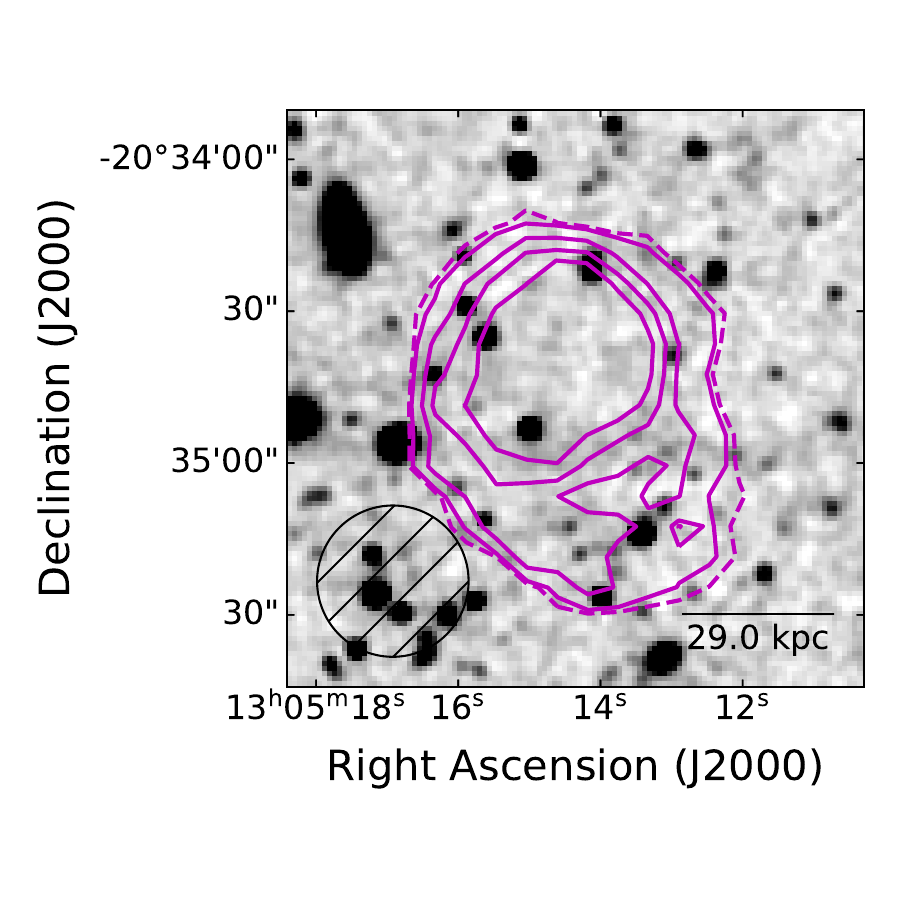}
         \caption{WALLABY J130514-203447}
     \end{subfigure}
     \begin{subfigure}[b]{0.24\textwidth}
         \centering
         \includegraphics[width=\textwidth]{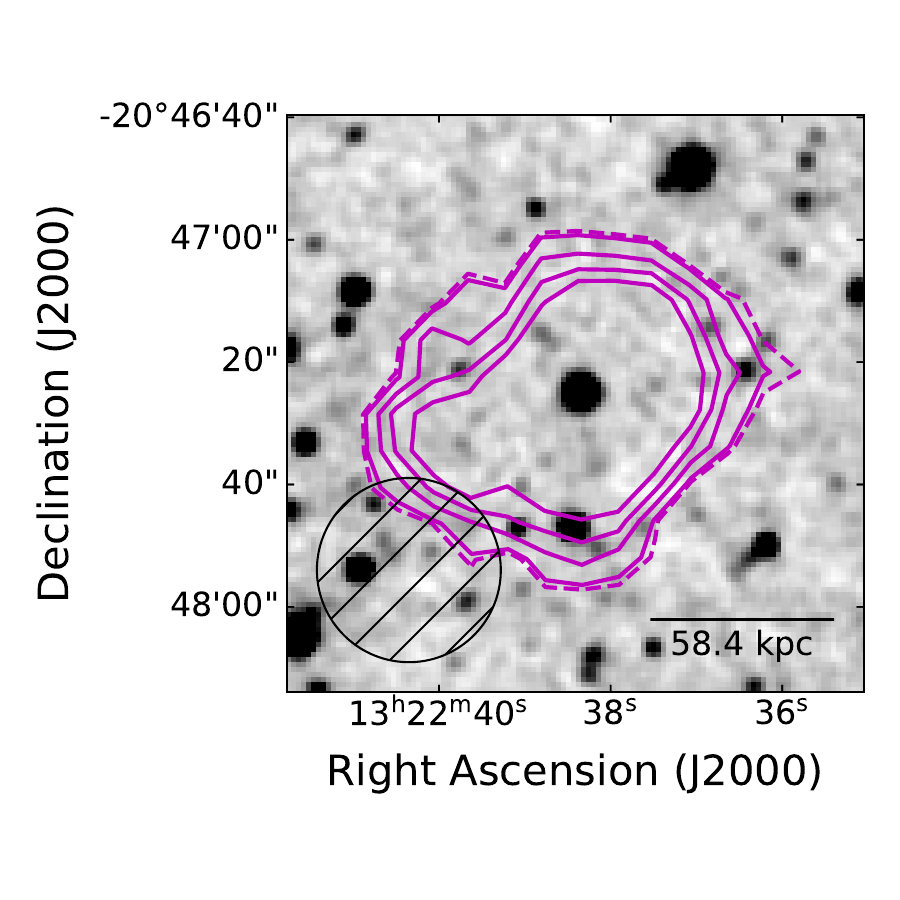}
         \caption{WALLABY J132238-204726}
     \end{subfigure}
     \begin{subfigure}[b]{0.24\textwidth}
         \centering
         \includegraphics[width=\textwidth]{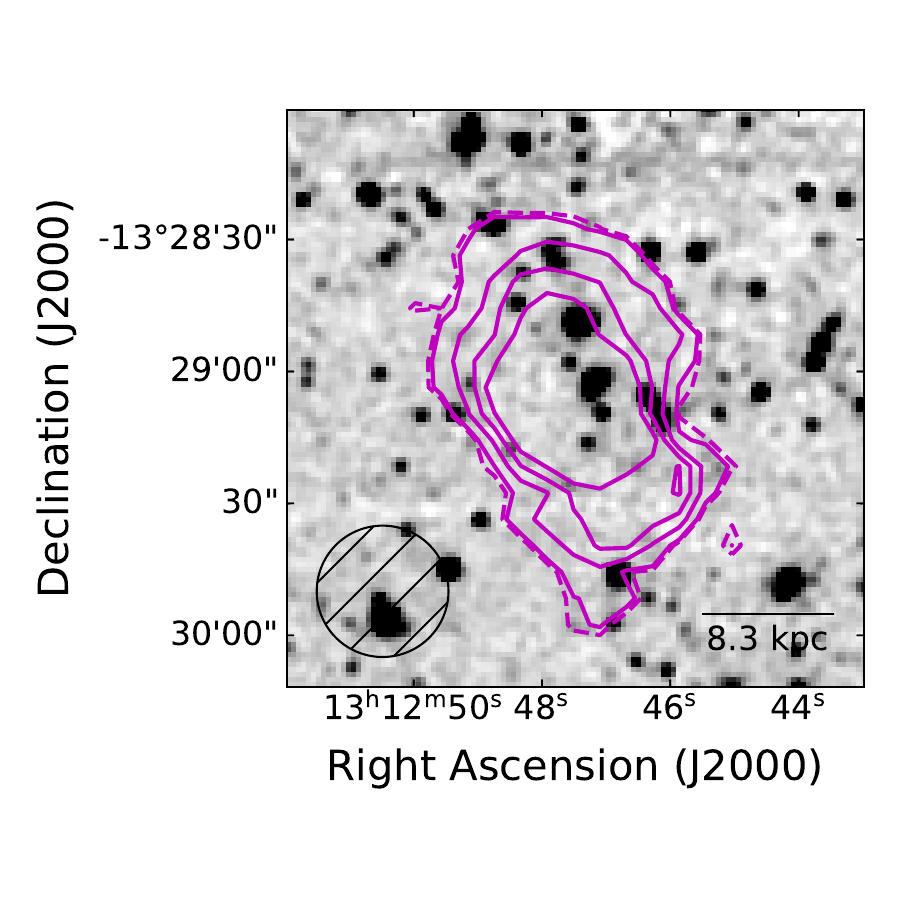}
         \caption{WALLABY J131247-132906}
     \end{subfigure}

    \begin{subfigure}[b]{0.24\textwidth}
         \centering
         \includegraphics[width=\textwidth]{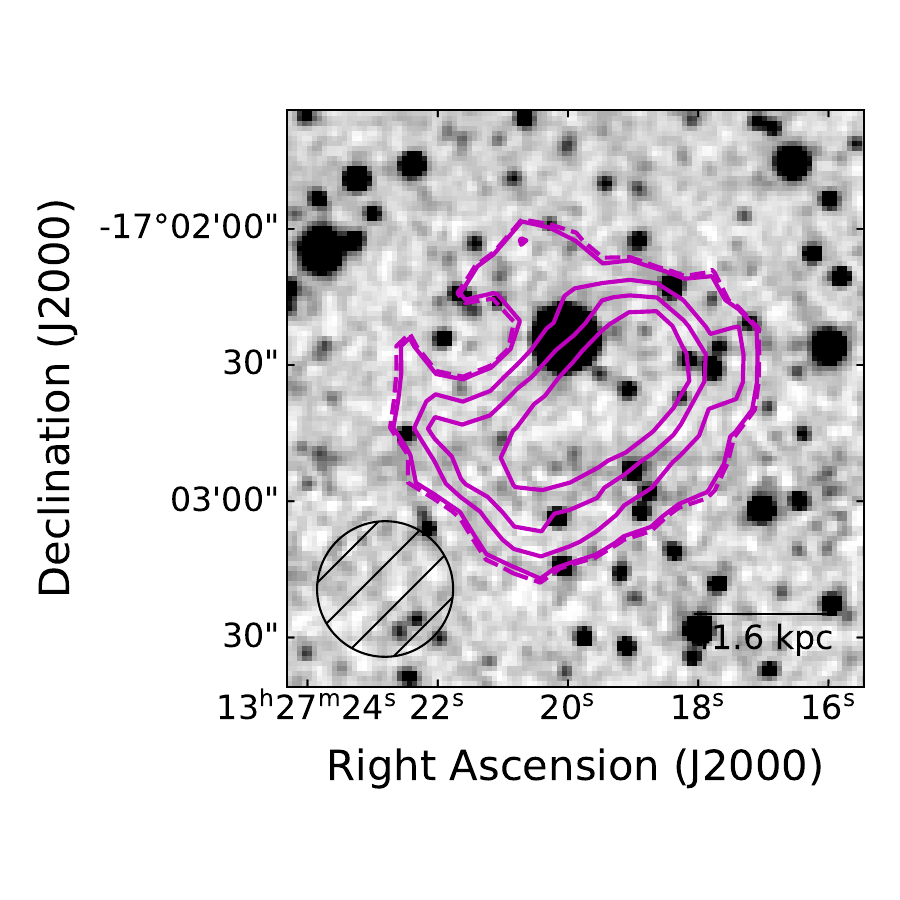}
         \caption{WALLABY J132719-170237}
     \end{subfigure}
     \begin{subfigure}[b]{0.24\textwidth}
         \centering
         \includegraphics[width=\textwidth]{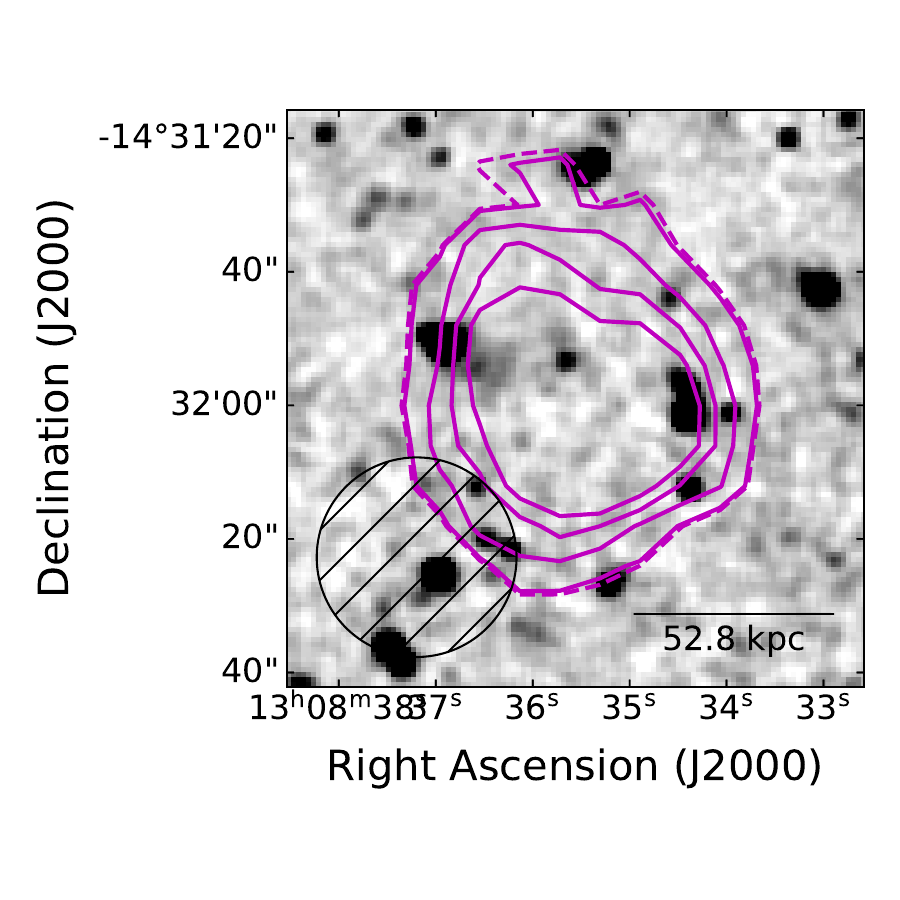}
         \caption{WALLABY J130835-143159}
     \end{subfigure}
     \begin{subfigure}[b]{0.24\textwidth}
         \centering
         \includegraphics[width=\textwidth]{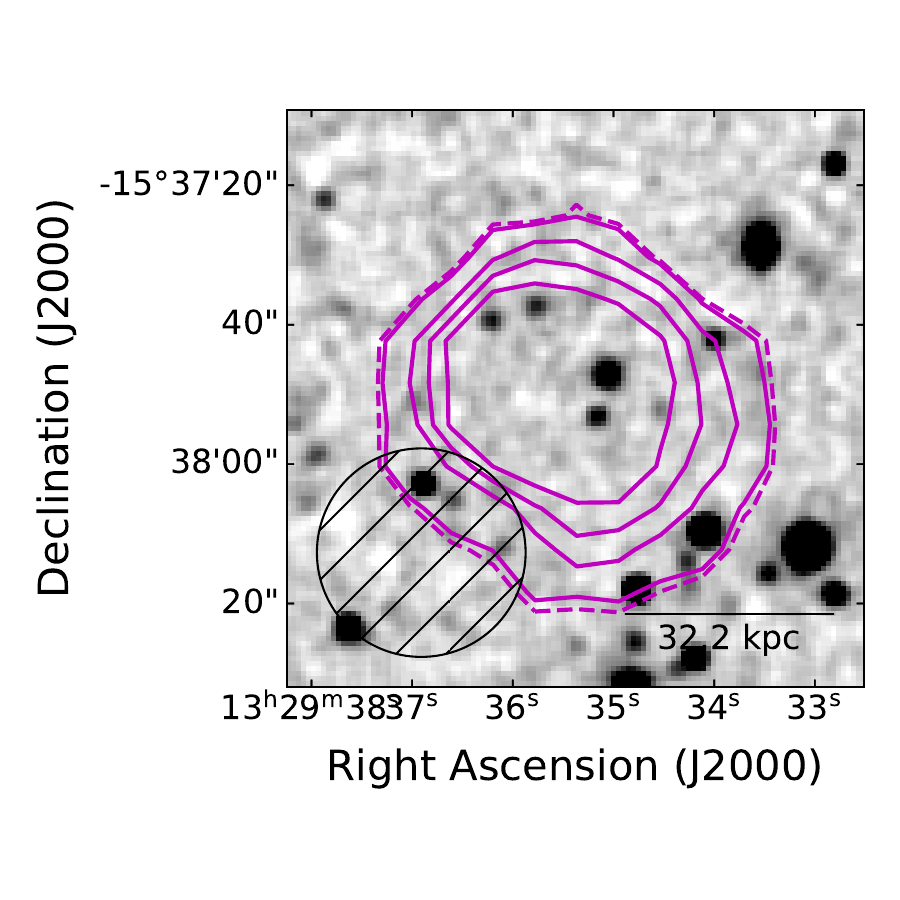}
         \caption{WALLABY J132935-153750}
     \end{subfigure}
     \begin{subfigure}[b]{0.24\textwidth}
         \centering
         \includegraphics[width=\textwidth]{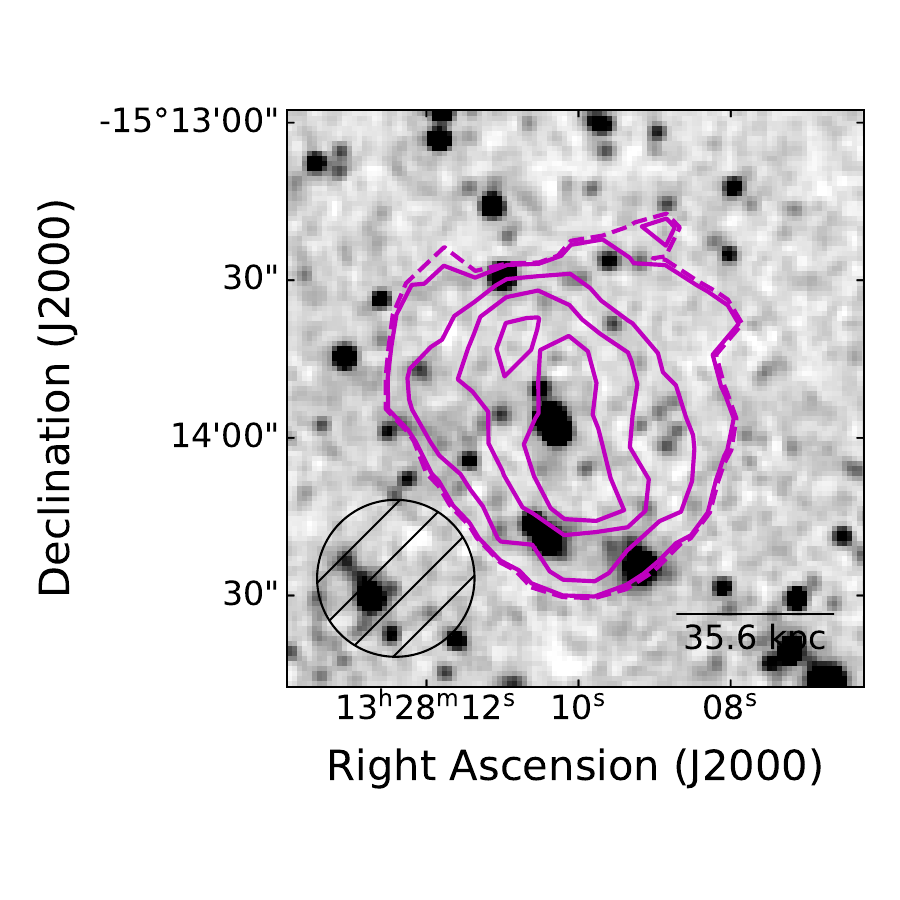}
         \caption{WALLABY J132810-151352}
     \end{subfigure}

     \begin{subfigure}[b]{0.24\textwidth}
         \centering
         \includegraphics[width=\textwidth]{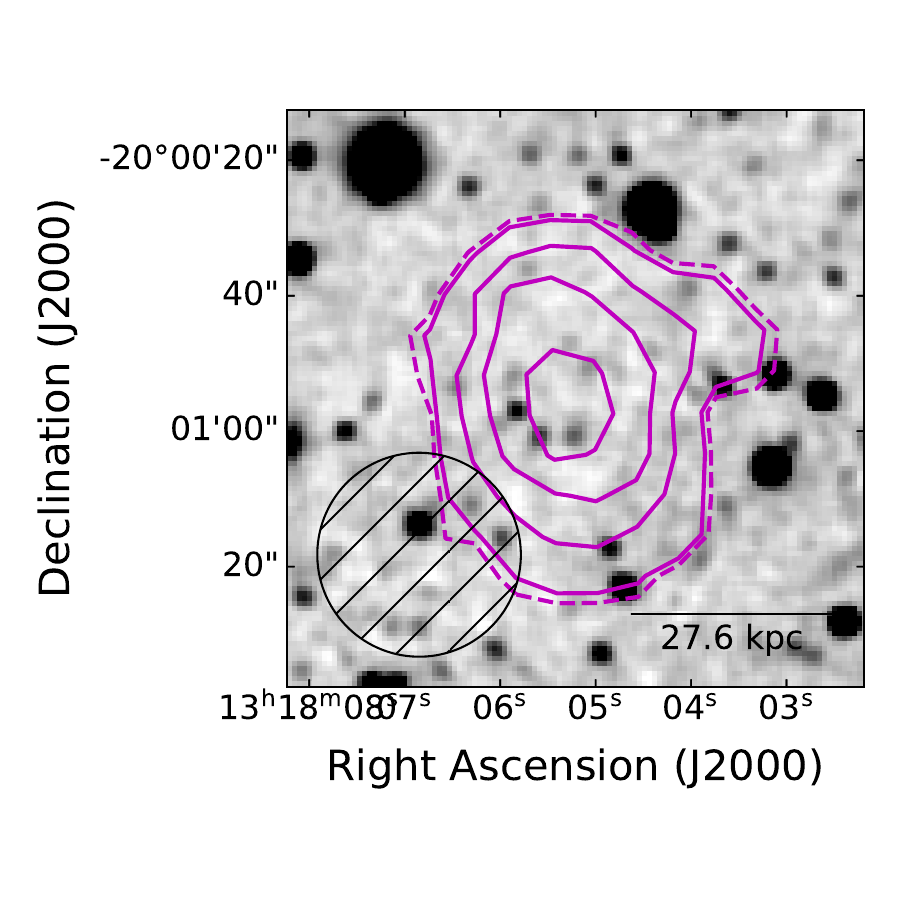}
         \caption{WALLABY J131805-200055}
     \end{subfigure}

    \caption{Co-added images of uncertain dark sources}
    \label{fig:coadd_u}
\end{figure*}

\end{document}